\documentclass{aa}
\usepackage[multidot]{grffile}
\usepackage{graphicx}
\usepackage{txfonts}
\usepackage{natbib}
\usepackage[colorlinks,allcolors=blue,bookmarks=false]{hyperref} 
\pdfoutput=1
\bibpunct{(}{)}{;}{a}{}{,}
\begin{document}
\title
{Inaccuracies and biases of the Gaussian size deconvolution\\for extracted sources and filaments}


\author
{A.~Men'shchikov}


\institute
{
Universit\'{e} Paris-Saclay, Universit\'{e} Paris Cit\'{e}, CEA, CNRS, AIM, 91191, Gif-sur-Yvette, France\\
\email{alexander.menshchikov@cea.fr}
}
\date{Received 16 February 2023 / Accepted 2 June 2023 }
\offprints{Alexander Men'shchikov}
\titlerunning{Inaccuracies and biases of Gaussian size deconvolution}
\authorrunning{A.~Men'shchikov}


\abstract
{ 
A simple Gaussian size deconvolution method is routinely used to remove the blur of observed images caused by insufficient angular
resolutions of existing telescopes, thereby to estimate the physical sizes of extracted sources and filaments. To ensure that the
physical conclusions derived from observations are correct, it is necessary to know the inaccuracies and biases of the size
deconvolution method, which is expected to work when the structures, as well as the telescope beams, have Gaussian shapes. This
study employed model images of the spherical and cylindrical objects with Gaussian and power-law shapes, representing the dense
cores and filaments observed in star-forming regions. The images were convolved to a wide range of angular resolutions to probe
various degrees of resolvedness of the model objects. Simplified shapes of the flat, convex, and concave backgrounds were added to
the model images, then planar backgrounds across the footprints of the structures are subtracted and sizes of the sources and
filaments were measured and deconvolved. When background subtraction happens to be inaccurate, the observed structures acquire
profoundly non-Gaussian profiles. The deconvolved half maximum sizes can be strongly under- or overestimated, by factors of up to
$\sim$\,$20$ when the structures are unresolved or partially resolved. For resolved structures, the errors are generally within a
factor of $\sim$\,$2$; although, the deconvolved sizes can be overestimated by factors of up to $\sim$\,$6$ for some power-law
models. The results show that Gaussian size deconvolution cannot be applied to unresolved structures, whereas it can only be
applied to the Gaussian-like structures, including the critical Bonnor-Ebert spheres, when they are at least partially resolved.
The deconvolution method must be considered inapplicable for the power-law sources and filaments with shallow profiles. This work
also reveals subtle properties of convolution for structures of different geometry. When convolved with different kernels,
spherical objects and cylindrical filaments with identical profiles obtain different widths and shapes. In principle, a physical
filament, imaged by the telescope with a non-Gaussian point-spread function, could appear substantially shallower than the
structure is in reality, even when it is resolved.
} 
\maketitle


\section{Introduction}
\label{introduction}

Astronomical imaging with modern telescopes and interferometers provide ever increasing angular resolutions and sensitivities.
Observed images contain distinct structural components, such as interstellar clouds, filaments, and sources. All of those
structures are the integral parts of the dynamically interacting interstellar matter, controlled by the gravity and other physical
processes, which play important roles in the formation of stars. Observationally, the filaments are significantly elongated
structures and the sources are relatively round peaks in the images, which stand out of (and blend with) the complex fluctuations
of the clouds and small-scale instrumental noise. The observed sources and filaments are produced by various physical objects in
the interstellar clouds. Because of their integral nature, separation of the sources from the filaments and from the background
clouds is a nontrivial problem. In the regions that are not too distant from the Galactic plane, the backgrounds are quite bright,
have complex structures, and strongly fluctuate on all spatial scales, as demonstrated in the recent decades by the well-known
sensitive, high-resolution images obtained with \emph{Hubble}, \emph{Spitzer}, and \emph{Herschel}.

The currently available resolutions are rarely sufficient to fully resolve the imaged sources and filaments in distant regions --
relatively large point-spread functions (PSFs) of the telescopes blur the physical objects, artificially widening their true
angular sizes. Although the higher sensitivities (lower noise levels) of newer instruments enable detection of fainter structures,
they also reveal more complex fluctuating backgrounds that complicate the extraction and accurate measurements of the structures of
interest. In general, an extracted source or filament does not contain emission of just a single physical object, but very often it
is a blend of background fluctuations and other nearby objects. Assessing their physical properties requires accurate methods of
background subtraction, deblending, and deconvolution of the observed structures. Such reliable methods are presently unavailable,
as manifested by the major problems with the physical properties derived for the unresolved sources in both observations and
numerical simulations, reported by \cite{Louvet_etal2021}. They showed that the numbers, sizes, and masses of the sources extracted
at different angular resolutions linearly depend on the resolution and, as a result, the peak of the source mass function shifts to
lower masses. The effects were attributed to the heavy blending of the sources with filaments and a background cloud as well as to
an inaccurate background determination at lower angular resolutions.

A simple approach to deconvolving the half maximum size $s$ of a structure from its measured half maximum size $w$ in an image
convolved with a kernel of a half maximum size $k$ relies on the basic mathematical property of convolution ($w^2 = s^2 + k^2$),
which is valid only for the Gaussian shapes of both the kernel and the object. Because of its simplicity, the method is widely used
to deconvolve the sizes of both sources
\citep[e.g.,][]{Motte_etal2003,Bontemps_etal2010a,NguenLuong_etal2011,Ko"nyves_etal2015,Ladjelate_etal2020,Pouteau_etal2022} and
filaments \citep[e.g.,][]{Arzoumanian_etal2011,Palmeirim_etal2013,Andre_etal2016,Cox_etal2016,Arzoumanian_etal2019}. In the
astronomical studies, this simple Gaussian size deconvolution method implies strictly Gaussian shapes of the telescope PSFs, the
extracted (background-subtracted) structures, and the corresponding physical objects, constituting a strong set of conditions.

In reality, the PSFs of orbital telescopes are flatter and more complex than a Gaussian shape, affected by the diffraction rings
and spikes, at angular distances from their centers larger than roughly their half maximum widths. Even if the beams were Gaussian,
most astrophysical objects would have non-Gaussian (often power-law) shapes beyond their peaks, at least the objects of common
interest in star-forming molecular clouds, such as the dense filaments and prestellar or protostellar cores. Even if both the beams
and the objects had Gaussian shapes, just a slightly inaccurate subtraction of the fluctuating backgrounds of the structures would
make them have non-Gaussian profiles.

Backgrounds of observed sources and filaments are blends of the interstellar clouds and instrumental noise that display complex
fluctuations on the spatial scales of and above the observational beam. To make the problem tractable, it makes sense to somewhat
simplify the complexity of background shapes and assume that, in general, the structures have chances to be situated on the flat,
convex (hill-like), or concave (bowl-like) background fluctuations. An unknown background of a specific source or filament must be
subtracted, which has the potential to strongly affect the measured sizes, peak intensities, and integrated fluxes of the objects,
and hence their derived masses. Benchmark extractions indicate that errors in the integrated fluxes of faint sources within a
factor of $\sim$\,$2$ are fairly common, reaching even factors of up to an order of magnitude in some cases \citep[cf.
Figs.~6\,--\,10 and A.1 in][]{Men'shchikov2021b}.

From a physical point of view, the embedded dense cores and elongated structures, forming inside a molecular cloud, are expected to
spatially coincide with significant volume density enhancements because the objects are created by their self-gravity (along with
other processes). This implies higher chances for such sources and filaments of being observed on the local volume density peaks,
although the latter would not necessarily appear as the surface density peaks \citep[e.g.,][]{Padoan_etal2023}. Depending on the
angular resolution and projected volume density variations along the lines of sight, backgrounds of the observed structures may
have approximately convex or concave shapes and, in the simplest cases, the backgrounds might also be flat.

The structures with shallower power-law profiles, occupying wider image areas, have higher chances of blending with the nonuniform
backgrounds of complex shapes. It is nearly impossible to accurately deblend the background contribution, especially for the
power-law structures whose dimensions are much larger than the telescope beam. The inaccuracies of background determination
primarily affect (and quite significantly) the fainter sources and filaments with lower signal-to-noise ratios, and hence lower
masses \citep{Men'shchikov2021b}. Completeness limits for source extractions are usually found near the peaks of the differential
mass functions \citep[e.g.,][]{Ko"nyves_etal2015,Ladjelate_etal2020,Pouteau_etal2022}, which means that the peaks contain large
numbers of the faintest sources. Most of the mass of the power-law structures is contained in their outermost areas, which are
strongly affected by the inaccurate background subtraction. This indicates that large fractions of the extracted structures in
observed regions may have poorly determined backgrounds and measured physical properties.

Except for the brightest structures with large signal-to-noise ratios, it is nearly impossible to accurately determine and subtract
the combined effects of the background cloud and instrumental noise fluctuations because they are completely blended with the
structures of interest. The practical approaches used by existing extraction algorithms \citep[e.g.,][]{Men'shchikov2021a} include
a two-dimensional linear interpolation within the structure, based on the pixel values just outside it. Clearly, such interpolated
backgrounds are generally too simple to accurately represent the complex fluctuations within the entire area of the source or
filament. The problem is aggravated for the power-law filaments that have significant elongation, complex shapes, and variable
widths, and hence they have much higher chances of blending with other nearby structures and resulting inaccuracies in their
footprint determination. In the absence of any deblending scheme for filaments, this might substantially increase the interpolation
distances between the edges of the filament footprint. Depending on the complexity of the background cloud and filament shapes,
which are often quite complex, the filament backgrounds may be determined with larger errors than in the case of the relatively
round sources.

The above considerations put into question whether the simple deconvolution method can produce reliable estimates of the physical
dimensions of the real objects producing the observed emission of sources and filaments. This work puts the method of Gaussian size
deconvolution through several model-based quantitative tests. The simulated images are described in Sect.~\ref{simimages}, the
deconvolution results are presented in Sect.~\ref{results} and discussed in Sect.~\ref{discussion}, the conclusions are given
in Sect.~\ref{conclusions}, and supplementary results are found in Appendices~\ref{resultsmom} and \ref{tabresults}.

The two-dimensional model images are referred to with capitalized calligraphic characters (e.g., $\mathcal{A}$, $\mathcal{B}$) to
distinguish them from other one-dimensional functions or scalar quantities. The software names and numerical methods are typeset
slanted (e.g., \textsl{getsf}) to set them apart from other emphasized words, such as the names of telescopes or instruments. The
curly brackets are used to collectively refer to either of the characters, separated by vertical lines (e.g., $\{A|B\}$ refers to
$A$ or $B$). All profiles refer to the angular dependence of the intensity or surface density, that is, the quantities integrated
along the lines of sight passing through the centers of pixels ($\sigma \propto \theta^{-\beta}$), with an exception being the
radial volume density profile of the models ($\rho \propto r^{-\alpha}$).

\begin{figure}
\centering
\centerline{
  \resizebox{0.6427\hsize}{!}{\includegraphics{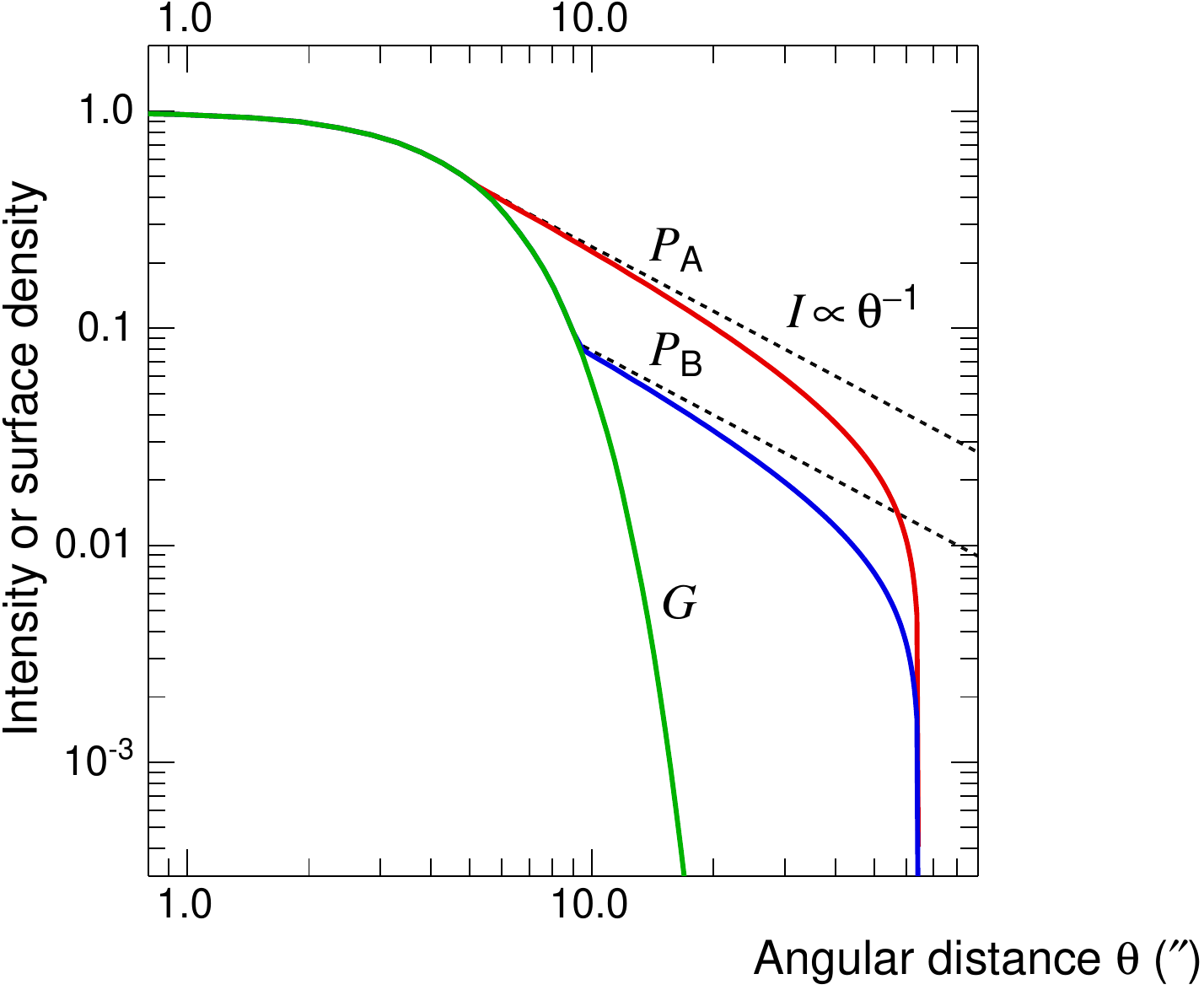}}   \hspace{-12mm}
  \resizebox{0.4926\hsize}{!}{\includegraphics{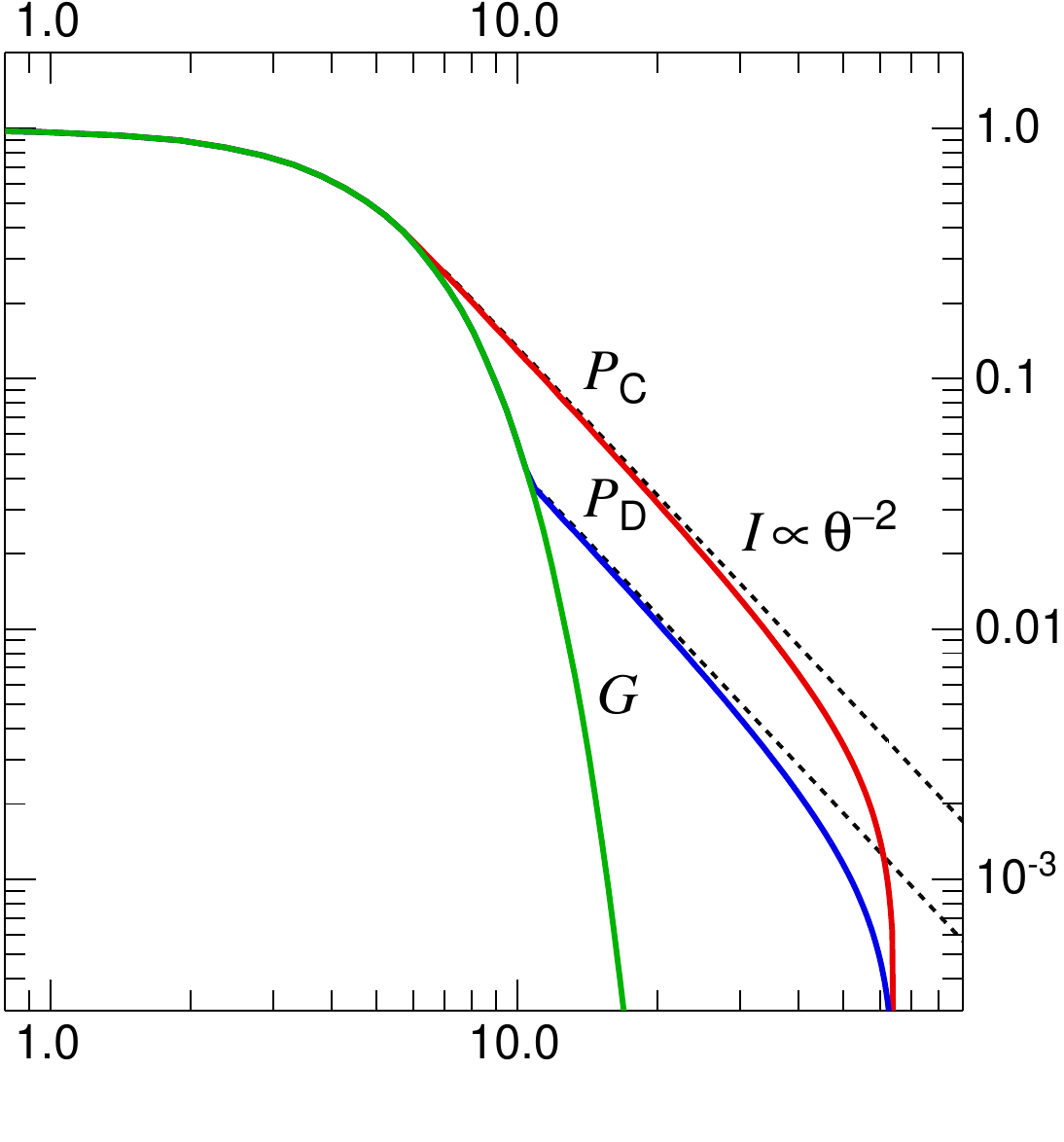}}}
\caption
{ 
Profiles of the spherical and cylindrical models. Model $\mathcal{G}$ has a Gaussian shape, whereas the power-law models
$\mathcal{P}_{\rm A}$, $\mathcal{P}_{\rm B}$, $\mathcal{P}_{\rm C}$, and $\mathcal{P}_{\rm D}$ have extended envelopes outside
their Gaussian core, with the volume densities $\rho \propto r^{-\alpha}$ and intensities $I \propto \theta^{-\beta}$ (Table
\ref{modeltable}). The profiles, differing in intensity (density) by a factor of $3$, fall off steeply at the outer boundary
$\Theta = 64${\arcsec} because of the adopted geometry of the finite models. The dashed profiles correspond to the ``infinite''
models (no outer edge within the entire image area), discussed in Sect.~\ref{convsrcfil}.
} 
\label{modelprofs}
\end{figure}

\begin{table} 
\caption{Properties of the Gaussian and power-law models $\mathcal{M}$ (Fig.~\ref{modelprofs}) in spherical and cylindrical
geometries. The columns list the model, exponent $\alpha_{\mathcal{M}}$ of the radial volume density profile, exponent
$\beta_{\mathcal{M}}$ of the projected intensity or surface density, true half maximum size $H_{\!\mathcal{M}}$ and moment size
$M_{\!\mathcal{M}}$ (for spherical models only), and their corresponding intensity levels $L_{H_{\!\mathcal{M}}}$ and
$L_{M_{\!\mathcal{M}}}$. The columns $M_{\!\mathcal{M}*}$ and $L_{M_{\!\mathcal{M}*}}$ give the moment sizes and levels for the
infinite models. Outer boundaries of the finite models are at $\Theta = 64${\arcsec} from the model peak or crest.
} 
\begin{tabular}{ccccccccc}
\hline\hline
\noalign{\smallskip}
$\mathcal{M}$ & \!$\alpha_{\mathcal{M}}$ & \!\!\!\!$\beta_{\mathcal{M}}$ & \!\!\!$H_{\!\mathcal{M}}$ & \!\!$M_{\!\mathcal{M}}$ & 
$L_{H_{\!\mathcal{M}}}$ & \!\!\!$L_{M_{\!\mathcal{M}}}$ &  $M_{\!\mathcal{M}*}$ & \!\!\!\!\!$L_{M_{\!\mathcal{M}*}}$\,\,\,\,\, \\
&  &  & \!\!(\arcsec) & (\arcsec) &  & & (\arcsec) & \\   
\noalign{\smallskip}
\hline
\noalign{\smallskip}
$\mathcal{G}$\,\,\,   &   &\!\!\!\!  &\!\!10.0 &\!\!10.0 & 0.50 & \!\!0.49\,\,\,&  \!10.0 &\!\!\!0.49\,\,\,\,\,\,\,\,\\
$\mathcal{P}_{\rm A}$ & 2 &\!\!\!\!1 &\!\!10.0 &\!\!51.7 & 0.50 & \!\!0.073     &\!\!1578 &\!\!\!2.9E$-$3 \\
$\mathcal{P}_{\rm B}$ & 2 &\!\!\!\!1 &\!\!10.0 &\!\!45.7 & 0.50 & \!\!0.029     &\!\!1572 &\!\!\!9.6E$-$4 \\
$\mathcal{P}_{\rm C}$ & 3 &\!\!\!\!2 &\!\!10.0 &\!\!36.5 & 0.50 & \!\!0.039     &   \,749 &\!\!\!9.1E$-$5 \\
$\mathcal{P}_{\rm D}$ & 3 &\!\!\!\!2 &\!\!10.0 &\!\!27.7 & 0.50 & \!\!0.023     &   \,637 &\!\!\!4.2E$-$5 \\
\noalign{\smallskip}
\hline
\label{modeltable}
\end{tabular}
\end{table} 

\begin{figure*}
\centering
\centerline{
  \resizebox{0.2622\hsize}{!}{\includegraphics{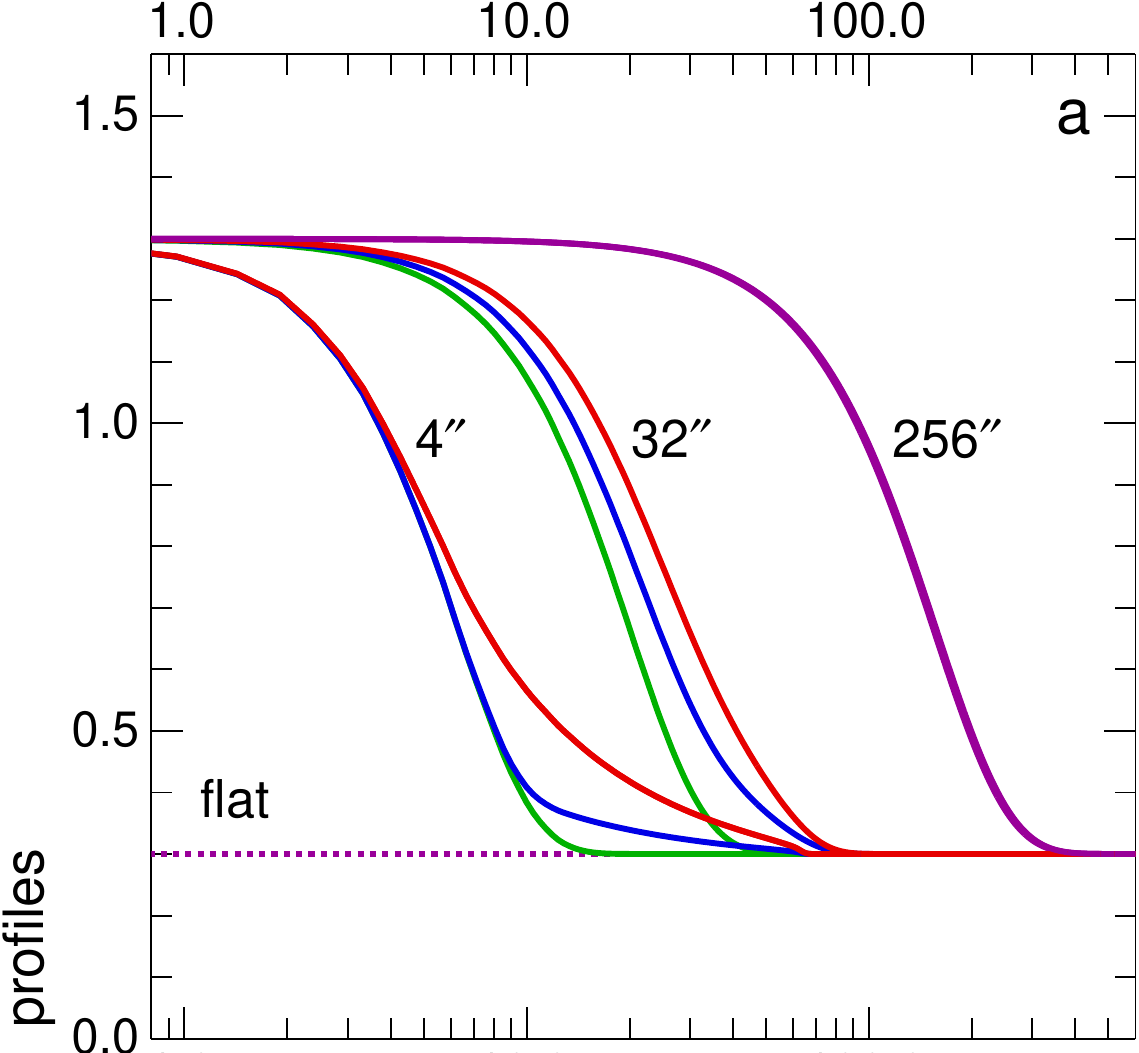}}    \hspace{-1.4mm}
  \resizebox{0.2287\hsize}{!}{\includegraphics{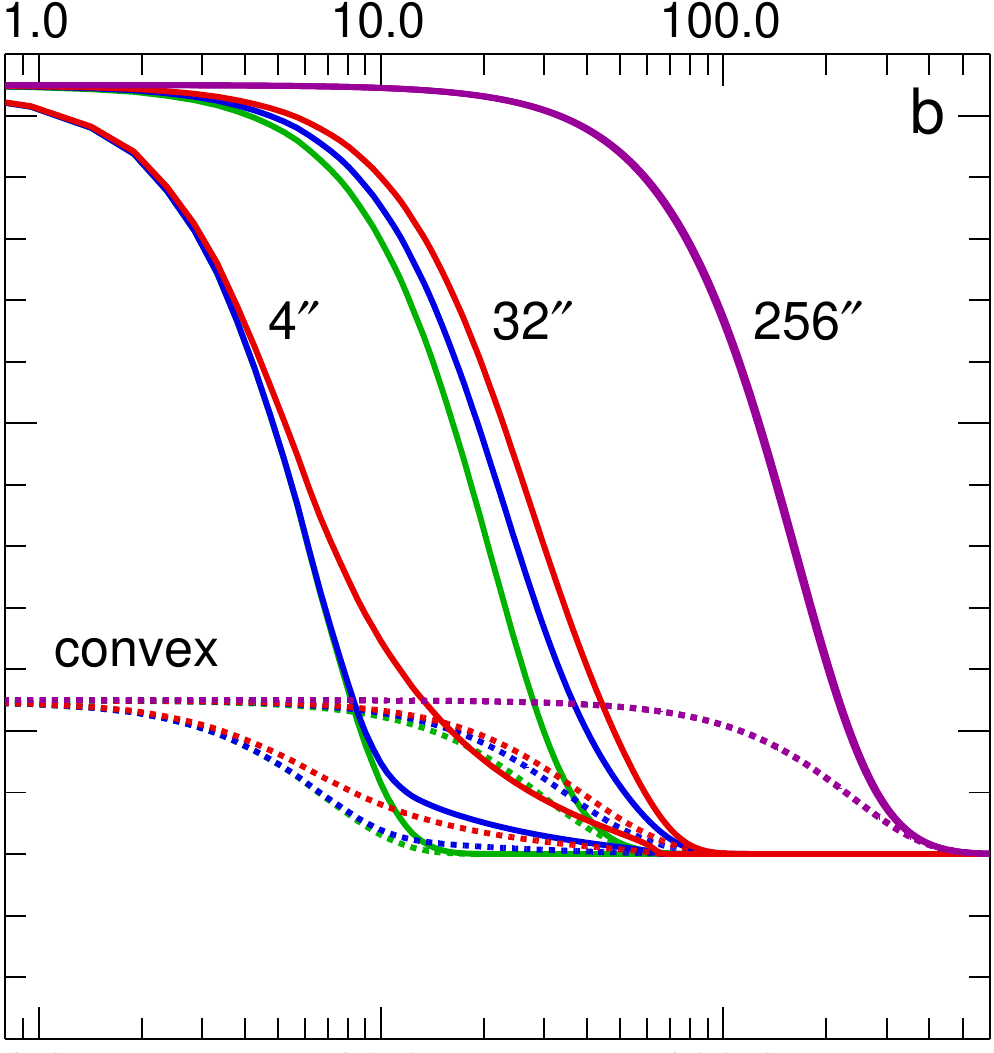}}  \hspace{-1.4mm}
  \resizebox{0.2527\hsize}{!}{\includegraphics{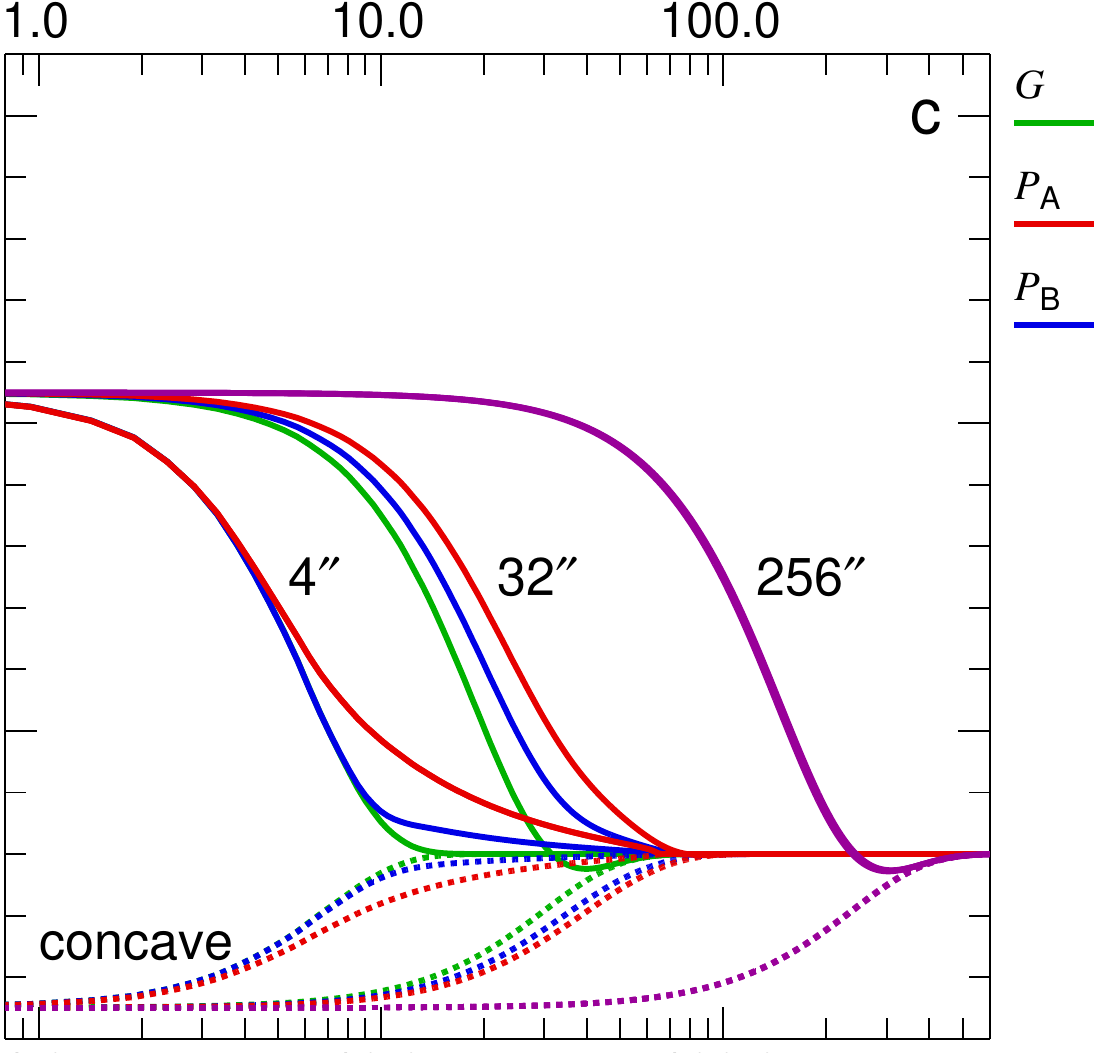}}}
\vspace{-0.45mm}
\centerline{
  \resizebox{0.2622\hsize}{!}{\includegraphics{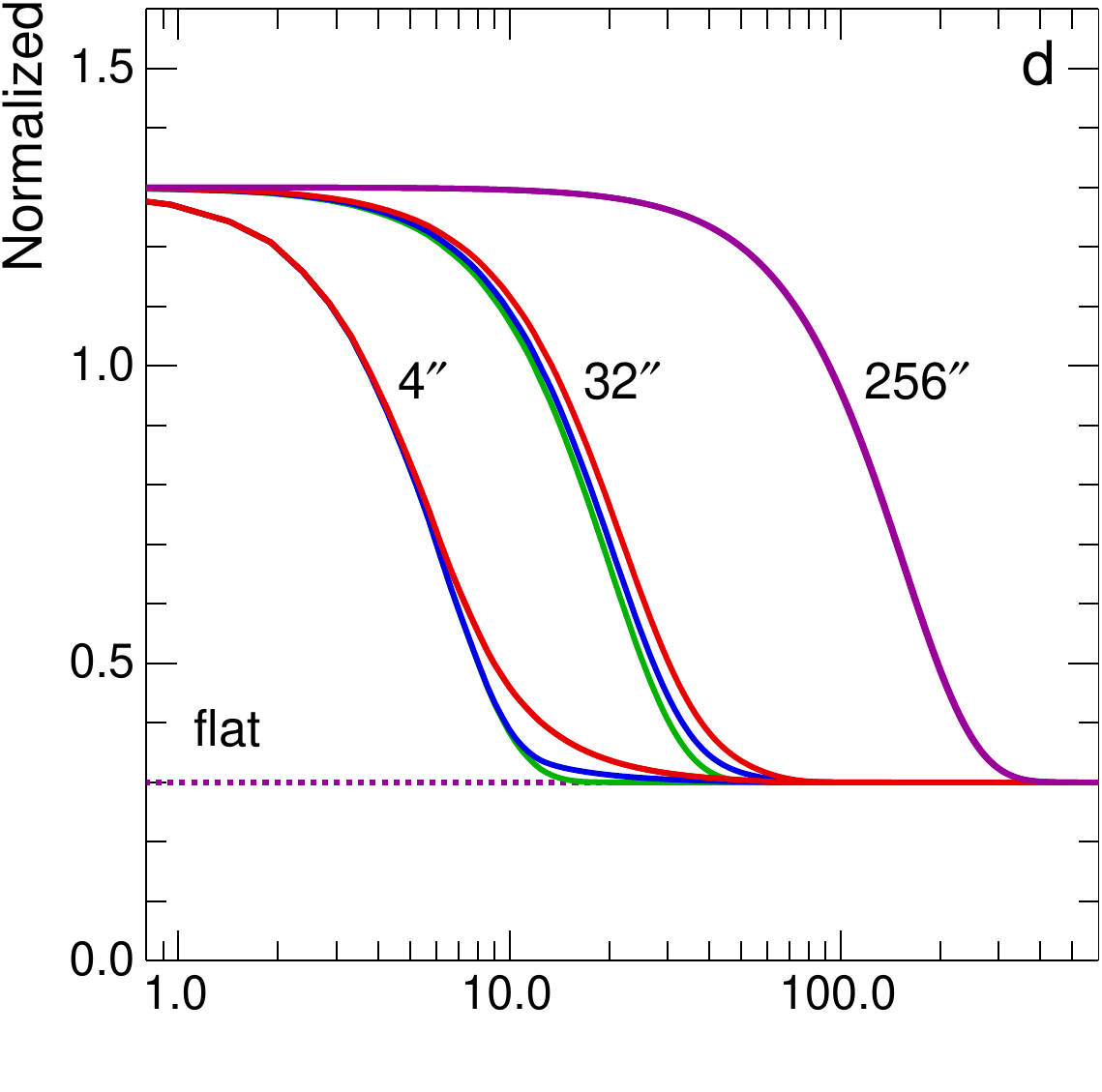}}    \hspace{-1.4mm}
  \resizebox{0.2287\hsize}{!}{\includegraphics{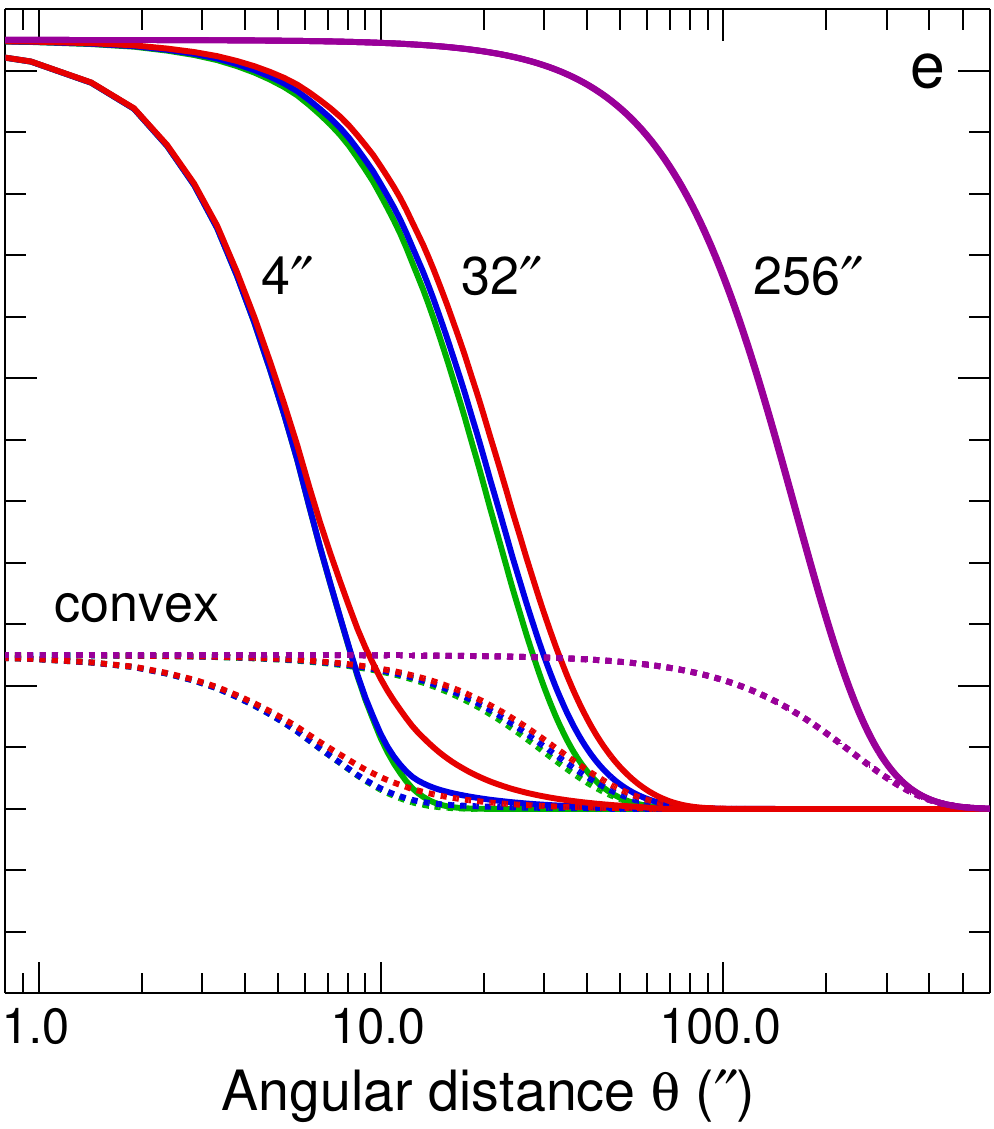}}  \hspace{-1.4mm}
  \resizebox{0.2527\hsize}{!}{\includegraphics{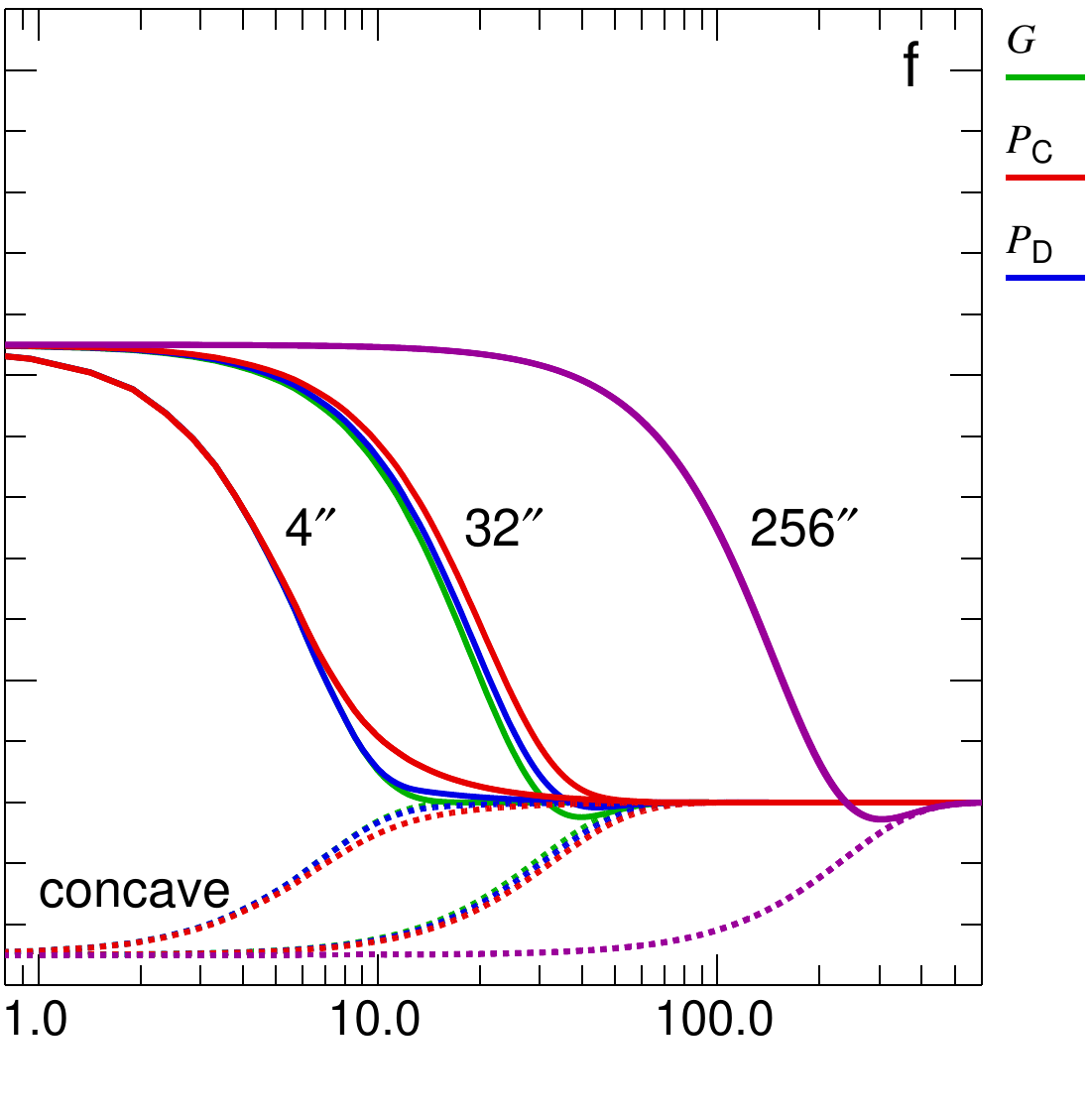}}}
\caption
{ 
Profiles of the simulated sources $\mathcal{S}_{{\mathcal{M}}{j}}$ and $\mathcal{S}_{{\mathcal{M}}{jk}{\pm}}$, defined as the
convolved spherical models $\mathcal{M}_{j}$ added to the flat, convex, and concave backgrounds (Eq.~(\ref{sources})). For an
illustration, the profiles correspond to three angular resolutions $O_{j} = \{4, 32$, $256$\}{\arcsec} (Eq.~(\ref{convol})) and the
background size factor $f_{k} = 1.54$ (Eq.~(\ref{beamback})). The profiles completely overlap with each other at $O_{j} =
256${\arcsec}.
} 
\label{modelsources}
\end{figure*}


\section{Simulated images}
\label{simimages}

This study investigates accuracy of the Gaussian size deconvolution on the basis of several spherical and cylindrical models with
profiles shown in Fig.~\ref{modelprofs} and parameters given in Table~\ref{modeltable}. Projections of each model onto the plane of
sky produce the images of a source or filament, with dimensions of $6079 \times 6079$ pixels and a pixel size of $0.47${\arcsec}.
The sources are placed at the image center, whereas the filaments are running through the center across the entire image.

\subsection{Models}
\label{models}

Model $\mathcal{G}$ has a Gaussian intensity distribution with a full width at half maximum (FWHM) of $10${\arcsec}
(Fig.~\ref{modelprofs}). Models $\mathcal{P}_{\rm A}$ and $\mathcal{P}_{\rm B}$ have power-law profiles with intensities (or
surface densities) $I \propto \theta^{-1}$ fitted to the Gaussian core, describing the cores and filaments with volume densities
$\rho \propto r^{-2}$. This choice is based on the fact that similar profiles between $r^{-1.5}$ and $r^{-2.5}$ (for large $r$) are
expected in both the Bonnor-Ebert spheres \citep{Bonnor1956} and collapsing protostars \citep{Larson1969}, as well as derived for
selected young stellar objects \citep[Sect.~4 in][]{Men'shchikovHenning1997,Men'shchikov_etal1999} and filaments
\citep{Arzoumanian_etal2011,Arzoumanian_etal2019}. Models $\mathcal{P}_{\rm C}$ and $\mathcal{P}_{\rm D}$ with steeper profiles $I
\propto \theta^{-2}$ ($\rho \propto r^{-3}$) are meant to test the dependence of size deconvolution results on different intensity
distributions. The fitting levels of the power-law profiles, differing by a factor of $3$, were chosen arbitrarily to test
different contrasts of the Gaussian core with respect to the power-law wings. All models have a peak intensity of unity and an
outer boundary at a radial distance $r_{2}$ from their centers, corresponding to the angular distance $\Theta = 64${\arcsec}, such
that $\rho = 0$ for $r > r_{2}$ and $I = 0$ for $\theta > \Theta$. As a result of the adopted spherical and cylindrical geometries,
the profiles of the power-law models fall down very steeply when $\theta \rightarrow \Theta$, whereas the intensities of model
$\mathcal{G}$ become negligible beyond $\theta \approx 20${\arcsec} (Fig.~\ref{modelprofs}).


It is convenient to collectively denote the power-law models as $\mathcal{P}$ and all five models ($\mathcal{G}$ and $\mathcal{P}$)
as $\mathcal{M}$. The images represent idealized sources and filaments with a high numerical resolution (small pixels), in the
absence of any background or noise. To simulate the finite angular resolutions (point-spread functions or PSFs) of astronomical
telescopes, the images are convolved with the circular Gaussian kernels $\mathcal{O}_{j}$,
\begin{equation} 
\mathcal{M}_{j} = \mathcal{O}_{j} * \mathcal{M}, \,\,\, j = 1, 2,\dots, J,
\label{convol}
\end{equation} 
where $J = 107$ and the FWHM sizes $O_{j}$ of the finely spaced set of kernels (beams, angular resolutions) between $O_{1} =
3.36{\arcsec}$ and $O_{J} = 332{\arcsec}$ are defined as
\begin{equation} 
O_{0} = 2^{-1/16} O_{1}, \, O_{j} = 2^{1/16} O_{j-1},
\label{beams}
\end{equation} 
and the resulting images $\mathcal{M}_{j}$ are renormalized to the peak intensity of the original models. The set of convolved
models, computed with \textsl{convolve} and \textsl{fftconv} from the \textsl{getsf} source and filament extraction
method\footnote{\url{http://irfu.cea.fr/Pisp/alexander.menshchikov/}} \citep{Men'shchikov2021a}, simulates the entire range of the
resolution states (unresolved to well resolved) of the model sources and filaments.

\subsection{Backgrounds}
\label{modbacks}

Most extraction methods cannot disentangle the complex backgrounds from the source or filament emission and apply simple (linear)
interpolation across the structures of interest, based on the pixel values just outside them \citep[e.g., Fig.~12
in][]{Men'shchikov_etal2012}. To investigate effects of the inaccurately determined backgrounds on the Gaussian size deconvolution,
it makes sense to reduce the observed complexity and adopt only three typical shapes for the models: the flat, convex, and concave
backgrounds (Fig.~\ref{modelsources}).

The flat backgrounds are assumed to always have a constant (arbitrary) value $\mathcal{B} = 0.3$. The two other types of
backgrounds are derived, for simplicity, from the shapes of the model sources and filaments at each angular resolution,
\begin{equation} 
\mathcal{B}_{{\mathcal{M}}{jk}{\pm}} = \mathcal{B} \pm 0.25\,\left(\mathcal{O}_{jk} * \mathcal{M}\right),
\label{modelback}
\end{equation} 
where ``$+$'' and ``$-$'' correspond to the convex and concave backgrounds, respectively, and the FWHM sizes of the Gaussian
kernels $\mathcal{O}_{jk}$ are defined by
\begin{equation} 
O_{jk} = f_{k} O_{j}, \,\,\, k = 1, 2,\dots, K,
\label{beamback}
\end{equation} 
where $K = 9$ and $f_{k}$ are the background size factors, defined as
\begin{equation} 
f_{0} = 2^{-1/8} f_{1}, \, f_{k} = 2^{1/8} f_{k-1},
\label{backfacts}
\end{equation} 
in the range between $1$ and $2$. For $f_{1} = 1$, the convex or concave background has the same width as that of the model at
resolution $O_{j}$, whereas for $f_{K} = 2$, the background has a wider shape, corresponding to the model convolved to a twice
lower resolution. With $f_{k} > 1$, the convex backgrounds tend to increase both the heights and widths of the sources, whereas the
concave backgrounds contribute to their decrease. 

For brevity, it is convenient to denote $\{\mathcal{S}|\mathcal{F}\}$ the images of spherical or cylindrical models
(Table~\ref{modeltable}, Fig.~\ref{modelsources}) and refer to either of the images as $\mathcal{I}=\{\mathcal{S}|\mathcal{F}\}$. A
superposition of the models and backgrounds completes the simulated images,
\begin{eqnarray} 
\left.\begin{aligned}
&\mathcal{I}_{{\mathcal{M}}{j}} = \mathcal{M}_{j} + \mathcal{B}, \\
&\mathcal{I}_{{\mathcal{M}}{jk}{\pm}} = \mathcal{M}_{j} + \mathcal{B}_{{\mathcal{M}}{jk}{\pm}}, 
\end{aligned}\right.
\label{sources} 
\end{eqnarray} 
representing the idealized sources ($\mathcal{S}_{{\mathcal{M}}{j}}$, $\mathcal{S}_{{\mathcal{M}}{jk}{\pm}}$) and filaments
($\mathcal{F}_{{\mathcal{M}}{j}}$, $\mathcal{F}_{{\mathcal{M}}{jk}{\pm}}$) with a FWHM size of $10${\arcsec}, blended with the
three different backgrounds and observed with the wide range of angular resolutions $O\approx 3{-}300${\arcsec}. This set of images
is sufficient to study effects of the complex, inaccurately subtracted backgrounds on the results of measurement and size
deconvolution in real observations.

\begin{figure*}
\centering
\centerline{
  \resizebox{0.2680\hsize}{!}{\includegraphics{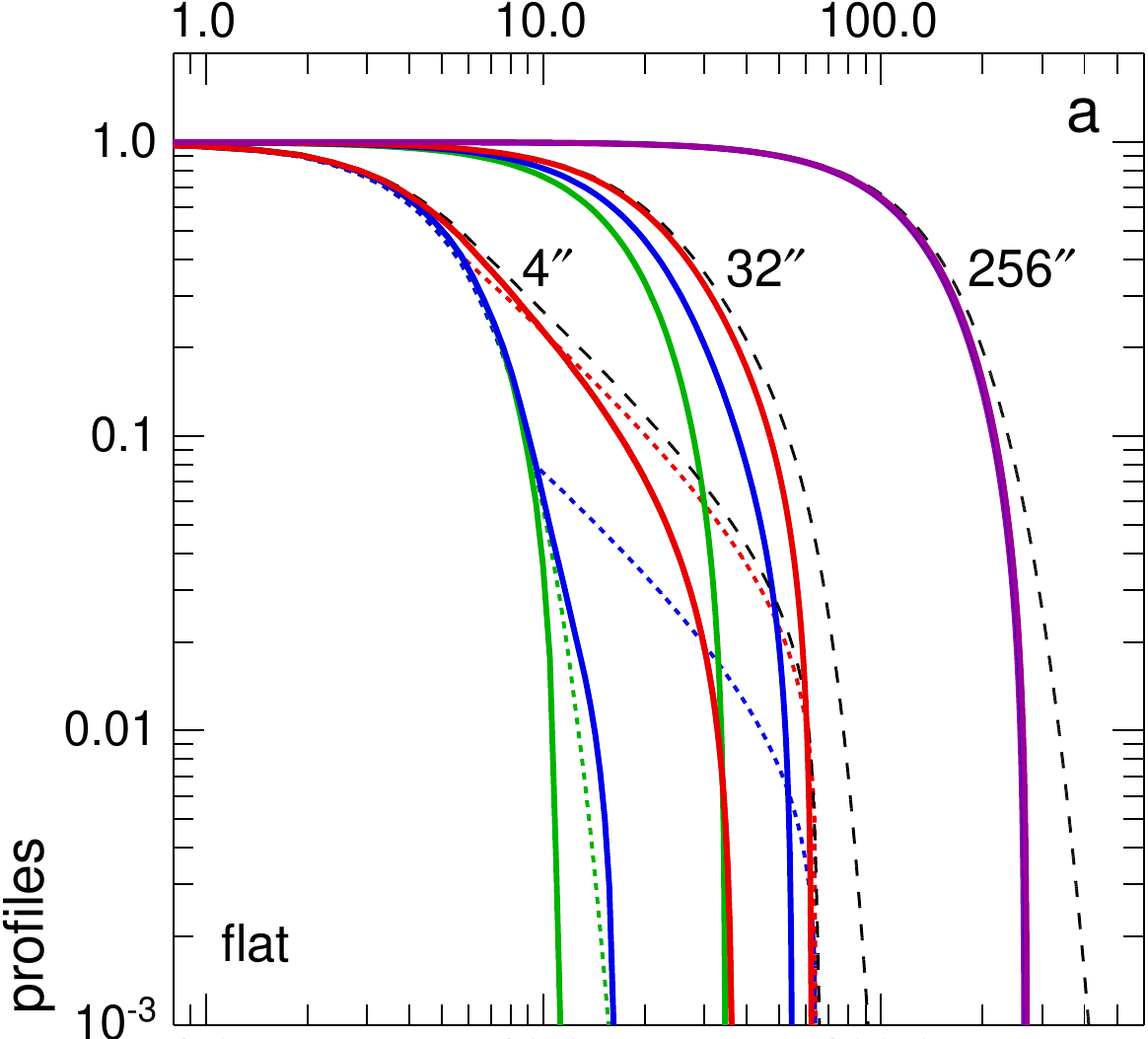}}    \hspace{-1.4mm}
  \resizebox{0.2287\hsize}{!}{\includegraphics{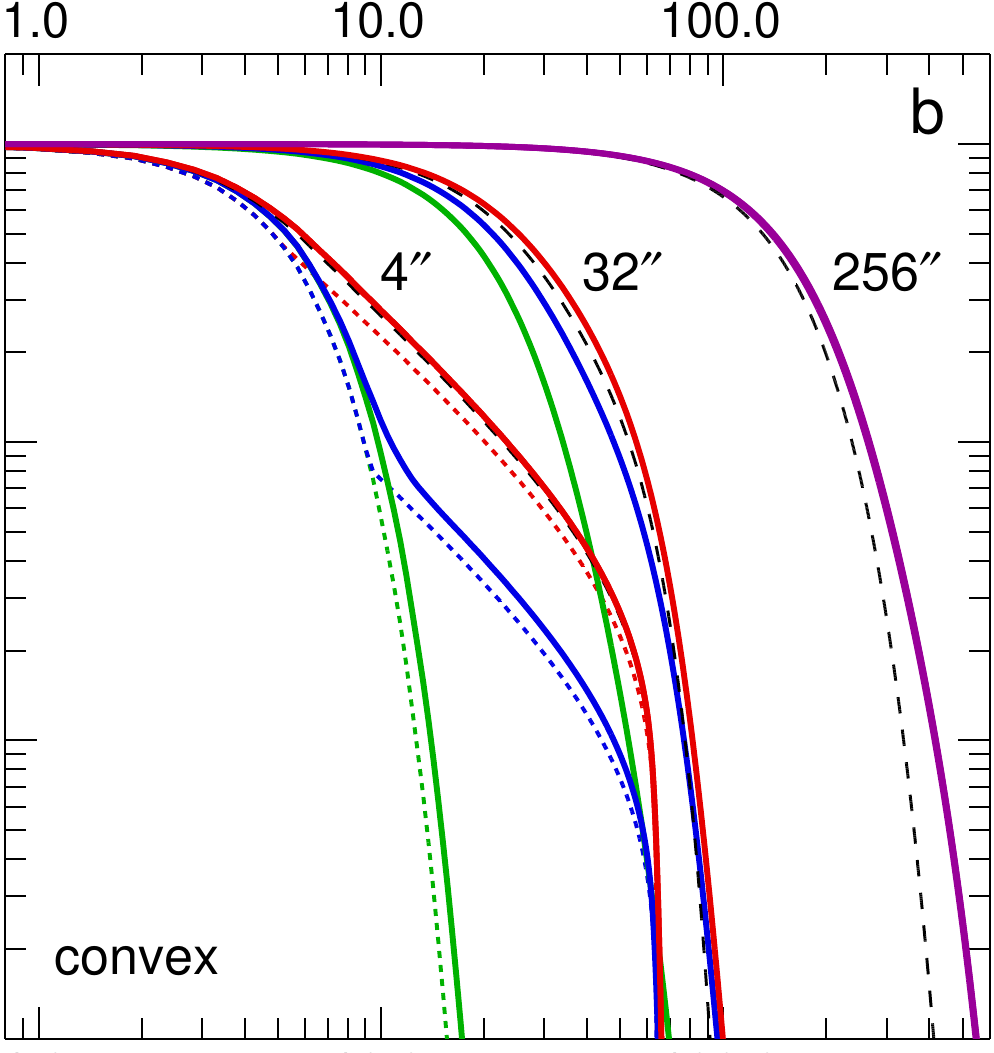}}  \hspace{-1.4mm}
  \resizebox{0.2527\hsize}{!}{\includegraphics{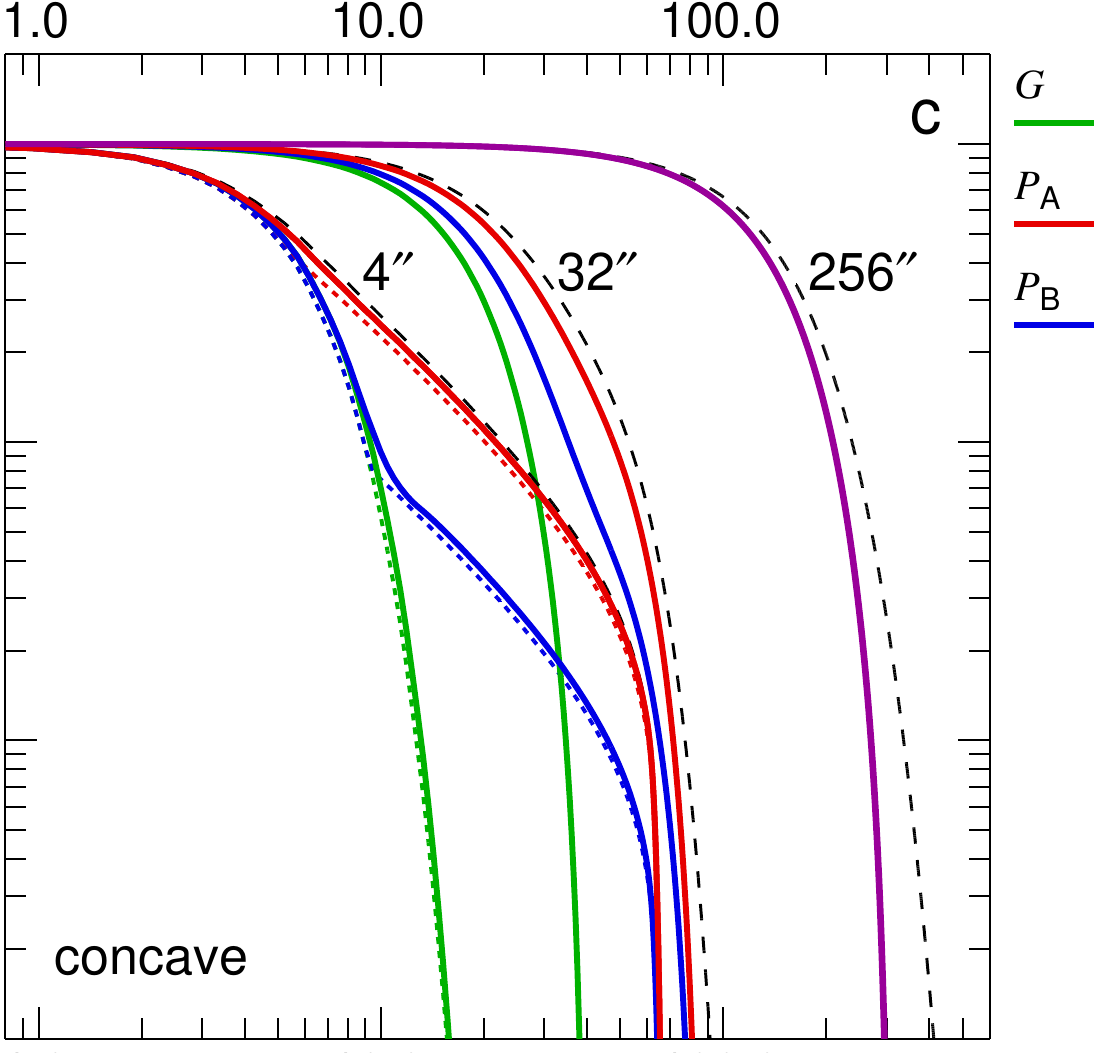}}}
\vspace{-0.45mm}
\centerline{
  \resizebox{0.2680\hsize}{!}{\includegraphics{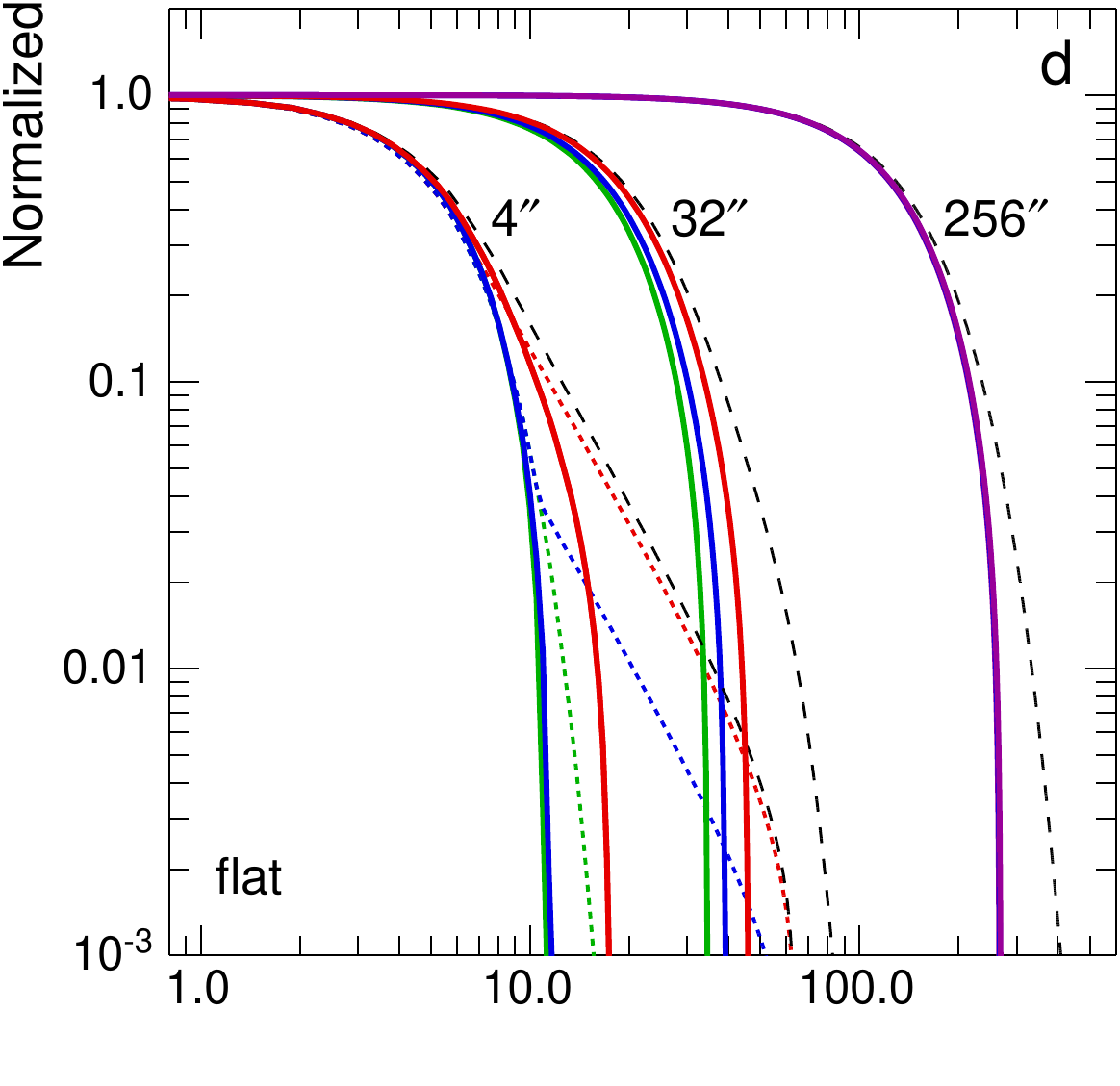}}    \hspace{-1.4mm}
  \resizebox{0.2287\hsize}{!}{\includegraphics{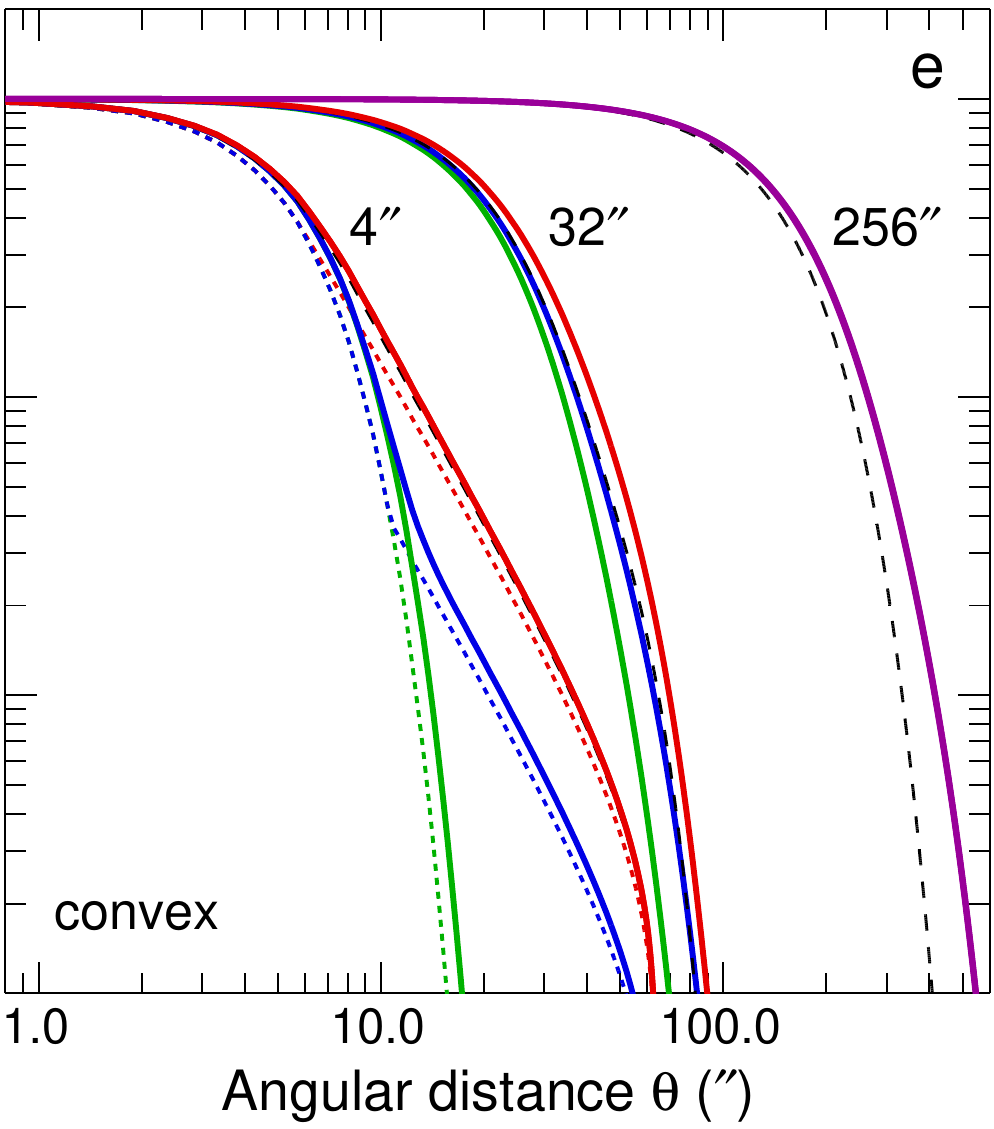}}  \hspace{-1.4mm}
  \resizebox{0.2527\hsize}{!}{\includegraphics{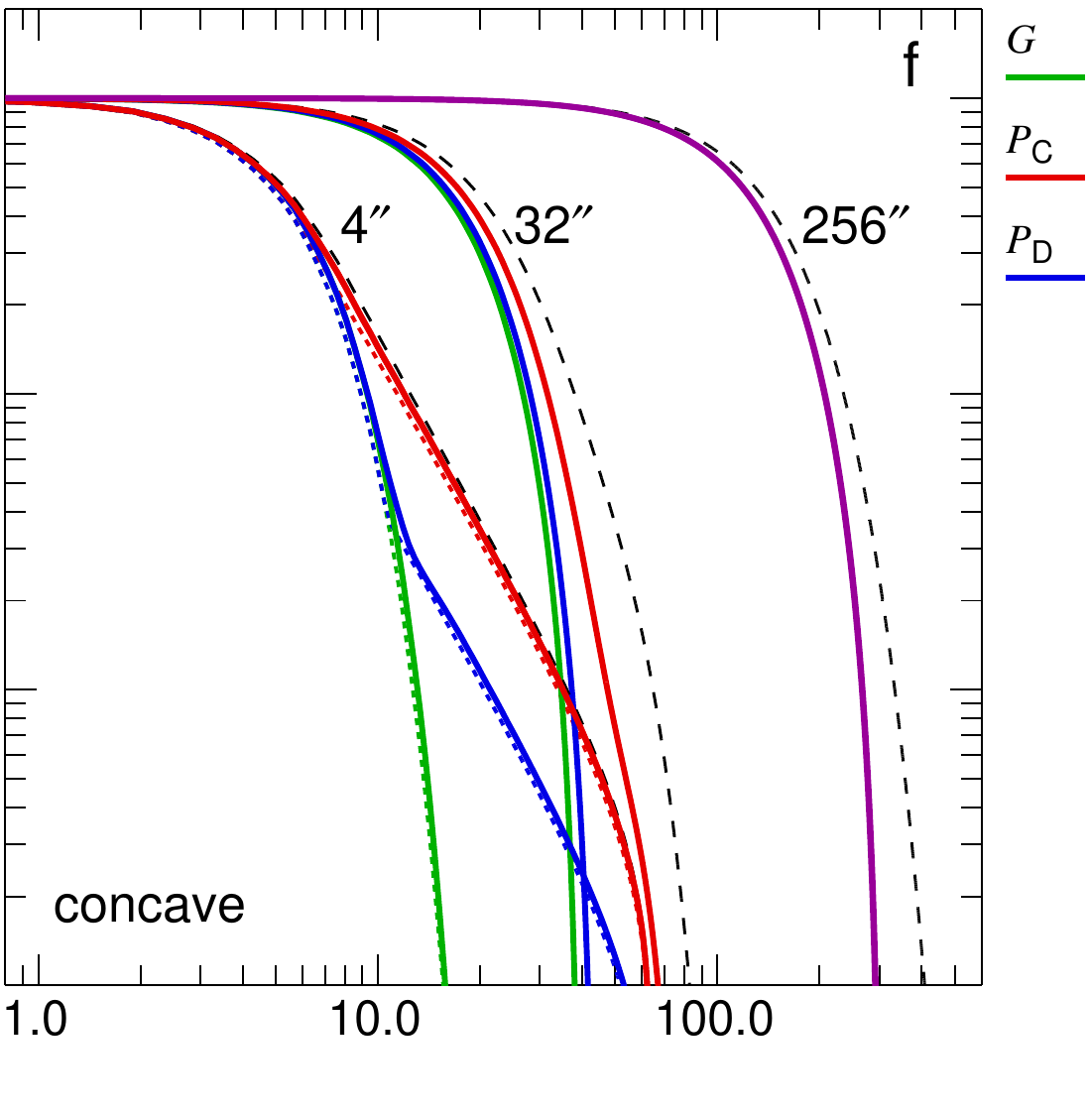}}}
\caption
{ 
Profiles of the sources $\mathcal{\tilde{S}}_{{\mathcal{M}}{jl}}$ and $\mathcal{\tilde{S}}_{{\mathcal{M}}{jk}{\pm}}$
(Fig.~\ref{modelsources}), separated from the flat, convex, and concave backgrounds by subtraction of their interpolated planar
backgrounds, according to Eq.~(\ref{bgsubtraction}). The profiles (solid lines) correspond to the angular resolutions $O_{j} = \{4,
32$, $256$\}{\arcsec}, background error $\epsilon_{\,l} = 0.05$, and background size factor $f_{k} = 1.54$ used in
Fig.~\ref{modelsources}. The profiles completely overlap with each other at $O_{j} = 256${\arcsec}. For reference, the dotted
curves reproduce the true model profiles from Fig.~\ref{modelprofs}. For comparison, the black dashed curves for models
$\mathcal{P}_{\rm \{A|C\}}$ with $\epsilon_{\,l} = 0$ and $f_{k} = 1$ display the background-separated profiles that get almost no
distortions of their shapes for $O_{j} = 4${\arcsec} in the process of background subtraction. To preserve clarity of the plots,
similar profiles for models $\mathcal{G}$ and $\mathcal{P}_{\rm \{B|D\}}$ are not shown.
} 
\label{sourcesbgs}
\end{figure*}

\subsection{Background subtraction}
\label{backsubtract}

In this model-based study, the general, nontrivial problem of source or filament extraction in complex observed images
\citep[e.g.,][]{Men'shchikov2021a} reduces to the separation of sources or filaments from their backgrounds and measurement of
their sizes. It is assumed that the structure has been detected and its location is precisely known.

To follow the standard procedure used in the extraction methods \citep[e.g., Sect.~3.4.6 in][]{Men'shchikov2021a}, it is necessary
to determine the source or filament footprint, defined as the image area, supposedly containing the entire emission of the
structure of interest, and subtract a planar (constant) background defined by the intensity at the footprint edge. This can be
accomplished by finding the angular distance $\Theta$ from the source peak or filament crest, at which the slope of intensity
distribution becomes sufficiently small:
\begin{equation} 
\left|\frac{{\rm d}\log{I(\theta)}}{{\rm d}\log{\theta}}\right|_{\Theta} = 10^{-4},
\label{footprint}
\end{equation} 
where the adopted (arbitrary) value is appropriate for the perfectly smooth model images. With this definition, the simulated (not
convolved) flat-background structures $\mathcal{I}_{{\mathcal{M}}} = \mathcal{M} + \mathcal{B}$ have the derived footprint radii of
$\Theta_{\mathcal{G}} = 21.8${\arcsec} and $\Theta_{\mathcal{P}} = 64.09{\arcsec}$. The footprints expand to larger distances
$\Theta_{{\mathcal{M}}{j}}$ and $\Theta_{{\mathcal{M}}{jk}{\pm}}$ at lower angular resolutions, when the images are convolved with
increasingly larger beams $O_{j}$ (Fig.~\ref{modelsources}). For models $\mathcal{I}_{{\mathcal{M}}{jk}-}$ on concave
backgrounds, the footprints $\Theta_{{\mathcal{M}}{jk}-}$ correspond to the minima that develop in their intensity profiles with
$f_{k} > 1$, hence they become smaller than the footprints for the flat backgrounds (Fig.~\ref{modelsources}). In contrast, the
models $\mathcal{I}_{{\mathcal{M}}{jk}+}$ on convex backgrounds with $f_{k} > 1$ have footprints $\Theta_{{\mathcal{M}}{jk}+}$ that
are larger than those for the flat backgrounds (Fig.~\ref{modelsources}).

When the footprints of the structures ($\mathcal{I}_{{\mathcal{M}}{j}}$, $\mathcal{I}_{{\mathcal{M}}{jk}{\pm}}$) are determined,
the background-subtracted intensity distributions, are obtained by subtracting intensities of their (interpolated) constant
backgrounds,
\begin{eqnarray} 
\left.\begin{aligned}
&\mathcal{\tilde{I}}_{{\mathcal{M}}{jl}} = \max \left(\mathcal{I}_{{\mathcal{M}}{j}} - 
\left(I(\Theta_{{\mathcal{M}}{j}}) + \epsilon_{\,l}\right), 0\right), \\
&\mathcal{\tilde{I}}_{{\mathcal{M}}{jk}{\pm}} = \max \left(\mathcal{I}_{{\mathcal{M}}{jk}{\pm}} - 
I(\Theta_{{\mathcal{M}}{jk}{\pm}}), 0\right), 
\end{aligned}\right.
\label{bgsubtraction} 
\end{eqnarray} 
where $\epsilon_{\,l}$ accounts for possible background errors, in order to test, how overestimated backgrounds affect the measured
sizes and their deconvolution,
\begin{equation} 
\epsilon_{\,l} = 0.05 \left(l - 1\right), \,\,\, l = 1, 2,\dots, L,
\label{backerrs}
\end{equation} 
where $L = 7$. The background errors $\epsilon_{\,l}$ sample values within $30$\% of the model maximum intensity ($I(0) = 1$),
which is a typical range of the measurement errors for the peak intensities of faint sources, as demonstrated by source extraction
benchmarks \citep[Figs.~6\,--\,10 in][]{Men'shchikov2021b}. Subtraction of the overestimated backgrounds
$I(\Theta_{{\mathcal{M}}{j}}) + \epsilon_{\,l}$, instead of the accurate background $I(\Theta_{{\mathcal{M}}{j}})$, cuts off the
low-intensity pedestal of the source shape and shrinks its footprint (Fig.~\ref{sourcesbgs}), leading to underestimated sizes, peak
intensities, and integrated fluxes. It makes no sense to similarly test the underestimated backgrounds for the flat-background
models, because their shapes would not change, if $\epsilon_{\,l}$ were subtracted from the background
$I(\Theta_{{\mathcal{M}}{j}})$ in Eq.~(\ref{bgsubtraction}). The errors are applied only to the flat-background models, because the
nonflat backgrounds $\mathcal{B}_{{\mathcal{M}}{jk}{\pm}}$ lead to under- or overestimation of the source or filament backgrounds
on their own and it is important to isolate the different effects.

\subsection{Measurements}
\label{measurements}

This study focuses on the measurements of half maximum sizes and their deconvolution, therefore it does not discuss measurements of
other properties of extracted sources and filaments. In most algorithms, the half maximum sizes of sources are obtained assuming a
Gaussian shape of their intensity distribution, by fitting two-dimensional Gaussians \citep[e.g., in
\textsl{cutex},][]{Molinari_etal2011} or employing intensity moments \citep[e.g., in \textsl{getsources}][]{Men'shchikov_etal2012}.
The sizes, estimated from the second moments of intensities \citep[e.g., Appendix F in][]{Men'shchikov_etal2012}, are referred to
as the moment sizes in this paper.

\begin{figure}
\centering
\centerline{
  \resizebox{0.6100\hsize}{!}{\includegraphics{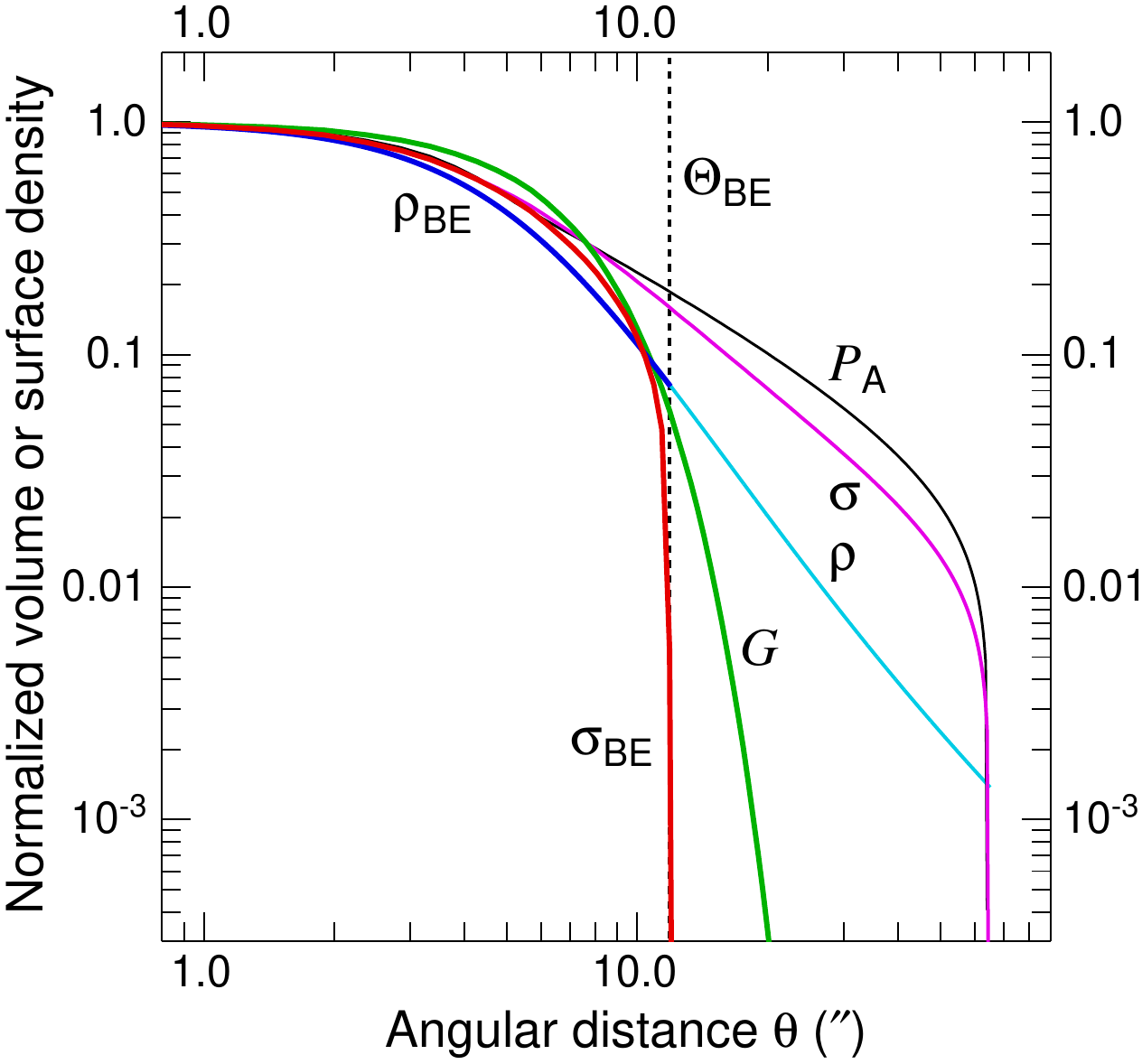}}}
\caption
{ 
Profiles of the volume density $\rho_{\rm BE}$ (\emph{blue}) and surface density $\sigma_{\rm BE}$ (\emph{red}) of a critical
Bonnor-Ebert sphere with an outer boundary at $\Theta_{\rm BE} = 11.9${\arcsec} and a surface-density half maximum width $H_{\rm
BE} = 10${\arcsec}. For comparisons, the truncated $\rho_{\rm BE}$ continues at $\theta > \Theta_{\rm BE}$ as $\rho$ (\emph{cyan})
and the resulting $\sigma$ (\emph{magenta}) corresponds to the outer boundary $\Theta = 64${\arcsec} of the finite spherical models
(Table~\ref{modeltable}). It can be easily estimated from the numerical data that the Bonnor-Ebert sphere has the power-law
profiles $\rho \propto r^{-2.43}$ and $\sigma \propto \theta^{-1.6}$ at the boundary $\Theta_{\rm BE}$. For reference, the profile
of model $\mathcal{P}_{\rm A}$ (\emph{black}) is also shown.
} 
\label{besdens}
\end{figure}

A serious drawback of such approaches is that the assumption of Gaussian profiles is very idealized and simplistic. Various
physical objects, such as the extended envelopes of protostellar cores \citep[e.g.,][]{Larson1969} and evolved stars
\citep[e.g.,][]{Men'shchikov_etal2001}, are well described by the power-law distributions of volume densities ($\rho \propto
r^{-2}$) and surface densities ($\sigma \propto \theta^{-1}$), markedly different from a Gaussian. Prestellar cores, modeled as the
critical Bonnor-Ebert spheres \citep{Bonnor1956}, are also expected to have a power-law volume densities at their boundaries ($\rho
\propto r^{-2.43}$, Fig.~\ref{besdens}). However, their sphericity and compactness make their surface densities appear similar to a
Gaussian with a half maximum size of $H_{\rm G} = \Theta_{\rm BE}$ above $10${\%} of the peak, but strongly dissimilar below that
level (Fig.~\ref{besdens}).


Moreover, even an imaginary purely Gaussian structure acquires a non-Gaussian shape in the extraction process, when its background
happens to be under- or overestimated, hence inaccurately subtracted (Fig.~\ref{sourcesbgs}). As a consequence of substantial
deviations of the extracted sources from a Gaussian shape, their moment sizes do not correspond to half maximum intensity. For the
power-law sources $\mathcal{P}$, Table~\ref{modeltable} demonstrates that the intensity levels corresponding to their moment sizes
($L_{M_{\!\mathcal{P}}}$), are below the actual half maximum level by very large factors. To avoid misunderstanding and confusion,
the sizes computed from the source intensity moments should not be called the FWHM sizes.

The approach of fitting a Gaussian shape to the background-subtracted source to measure its FWHM, assumes that a Gaussian shape can
be found by the fitting algorithm that closely approximates the intensity distribution. Replacing the actual source shape with a
fitted Gaussian gives the size estimates that do correspond to the source half maximum intensity, eliminating the problem of
incompatible intensity levels of the moment sizes for different sources. A similar approach is also used to estimate the widths of
filaments, when one-dimensional Gaussian is fitted to the filament profile, usually averaged along the filament length
\citep[e.g.,][]{Palmeirim_etal2013}. However, the fitting method measures properties of the fitted Gaussians, not those of the
observed (non-Gaussian) sources. As a consequence, the derived half maximum sizes may be incorrect, depending on the inaccuracies
of the data fitting procedure in the presence of the background and noise fluctuations and on the properties of the object that
produced the observed structure shape. Fitting a Gaussian to an average filament profile ignores the important fact that filaments
often have substantially different widths along their crests \citep[e.g.,][]{Arzoumanian_etal2019}.

Both the half maximum and moment sizes are investigated in this work, following the algorithms applied by the \textsl{getsf}
extraction method \citep{Men'shchikov2021a}. The half maximum sizes are determined by a direct interpolation of intensity profiles
(Eq.~(\ref{bgsubtraction})) to the half maximum level \citep[Sect.~3.4.6 in][]{Men'shchikov2021a}, whereas the moment sizes (of
sources) are computed from the second moments of intensities \citep[Appendix F in][]{Men'shchikov_etal2012}. The two definitions
are implemented in the \textsl{imgstat} utility from \textsl{getsf} and they give identical results for the Gaussian intensity
distribution (Table~\ref{modeltable}), as expected.

It is convenient to collectively denote $\{H|M\}$ the true model sizes (Table~\ref{modeltable}, Fig.~\ref{modelprofs}), $\{H|M\}_j$
the sizes of the convolved models at a resolution $O_j$, and $\{\tilde{H}|\tilde{M}\}_j$ the sizes of the background-subtracted
sources. The obvious indices $\mathcal{M}$, $l$, and $k{\pm}$, related to the models and backgrounds (Eq.~(\ref{bgsubtraction})),
are not given explicitly in $\{\tilde{H}|\tilde{M}\}_j$, to simplify the notation.

\begin{figure*}
\centering
\centerline{
  \resizebox{0.2675\hsize}{!}{\includegraphics{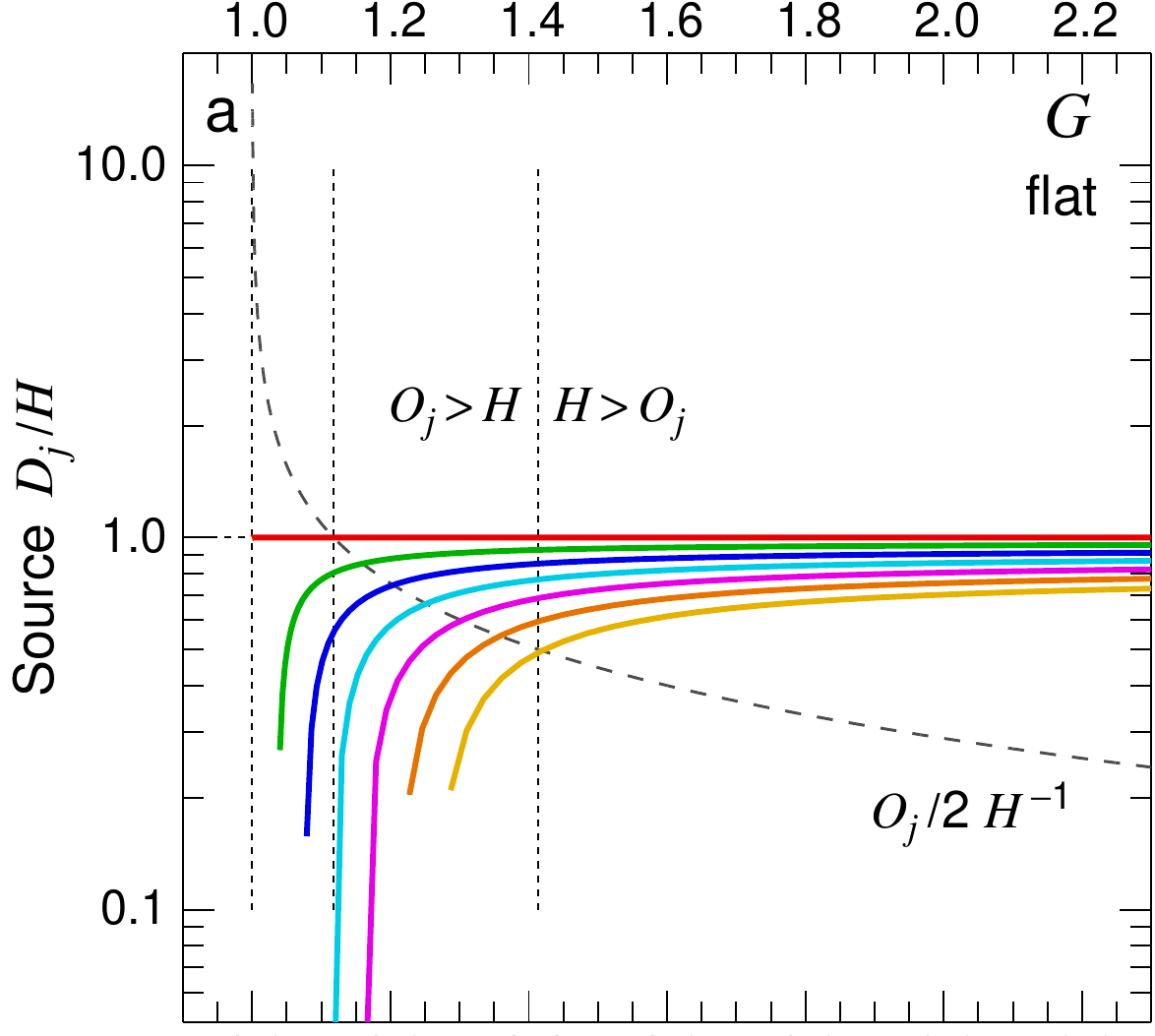}}     \hspace{-1.8mm}
  \resizebox{0.2755\hsize}{!}{\includegraphics{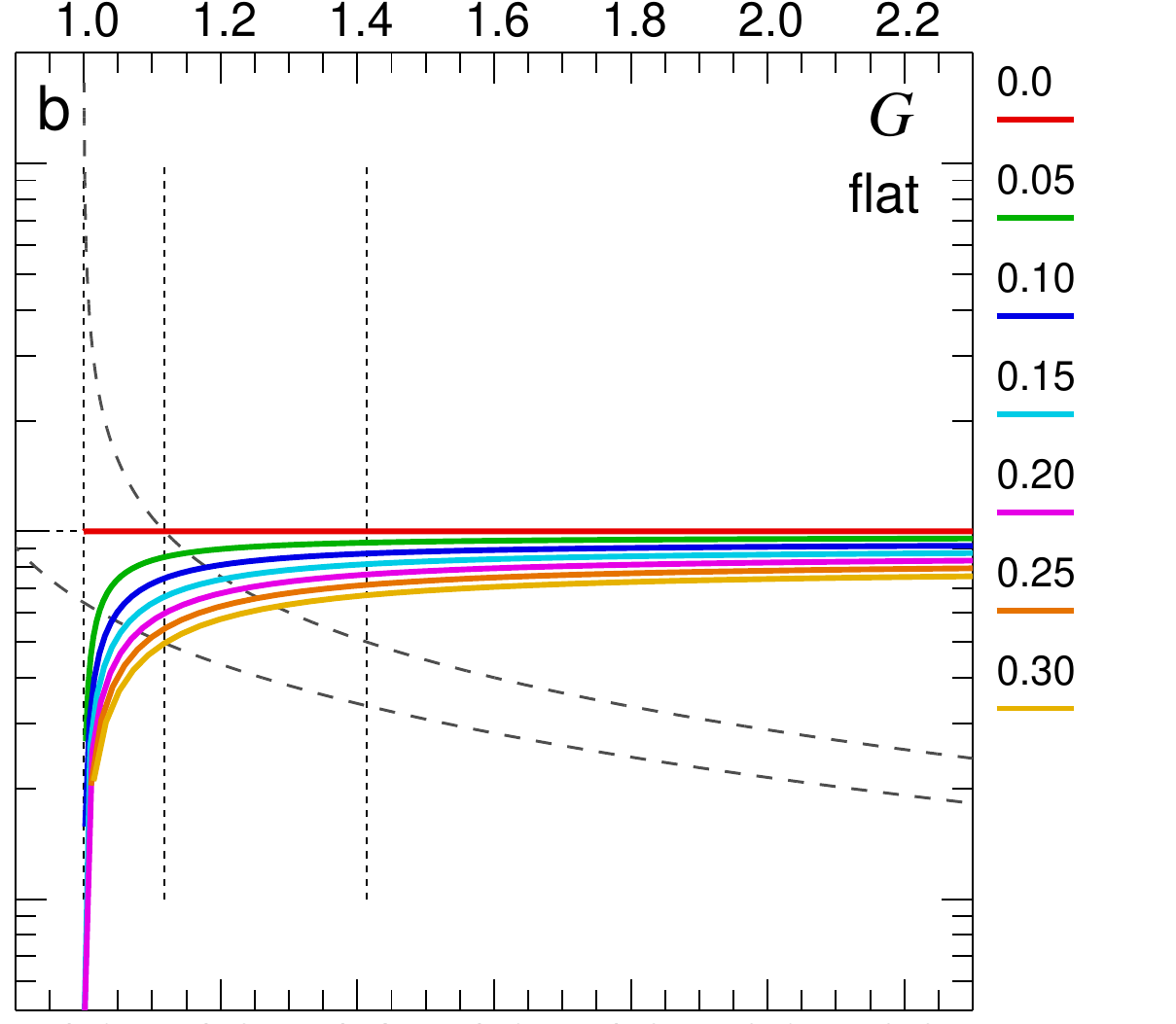}}       \hspace{-1.8mm}
  \resizebox{0.2912\hsize}{!}{\includegraphics{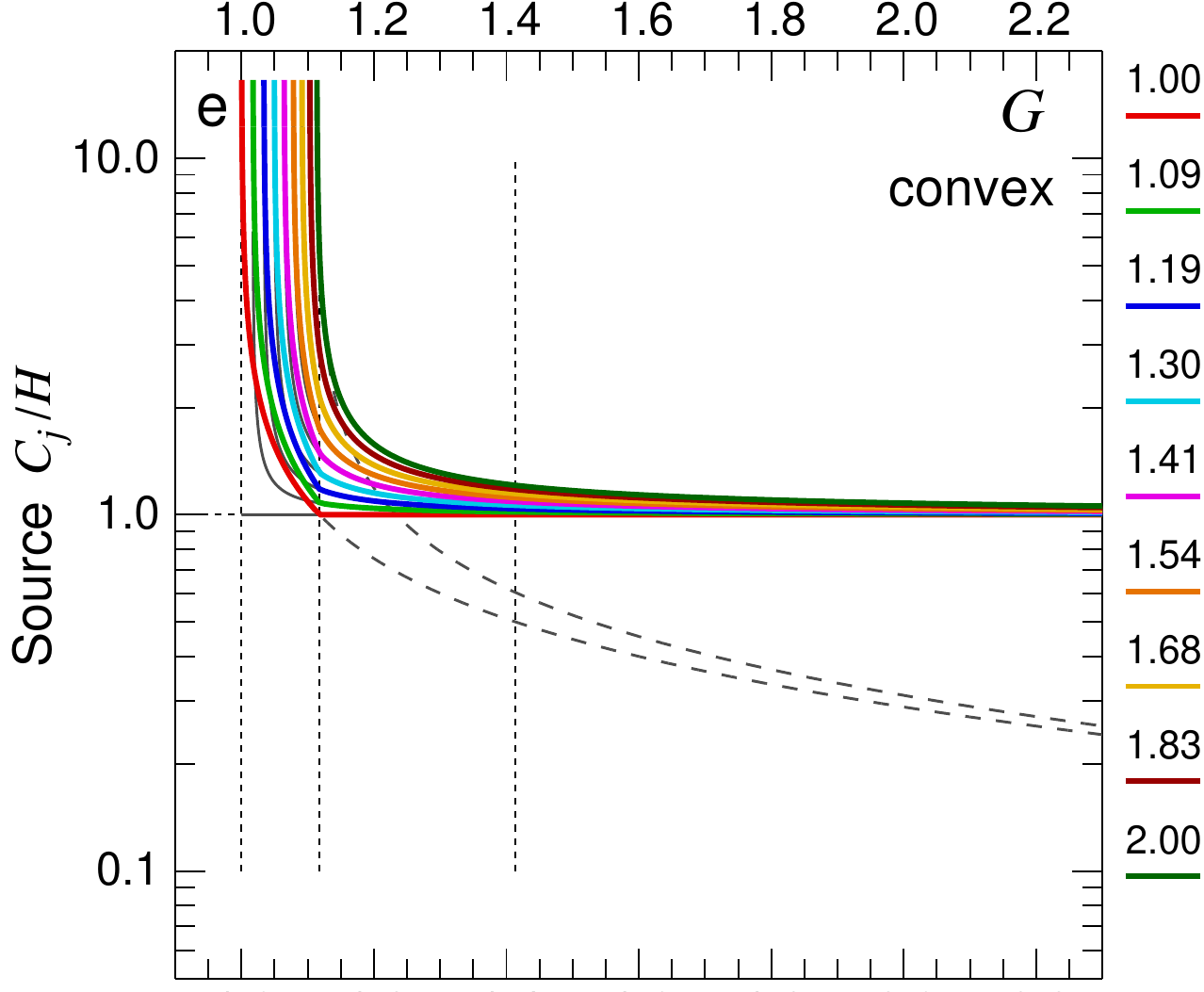}}}
\vspace{-2.54mm}
\centerline{
  \resizebox{0.2675\hsize}{!}{\includegraphics{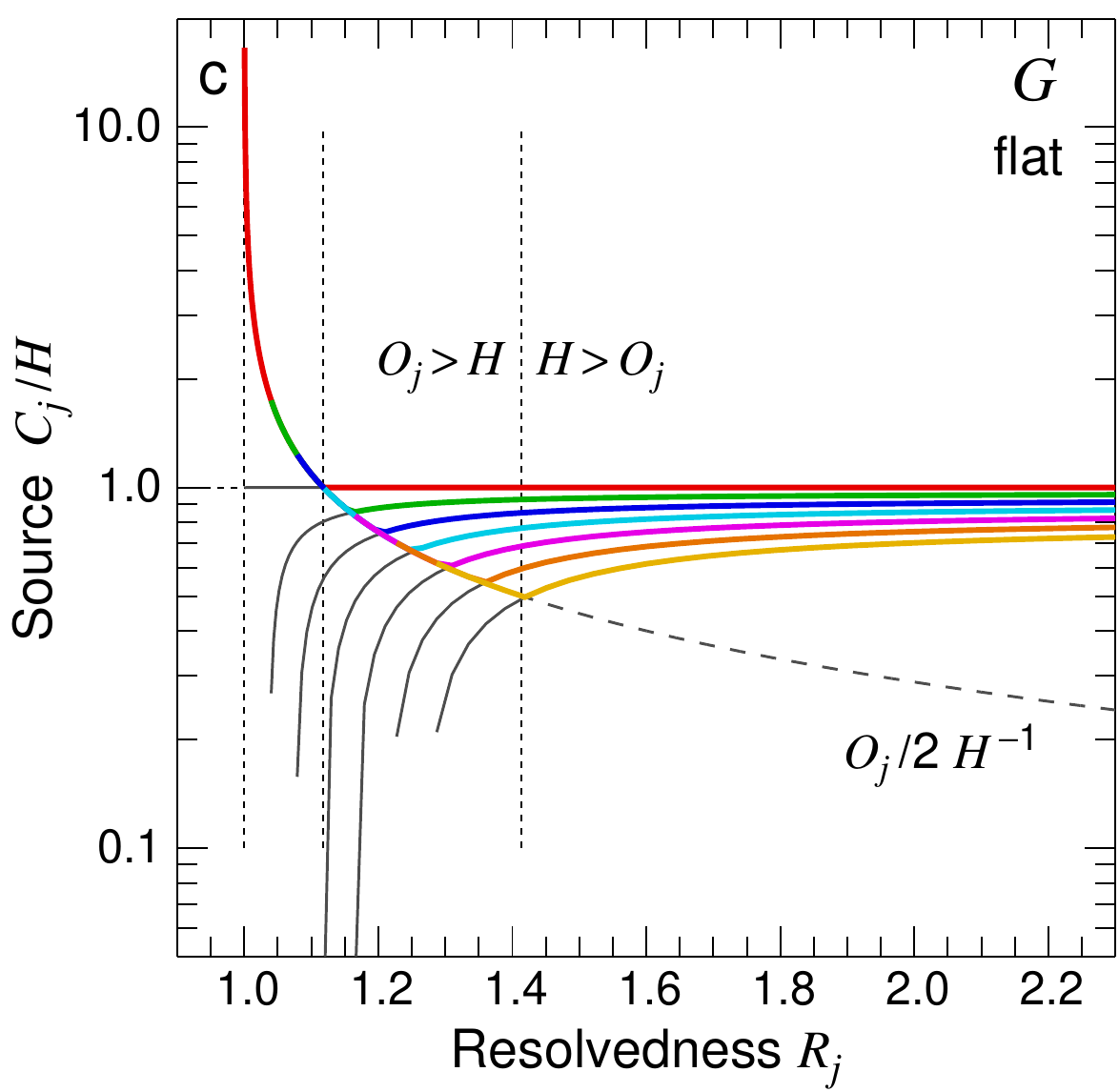}}    \hspace{-1.8mm}
  \resizebox{0.2755\hsize}{!}{\includegraphics{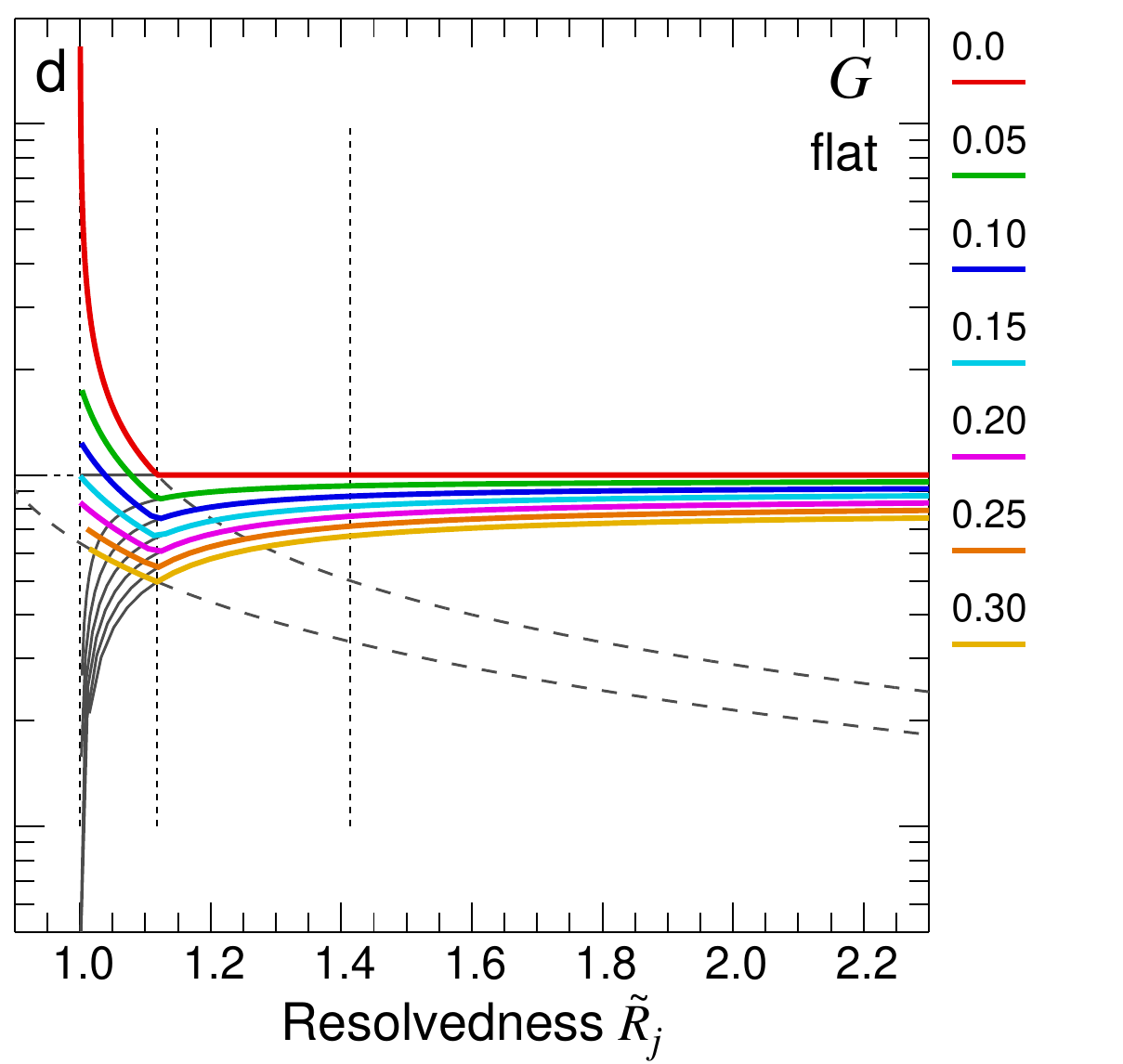}}      \hspace{-1.8mm}
  \resizebox{0.2912\hsize}{!}{\includegraphics{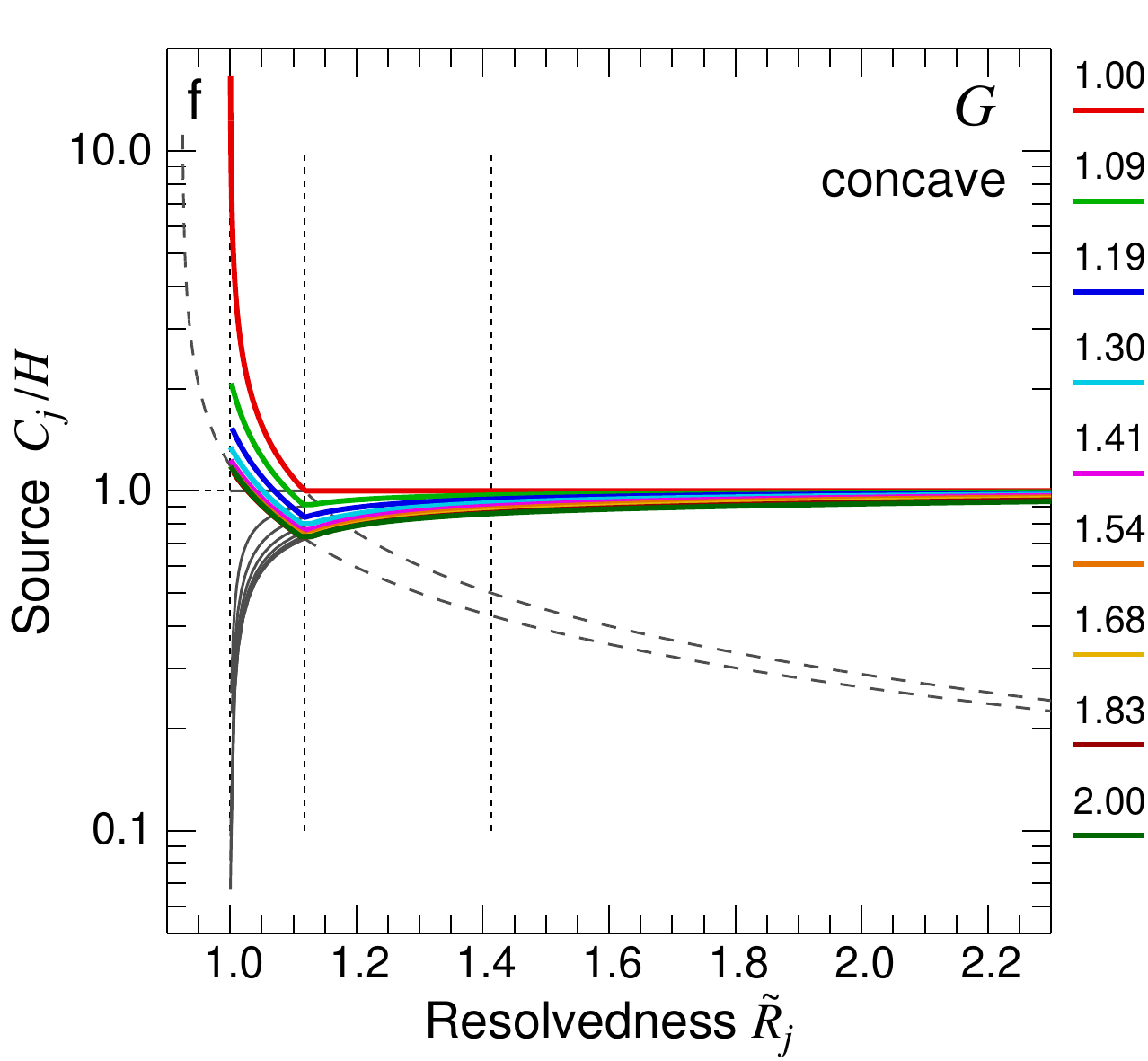}}}
\caption
{ 
Deconvolution accuracy of the half maximum sizes $\tilde{H}_{j}$ for the Gaussian sources $\mathcal{\tilde{S}}_{{\mathcal{G}}{jl}}$
(Eq.~(\ref{bgsubtraction})), separated from the flat background $\mathcal{B}$, for background over-subtraction levels $0 \le
\epsilon_{\,l} \le 0.3$ (\emph{left}) and from the convex and concave backgrounds $\mathcal{B}_{{\mathcal{G}}{jk}{\pm}}$, for
background size factors $1 \le f_{k} \le 2$ (\emph{right}). The ratios of the deconvolved sizes $D_{j}$ and $C_{j}$ to the true
model size $H$ (Table~\ref{modeltable}) are plotted as functions of the model and source resolvedness $R_{j}$ and $\tilde{R}_{j}$.
For reference, the thin black curves display $D_{j}/H$ from panels \emph{a} and \emph{b} and the dashed curves visualize
$(O_{j}/2)/H$ for $\epsilon_{\,l} = \{0, 0.3\}$. The dashed vertical lines divide the horizontal axis into the unresolved,
partially resolved, and resolved domains (Eq.~(\ref{resolstate})). All curves for $\epsilon_{\,l} > 0$ in panels \emph{b} and
\emph{d} are shifted to the left in comparison with panels \emph{a} and \emph{c}, because the resolvedness is underestimated
($\tilde{R}_{j}\!< R_{j}$). Corresponding plots for the deconvolved moment sizes $\tilde{M}_{j}$ are presented in
Fig.~\ref{decgaussflathillhellmom}.
} 
\label{decgaussflathillhell}
\end{figure*}


\subsection{Resolvedness}
\label{resolve}

To quantify the degree to which a source or filament is resolved (extended), it is useful to define the resolvedness $R_j$ of the
convolved model objects and the measured resolvedness $\tilde{R}_j$ of the background-subtracted structure,
\begin{equation} 
R_{j} = \{H|M\}_j / O_j, \,\,\, \tilde{R}_{j} = \max \left(\{\tilde{H}|\tilde{M}\}_j, O_{j}\right) / O_{j}.
\label{resolvedness}
\end{equation} 
The model resolvedness $R_{j}$ is known exclusively for the simulated models, without any background (cf. Table~\ref{resolvtable}).
Only the quantity $\tilde{R}_j$ is available from the observed images, whose value is generally affected by the errors made in the
background subtraction and size measurements. Examples of the relationships between $R_{j}$ and $\tilde{R}_{j}$ for the power-law
models are presented in Fig.~\ref{resolvednM}. It follows from the above definition, that only the point-like models whose smallest
angular dimension fits entirely within a single pixel can have $R_j = 1$. All models $\mathcal{M}$, considered in this paper,
extend over more than one pixel, therefore they have $R_j > 1$. Measured sizes $\{\tilde{H}|\tilde{M}\}_j < O_{j}$ would be a clear
signal that they are significantly underestimated, whereas overestimated sizes do not have such obvious indicators.


It makes sense to adopt the following definitions of the state of resolvedness, based on the relationship between the beam $O_{j}$
and the true object size $\{H|M\}$. When the beam is narrower than the object size, the object is considered to be resolved and
when the beam is wider than twice the object size, the object is regarded as unresolved. In the intermediate range, the structure
is deemed to be partially resolved. These definitions, used throughout this paper, can be summarized as
\begin{eqnarray} 
\left.\begin{aligned}
      \{R|\tilde{R}\}_j & >  1.4 \,\,\,(\mathrm{resolved{:}} \,\,O_{j} < \{H|M\}), \\
1.1 < \{R|\tilde{R}\}_j &\le 1.4 \,\,\,(\mathrm{partially{:}} \,\,\{H|M\} \le O_{j} < 2\{H|M\}), \\
1 \le \{R|\tilde{R}\}_j &\le 1.1 \,\,\,(\mathrm{unresolved{:}} \,\,2\{H|M\} \le O_{j}),
\end{aligned}\right.
\label{resolstate} 
\end{eqnarray} 
where the limits of $(5/4)^{1/2}\!\approx 1.1$ and $2^{1/2}\!\approx 1.4$ can be readily obtained from Eqs.~(\ref{resolvedness})
and (\ref{deconvolution}), assuming that deconvolution recovers the true size $\{H|M\}$. In Eq.~(\ref{modelback}), the small beams
($3.34{\arcsec}\!\leftarrow O_j$) are used to simulate the resolved structures ($R_j\!\rightarrow 3.14$ or above), whereas the
large beams ($O_j\!\rightarrow 332${\arcsec}) are used to simulate the unresolved sources ($1\!\leftarrow R_j$). Examples of the
resolvedness values for all models are given in Table~\ref{resolvtable}.
The concept of resolvedness may be ambiguous for the structures with extended power-law profiles: there exist angular resolutions
$O_{j}$, at which their interiors appear completely unresolved ($2\{H|M\} \ll O_{j}$), while their extended power-law areas are
well resolved ($O_{j} \ll 2 \Theta$).


\section{Gaussian size deconvolution}
\label{results}

The simulated models on different backgrounds in Eq.~(\ref{sources}) and the corresponding background-subtracted sources and
filaments in Eq.~(\ref{bgsubtraction}) represent wide ranges of the observable structures, from the unresolved to the well resolved
ones (Table~\ref{resolvtable}). As can be seen from Eq.~(\ref{resolvedness}) and Table~\ref{modeltable}, the maximum range of
resolvedness in this study is $1 \la R_1 \la \{H|M\}_{\,\mathcal{P}_{\rm A}} / O_{1} \approx \{3|15\}$. Assuming a Gaussian
shape for the source or filament intensity profile and telescope PSF, the deconvolved sizes can be conveniently derived from the
basic property of convolution,
\begin{equation} 
D_{j} = \left(\{\tilde{H}|\tilde{M}\}_{j}^2 - O_{j}^2\right)^{1/2}, \,\,\, \{\tilde{H}|\tilde{M}\}_{j} \ge O_{j},
\label{deconvolution}
\end{equation} 
where $D_{j}$ collectively denotes the deconvolved half maximum and moment sizes. For the round sources and straight filaments,
convolution and deconvolution produce identical results, only when both structures have Gaussian profiles with the same width (cf.
Sect.~\ref{convsrcfil}). In practice, some observational studies \citep[e.g.,][]{Pouteau_etal2022} use the modified (``corrected'')
deconvolved sizes,
\begin{equation} 
C_{j} = \max \left(D_{j}, O_{j}/2\right),
\label{maximized}
\end{equation} 
to adjust the values $D_{j}$ that happened to be exceptionally small (hence considered unrealistic) to a more acceptable range. The
deconvolved sizes $D_{j}$ or $C_{j}$ are usually interpreted in terms of the real transverse widths of the physical objects that
produce the observed emission. It can be seen from Eqs.~(\ref{resolvedness}) and (\ref{deconvolution}) that the inequalities $O_{j}
< \{H|M\}$ and $O_{j} \ge 2\{H|M\}$, used in the definitions of the resolvedness domains of Eq.~(\ref{resolstate}), correspond to
$D_{j} \le O_{j}/2$ (or $C_{j} > D_{j}$) and $D_{j} > O_{j}$, respectively. For simplicity, next sections present results for only
the half maximum sizes, whereas results for the moment sizes are described in Appendix~\ref{resultsmom}.




\begin{figure*}
\centering
\centerline{
  \resizebox{0.2675\hsize}{!}{\includegraphics{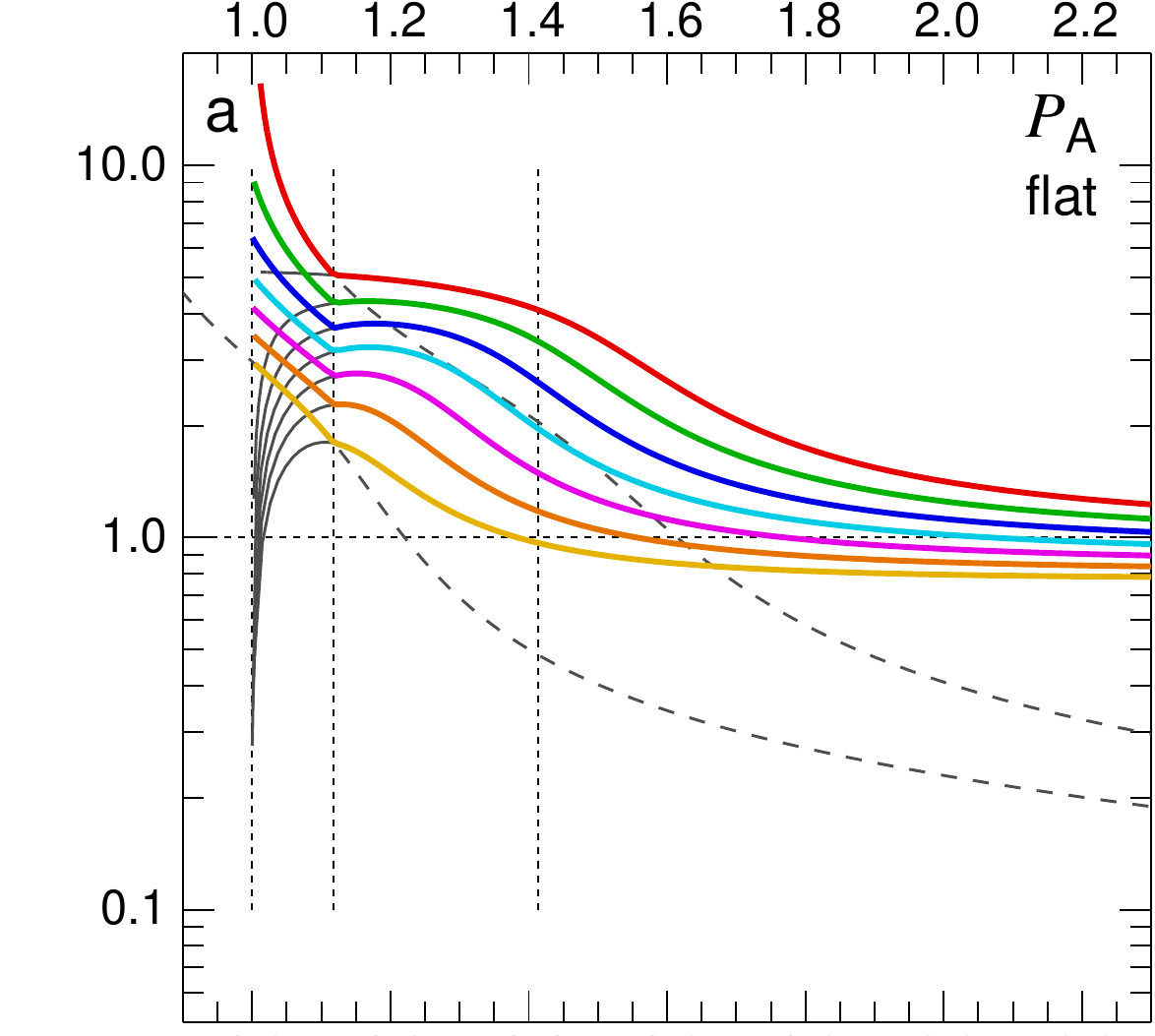}}  \hspace{-1.8mm}
  \resizebox{0.2287\hsize}{!}{\includegraphics{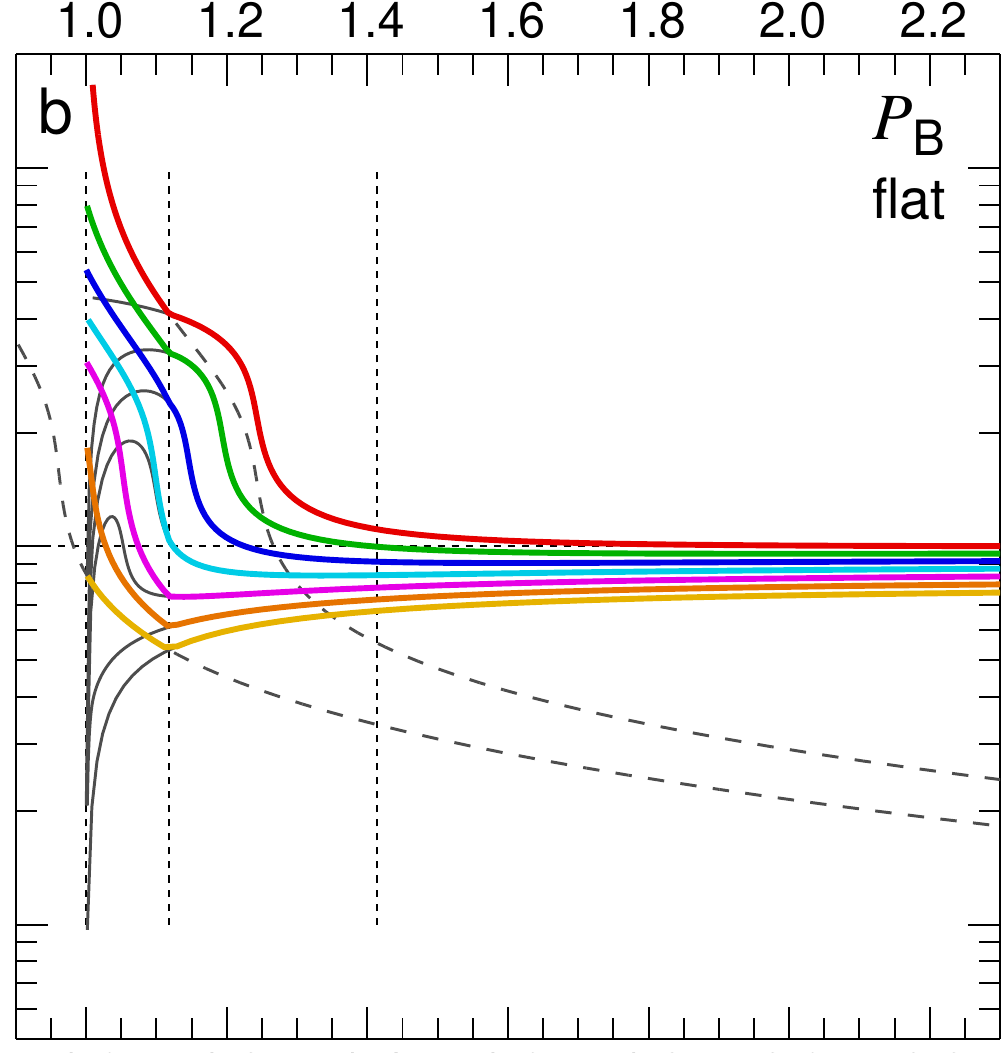}}  \hspace{-1.8mm}
  \resizebox{0.2287\hsize}{!}{\includegraphics{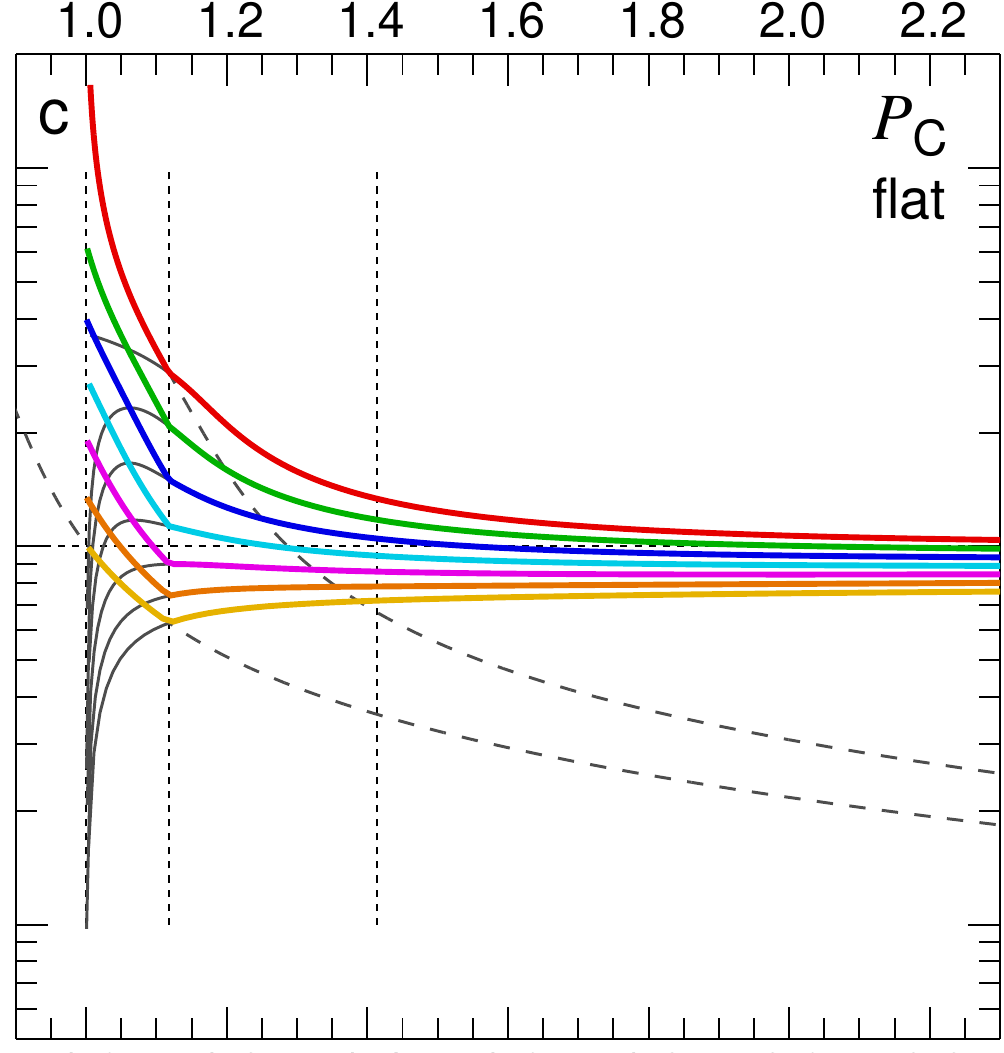}}  \hspace{-1.8mm}
  \resizebox{0.2755\hsize}{!}{\includegraphics{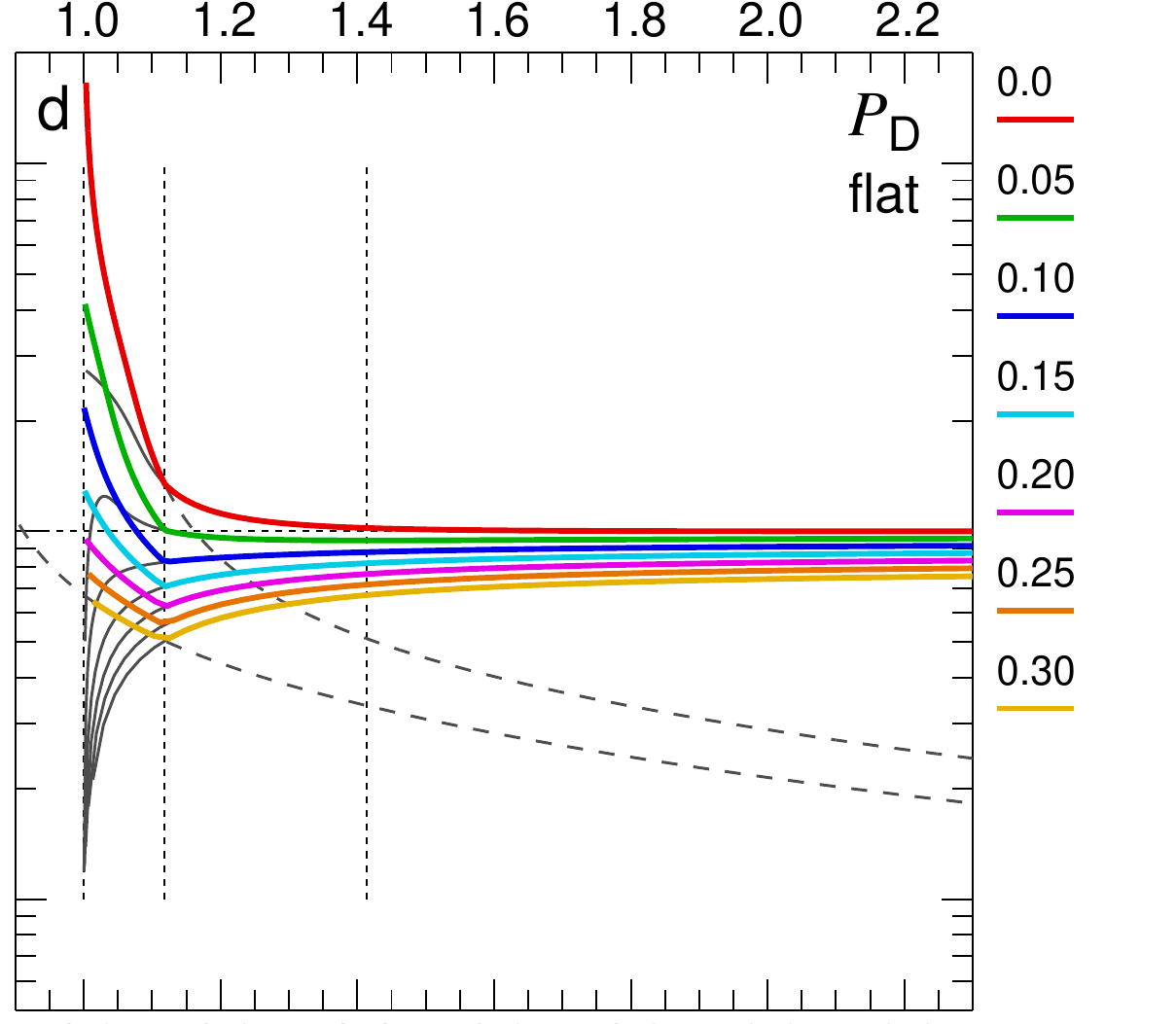}}}
\vspace{-0.94mm}
\centerline{
  \resizebox{0.2675\hsize}{!}{\includegraphics{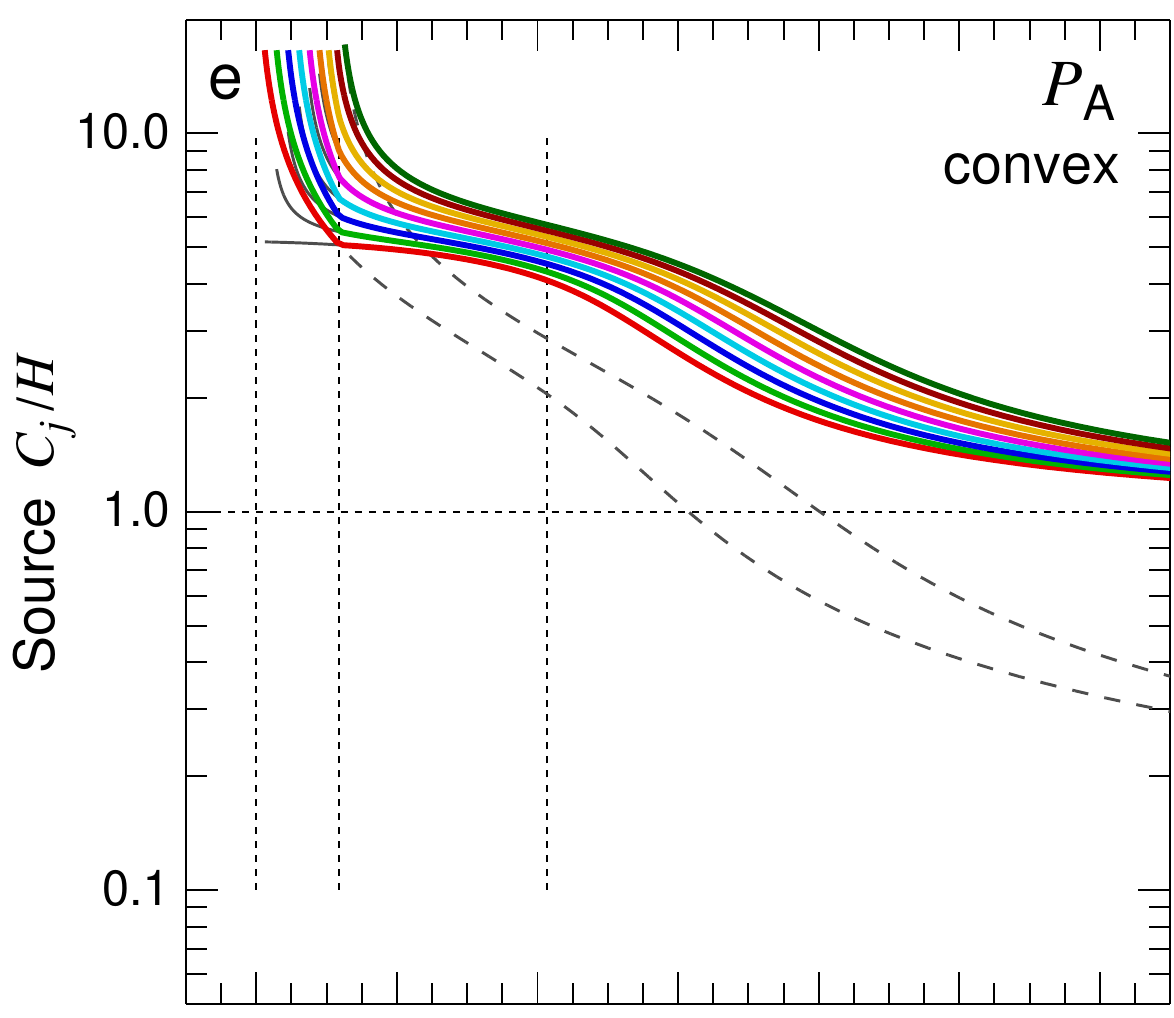}}  \hspace{-1.8mm}
  \resizebox{0.2287\hsize}{!}{\includegraphics{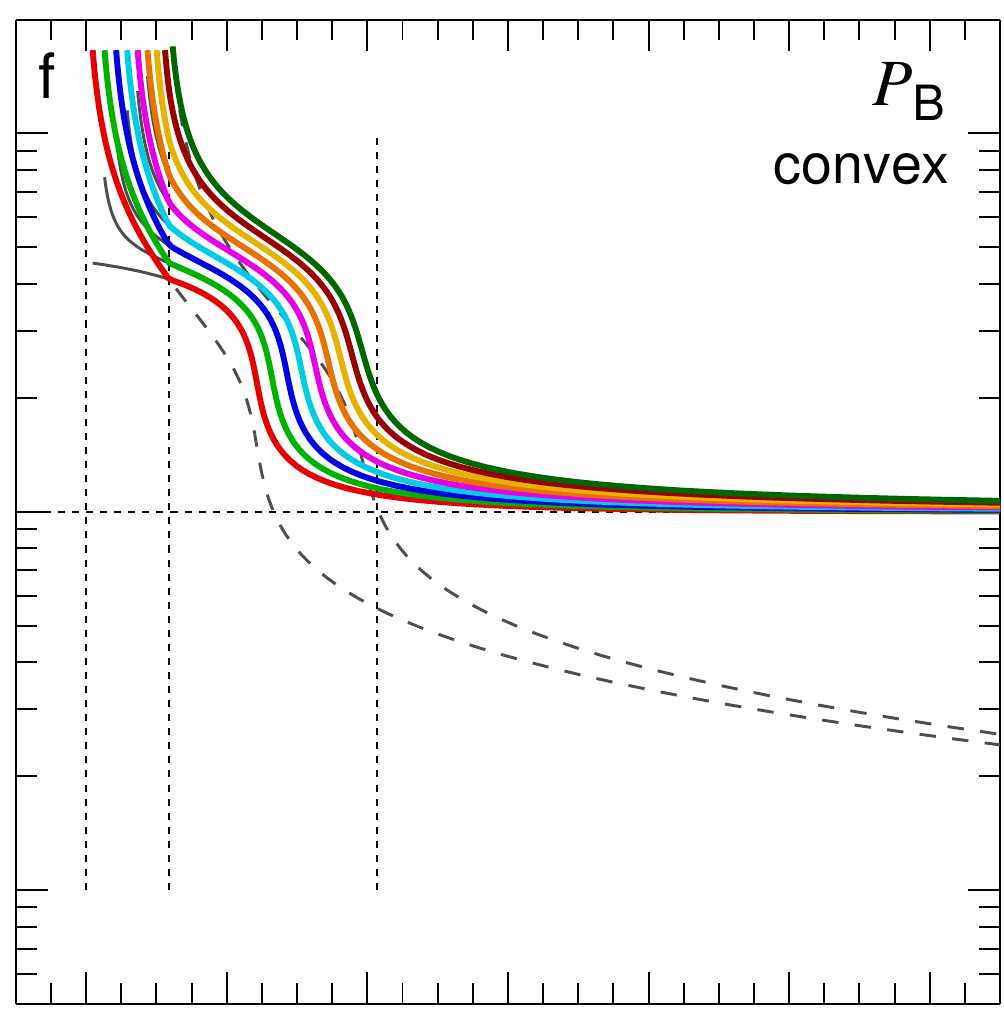}}  \hspace{-1.8mm}
  \resizebox{0.2287\hsize}{!}{\includegraphics{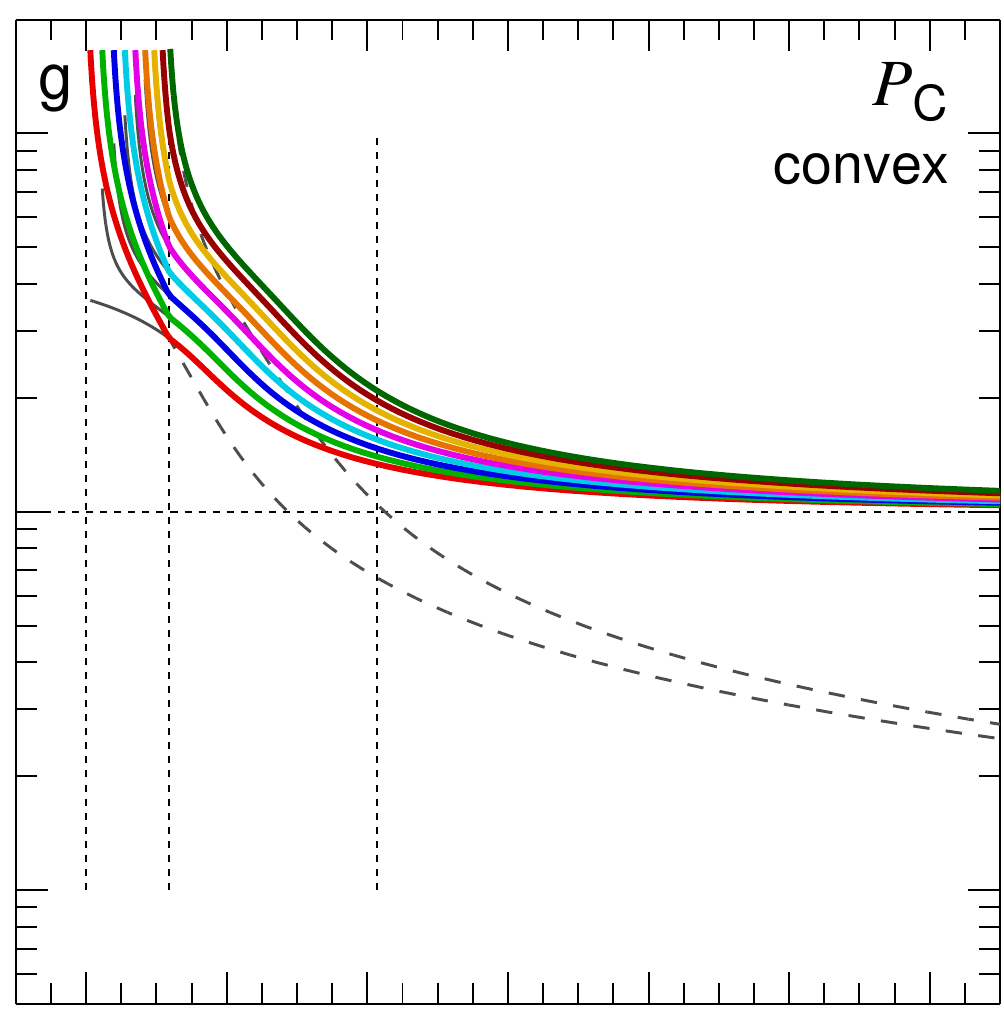}}  \hspace{-1.8mm}
  \resizebox{0.2755\hsize}{!}{\includegraphics{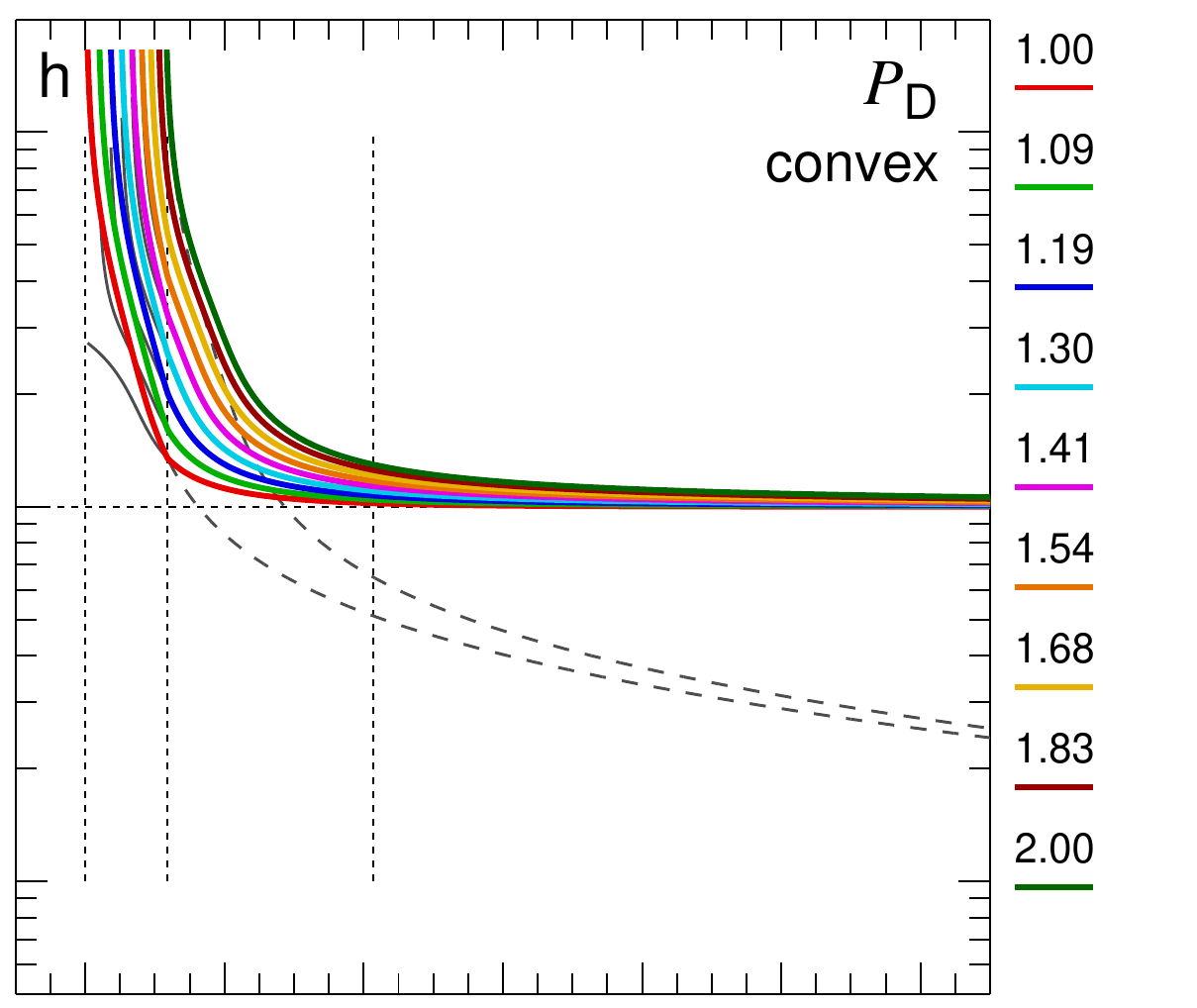}}}
\vspace{-0.94mm}
\centerline{
  \resizebox{0.2675\hsize}{!}{\includegraphics{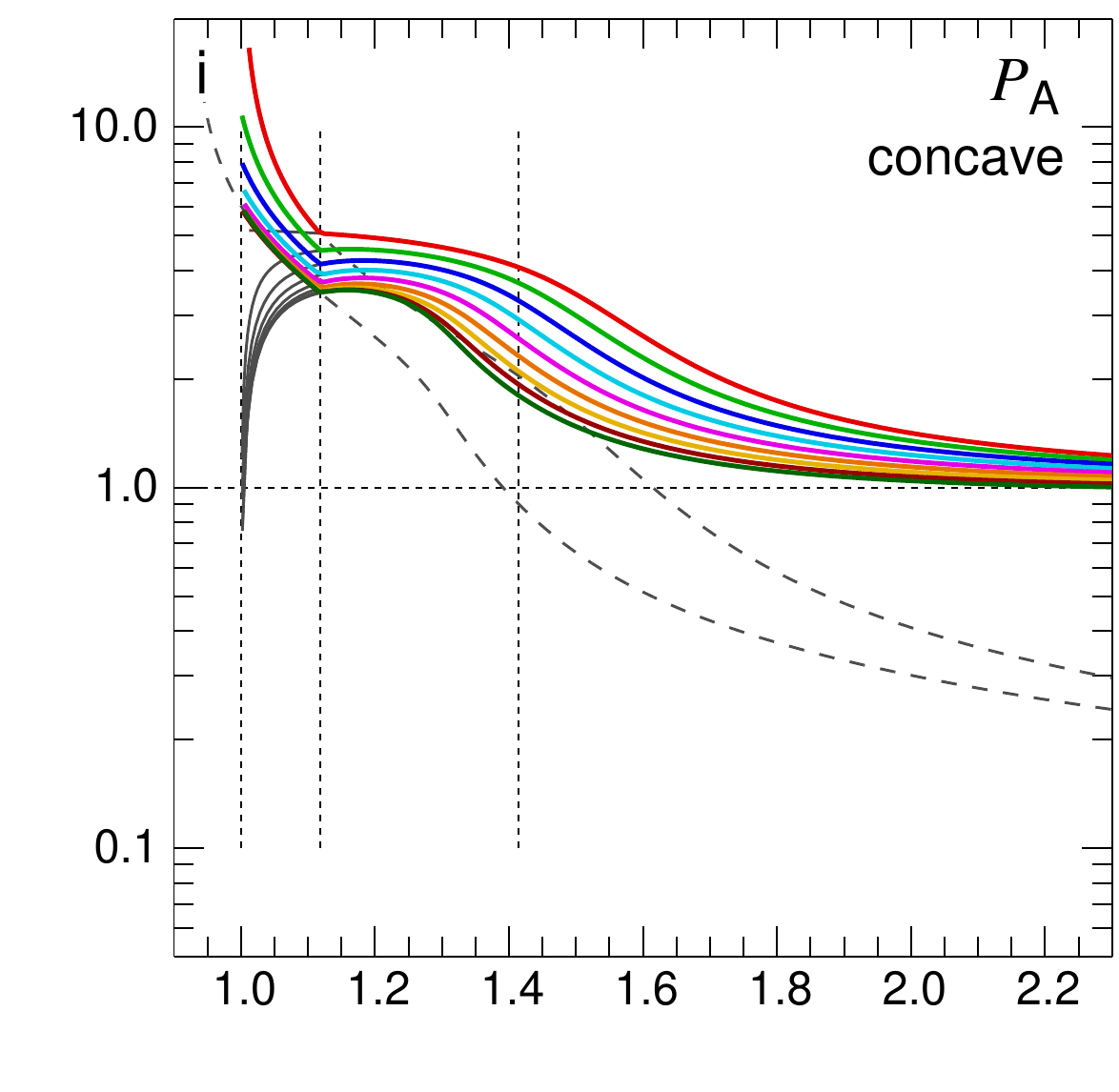}}  \hspace{-1.9mm}
  \resizebox{0.2832\hsize}{!}{\includegraphics{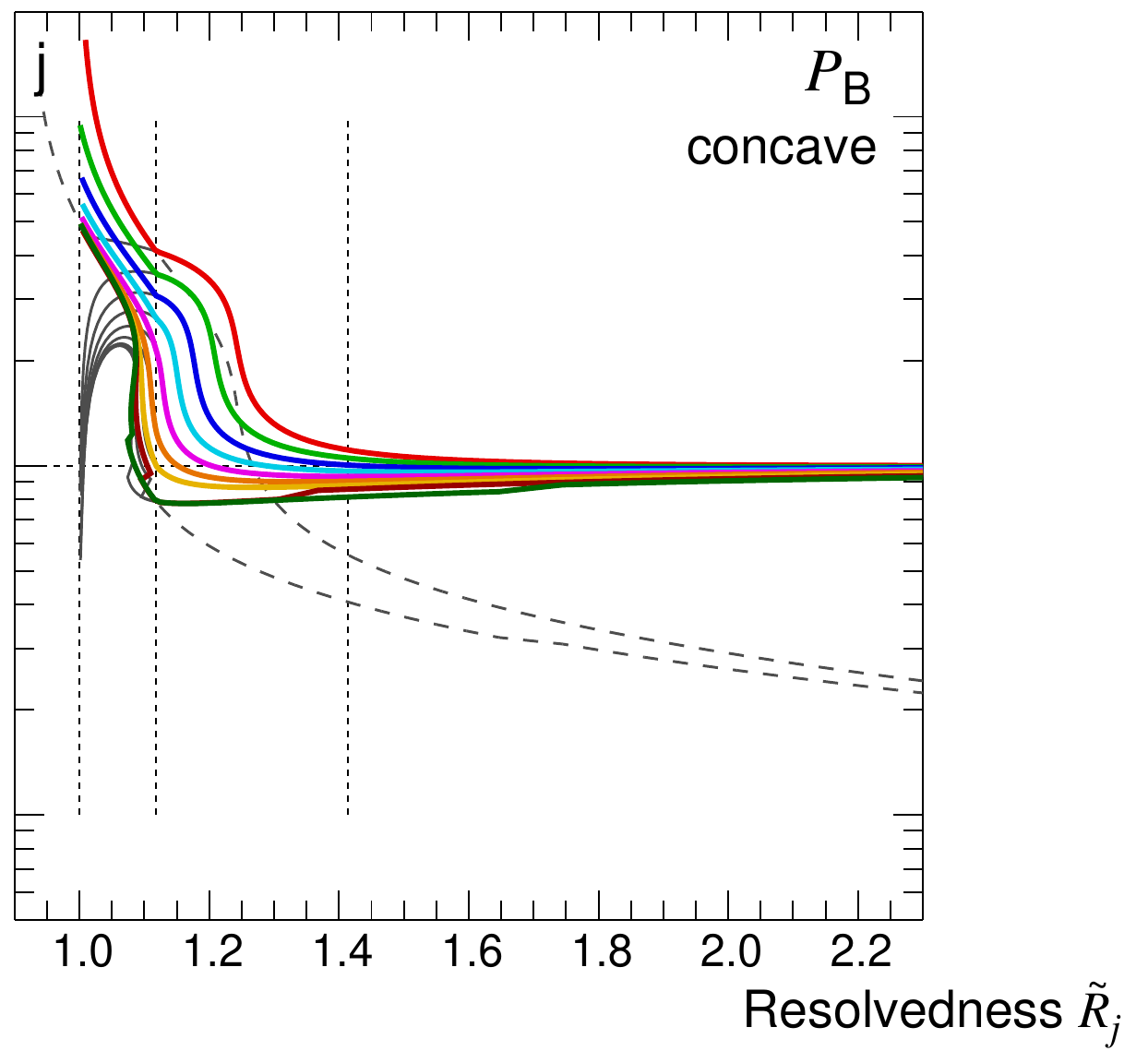}}  \hspace{-11.85mm}
  \resizebox{0.2287\hsize}{!}{\includegraphics{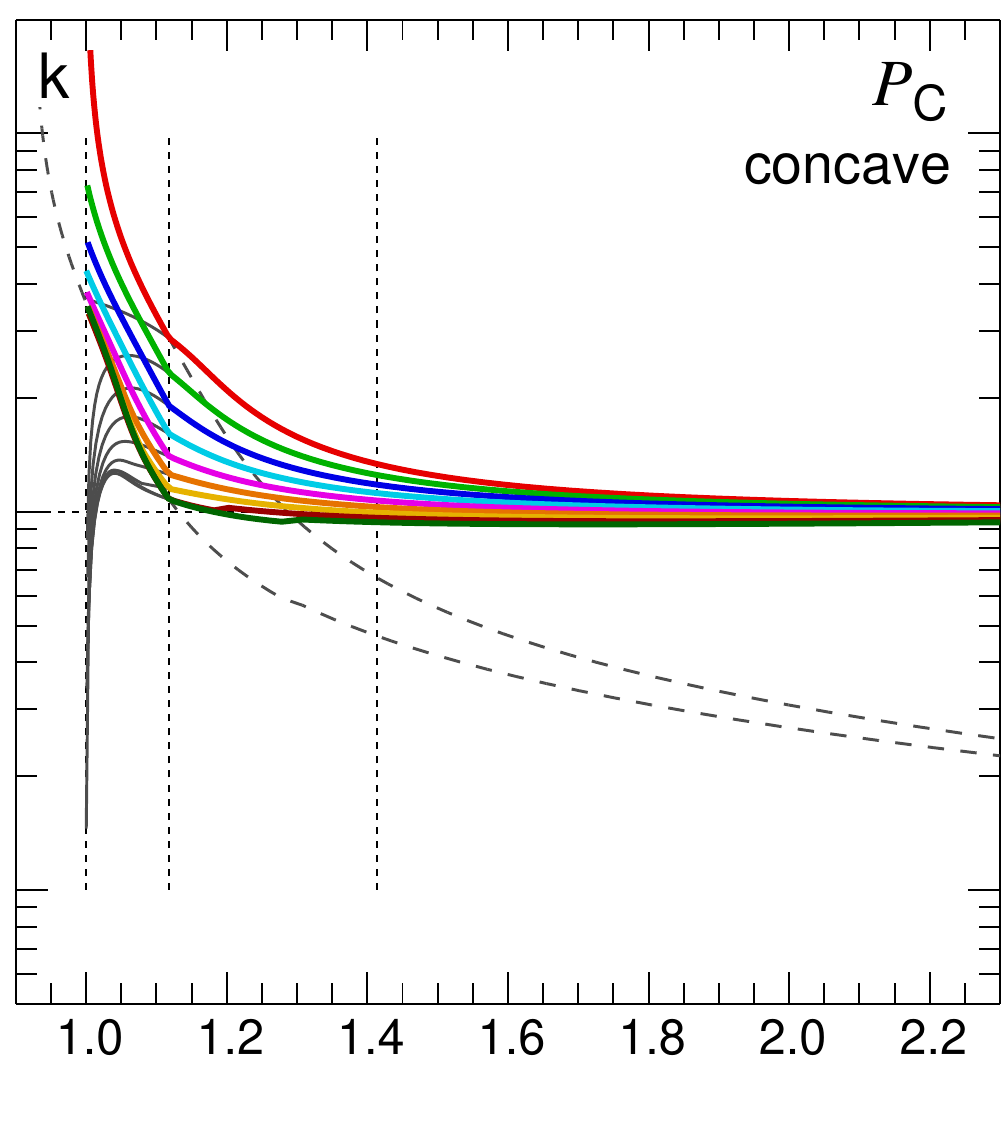}}  \hspace{-1.8mm}
  \resizebox{0.2738\hsize}{!}{\includegraphics{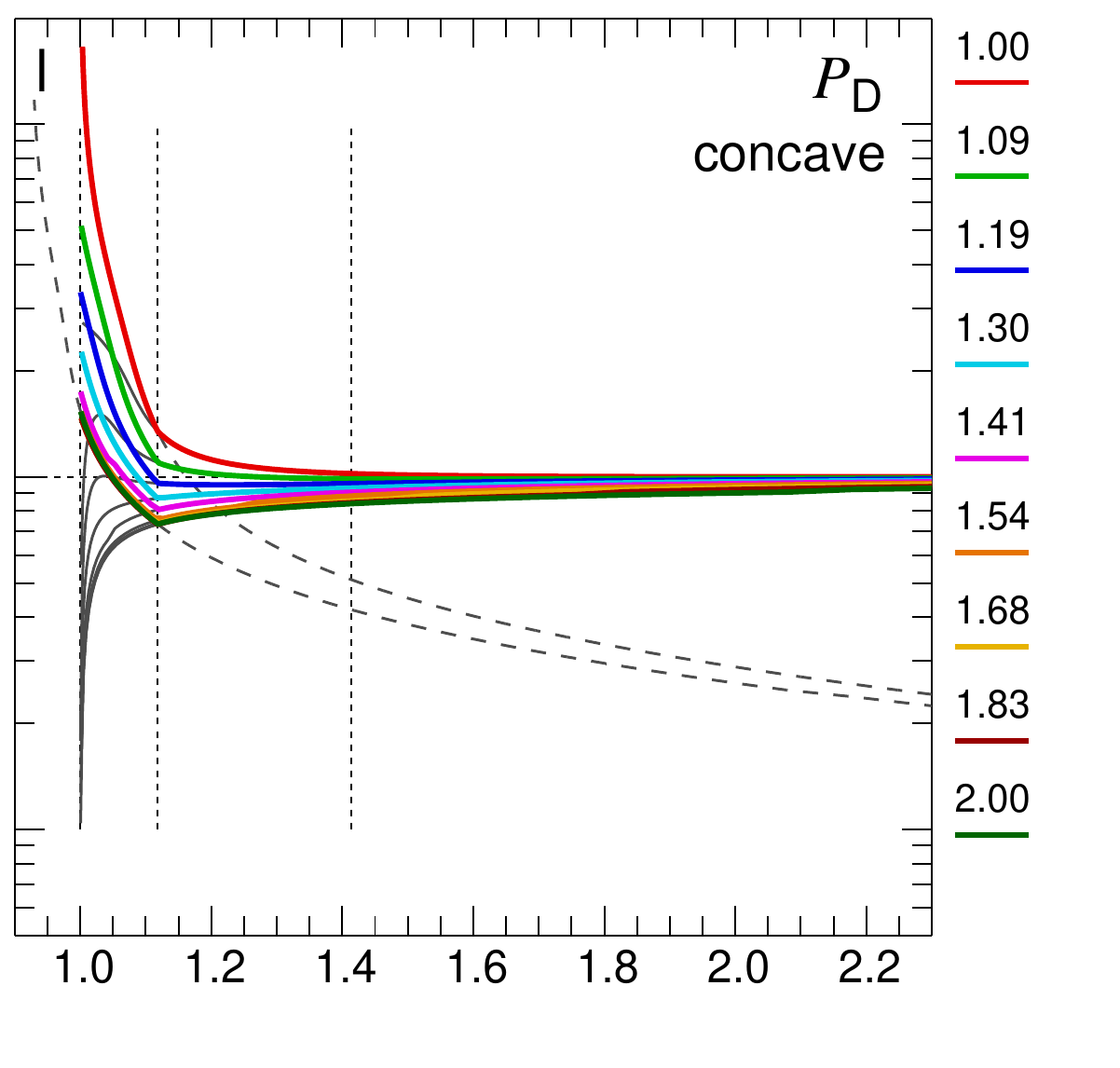}}}
\caption
{ 
Deconvolution accuracy of the half maximum sizes $\tilde{H}_{j}$ for the power-law sources
$\mathcal{\tilde{S}}_{{\mathcal{P}}{jl}}$ and $\mathcal{\tilde{S}}_{{\mathcal{P}}{jk}{\pm}}$ (Eq.~(\ref{bgsubtraction})), separated
from the flat (\emph{top}), convex (\emph{middle}), and concave (\emph{bottom}) backgrounds, for background over-subtraction levels
$0 \le \epsilon_{\,l} \le 0.3$ and size factors $1 \le f_{k} \le 2$. The ratio of the modified deconvolved sizes $C_{j}$ to the
true model size $H$ (Table~\ref{modeltable}) is plotted as a function of the source resolvedness $\tilde{R}_{j}$. For reference,
the thin black curves display $D_{j}/H$ and the dashed curves visualize $(O_{j}/2)/H$ for $\epsilon_{\,l} = \{0, 0.3\}$ and $f_{k}
= \{1, 2\}$. Corresponding plots for the deconvolved moment sizes $\tilde{M}_{j}$ are presented in Fig.~\ref{decgaussplawM}.
} 
\label{decgaussplawH}
\end{figure*}

\subsection{Spherical Gaussian models}
\label{deconvgauss}

The Gaussian sources $\mathcal{S}_\mathcal{G}$ comply with the assumptions made in Eq.~(\ref{deconvolution}). The size
deconvolution results for the background-subtracted sources $\mathcal{\tilde{S}}_{{\mathcal{G}}{jl}}$ are shown in
Fig.~\ref{decgaussflathillhell}. The simplest (flat) background $\mathcal{B}$ with $\epsilon_{\,l} = 0$ and the more complex
(convex and concave) backgrounds $\mathcal{B}_{{\mathcal{G}}{jk}{\pm}}$ with $f_{k} = 1$ do not distort Gaussian intensity
distributions, therefore they lead to accurate deconvolved sizes $D_{j}$ for all angular resolutions.

Figure~\ref{decgaussflathillhell} (\emph{a}, \emph{b}) demonstrates that the increasingly overestimated flat backgrounds
$\mathcal{B}$ ($\epsilon_{\,l} > 0$) lead to the progressively underestimated sizes $D_{j}$ that steeply drop to zero toward the
unresolved structures ($1\!\leftarrow \tilde{R}_{j}$), even for slightly overestimated backgrounds. In contrast,
Fig.~\ref{decgaussflathillhell} (\emph{c}, \emph{d}) reveals that the modified deconvolved sizes $C_{j}$ become greatly
overestimated (more than by an order of magnitude) toward the unresolved sources ($1\!\leftarrow \tilde{R}_{j}$), when the
background errors are small ($\epsilon_{\,l}\!\rightarrow 0$).

Figure~\ref{decgaussflathillhell} (\emph{e}) shows that the convex backgrounds $\mathcal{B}_{{\mathcal{G}}{jk}+}$ that happen to be
wider than the sources ($f_{k} > 1$) give rise to the deconvolved sizes $\{D|C\}_{j}$ that become steeply overestimated at small
$\tilde{R}_{j}$, even for the convex backgrounds that are just slightly wider than the sources. With the size factors
$f_{k}\!\rightarrow 2$, the overestimation becomes enormously biased toward the unresolved sources, where the $\{D|C\}_{j}/H$
curves get extremely steep on their left ends.

Figure~\ref{decgaussflathillhell} (\emph{f}) reveals that the concave backgrounds $\mathcal{B}_{{\mathcal{G}}{jk}-}$ that happen to
be wider than the sources ($f_{k} > 1$) result in the deconvolved sizes $D_{j}$ that are substantially underestimated, steeply
dropping to zero at $1\!\leftarrow \tilde{R}_{j}$, whereas the modified sizes $C_{j}$ tend to largely compensate that deficiency at
$\tilde{R}_{j} \la 1.1$. However, the correction creates a bias toward overestimated sizes of the unresolved sources, while not
correcting the partially resolved ones. The bias becomes strong for the backgrounds that are almost as wide as the sources ($1 <
f_{k} \la 1.1$). When the sources are not distorted by background subtraction ($f_{k} = 1$), the sizes $C_{j}$ turn out to be
intolerably overestimated. For the wider backgrounds, the lower limit on $D_{j}$ in Eq.~(\ref{maximized}) combines with the
increasingly underestimated $\tilde{H}_{j}$ to make $C_{j}$ have errors within $30$\,--\,$40{\%}$. Such errors might be deemed
acceptable for some applications, in comparison with the overestimations in excess of an order of magnitude. In the real
extractions, however, backgrounds are completely unknown and it is impossible to predict or control the actual magnitudes of the
deconvolution errors.

\subsection{Spherical power-law models}
\label{deconvplaws}

The power-law models $\mathcal{S}_\mathcal{P}$ violate the assumptions of Eq.~(\ref{deconvolution}), therefore substantial errors
in size deconvolution are expected. The deconvolution results for the background-subtracted power-law sources are displayed in
Fig.~\ref{decgaussplawH}. Even the simplest background $\mathcal{B}$ with $\epsilon_{\,l} = 0$ leads to distortions of the
background-subtracted source shapes. The convex and concave backgrounds $\mathcal{B}_{{\mathcal{P}}{jk}{\pm}}$, blended with the
intrinsic source shapes, alter their intensity distributions, making the background-subtracted sources wider or narrower
(Fig.~\ref{sourcesbgs}).

Larger convolution beams, used to simulate the unresolved sources ($1\!\leftarrow \tilde{R}_{j}$), cause greater overestimations of
$D_{j}$ (Fig.~\ref{decgaussplawH}), because they incorporate more significant contributions of the power-law shapes into the source
peak within its half maximum radius. Strongly overestimated deconvolved sizes $D_{j}$ are obtained toward the unresolved sources,
identical for all three types of backgrounds, when the background-subtracted source shapes are not distorted ($\epsilon_{\,l} = 0$,
$f_{k} = 1$). For the flat and concave backgrounds with $\epsilon_{\,l} > 0$ and $f_k > 1$, the deconvolved sizes $D_{j}$ turn out
to be underestimated again in the limit $1\!\leftarrow \tilde{R}_{j}$, hence the overestimation gets localized to intermediate
ranges of $\tilde{R}_{j}$ values (Fig.~\ref{decgaussplawH}). In addition, the modified sizes $C_{j}$ become steeply biased and
overestimated at $\tilde{R}_{j} < 1.1$.

\begin{figure*}
\centering
\centerline{
  \resizebox{0.2675\hsize}{!}{\includegraphics{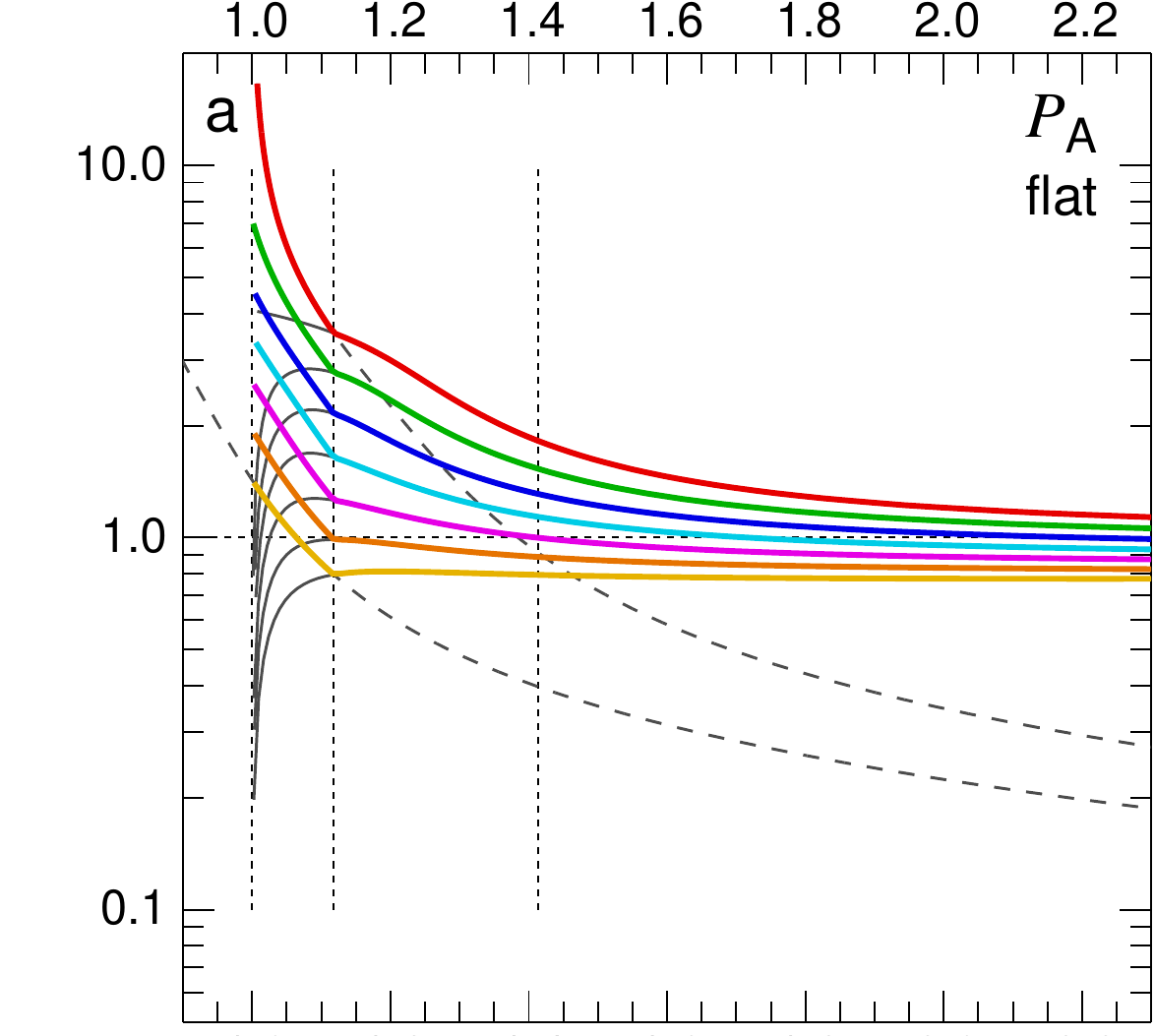}}  \hspace{-1.8mm}
  \resizebox{0.2287\hsize}{!}{\includegraphics{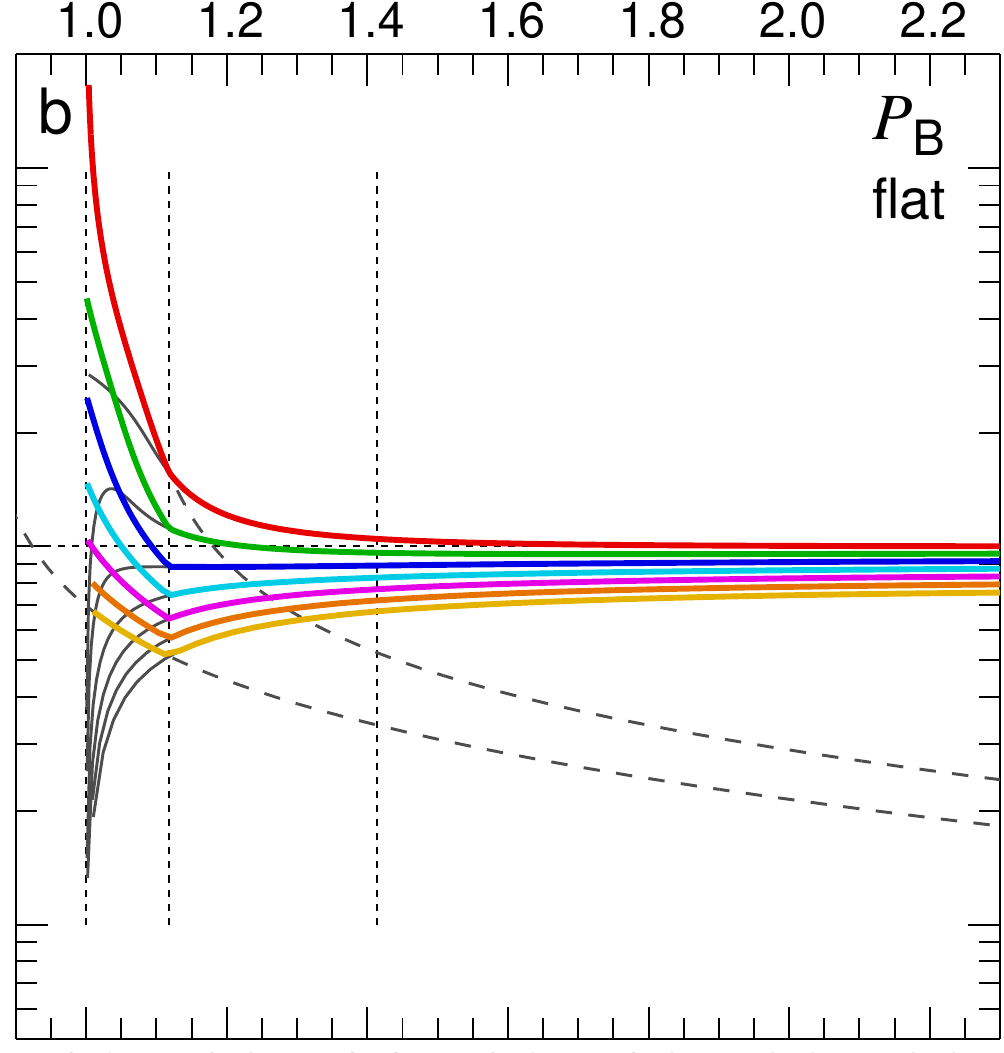}}  \hspace{-1.8mm}
  \resizebox{0.2287\hsize}{!}{\includegraphics{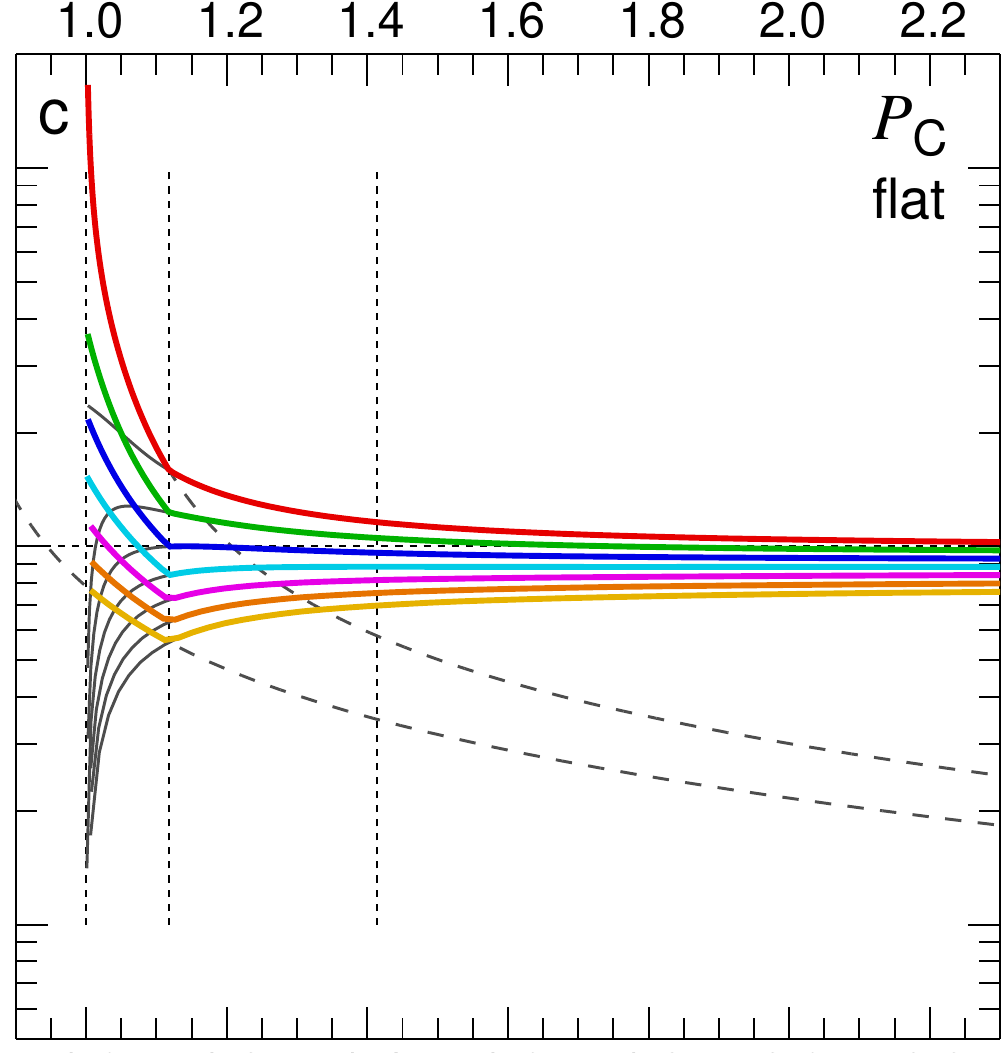}}  \hspace{-1.8mm}
  \resizebox{0.2755\hsize}{!}{\includegraphics{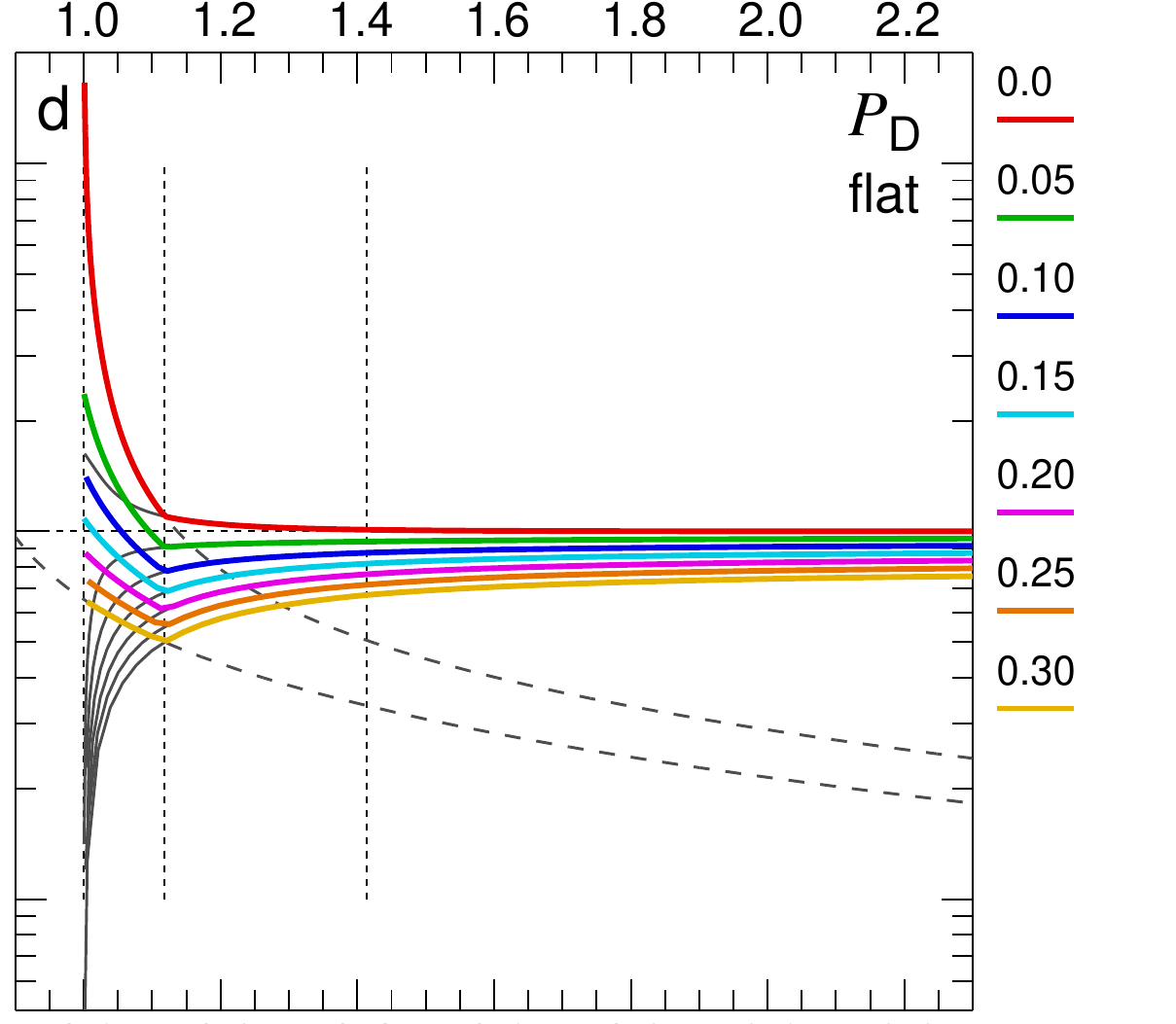}}}
\vspace{-0.94mm}
\centerline{
  \resizebox{0.2675\hsize}{!}{\includegraphics{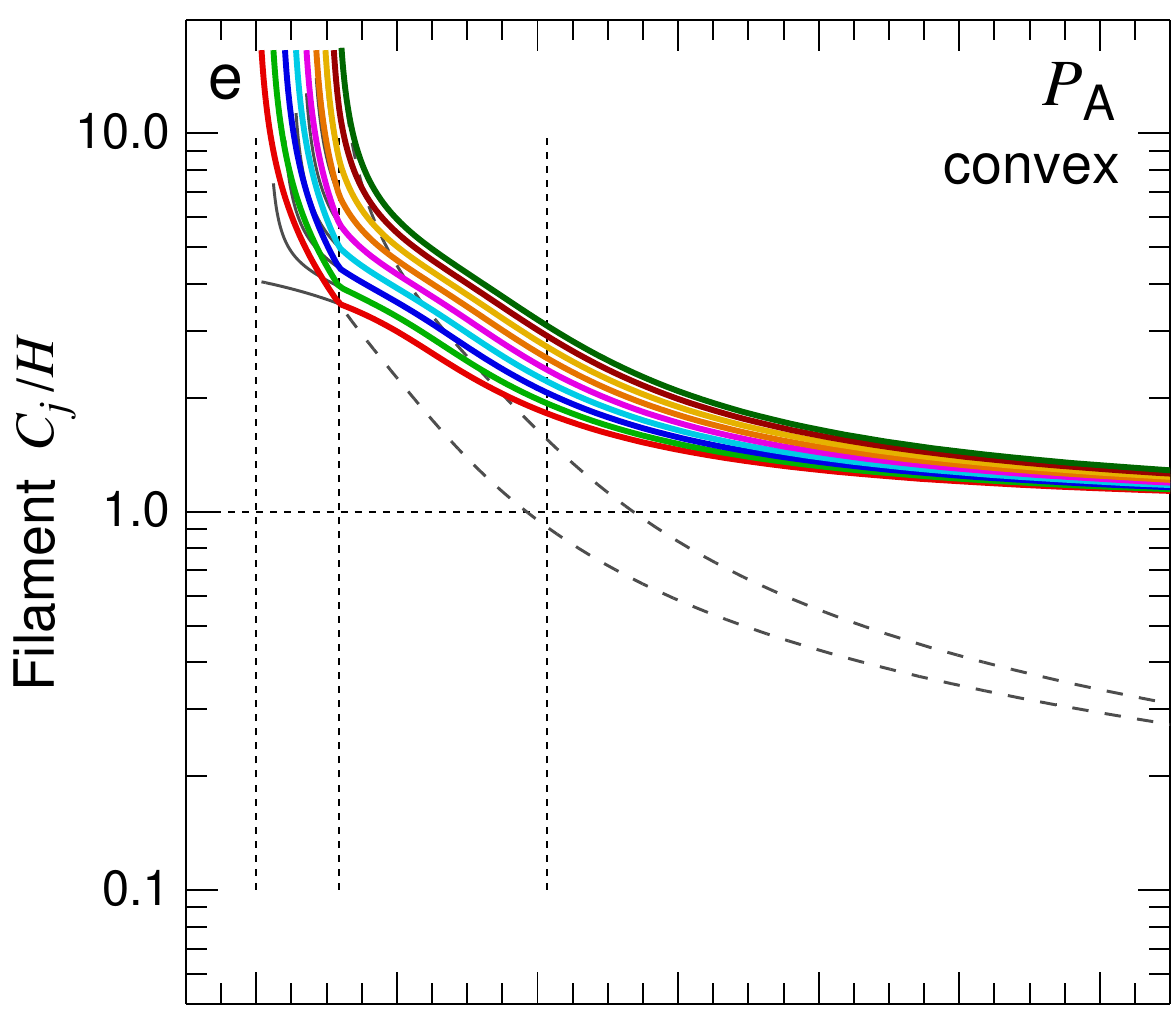}}  \hspace{-1.8mm}
  \resizebox{0.2287\hsize}{!}{\includegraphics{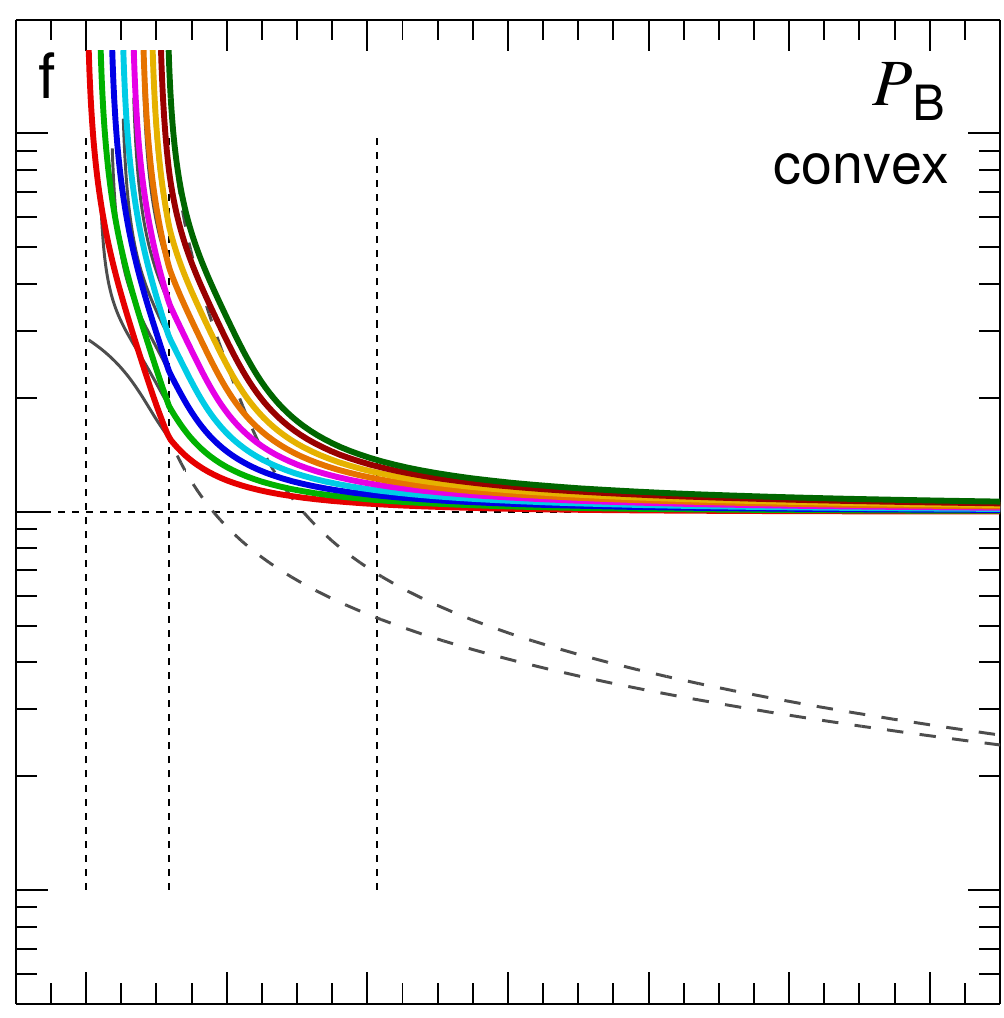}}  \hspace{-1.8mm}
  \resizebox{0.2287\hsize}{!}{\includegraphics{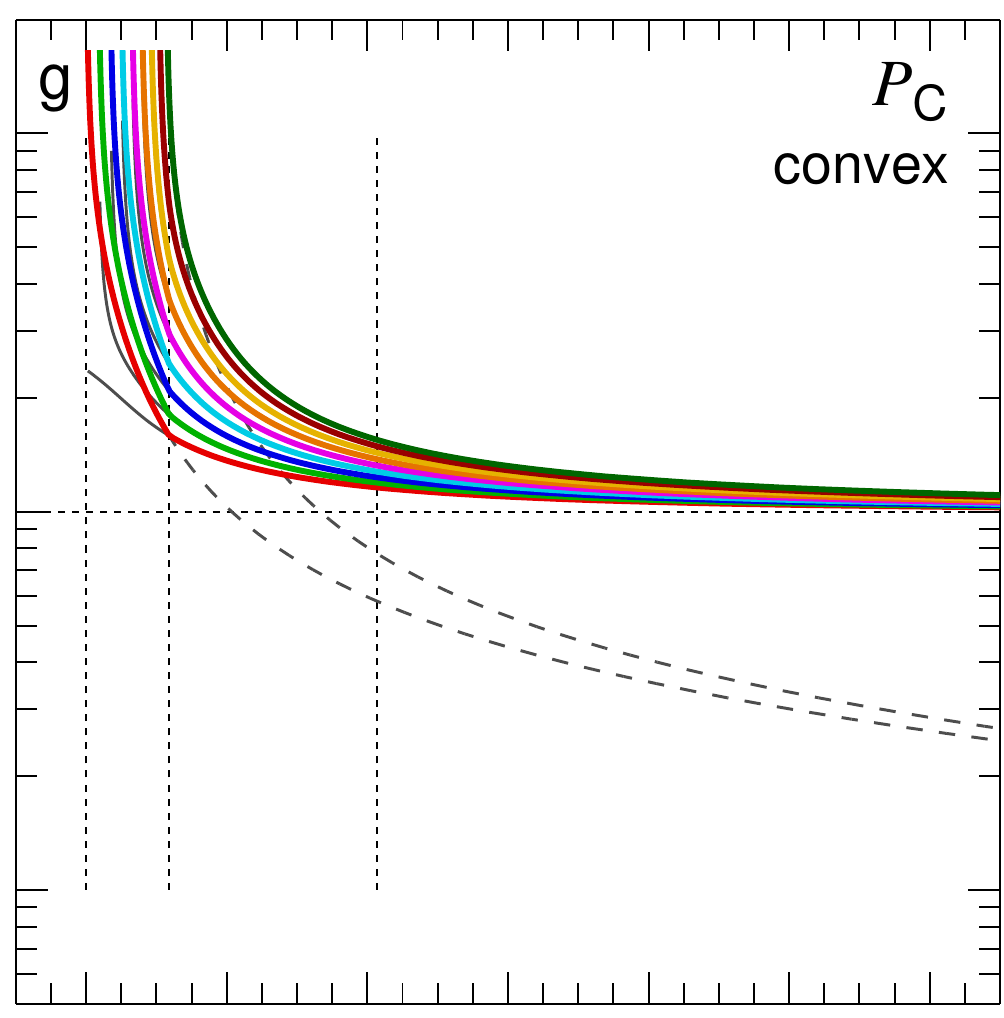}}  \hspace{-1.8mm}
  \resizebox{0.2755\hsize}{!}{\includegraphics{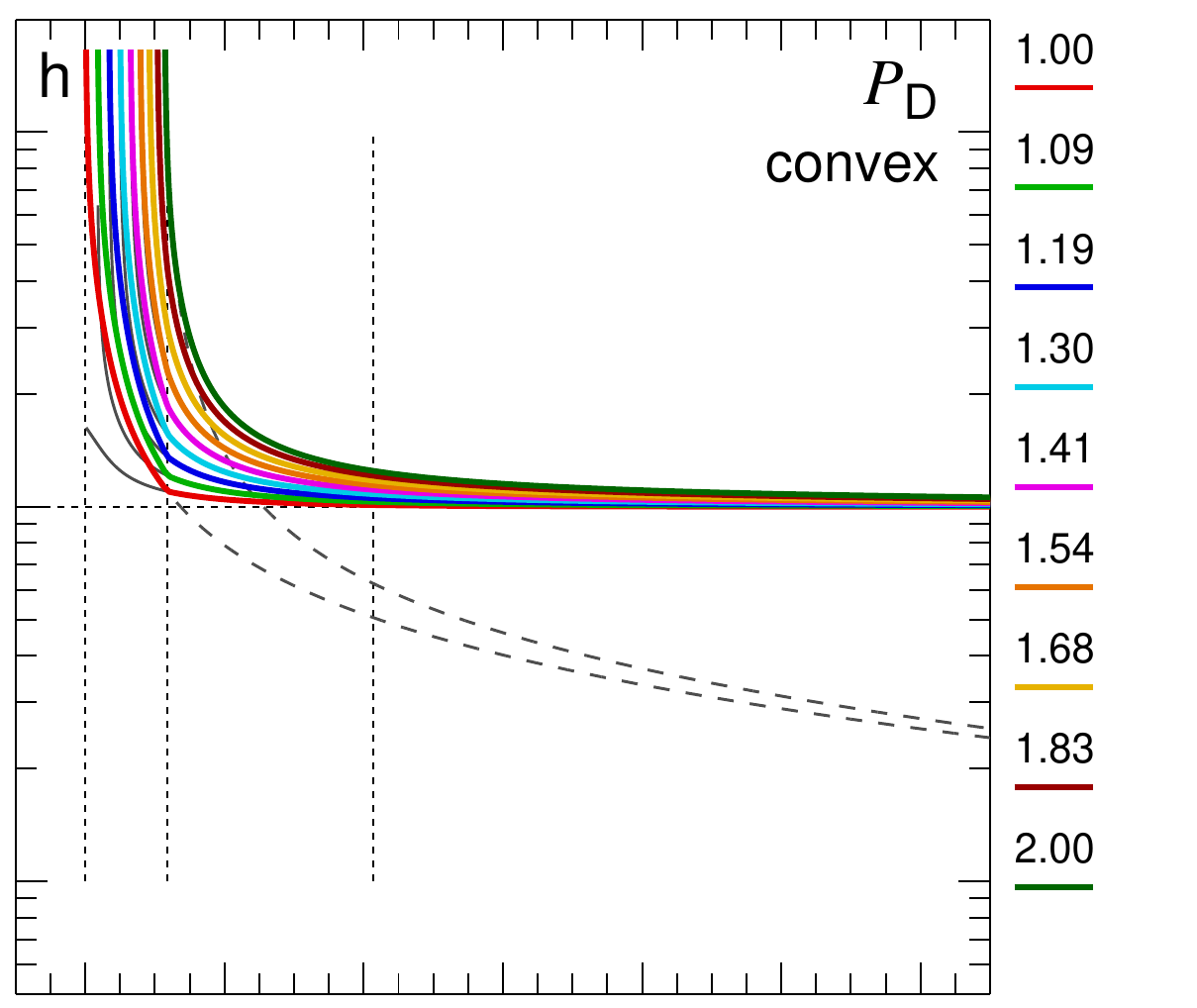}}}
\vspace{-0.94mm}
\centerline{
  \resizebox{0.2675\hsize}{!}{\includegraphics{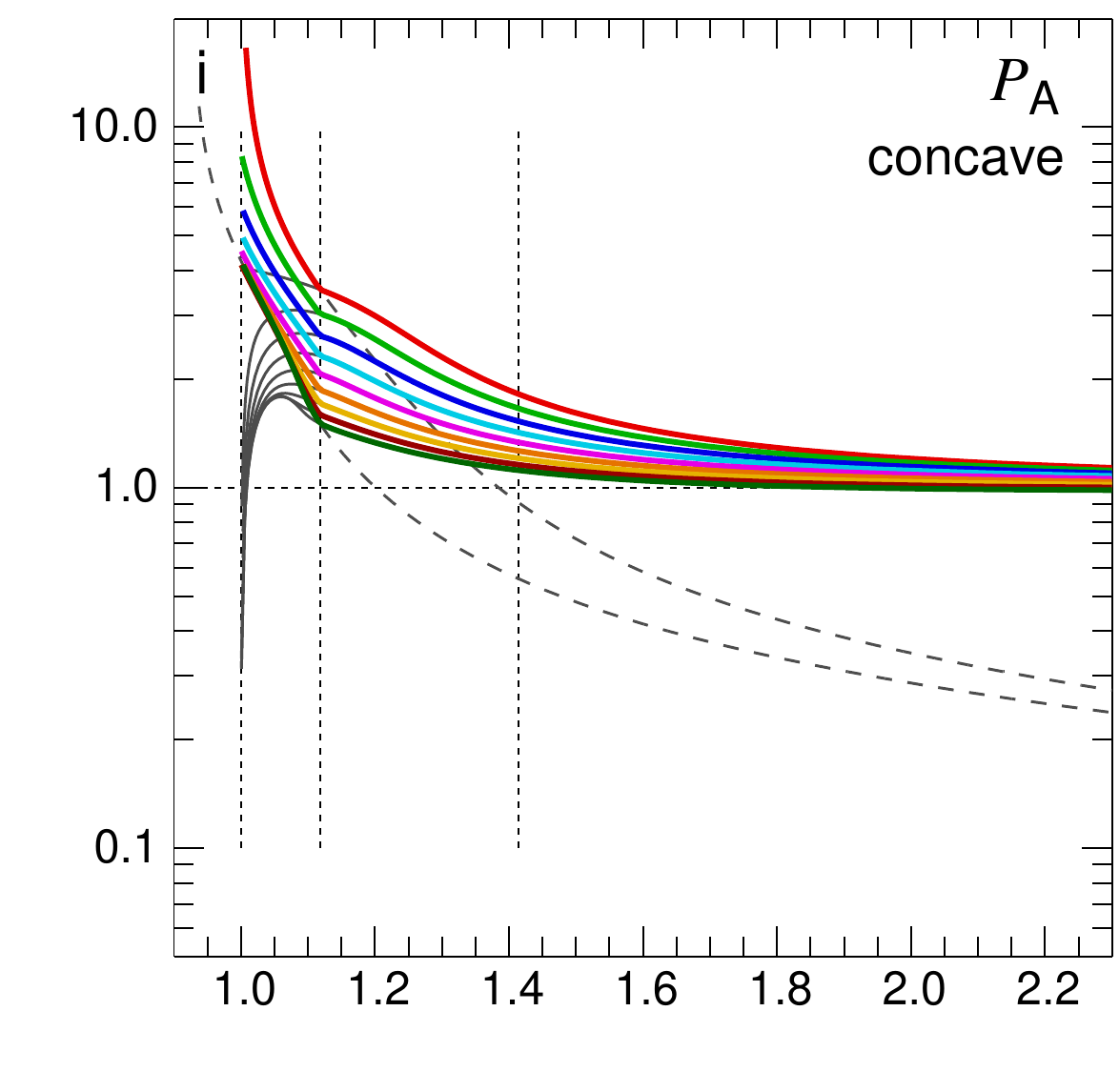}}  \hspace{-1.9mm}
  \resizebox{0.2832\hsize}{!}{\includegraphics{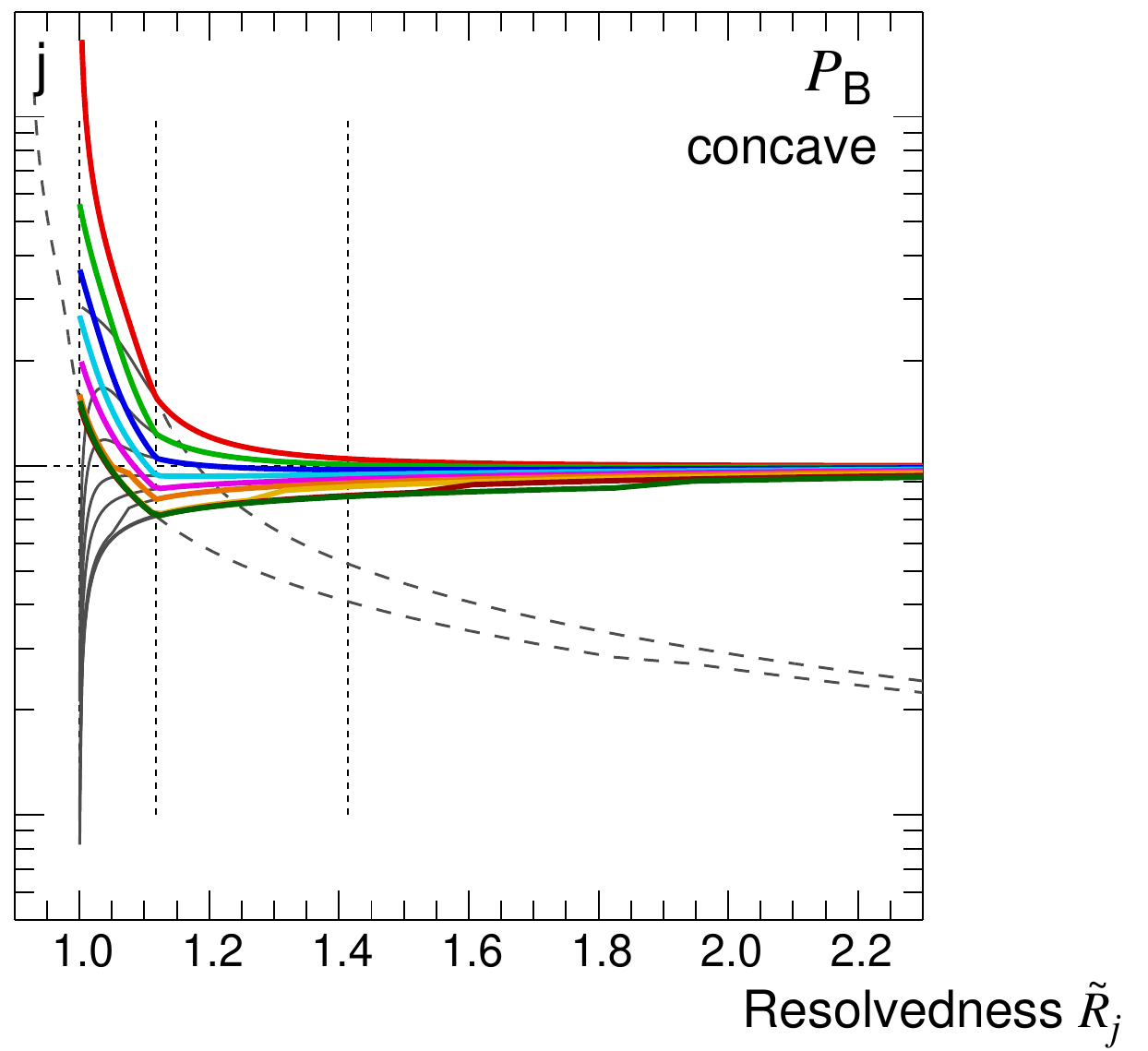}}  \hspace{-11.85mm}
  \resizebox{0.2287\hsize}{!}{\includegraphics{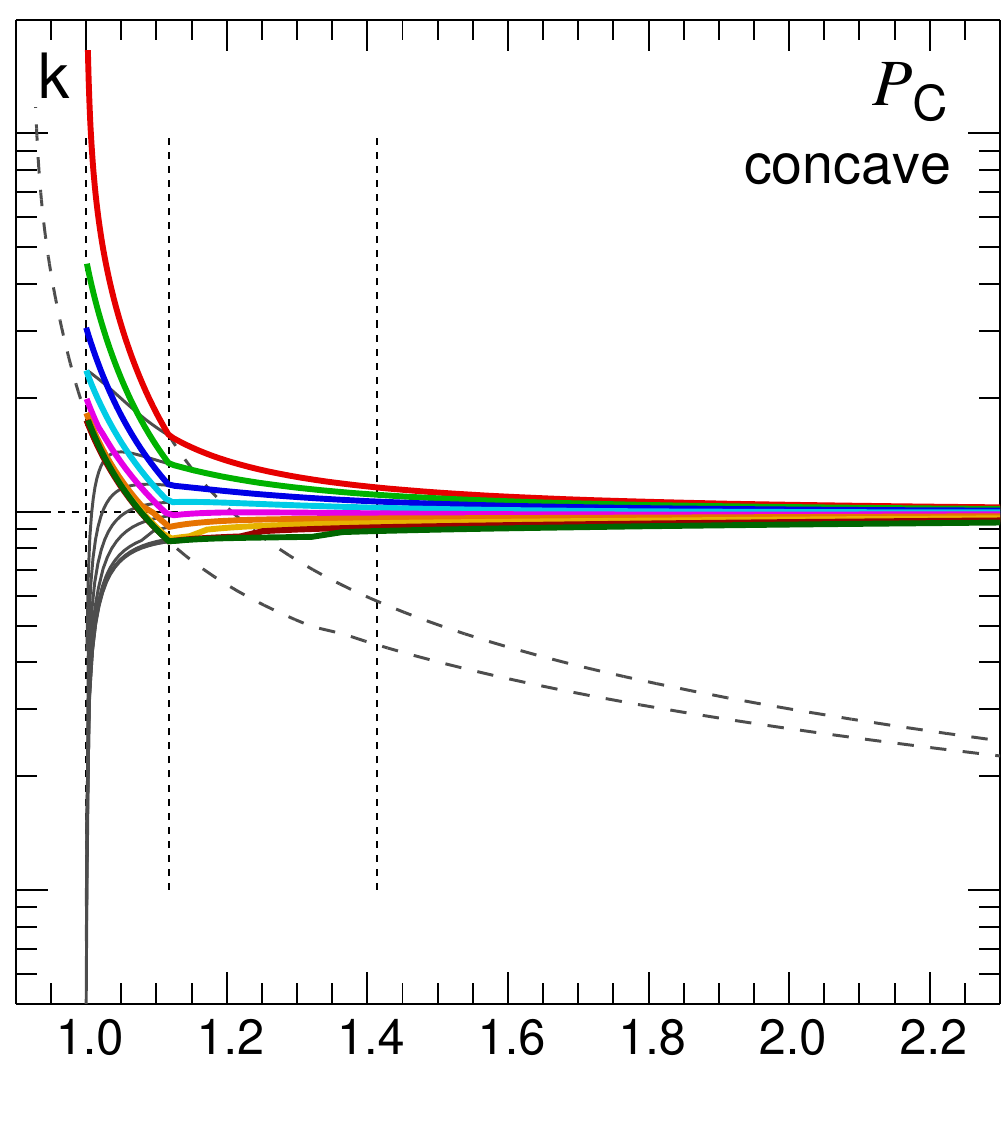}}  \hspace{-1.8mm}
  \resizebox{0.2738\hsize}{!}{\includegraphics{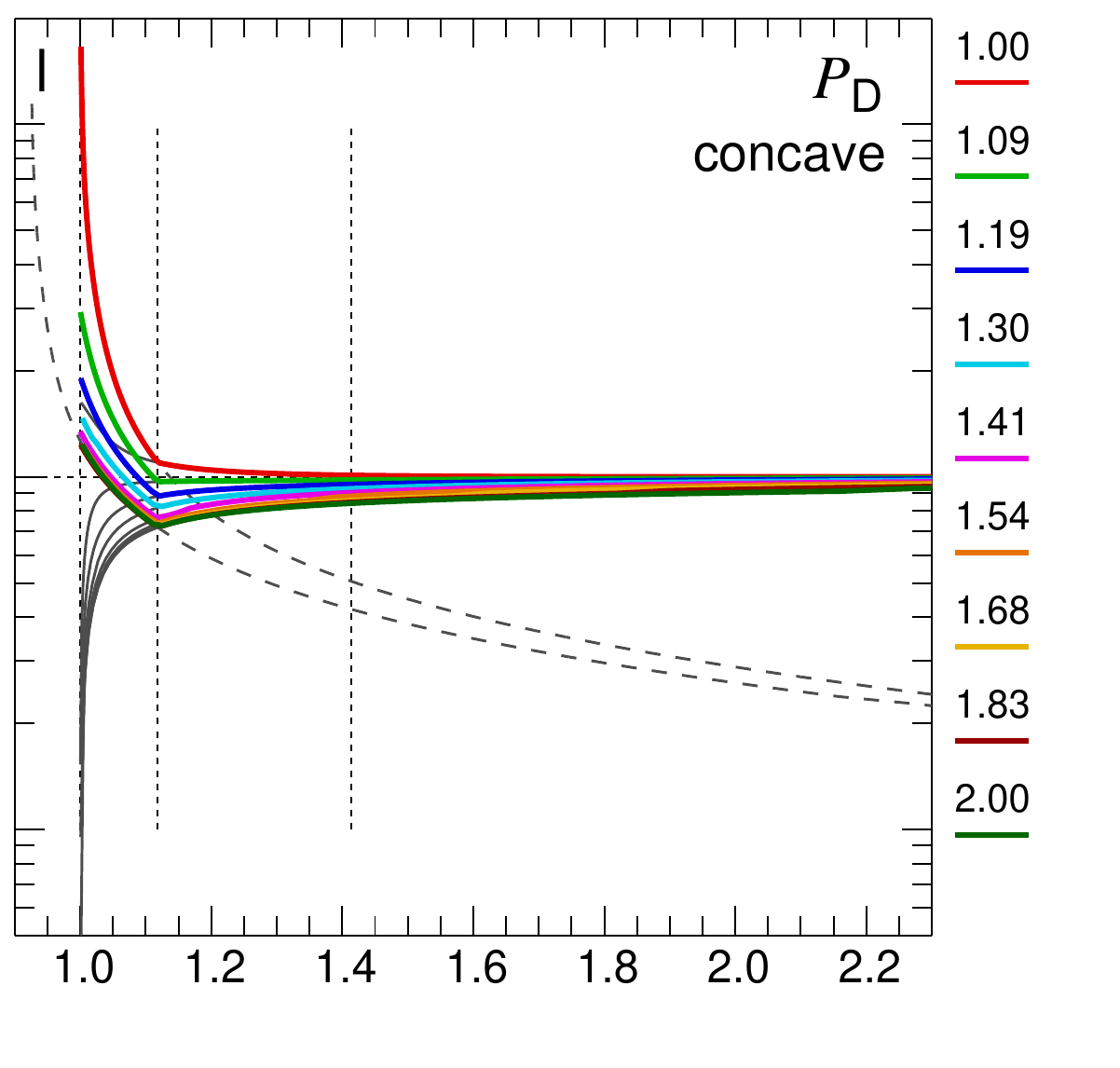}}}
\caption
{ 
Deconvolution accuracy of the half maximum sizes $\tilde{H}_{j}$ for the power-law filaments
$\mathcal{\tilde{F}}_{{\mathcal{P}}{jl}}$ and $\mathcal{\tilde{F}}_{{\mathcal{P}}{jk}{\pm}}$ (Eq.~(\ref{bgsubtraction})), separated
from the flat (\emph{top}), convex (\emph{middle}), and concave (\emph{bottom}) backgrounds, for background over-subtraction levels
$0 \le \epsilon_{\,l} \le 0.3$ and size factors $1 \le f_{k} \le 2$. The ratio of the modified deconvolved sizes $C_{j}$ to the
true model size $H$ (Table~\ref{modeltable}) is plotted as a function of the source resolvedness $\tilde{R}_{j}$. For reference,
the thin black curves display $D_{j}/H$ and the dashed curves visualize $(O_{j}/2)/H$ for $\epsilon_{\,l} = \{0, 0.3\}$ and $f_{k}
= \{1, 2\}$.
} 
\label{decgaussplawHfils}
\end{figure*}

Figure~\ref{decgaussplawH} (\emph{a}) demonstrates that model $\mathcal{P}_{\rm A}$ with the most intense power-law profile on the
flat background $\mathcal{B}$ has the largest overestimation of the deconvolved sizes $\{D|C\}_{j}$ at the lowest angular
resolutions ($1\!\leftarrow \tilde{R}_{j}$). Subtraction of the increasingly overestimated backgrounds ($\epsilon_{\,l} > 0$) leads
to the steeper profiles (Fig.~\ref{sourcesbgs}) and smaller $\{D|C\}_{j}$ values, thereby somewhat compensating the errors and
producing better accuracies. Model $\mathcal{P}_{\rm B}$, with a scaled-down power-law profile, displays a smaller overestimation
of the deconvolved sizes $\{D|C\}_{j}$, especially for the resolved structures ($\tilde{R}_{j} \ga 1.4$), because of the lesser
contribution of the fainter power-law wings to the source peak (Fig.~\ref{decgaussplawH} (\emph{b})). In models $\mathcal{P}_{\rm
\{B|C|D\}}$, with a greater overestimation of the background ($\epsilon_{\,l}\!\rightarrow 0.3$), the deconvolved sizes become
considerably underestimated (up to $40${\%}), especially at the border between the partially resolved and unresolved sources
($\tilde{R}_{j} \approx 1.1$).

Figure~\ref{decgaussplawH} (\emph{e}) demonstrates that the strongest power-law model $\mathcal{P}_{\rm A}$ on the convex
backgrounds $\mathcal{B}_{{\mathcal{P}}{jk}+}$ with $f_{k} = 1$ leads to the identical large overestimation of the deconvolved
sizes $\{D|C\}_{j}$, like for the flat backgrounds with $\epsilon = 0$. The wider backgrounds ($f_{k} > 1$) completely blend with
the sources, widening the source peak further and causing even greater overestimation of $\{D|C\}_{j}$. Models $\mathcal{P}_{\rm
\{B|C|D\}}$ with fainter power-law profiles produce results that become more resembling those for the Gaussian sources on convex
backgrounds (Fig.~\ref{decgaussflathillhell} (\emph{e})).

Figure~\ref{decgaussplawH} (\emph{i}\,--\,\emph{l}) shows that the power-law sources $\mathcal{S}_{{\mathcal{P}}{jk}-}$ on concave
backgrounds with $f_{k} = 1$ have the same large overestimation of $\{D|C\}_{j}$ as that for the flat and convex backgrounds,
because the source intensity distribution remains unchanged. The concave backgrounds that are wider than the sources ($f_{k} > 1$)
lead to less overestimated or even underestimated sizes. In effect, this slightly reduces the inaccuracy of the strongly
overestimated $\{D|C\}_{j}$. However, the errors in the modified sizes $C_{j}$ remain sharply rising toward the unresolved sources
($1\!\leftarrow \tilde{R}_{j}$). The power-law profile is much fainter in model $\mathcal{P}_{\rm D}$, hence the results resemble
those for the Gaussian sources in Fig.~\ref{decgaussflathillhell} (\emph{f}).

\begin{figure}
\centering
\centerline{
  \resizebox{0.6100\hsize}{!}{\includegraphics{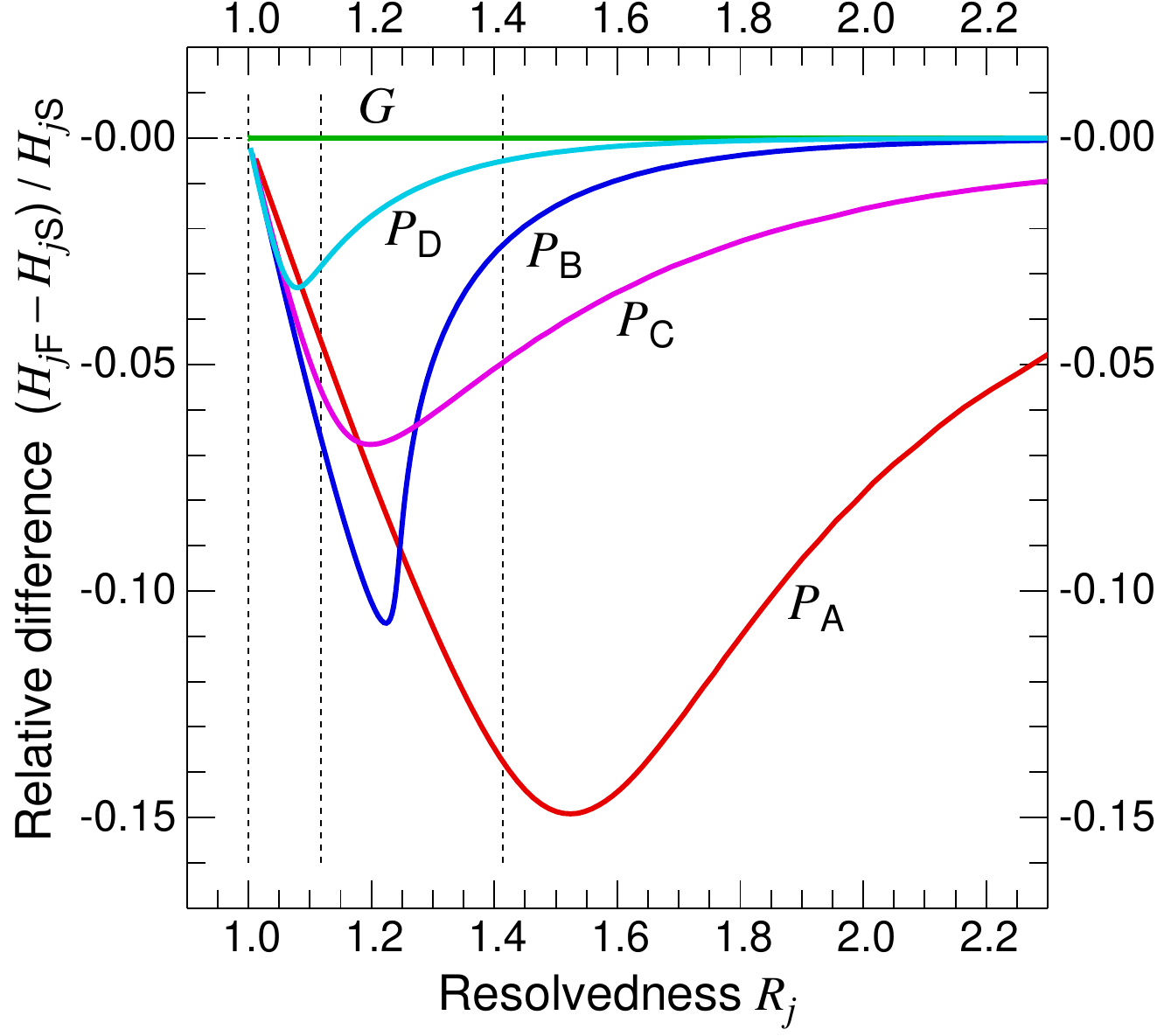}}}
\caption
{ 
Dependence of the convolution results on the model geometry. Shown are the differences between the models $\mathcal{M}_{j}$ of
filaments and sources with identical radial profiles (Fig.~\ref{modelprofs}), convolved with the Gaussian kernels $O_{j}$. The
relative differences $(H_{{j}{\,\rm F}} - H_{{j}{\,\rm S}}) / H_{{j}{\,\rm S}}$ of the half maximum widths of the structures are
plotted as a function of the model resolvedness $R_{j}$. The corresponding deconvolution accuracies for the structures are
displayed in Figs.~\ref{decgaussflathillhell}\,--\,\ref{decgaussplawHfils}.
} 
\label{comparison}
\end{figure}

\begin{figure*}
\centering
\centerline{
  \resizebox{0.2675\hsize}{!}{\includegraphics{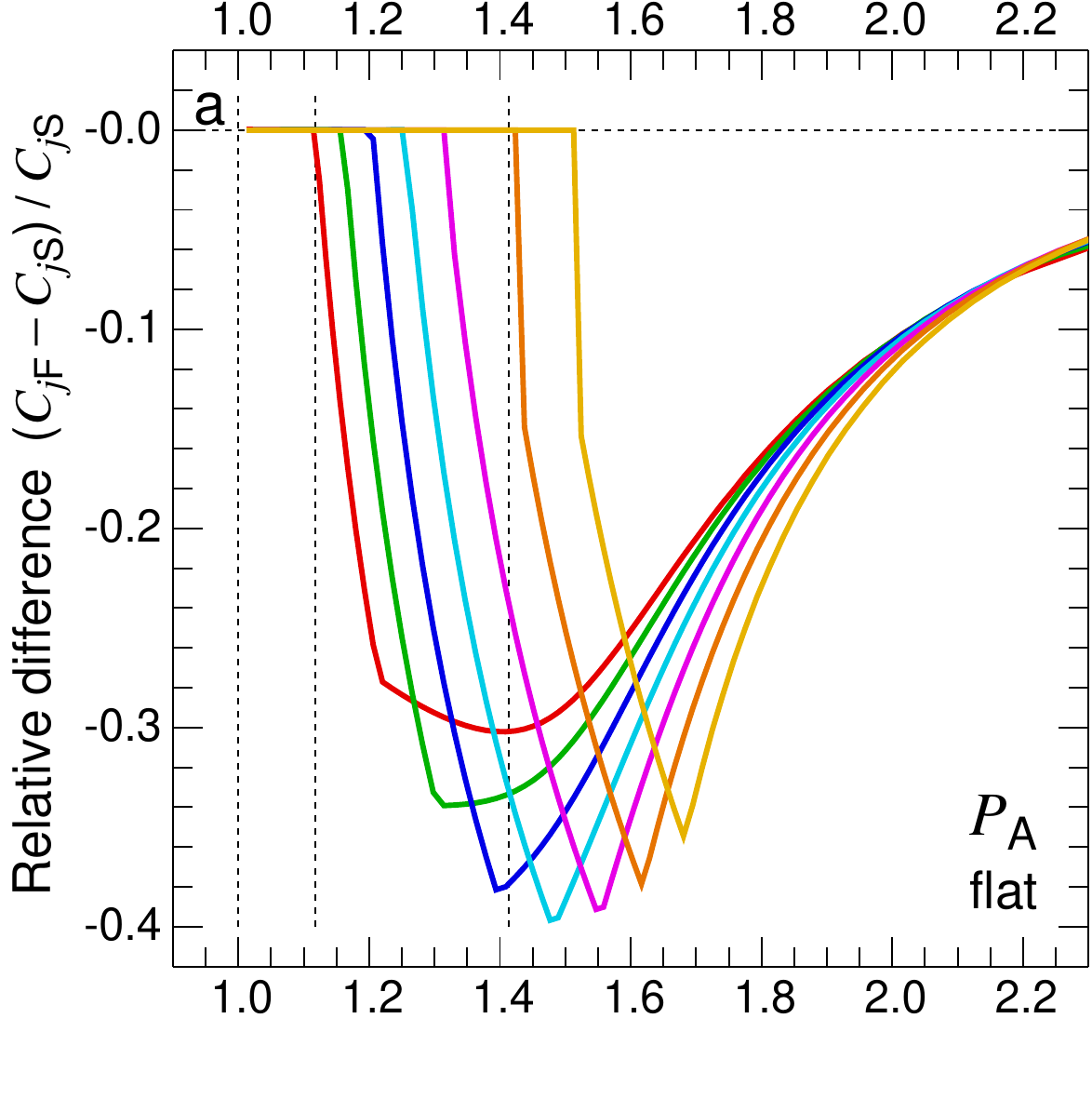}}  \hspace{-1.9mm}
  \resizebox{0.2832\hsize}{!}{\includegraphics{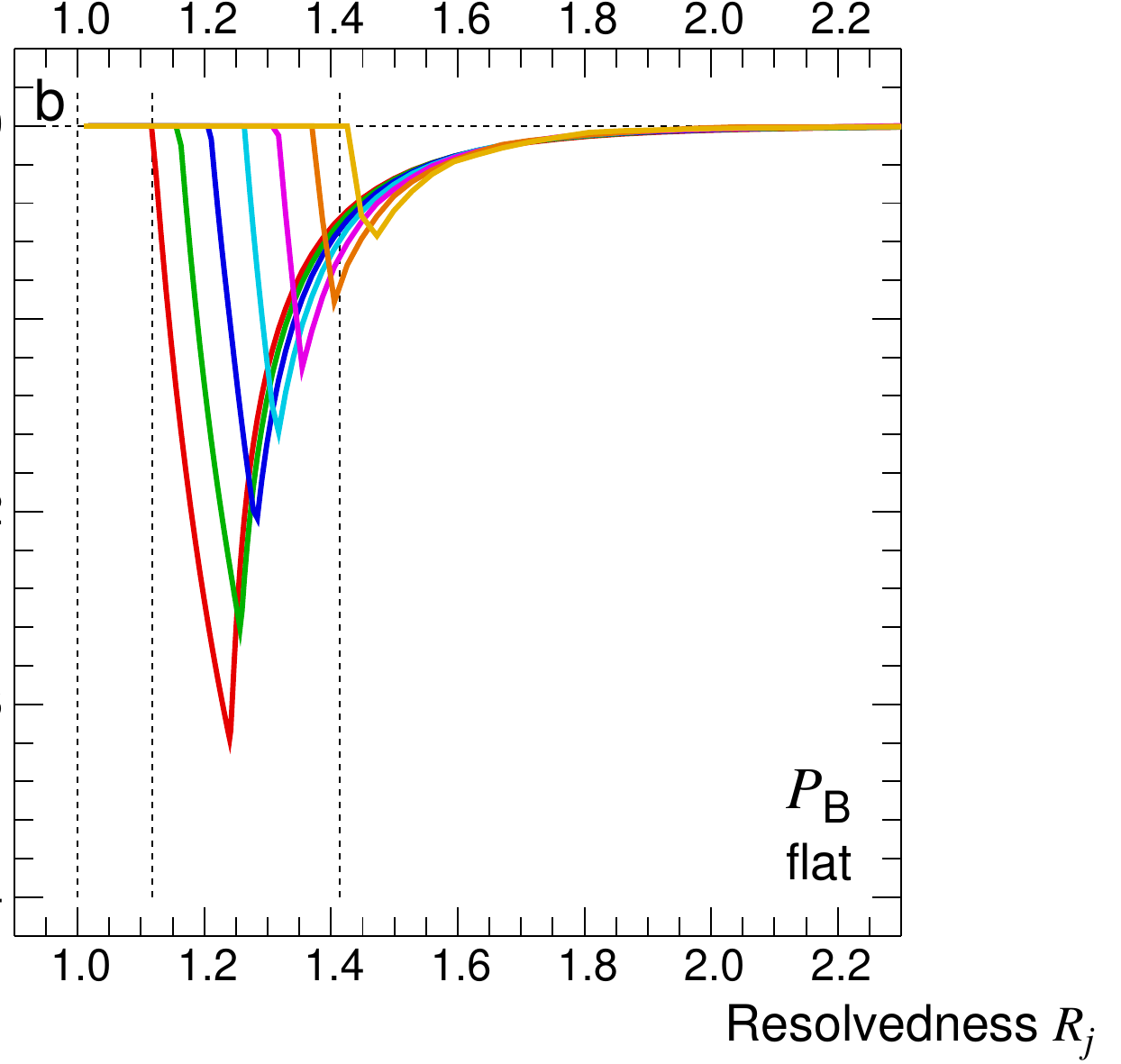}}  \hspace{-11.85mm}
  \resizebox{0.2287\hsize}{!}{\includegraphics{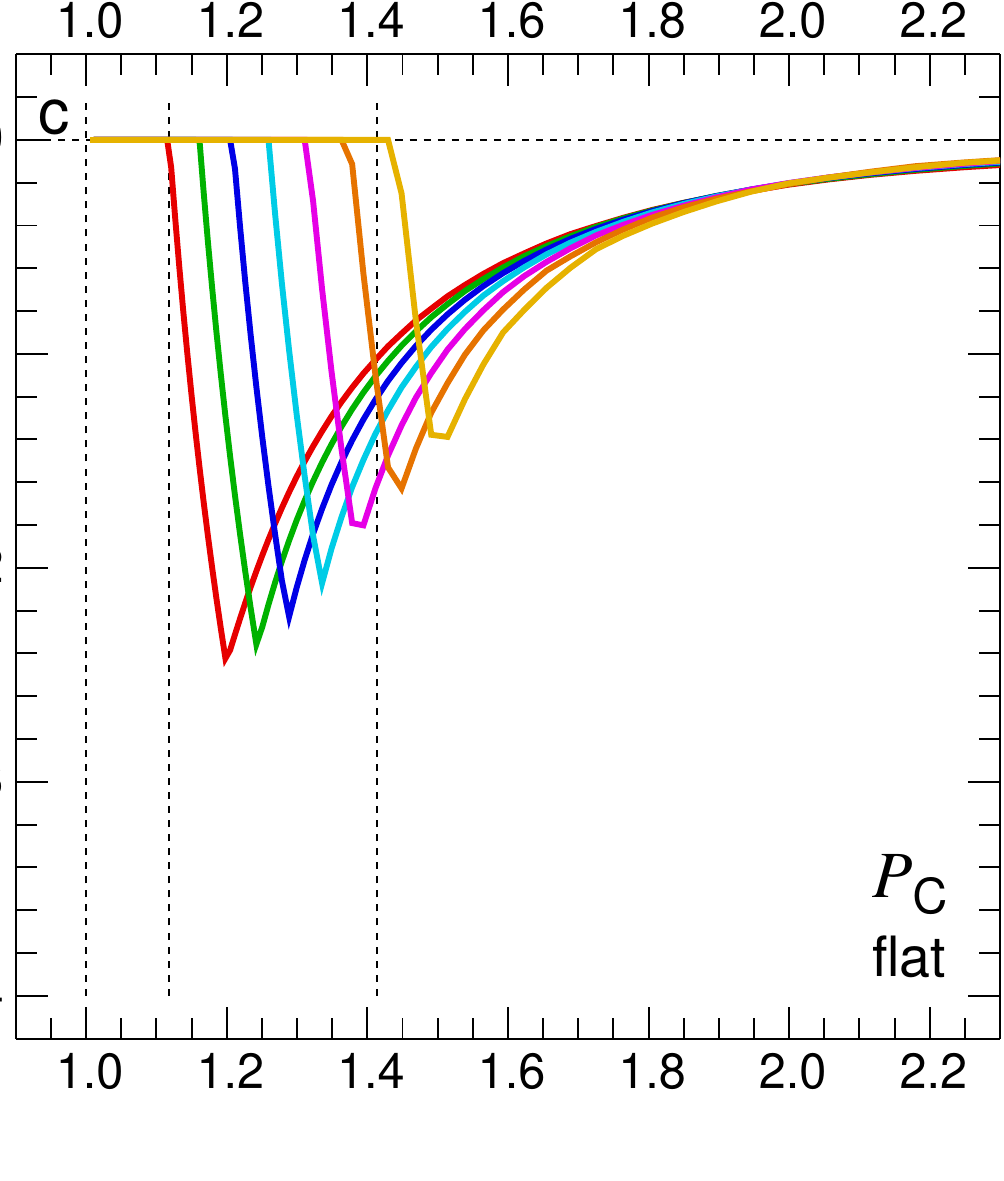}}  \hspace{-1.8mm}
  \resizebox{0.2738\hsize}{!}{\includegraphics{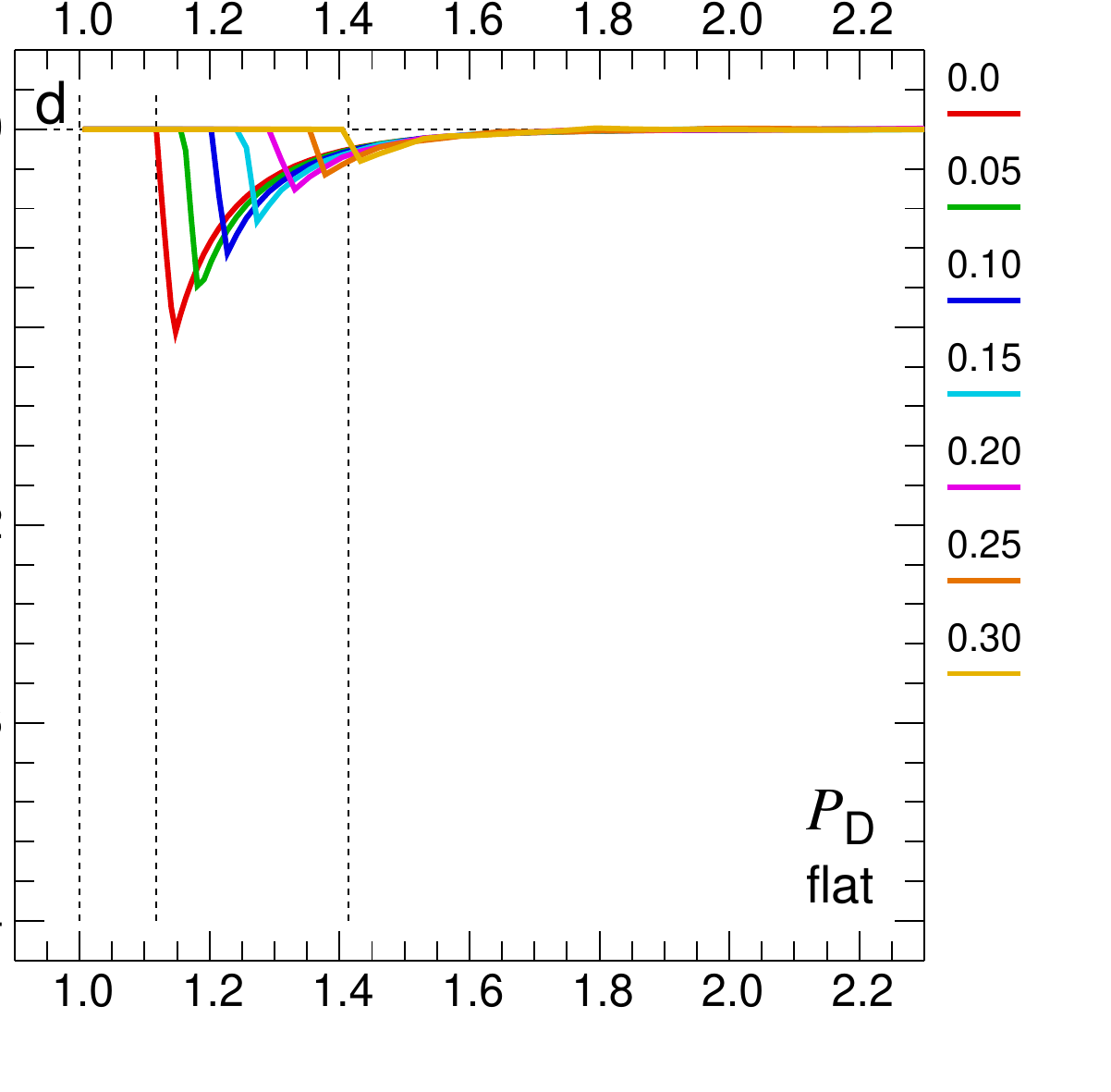}}}
\caption
{ 
Comparison of the deconvolved half-maximum sizes for the power-law filaments $\mathcal{\tilde{F}}_{{\mathcal{P}}{jl}}$ and sources
$\mathcal{\tilde{S}}_{{\mathcal{P}}{jl}}$ (with identical profiles) separated from the flat background $\mathcal{B}$ with the
over-subtraction levels $0 \le \epsilon_{\,l} \le 0.3$ (Figs.~\ref{decgaussplawH} and \ref{decgaussplawHfils}). The relative
differences between the deconvolved half maximum sizes $C_{{j}{\,\rm F}}$ for the filaments and $C_{{j}{\,\rm S}}$ for the sources
are plotted as a function of the model resolvedness $R_{j}$. For the convex and concave backgrounds, results are qualitatively
similar.
} 
\label{decgaussplawHreldiffs}
\end{figure*}

\subsection{Cylindrical Gaussian models}
\label{deconvgaussfils}

The Gaussian filaments $\mathcal{F}_\mathcal{G}$ are consistent with the assumptions made in Eq.~(\ref{deconvolution}). Despite
their cylindrical geometry, convolution of the filaments with the Gaussian kernels $O_{j}$ produces the radial profiles and sizes
$H_{j}$ identical to those of the equivalent sources $\mathcal{S}_\mathcal{G}$ (Fig.~\ref{comparison}). In other words, the size
deconvolution results and inaccuracies for the Gaussian filaments, for all angular resolutions and background types, are identical
to those presented in Sect.~\ref{deconvgauss} (Fig.~\ref{decgaussflathillhell}).

\subsection{Cylindrical power-law models}
\label{deconvplawsfils}

The power-law filaments $\mathcal{F}_\mathcal{P}$ are inconsistent with the assumptions made in Eq.~(\ref{deconvolution}).
Convolution of such filaments with the Gaussian kernels creates narrower radial profiles than those of the equivalent sources
$\mathcal{S}_\mathcal{P}$ (Fig.~\ref{comparison}, Sect.~\ref{convsrcfil}), depending on the model resolvedness $\tilde{R}_{j}$.

Figure~\ref{decgaussplawHfils} shows that the accuracies of the deconvolved sizes $\{C|D\}_{j}$ for the power-law filaments,
separated from the flat, convex, and concave backgrounds, are qualitatively similar to those presented in Sect.~\ref{deconvplaws}
for the power-law sources (Fig.~\ref{decgaussplawH}). The differences between the filaments and sources for the flat background are
displayed in Fig.~\ref{decgaussplawHreldiffs}. When filaments are well resolved ($\tilde{R}_{j} \gg 1.4$) or unresolved
($1\!\leftarrow \tilde{R}_{j} < 1.1$), they tend to have the same sizes as the sources, whereas in the intermediate range of $1.1
\la \tilde{R}_{j} \la 2$ the filaments become substantially narrower than the sources. As expected, the largest differences (up to
$30$\,--\,$40${\%}) are found for the strongest power-law model $\mathcal{P}_{\rm A}$ and the smallest deviations (within $10${\%})
are shown by model $\mathcal{P}_{\rm D}$, resembling the Gaussian model $\mathcal{G}$ most closely. Within such discrepancies, the
size deconvolution results for the power-law sources $\mathcal{S}_\mathcal{P}$ (Fig.~\ref{decgaussplawH}) are also applicable to
the filaments $\mathcal{F}_\mathcal{P}$ (Fig.~\ref{decgaussplawHfils}).



\section{Discussion}
\label{discussion}

\subsection{Pixel sizes}

This study was done using the model images with small $0.47${\arcsec} pixels (Sect.~\ref{simimages}). As a result, the convolved
images that simulated the wide range of angular resolutions were progressively smoother and oversampled toward the low-resolution
end, with up to $700$ pixels per largest beam $O_{J}$. To verify robustness of the results (Sect.~\ref{results}), the calculations
were repeated with variable pixel sizes, providing the standard sampling of the astronomical images. At each angular resolution
$O_{j}$, the smooth images were resampled \citep[using \textsl{swarp},][]{Bertin_etal2002} to always have three pixels per beam.
The tests showed mostly insignificant differences with respect to the original size deconvolution results, increasing to
approximately $20${\%} for the partially resolved and unresolved structures ($\tilde{R}_{j} < 1.4$).

\begin{figure*}
\centering
\centerline{
  \resizebox{0.2675\hsize}{!}{\includegraphics{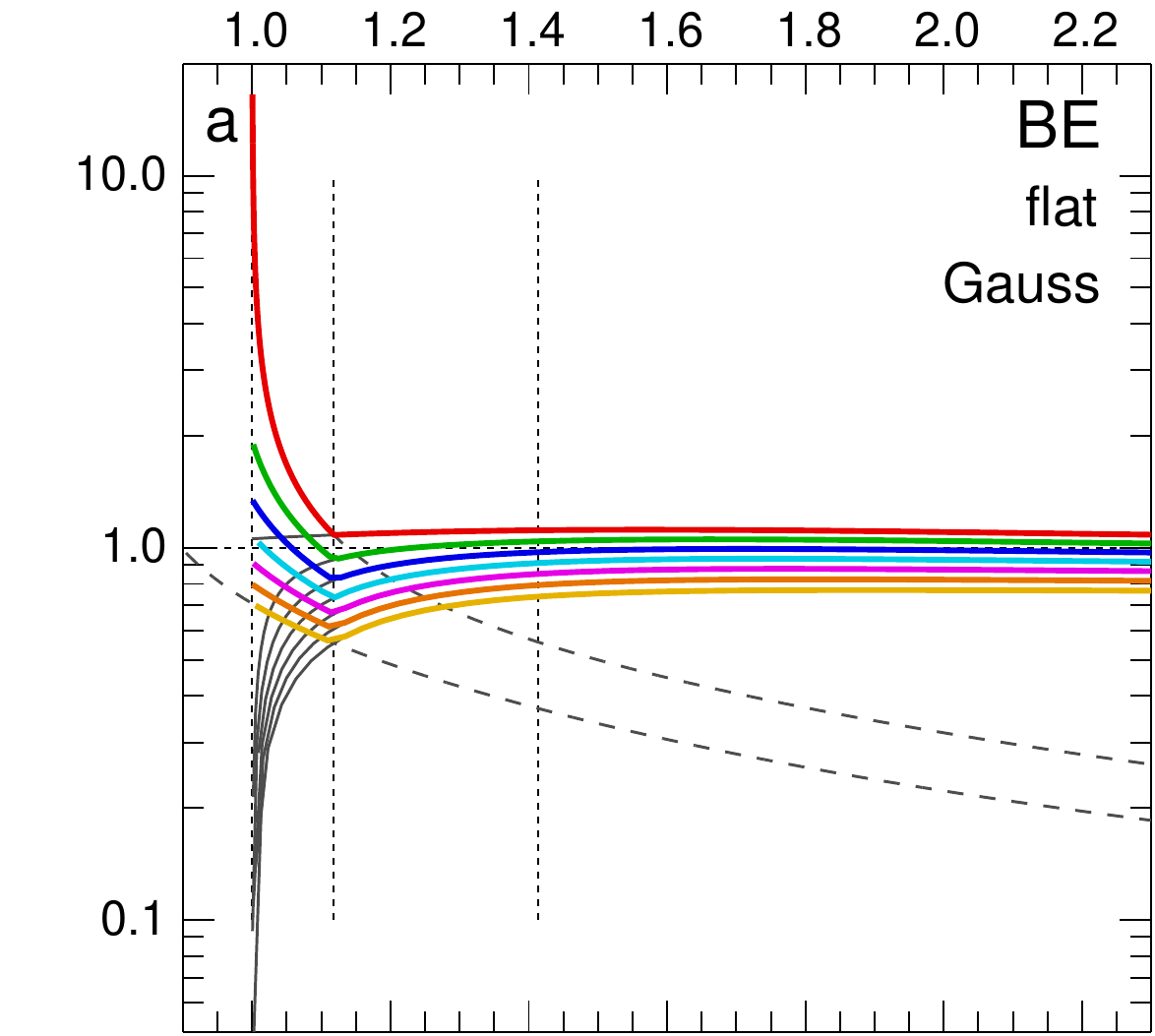}}      \hspace{-1.8mm}
  \resizebox{0.2287\hsize}{!}{\includegraphics{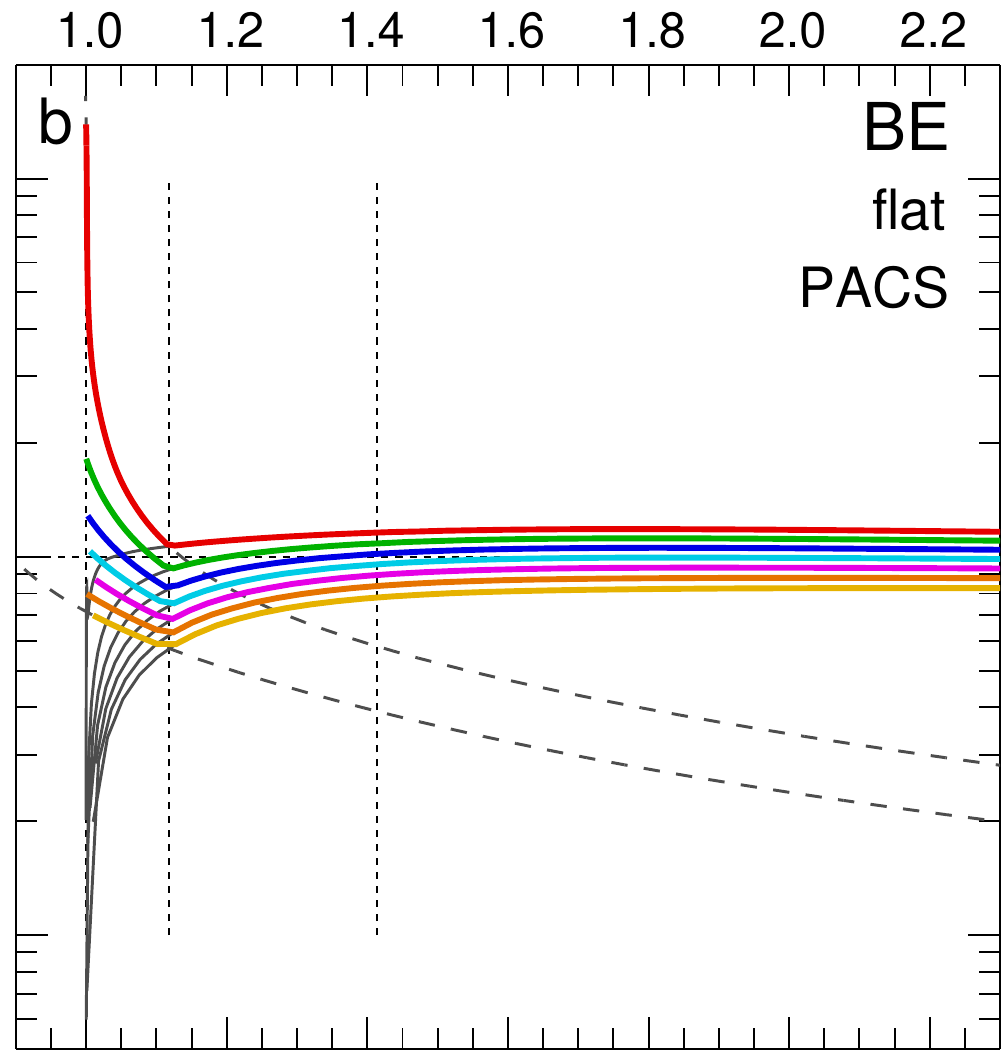}}  \hspace{-1.8mm}
  \resizebox{0.2287\hsize}{!}{\includegraphics{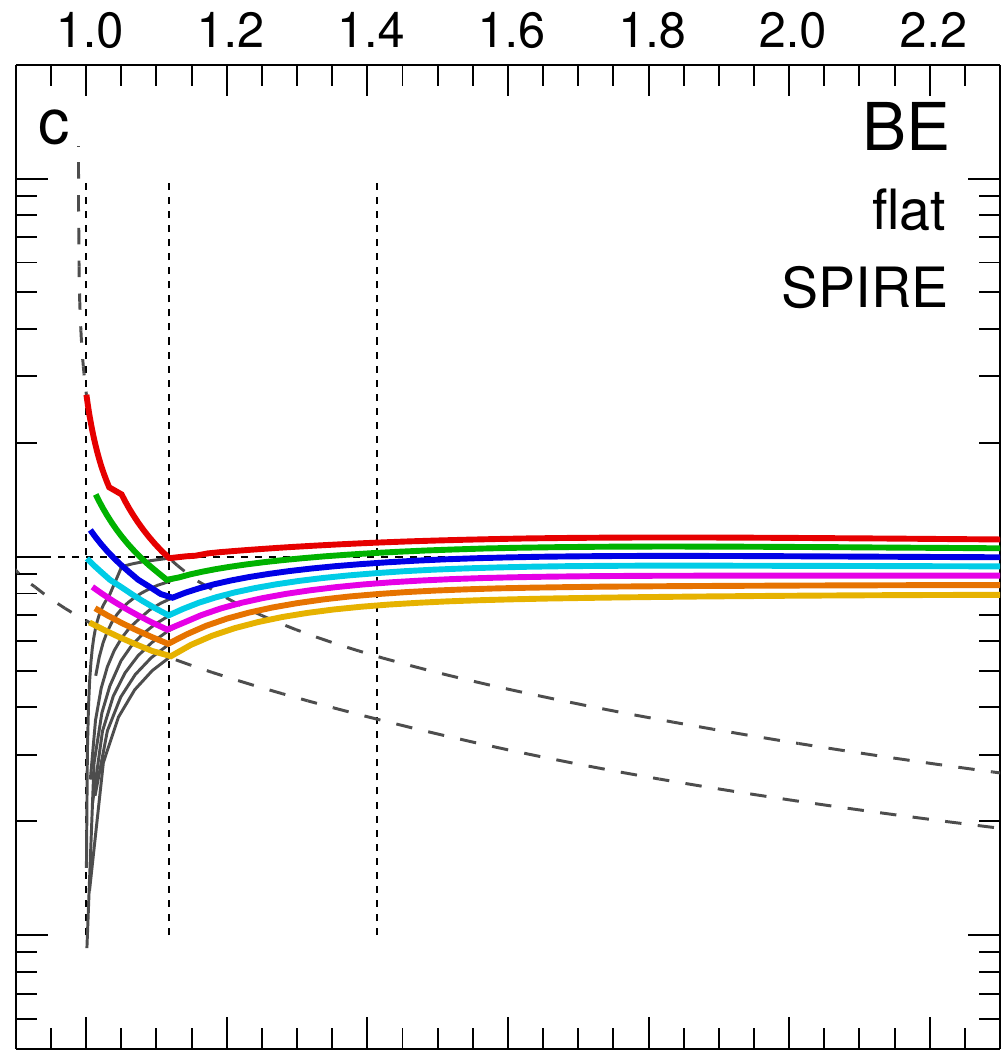}}  \hspace{-1.8mm}
  \resizebox{0.2675\hsize}{!}{\includegraphics{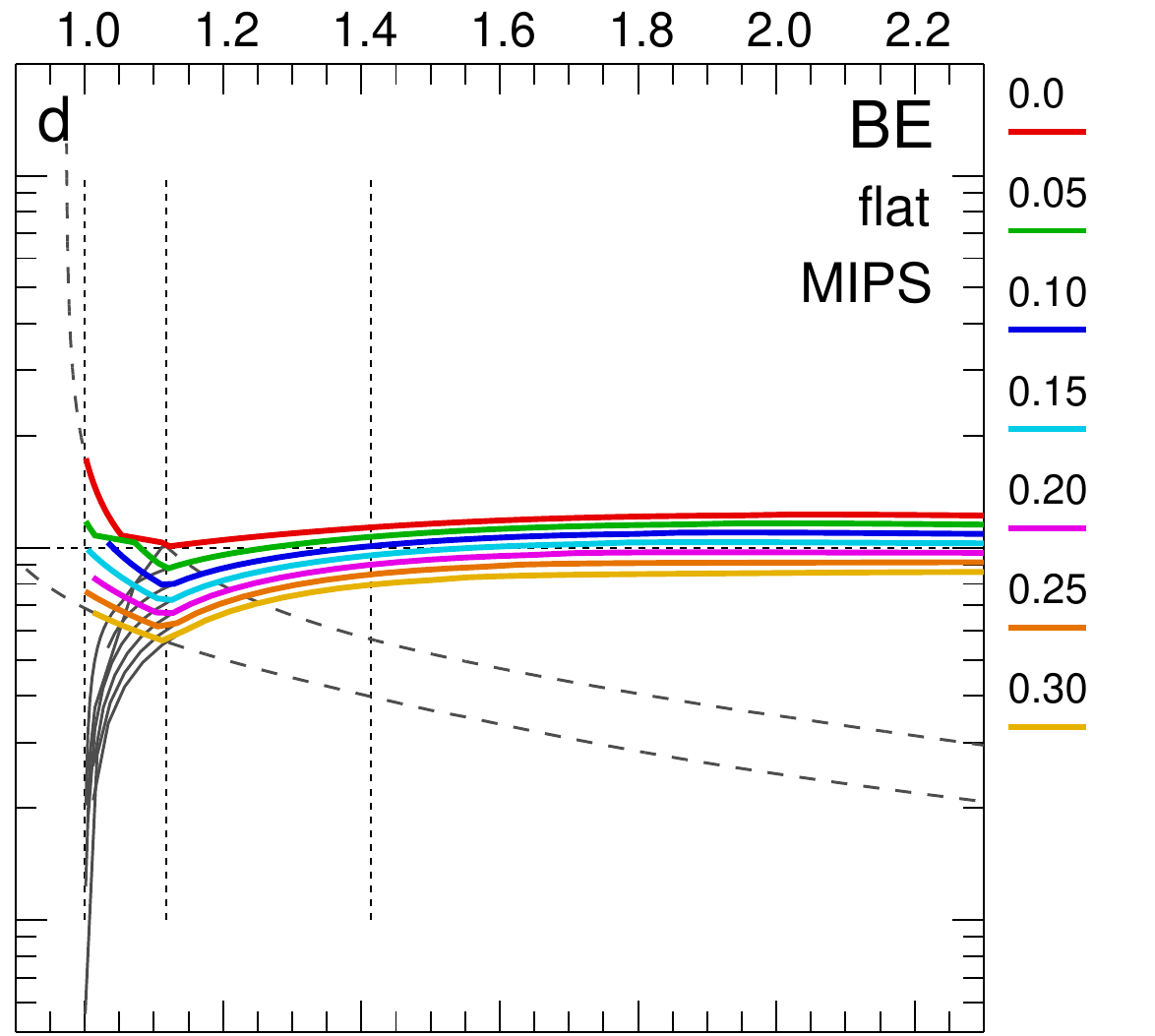}}}
\vspace{-2.40mm}
\centerline{
  \resizebox{0.2675\hsize}{!}{\includegraphics{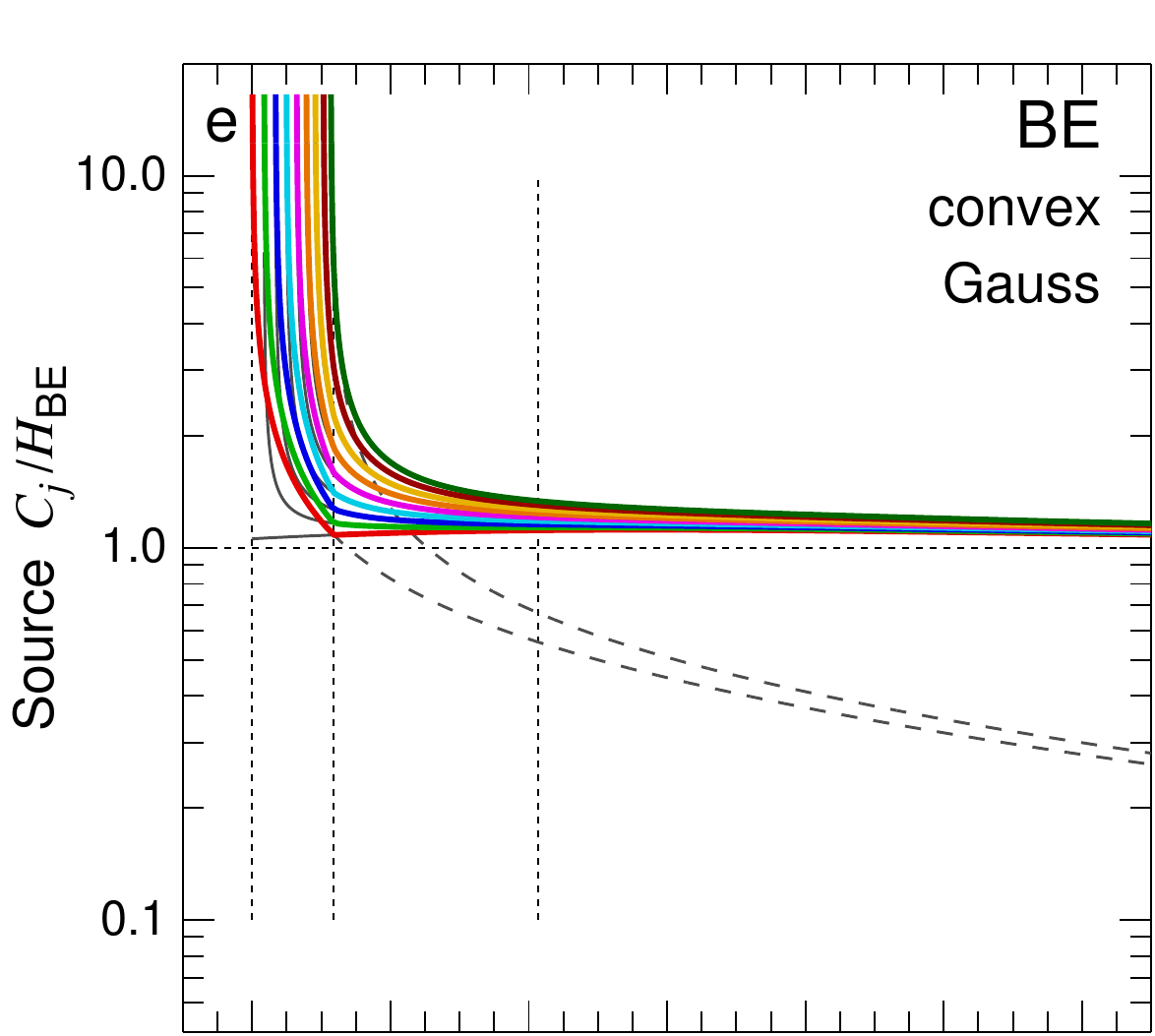}}      \hspace{-1.8mm}
  \resizebox{0.2287\hsize}{!}{\includegraphics{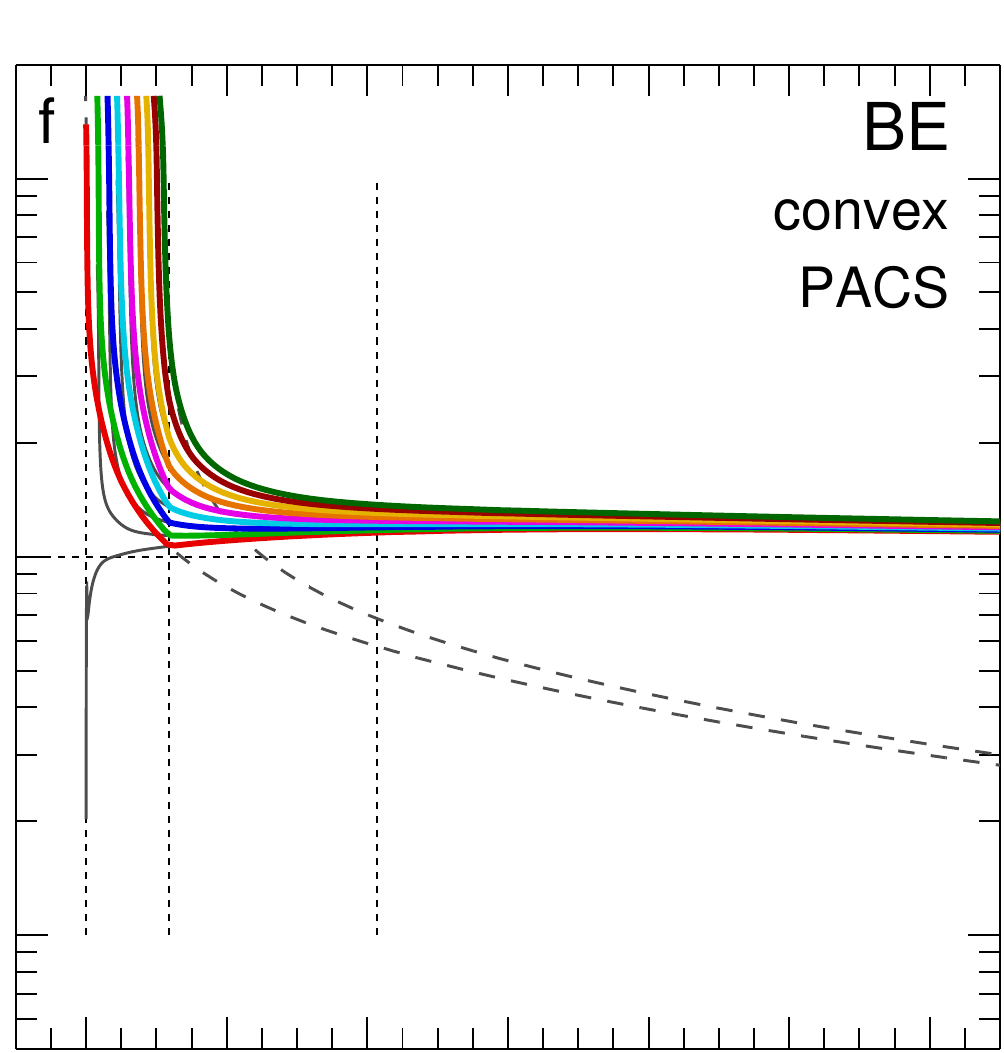}}  \hspace{-1.8mm}
  \resizebox{0.2287\hsize}{!}{\includegraphics{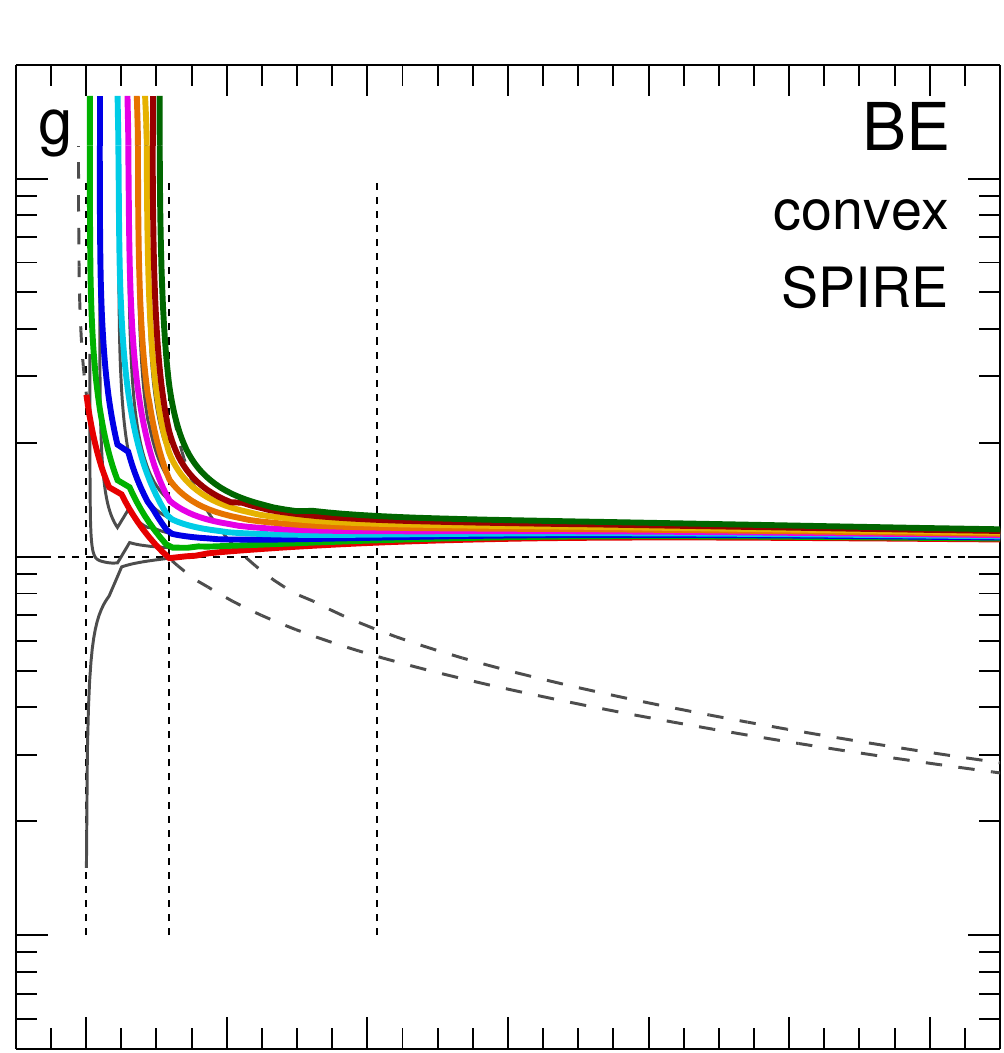}}  \hspace{-1.8mm}
  \resizebox{0.2675\hsize}{!}{\includegraphics{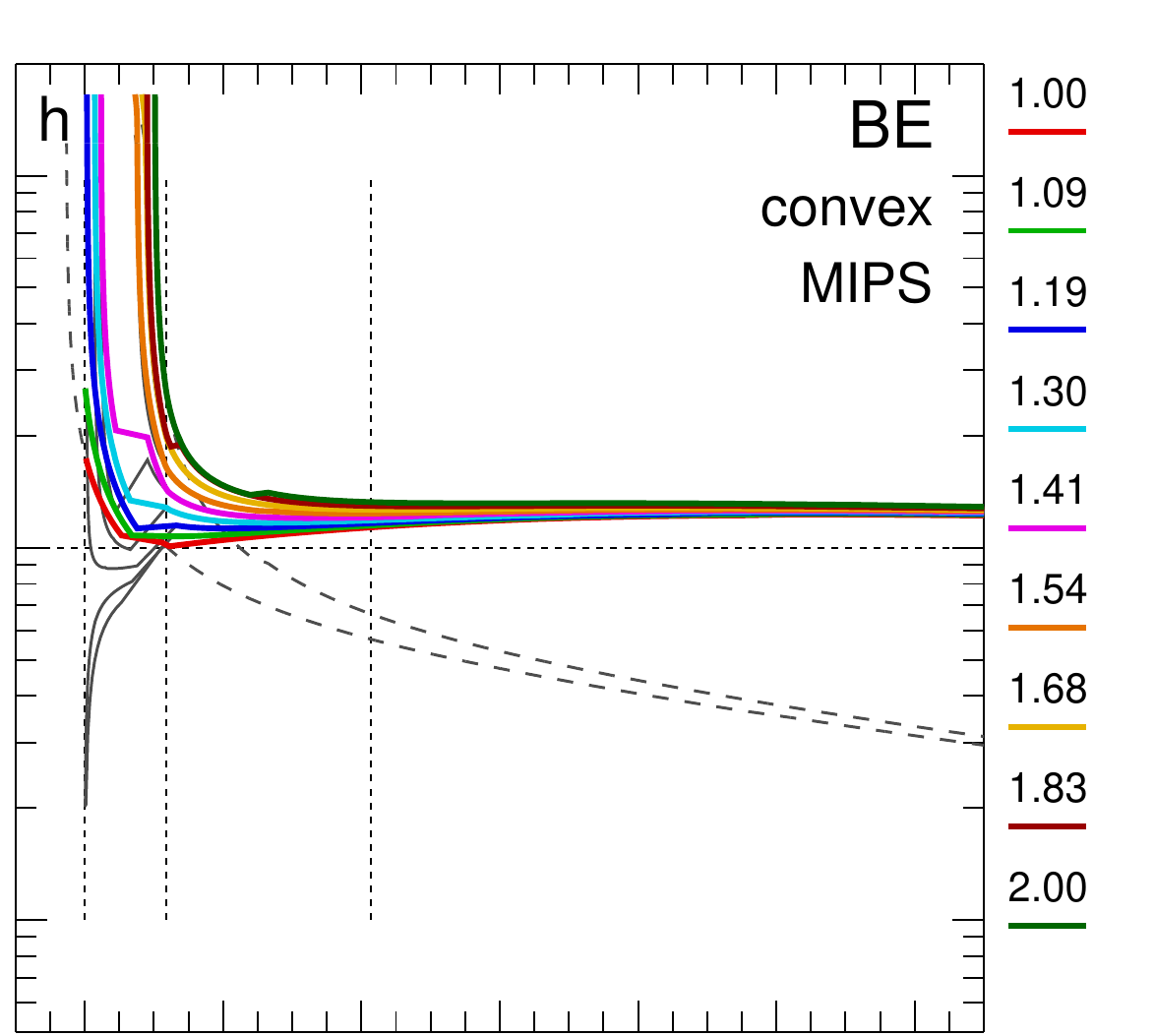}}}
\vspace{-0.5mm}
\centerline{
  \resizebox{0.2675\hsize}{!}{\includegraphics{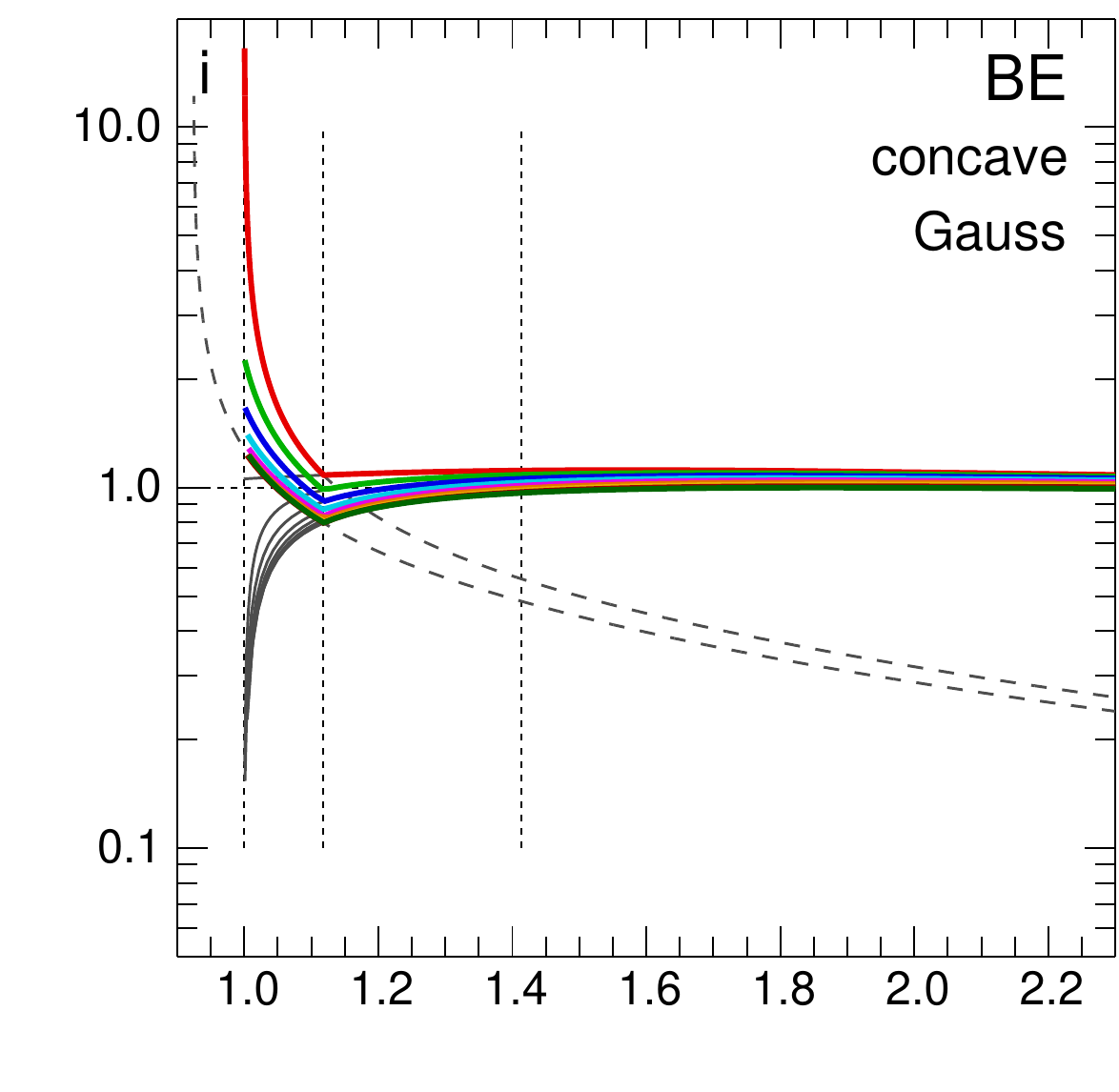}}      \hspace{-1.8mm}
  \resizebox{0.2832\hsize}{!}{\includegraphics{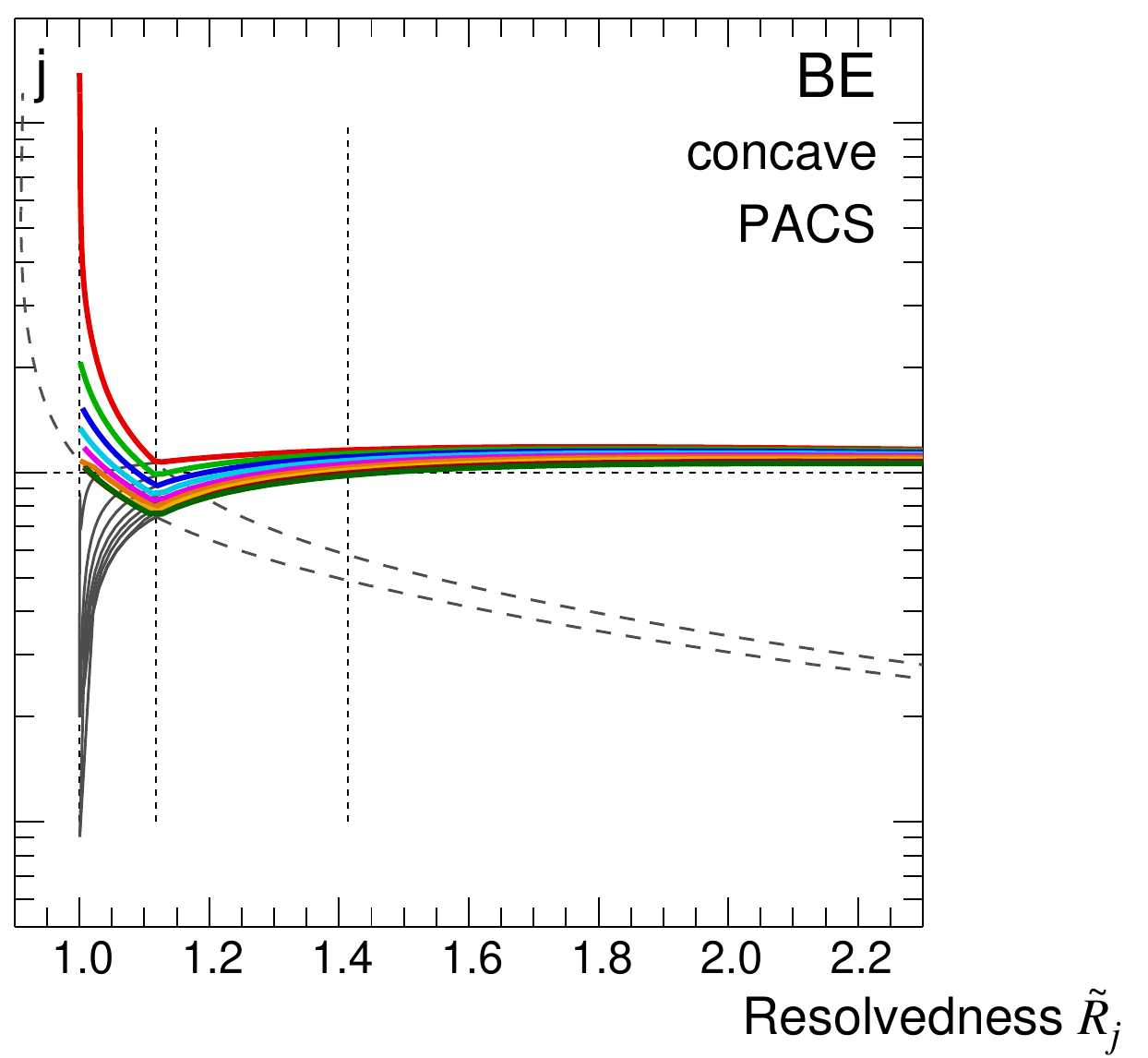}}  \hspace{-11.85mm}
  \resizebox{0.2287\hsize}{!}{\includegraphics{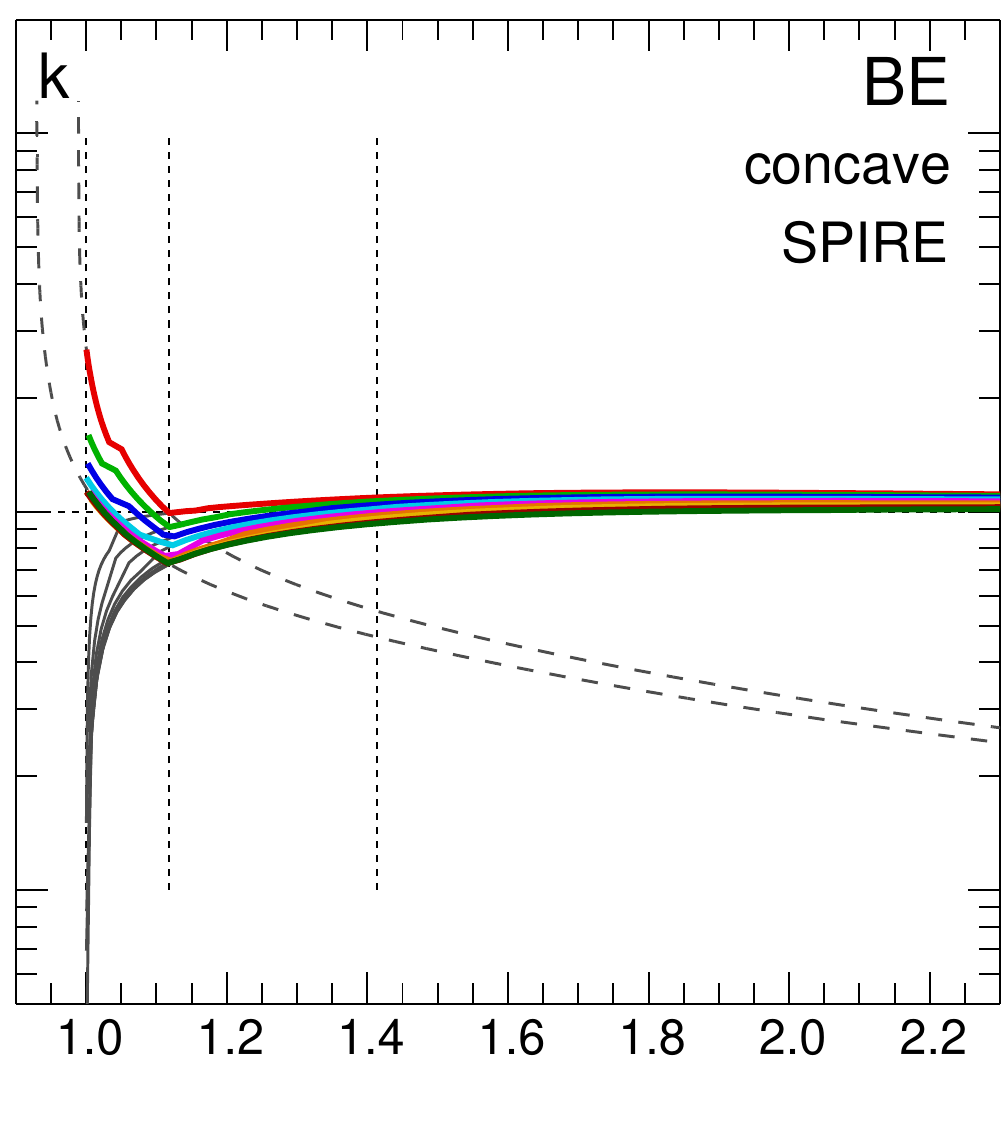}}  \hspace{-1.8mm}
  \resizebox{0.2675\hsize}{!}{\includegraphics{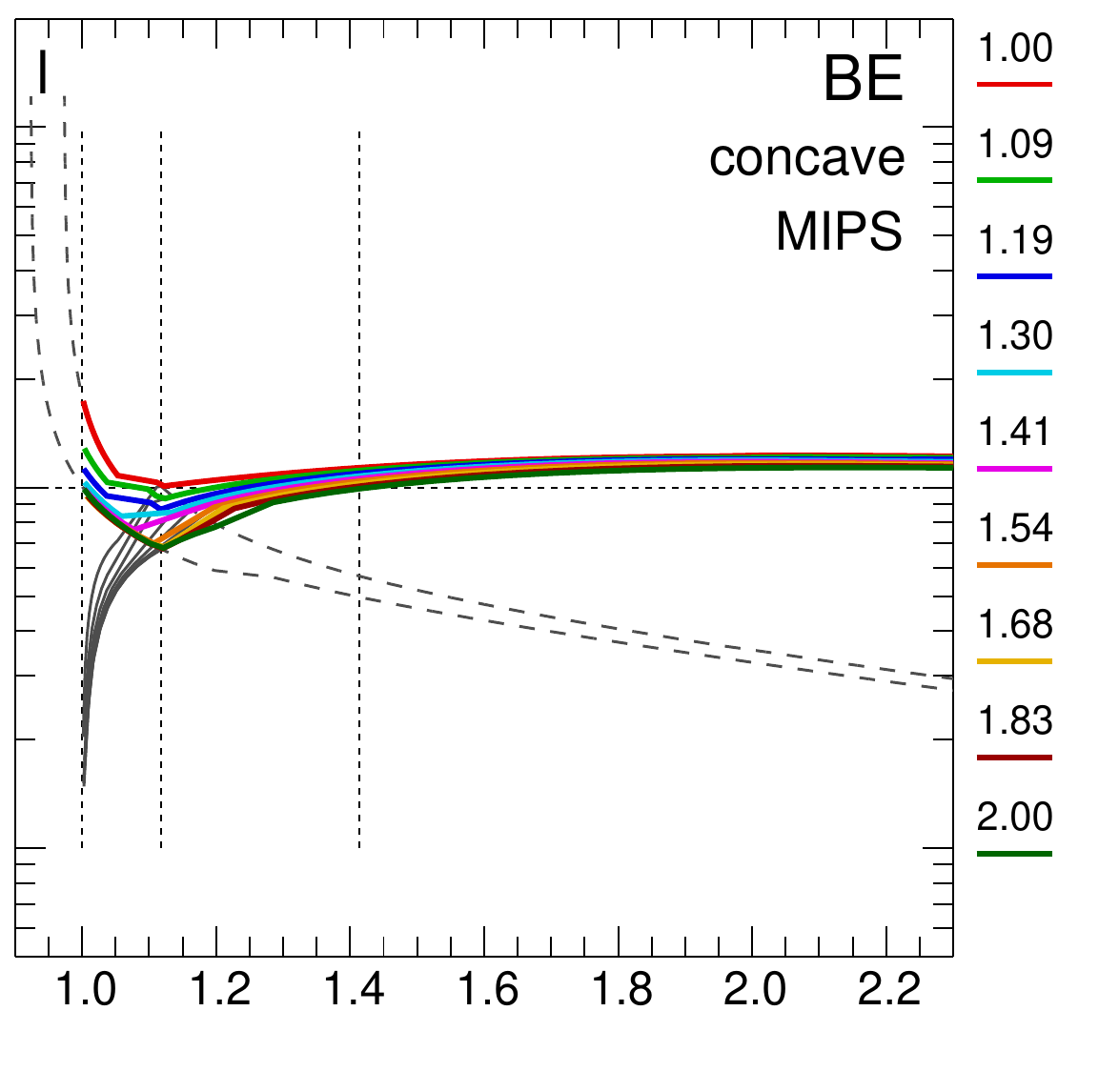}}}
\caption
{ 
Deconvolution accuracy of the half maximum sizes $\tilde{H}_{j}$ for the critical Bonnor-Ebert sphere (Fig.~\ref{besdens}),
separated from the flat (\emph{top}), convex (\emph{middle}), and concave (\emph{bottom}) backgrounds, for background
over-subtraction levels $0 \le \epsilon_{\,l} \le 0.3$ and size factors $1 \le f_{k} \le 2$. The convolution kernels, besides a
Gaussian shape, assumed the PSF shapes of \emph{Herschel} PACS at $160$\,$\mu$m, \emph{Herschel} SPIRE at $250$\,$\mu$m, and
\emph{Spitzer} MIPS at $160$\,$\mu$m \citep{Aniano_etal2011} with progressively larger deviations from the Gaussian shape. The
ratio of the modified deconvolved sizes $C_{j}$ to the true model size $H_{\rm BE} = 10${\arcsec} is plotted as a function
of the source resolvedness $\tilde{R}_{j}$. Some curves display small irregularities in their shapes, similar to the jumps
appearing in Fig.~\ref{decgaussplawM}, discussed in Sect.~\ref{deconvplaws}. They are also caused by the appearance of the
secondary minima in the source profiles, induced by the wavy profiles of the PSFs (Fig.~\ref{anianoconv}). For reference, the thin
black curves display $D_{j}/H_{\rm BE}$ and the dashed curves visualize $(O_{j}/2)/H_{\rm BE}$ for $\epsilon_{\,l} = \{0,
0.3\}$ and $f_{k} = \{1, 2\}$.
} 
\label{decbesH}
\end{figure*}

\subsection{Convolution kernels}
\label{convkernels}

The size deconvolution results (Sect.~\ref{results}), were presented for the models convolved with the Gaussian kernels
$\mathcal{O}_{j}$, which facilitated understanding and direct comparisons with the simplest reference models of the sources and
filaments with Gaussian profiles. To test robustness of the results, the model objects are also convolved to the angular
resolutions $O_{j}$ using the kernels with the PSF shape of the \emph{Herschel} SPIRE instrument at $250$ $\mu$m
\citep{Aniano_etal2011}. To create such set of kernels, the PSF is resampled (using \textsl{swarp}) to the pixel sizes that make
it correspond to the resolutions $O_{j}$.

Overall, the tests reveal minor differences, because the PSF resembles a Gaussian down to several percent of its peak (e.g.,
Fig.~\ref{anianoconv} (\emph{d})). For an illustration of these results (in the case of the Gaussian model $\mathcal{G}$), readers
are referred to Fig.~\ref{decbesH} (\emph{c}, \emph{g}, \emph{k}) with practically equivalent results. The differences in $D_{j}$
with respect to the original deconvolution results (Figs.~\ref{decgaussflathillhell}\,--\,\ref{decgaussplawHfils}) are mostly
within $10${\%} for model $\mathcal{G}$ and $20${\%} for the power-law models $\mathcal{P}$. The deviations are significant only
for the unresolved structures ($1\!\leftarrow \tilde{R}_{j} < 1.1$) on the simplest backgrounds ($\epsilon_{\,l} = 0$, $f_{k} =
1$). The deconvolved sizes $D_{j}$ become steeply underestimated ($D_{j}\!\rightarrow 0$), whereas the corrected sizes $C_{j}$ are
less severely overestimated than previously (by a factor of $2.5$ for model $\mathcal{G}$ with $\epsilon_{\,l} = 0$). The
$\epsilon_{\,l} > 0$ and $f_{k} > 1$ cases display minor differences, because the low-level intensities, affected by the
non-Gaussian PSF shape, are largely removed together with the subtracted backgrounds.

The differences are not surprising, because the PSF deviates from the Gaussian shape, violating the assumptions of the
deconvolution method (Eq.~(\ref{deconvolution})). As a result, the sizes become inaccurate even in the simplest case of the
Gaussian model $\mathcal{G}$, when the background subtraction does not modify the source shapes ($\epsilon_{\,l} = 0$, $f_{k} =
1$), because the shape is altered by the convolution with the non-Gaussian PSF.


Additional tests with the PSF shapes of the \emph{Spitzer} and \emph{Herschel} instruments were done for the model of a critical
Bonnor-Ebert (BE) sphere, which is often used in the studies of star formation to describe prestellar cores \citep{Bonnor1956}. The
surface densities of the critical BE sphere with a boundary radius $\Theta_{\rm BE}$ are usually assumed to resemble a Gaussian
with a half maximum size $H_{\rm G} \approx \Theta_{\rm BE}$ \citep[e.g.,][]{Ko"nyves_etal2015}. The approximate similarity of the
two shapes in their interiors (within $24${\%} at $\theta \approx \Theta_{\rm BE}/2$, Fig.~\ref{besdens}) is the consequence of the
model compactness and spherical geometry. The critical BE sphere becomes, however, much steeper than a Gaussian in the outermost
$10${\%} of the sphere ($0.9\,\Theta_{\rm BE} < \theta \rightarrow \Theta_{\rm BE}$), since its volume densities $\rho_{\rm BE}$
drop to zero at the outer boundary. Because of the differences and relevance for the star formation research, the size
deconvolution method is tested also for the BE model using the real-life convolution kernels.

The testing is done following the standard approach, by substituting the image of the Gaussian model (Fig.~\ref{modelprofs}) with
that of the critical BE sphere (Fig.~\ref{besdens}) having $\Theta_{\rm BE} = 11.9${\arcsec} and $H_{\rm BE} = 10${\arcsec} (the
same as $H_{\mathcal{M}}$, cf. Table~{\ref{modeltable}}). In addition to the Gaussian-shaped kernels $O_{j}$, three other sets of
kernels were derived from the PSF shapes of \emph{Herschel} PACS at $160$\,$\mu$m, \emph{Herschel} SPIRE at $250$\,$\mu$m, and
\emph{Spitzer} MIPS at $160$\,$\mu$m \citep{Aniano_etal2011} by resampling them to new pixels, thereby making their half maximum
widths correspond to $O_{j}$.

Figure~\ref{decbesH} presents the deconvolution results for the critical BE model. With the Gaussian kernels, the deconvolution
accuracies closely match (within $10${\%}) those of the Gaussian model $\mathcal{G}$ (Fig.~\ref{decgaussflathillhell}). The
realistic PSFs make the deconvolved sizes somewhat more overestimated (up to $25${\%}) for the resolved sources. The results, shown
in Fig.~\ref{decbesH} (\emph{c}, \emph{g}, \emph{k}), are very similar to those obtained with the SPIRE PSF for model
$\mathcal{G}$. Furthermore, the inaccuracies and irregularities in the $D_{j}/H_{\rm BE}$ curves for the unresolved sources
($1\!\leftarrow \tilde{R}_{j} < 1.1$) increase between the four columns of panels in Fig.~\ref{decbesH} (from left to right), along
the sequence of the PSFs with progressively stronger deviations from the Gaussian shape (cf. Fig.~\ref{anianoconv} (\emph{a},
\emph{d}, \emph{g})). As discussed above, the BE source shape gets altered by the convolution with the non-Gaussian kernels and,
therefore, the deconvolved sizes become inaccurate ($D_{j}\!\rightarrow 0$) even in the simplest case, when the background
subtraction does not modify the source shapes ($\epsilon_{\,l} = 0$, $f_{k} = 1$).


\subsection{Deconvolution inaccuracies}

This model-based study demonstrates that the Gaussian size deconvolution method works only when the extracted sources and filaments
fully comply with the strong assumptions of Eq.~(\ref{deconvolution}). The deconvolved sizes $D_{j}$ are accurate, when the
telescope PSF is nearly Gaussian, the object that produces the observed source or filament emission has a Gaussian shape, the
source or filament footprint fully includes the object's emission, and the subtracted background is correct. The (linearly)
interpolated background is accurate only when the actual background is flat. Furthermore, an accurately determined background
requires that the errors caused by the noise fluctuations in the image be very low, and hence the signal-to-noise ratio for the
structure must be high. Examples of the accurately deconvolved sizes $D_{j}$ for all angular resolutions $O_{j}$ are found in
Figs.~\ref{decgaussflathillhell}, \ref{decgaussflathillhellmom}, and \ref{decgaussplawM}, only for the simplest (unrealistic)
cases ($\epsilon_{\,l} = 0$ and $f_{k} = 1$), when the structure shape is not distorted by background subtraction.


For the simulated power-law sources $\mathcal{S}_\mathcal{P}$ and filaments $\mathcal{F}_\mathcal{P}$, the general trends in the
deconvolution results can be understood in terms of how much their profiles (Fig.~\ref{modelprofs}) deviate from the Gaussian shape
$\mathcal{G}$, when convolved to different angular resolutions. Models $\mathcal{P}_{\rm \{A|C\}}$, with their most intense
power-law profiles, display larger inaccuracies, than do the scaled-down models $\mathcal{P}_{\rm \{B|D\}}$ with smaller departures
from the Gaussian model. Similarly, models $\mathcal{P}_{\rm \{C|D\}}$, with their steeper power-law profiles, resemble
$\mathcal{G}$ better than $\mathcal{P}_{\rm \{A|B\}}$ do, hence their deconvolved sizes are less inaccurate.

The modified deconvolved sizes $C_{j}$ from Eq.~(\ref{maximized}) are accurate, when all assumptions of Eq.~(\ref{deconvolution})
are valid and the structures are not unresolved ($\tilde{R}_{j} \ga 1.1$). For the unresolved structures ($1\!\leftarrow
\tilde{R}_{j} < 1.1$) and correctly determined backgrounds, the modified sizes $C_{j}$ become greatly overestimated and steeply
biased (Figs.~\ref{decgaussflathillhell}\,--\,\ref{decgaussplawHfils}). The great inaccuracies in $C_{j}$ are especially obvious in
the simplest case of the Gaussian source in Fig.~\ref{decgaussflathillhell}, whose shape is not altered by background subtraction
($\epsilon_{\,l} = 0$, $f_{k} = 1$) and for which the measured size $\tilde{H}_{j}$ is correct and the standard deconvolution of
Eq.~(\ref{deconvolution}) gives the perfectly accurate $D_{j}$ values. When deconvolution gives more accurate or even perfectly
correct sizes $D_{j}$, the use of $C_{j}$ is evidently counterproductive.




\subsection{Background shapes}

Despite the simplicity of the models of the flat, convex, and concave backgrounds, they represent three typical forms of the
observed fluctuating backgrounds. The convex, concave, and more complex backgrounds totally blend with the source or filament
emission and cannot be correctly deblended (interpolated). Cubic splines or even more complex interpolation schemes that one might
think of are not necessarily better in practice than the simplest planar backgrounds used in the source or filament extractions.
The backgrounds of the observed structures are the fundamental unknowns and unlikely to be accurately reconstructable, especially
in the presence of noise and strong background fluctuations on all spatial scales. It is difficult to envision how the background
values and derivatives at the edges of the structures can be determined from the data to ensure an accurate approximation of the
true intensity or surface density distribution of the background cloud with cubic splines or more complex shapes. The edges of the
structure footprint are poorly determined and heavily affected by the fluctuations, and the background inside the footprint remains
completely unconstrained by the data.

The measured sizes and deconvolved sizes of the simulated sources and filaments do not depend on the background shape, when
background subtraction does not modify the structure intensity distribution, except by just a scaling factor. The intensity shape
is preserved in only two simplest cases, when the background is flat (and $\epsilon_{\,l} = 0$) or it is convex or concave with
exactly the same shape and width as the structure itself ($f_{k} = 1$). In practice, it is much more likely that subtraction of the
interpolated (planar) background does alter the structure intensity distribution, which leads to very large errors in the
deconvolved sizes of unresolved structures. 

For the simplistic Gaussian source or filament on a flat background, even a small overestimation of the background level causes
severely underestimated $D_{j}$ (Fig.~\ref{decgaussflathillhell} (\emph{d})). A similar behavior is obvious for the Gaussian
structures on the concave backgrounds that happen to be wider than the structure (Fig.~\ref{decgaussflathillhell} (\emph{f})),
because the interpolated backgrounds become overestimated. In contrast, the convex backgrounds that happen to be wider than the
structures, are always underestimated, which leads to the explosive overestimations of both deconvolved sizes $\{D|C\}_{j}$ for
even those structures that appear resolved ($\tilde{R}_{j} \la 1.2$ in Fig.~\ref{decgaussflathillhell} (\emph{e}), and
$\tilde{R}_{j} \la 1.7$ in Fig.~\ref{decgaussflathillhellmom}).

The simulated sources and filaments with the intense power-law profiles of models $\mathcal{P}_{\rm \{A|B\}}$ demonstrate a severe
overestimation of the deconvolved half maximum sizes $C_{j}$, independently of the type of their background, which is worse than
for the Gaussian structures (Figs.~\ref{decgaussplawH} and \ref{decgaussplawHfils}). At the same time, the deconvolved $D_{j}$
values display strongly nonmonotonic trends to either over- or underestimations at $\tilde{R}_{j} \la 1.3$ for the flat and
concave backgrounds. For filaments, the amplitude of the above behavior is diminished by the fact that their convolved widths are
substantially smaller than those of the sources. 

The moment sizes $M_{j}$ and their deconvolved values $\{D|C\}_{j}$ are especially badly affected by the background inaccuracies,
for all types of sources and backgrounds (Figs.~\ref{decgaussflathillhellmom} and \ref{decgaussplawM}). In general, the sizes
computed from the intensity moments over the entire source footprints are too sensitive to the inaccuracies of the interpolated
backgrounds, hence they are significantly less reliable than the half maximum sizes. Moreover, the intensity level, to which the
moment sizes correspond, depends on the profile of the background-subtracted intensity distribution (e.g., Table~\ref{modeltable}),
therefore it can strongly vary for different observed sources.

The complex observed backgrounds can only give rise to even larger inaccuracies than those reported in this paper. The wider the
footprints of structures are, the less reliable the interpolated backgrounds become. Therefore, the extracted sources and filaments
with strong and wide power-law wings must have less accurate backgrounds and, as a consequence, the measured and deconvolved sizes.
This is consistent with a recent analysis of the \emph{Herschel} filaments \citep{Andre_etal2022} that found the Gaussian size
deconvolution inaccurate and unreliable for the unresolved or power-law filaments.

\subsection{Physical interpretations}

Interpretation of $\{D|C\}_{j}$ in terms of the dimensions of the physical objects that produce the observed emission of sources
and filaments is questionable. The deconvolved half maximum sizes, even if they were perfectly accurate, describe only the
innermost bright peak, (much) smaller than the entire (footprint) diameters $2 \Theta$ of the objects (Table~\ref{modeltable}). In
general, it is impossible to deduce the full extent of the partially resolved or unresolved non-Gaussian objects from the
half maximum values without additional model assumptions, because of the unknown relationship between the $\{H|M\}$ values and the
footprints of the structures. The information is revealed in the observed images for only the well resolved sources or filaments,
in which case the deconvolution loses its value, because convolution has only a minor effect on their widths.

Filaments are more difficult structures for a proper analysis than the round sources. The implicit assumptions that each physical
filament lies entirely in the plane of sky and has a cylindrical geometry are unrealistic. In general, the observed filaments have
substantial variations of the physical conditions (profiles, widths, linear densities, etc.) in distant segments along their
crests. The usual approach to the analysis of their radial profiles and widths rests on the assumption that they have the same
cylindrical properties along the entire length. With this assumption, the individual filament profiles are averaged along the
entire crest \citep[e.g.,][]{Arzoumanian_etal2011} to dilute the background and noise fluctuations and enhance the signal-to-noise
ratio of the imaginary ``global'' profile. An average physical property of a long and variable filament is a simplified abstract
concept that needs to be replaced with the property being a function of the coordinate along its skeleton.

\begin{figure*}
\centering
\centerline{
  \resizebox{0.3228\hsize}{!}{\includegraphics{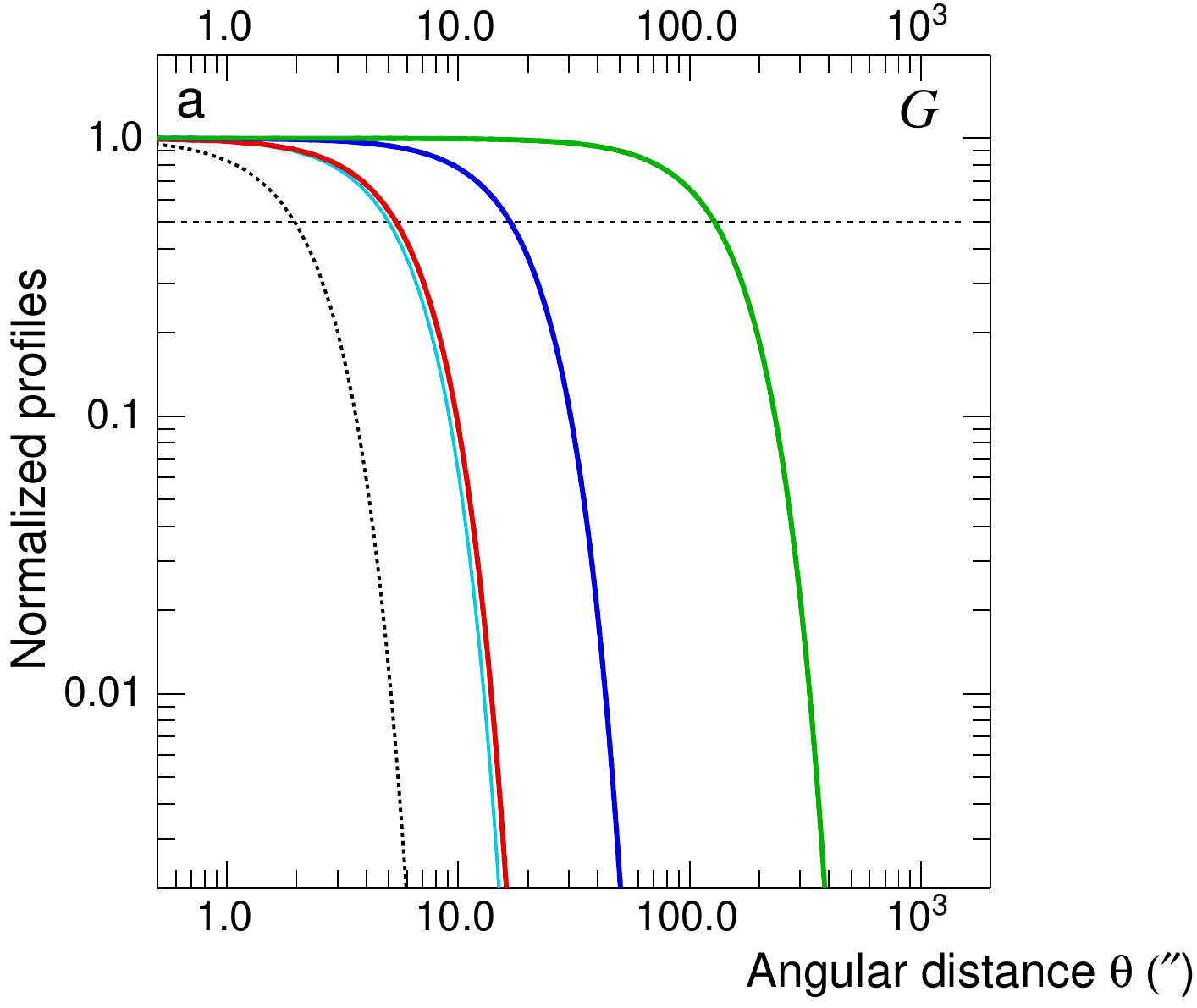}}  \hspace{-11.85mm}
  \resizebox{0.2760\hsize}{!}{\includegraphics{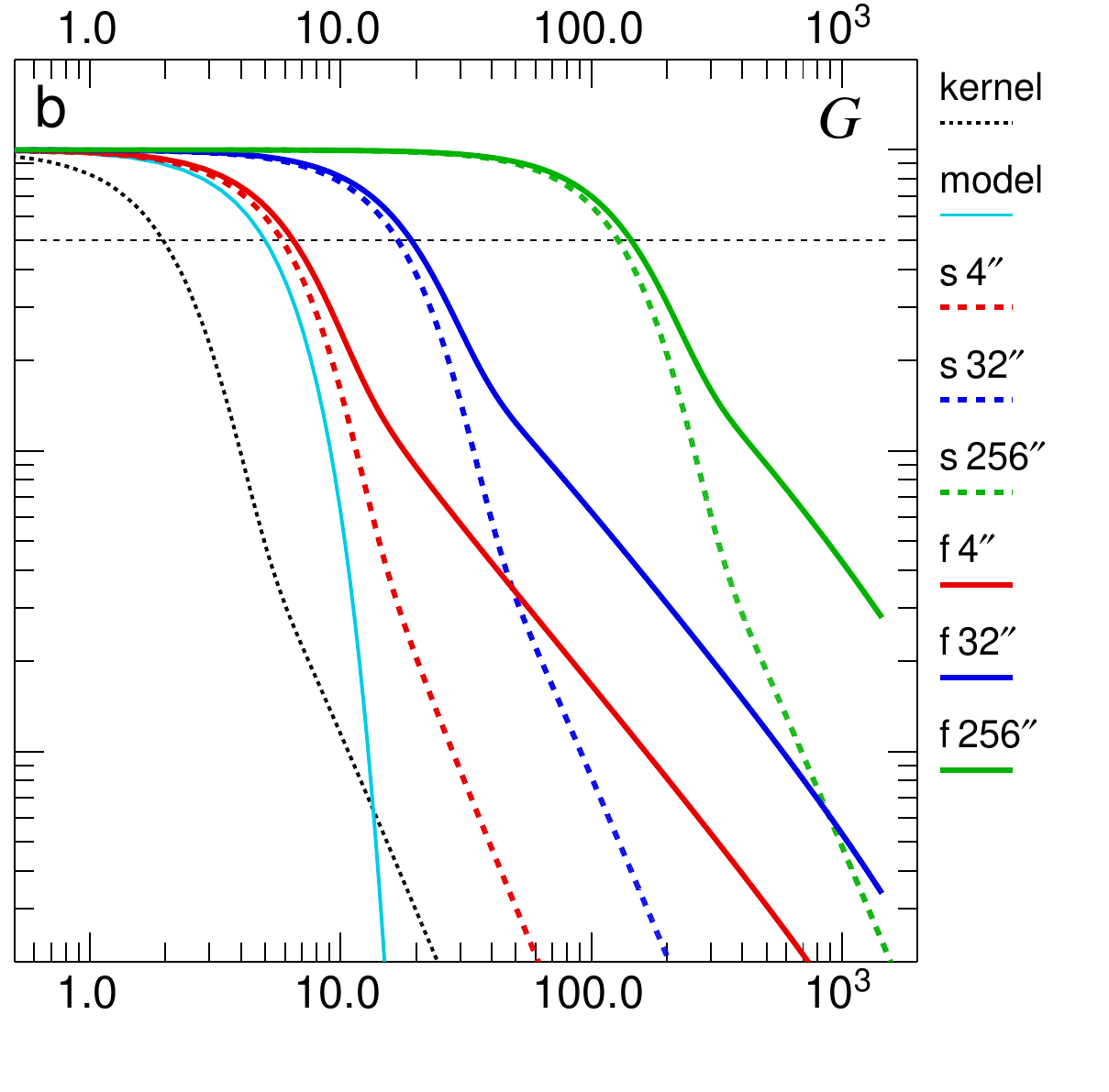}}}
\caption
{ 
Convolution of the Gaussian models. The spherical (dashed curves) and cylindrical (solid curves) models $\mathcal{G}$
(Fig.~\ref{modelprofs}) were convolved with two types of kernels (thin dotted curves) of the half maximum sizes of \{$4, 32$,
$256$\}{\arcsec}. Shown are the results for the pure Gaussian kernels $\mathcal{O}_{j}$ (\emph{left}) and power-law kernels
$\mathcal{K}_{j}$ (\emph{right}). The dashed horizontal lines indicate the half maximum level.
} 
\label{gausconv}
\end{figure*}

\begin{figure*}
\centering
\centerline{
  \resizebox{0.2675\hsize}{!}{\includegraphics{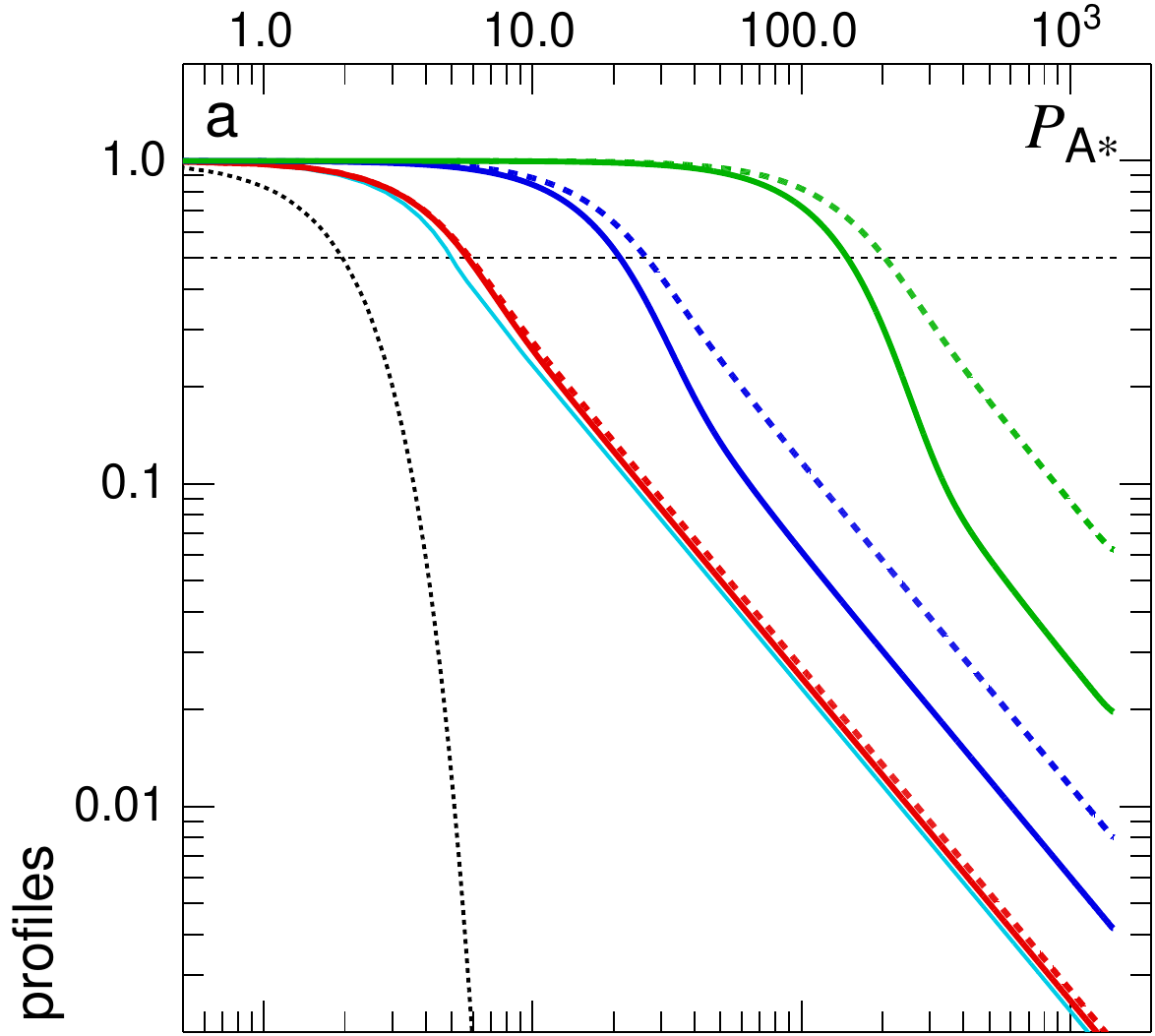}}  \hspace{-1.8mm}
  \resizebox{0.2287\hsize}{!}{\includegraphics{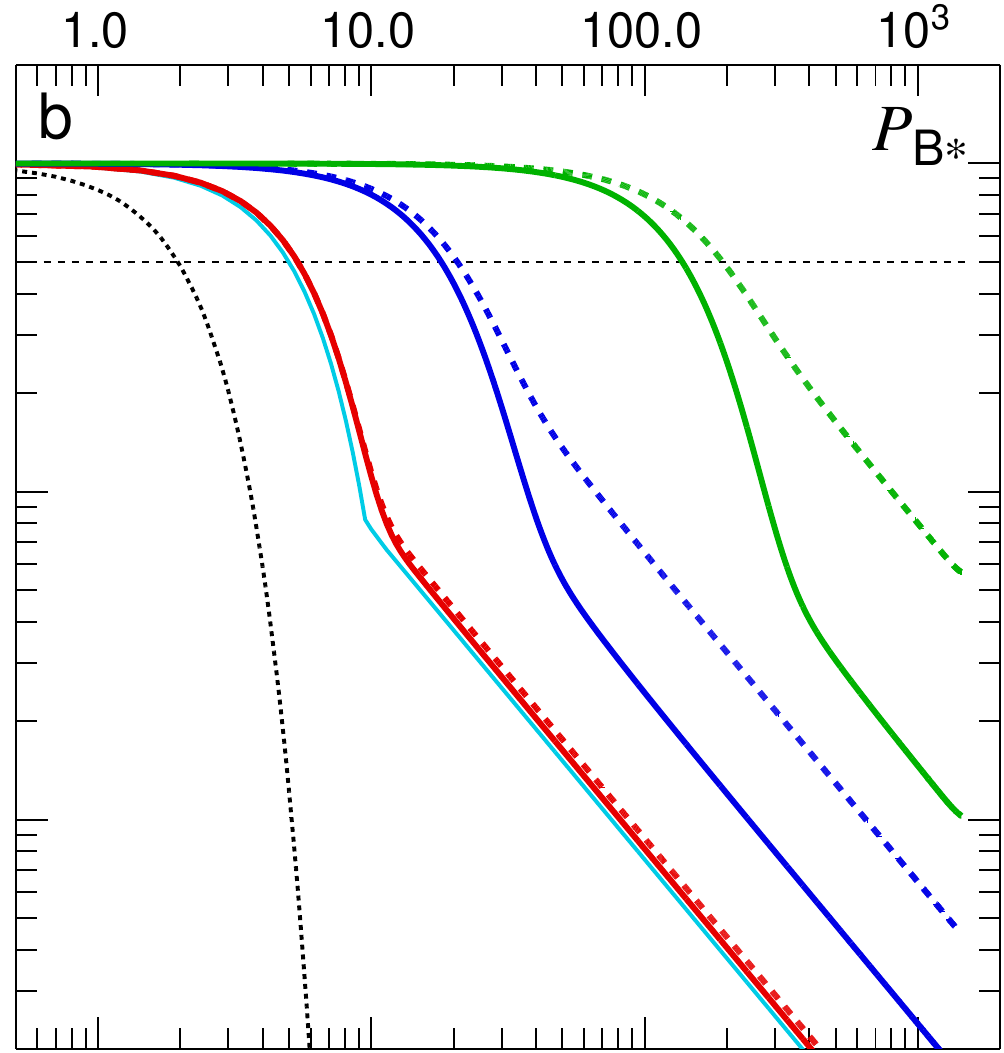}}  \hspace{-1.8mm}
  \resizebox{0.2287\hsize}{!}{\includegraphics{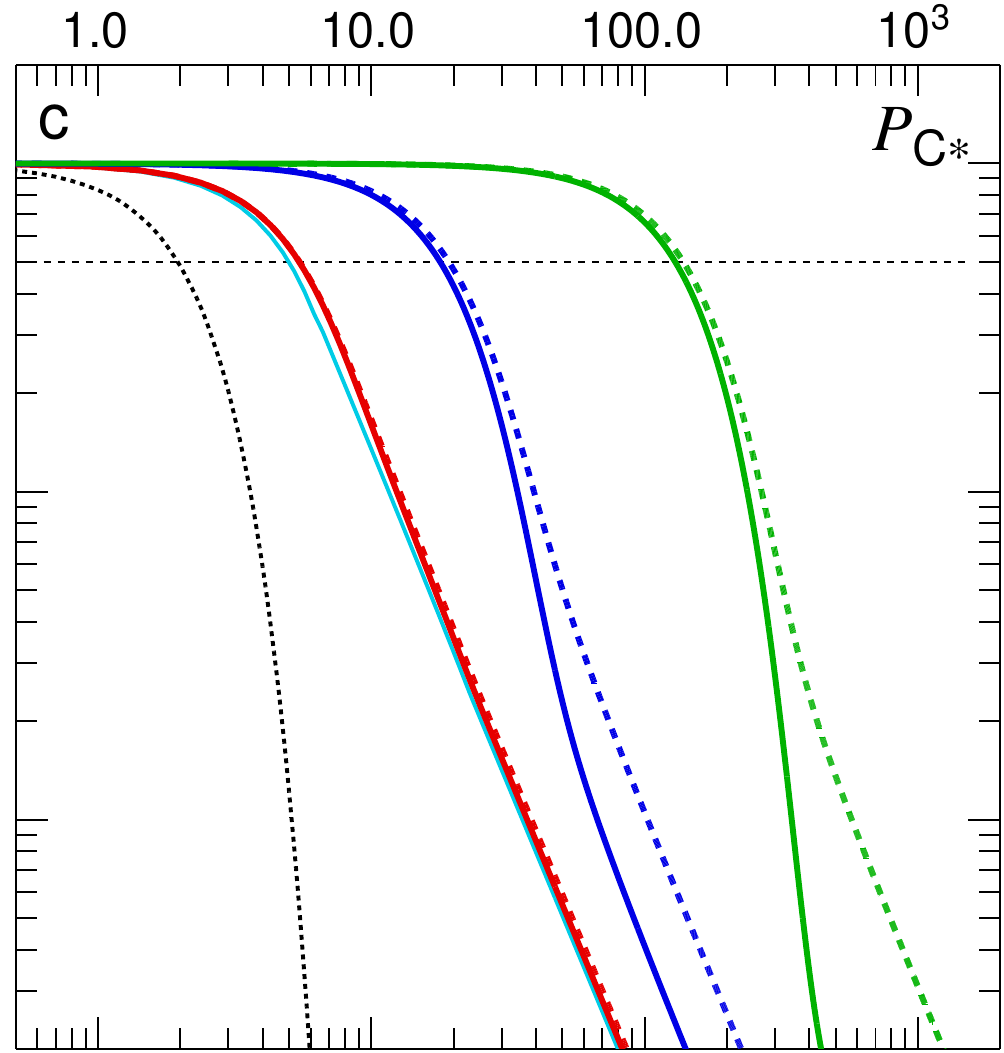}}  \hspace{-1.8mm}
  \resizebox{0.2675\hsize}{!}{\includegraphics{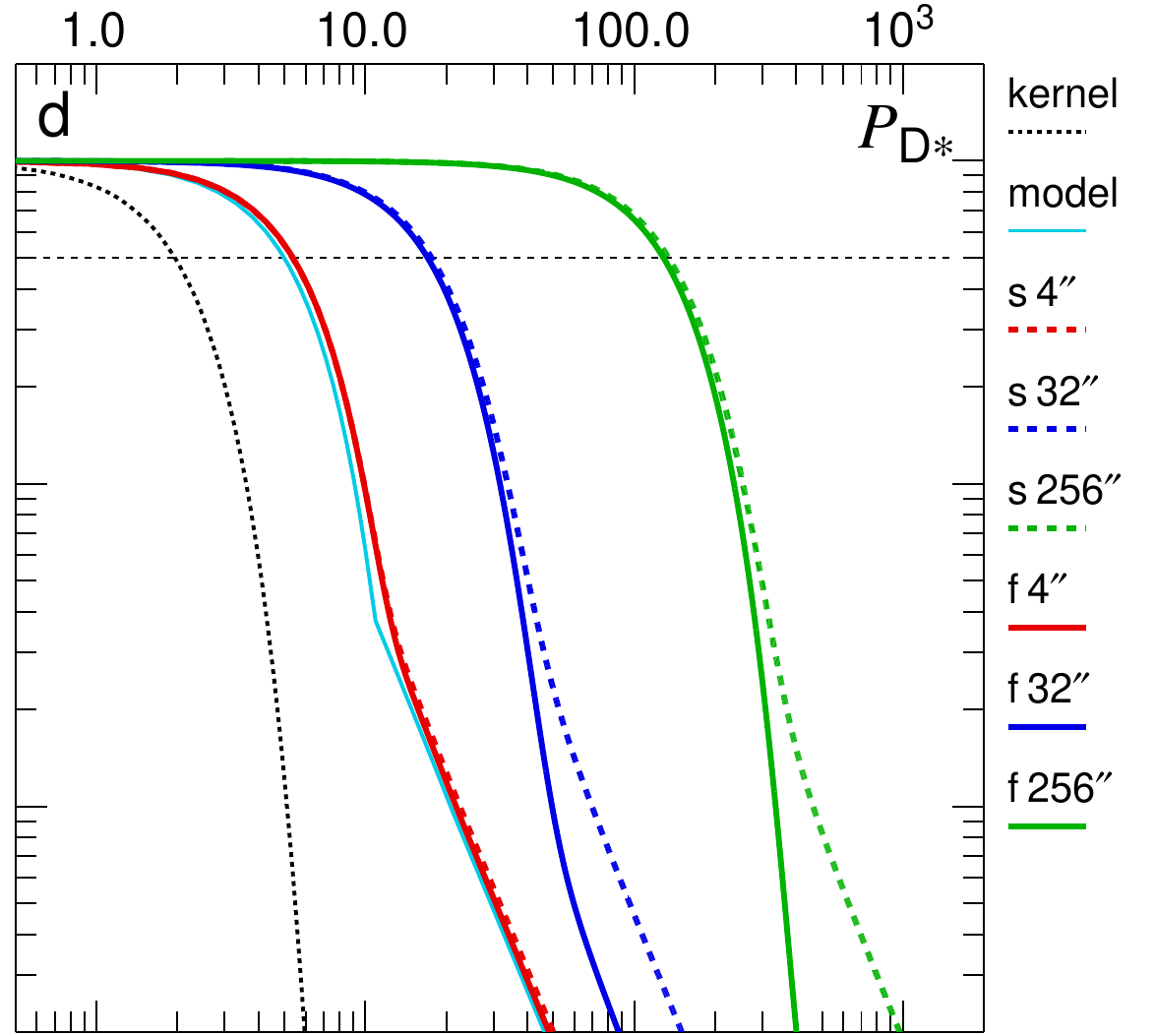}}}
\vspace{-0.46mm}
\centerline{
  \resizebox{0.2675\hsize}{!}{\includegraphics{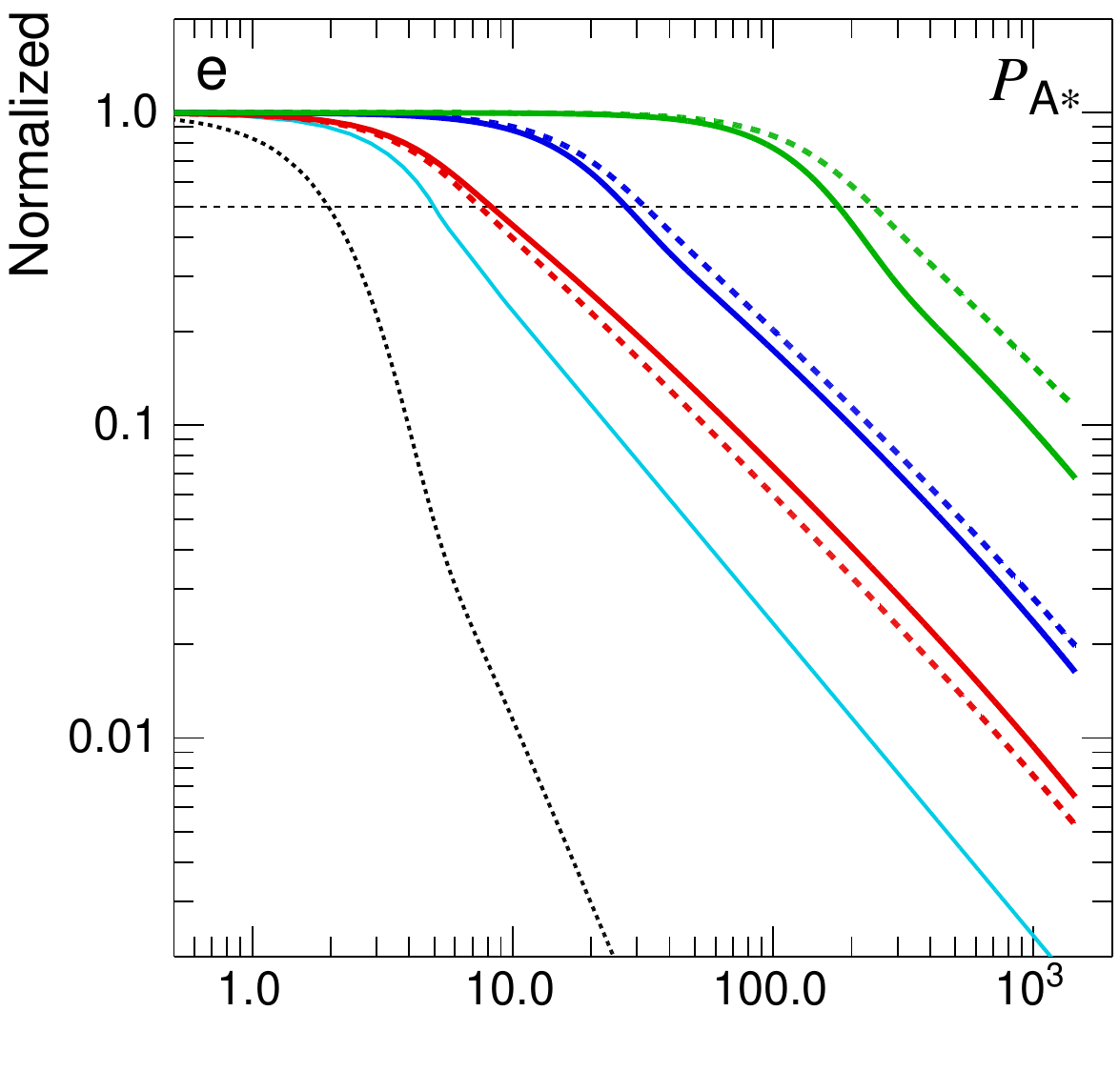}}  \hspace{-1.9mm}
  \resizebox{0.2832\hsize}{!}{\includegraphics{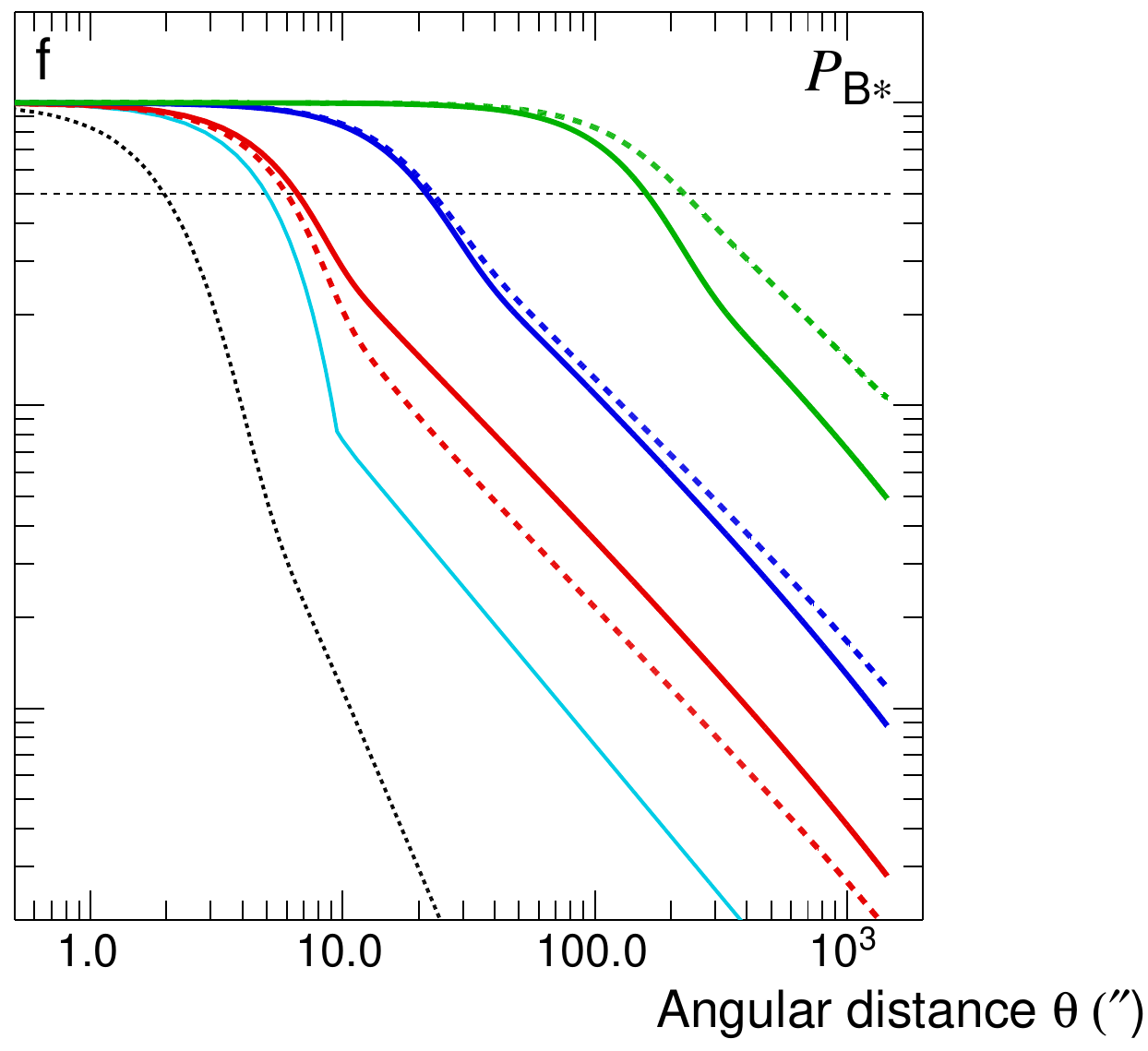}}  \hspace{-11.85mm}
  \resizebox{0.2287\hsize}{!}{\includegraphics{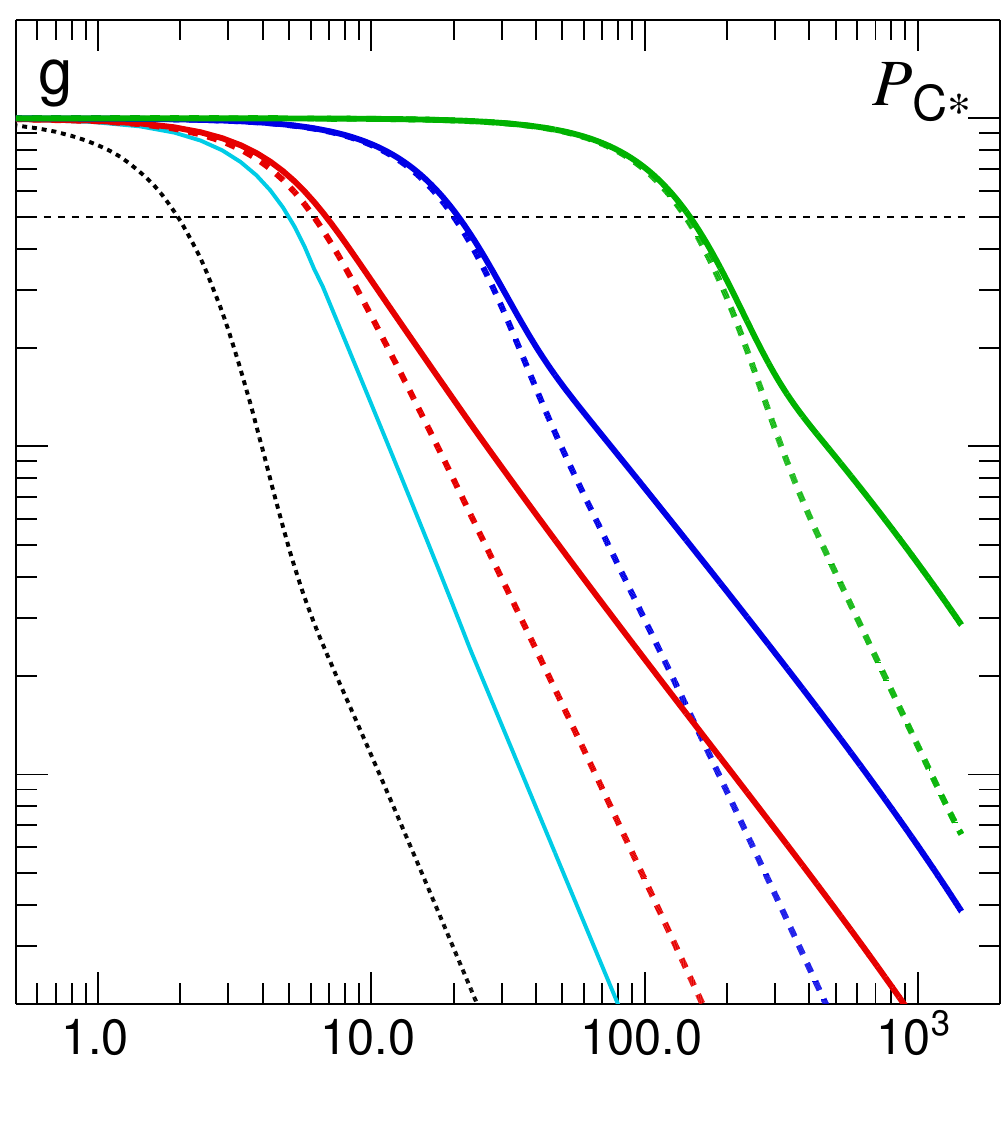}}  \hspace{-1.8mm}
  \resizebox{0.2675\hsize}{!}{\includegraphics{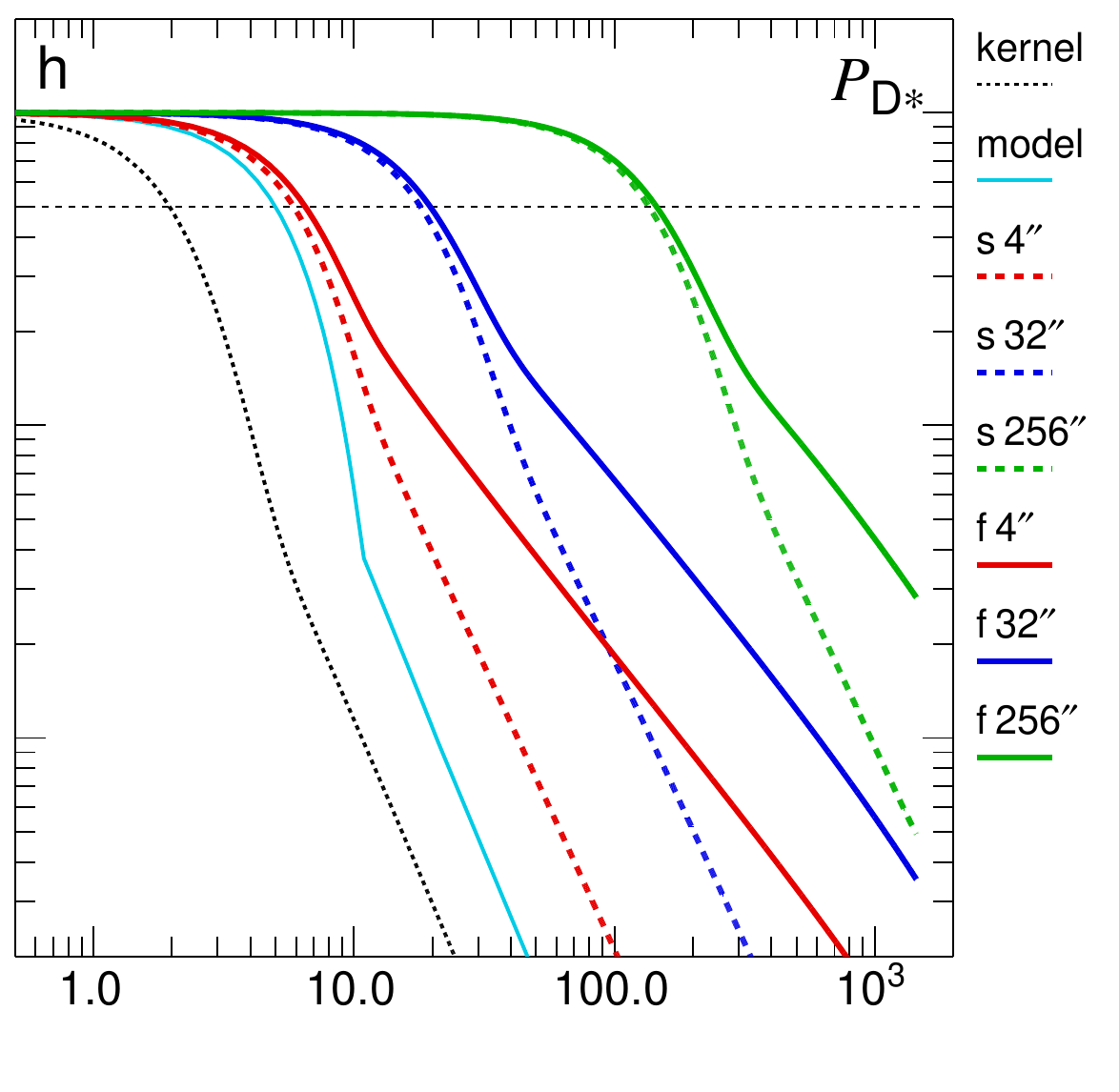}}}
\caption
{ 
Convolution of the infinite power-law models (no outer edge within the entire image area). The spherical (dashed curves) and
cylindrical (solid curves) models with the radial profiles, analogous to $\mathcal{P_{\rm \{A|B|C|D\}}}$ (Fig.~\ref{modelprofs}),
were convolved with two types of kernels (thin dotted curves) of the half maximum sizes of \{$4, 32$, $256$\}{\arcsec}. Shown are
the results for the pure Gaussian kernels $\mathcal{O}_{j}$ (\emph{top}) and power-law kernels $\mathcal{K}_{j}$ (\emph{bottom}).
The dashed horizontal lines indicate the half maximum level.
} 
\label{plawinfconv}
\end{figure*}

Most observed filaments appear significantly curved and (or) blended with a variety of other nearby structures. Convolution of a
wavy filament with a telescope PSF might produce asymmetric profiles in the segments of the filament, whose radii of curvature are
comparable to the angular resolution. When observed with telescope beams wider than the curvature radius of the filament segment,
the latter would appear as an elongated source. Observations indicate that filament profiles are often represented by shallow power
laws \citep[e.g.,][]{Palmeirim_etal2013,Arzoumanian_etal2019}, hence their footprints must have large and variable widths, which
implies large inaccuracies in the interpolated backgrounds and profiles. With higher angular resolutions, the apparently single
filaments may be expected to be resolved into thinner subfilaments \citep[e.g.,][]{Hacar_etal2018,Dewangan_etal2023}, like the
seemingly single sources in numerical simulations are resolved into clusters \citep{Louvet_etal2021}. Many unknowns and sources or
errors can make deconvolved sizes of the filaments, extracted from observations, fairly unreliable.

\begin{figure*}
\centering
\centerline{
  \resizebox{0.2675\hsize}{!}{\includegraphics{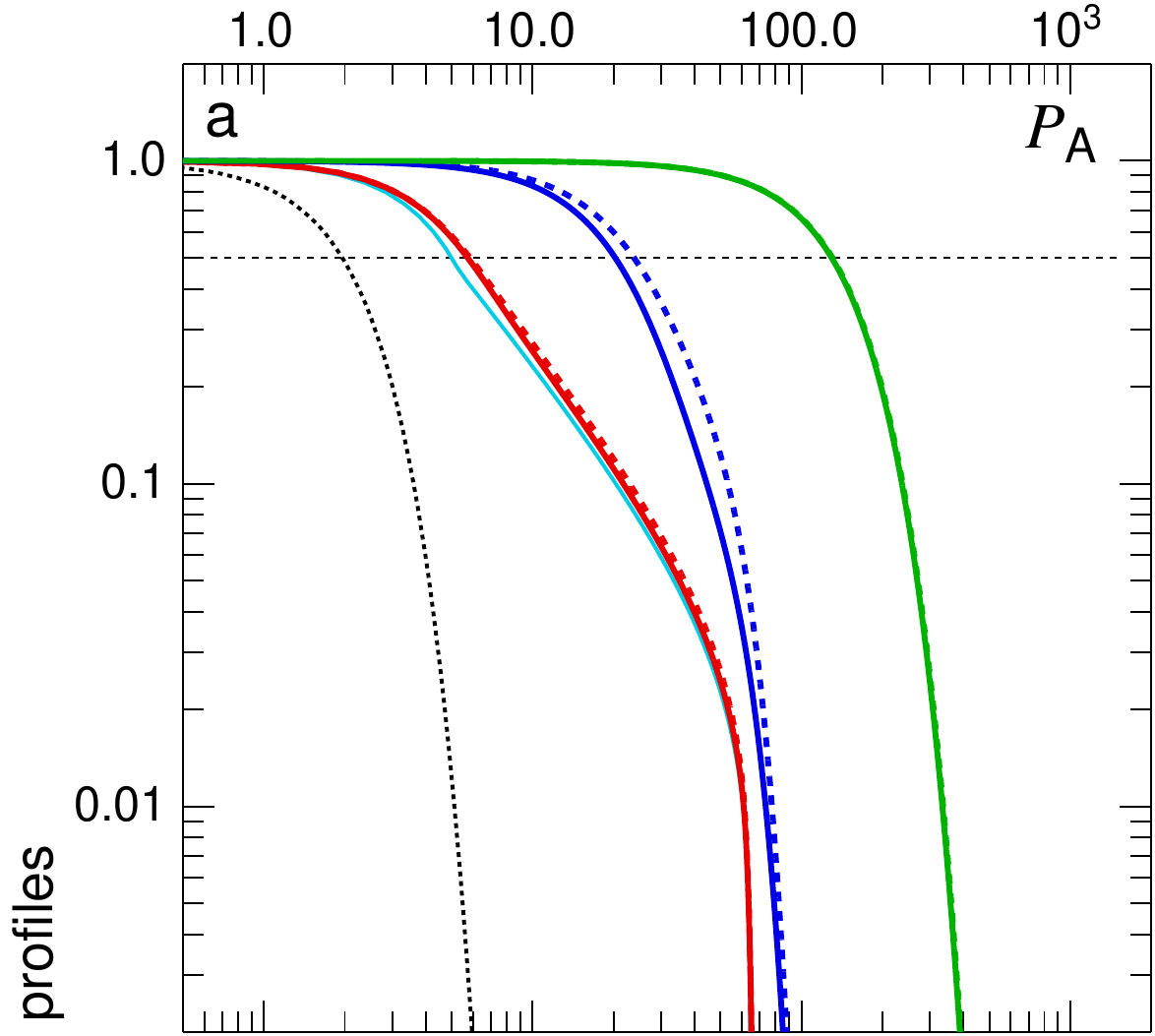}}  \hspace{-1.8mm}
  \resizebox{0.2287\hsize}{!}{\includegraphics{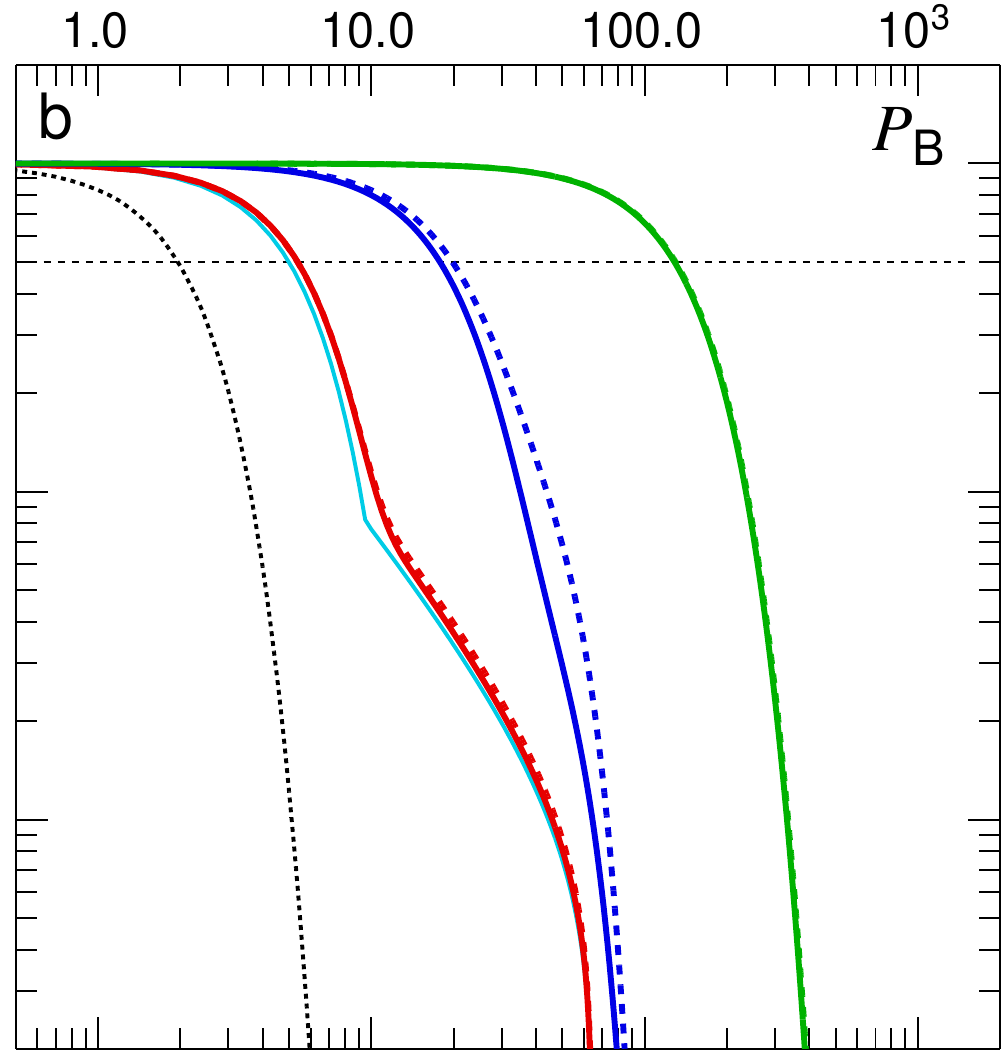}}  \hspace{-1.8mm}
  \resizebox{0.2287\hsize}{!}{\includegraphics{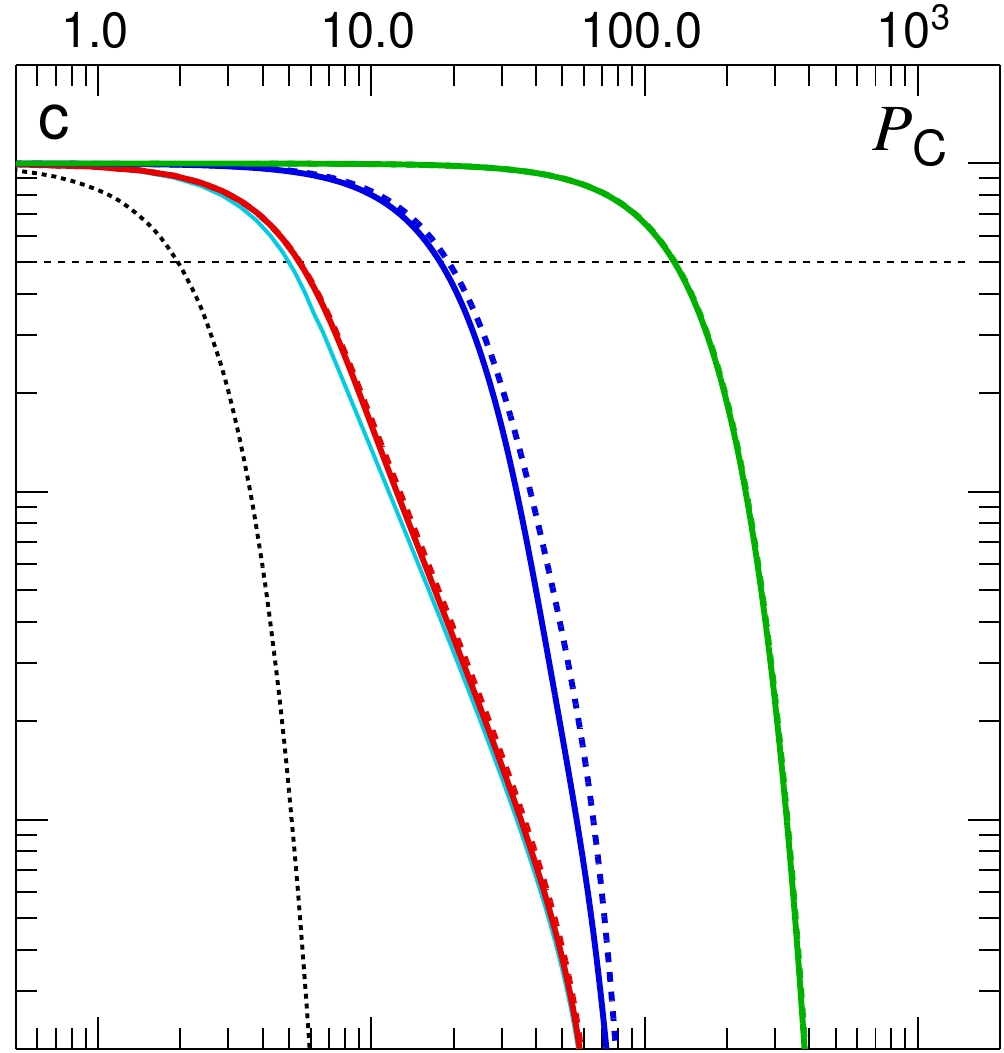}}  \hspace{-1.8mm}
  \resizebox{0.2675\hsize}{!}{\includegraphics{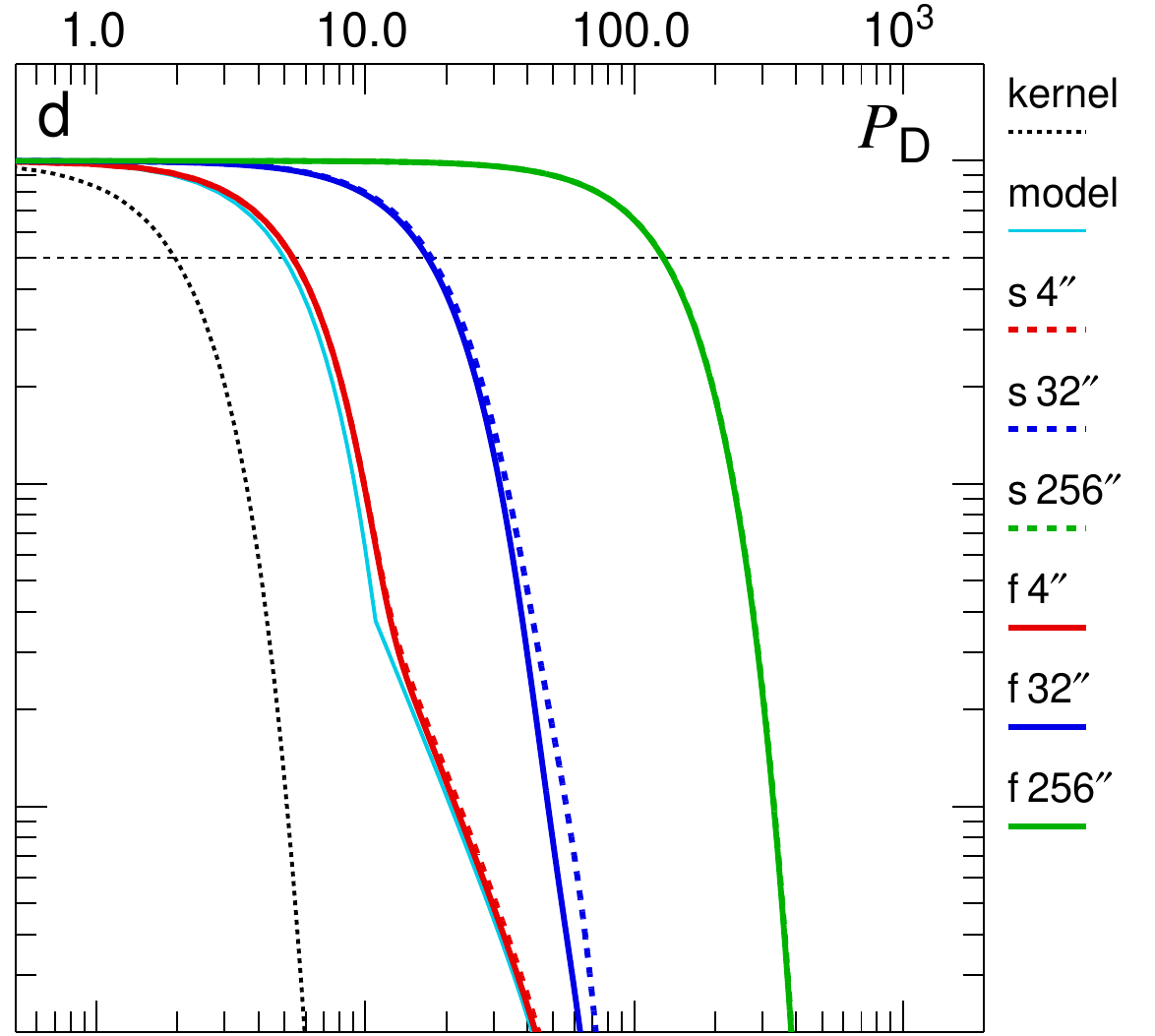}}}
\vspace{-0.46mm}
\centerline{
  \resizebox{0.2675\hsize}{!}{\includegraphics{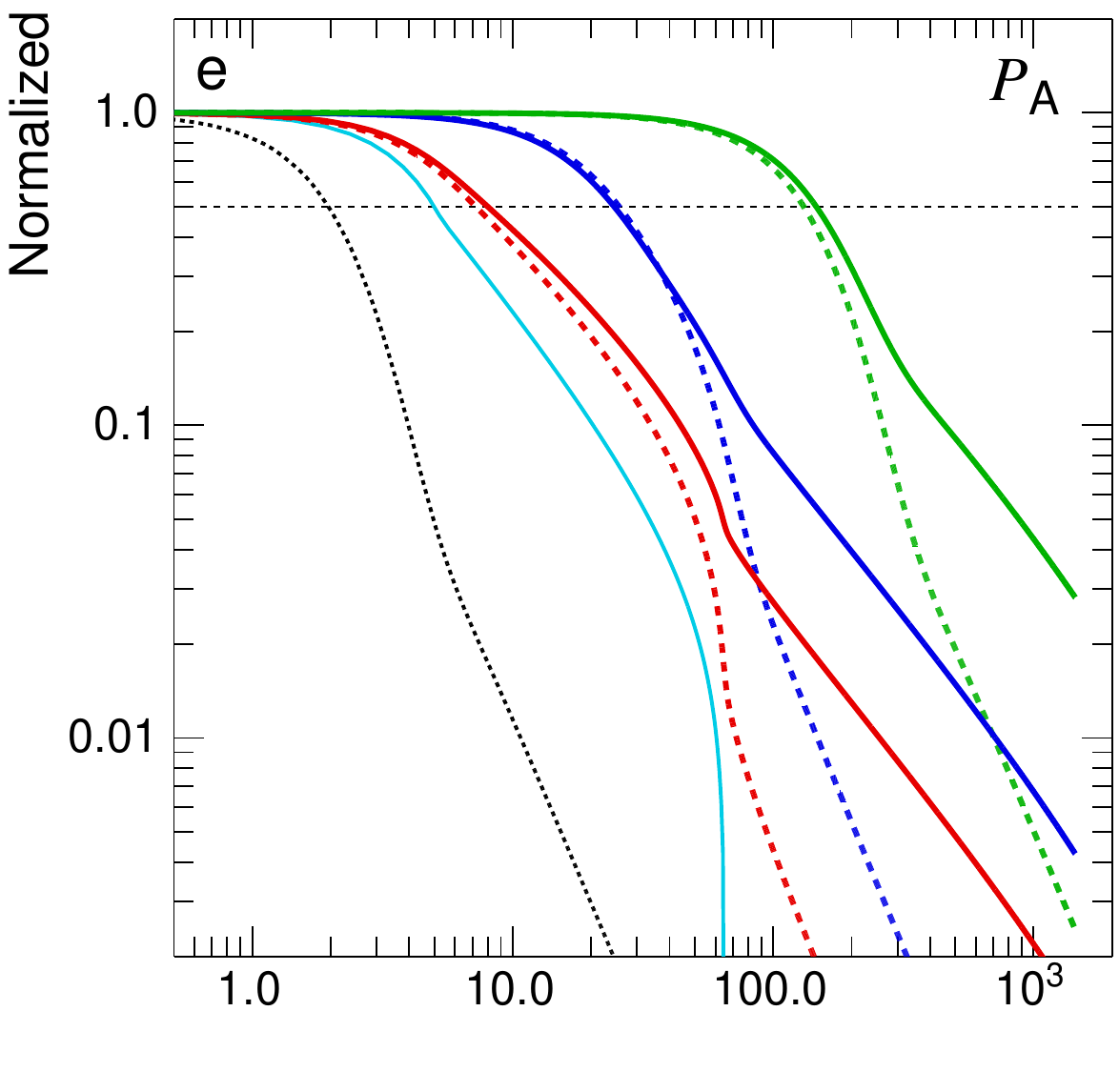}}  \hspace{-1.9mm}
  \resizebox{0.2832\hsize}{!}{\includegraphics{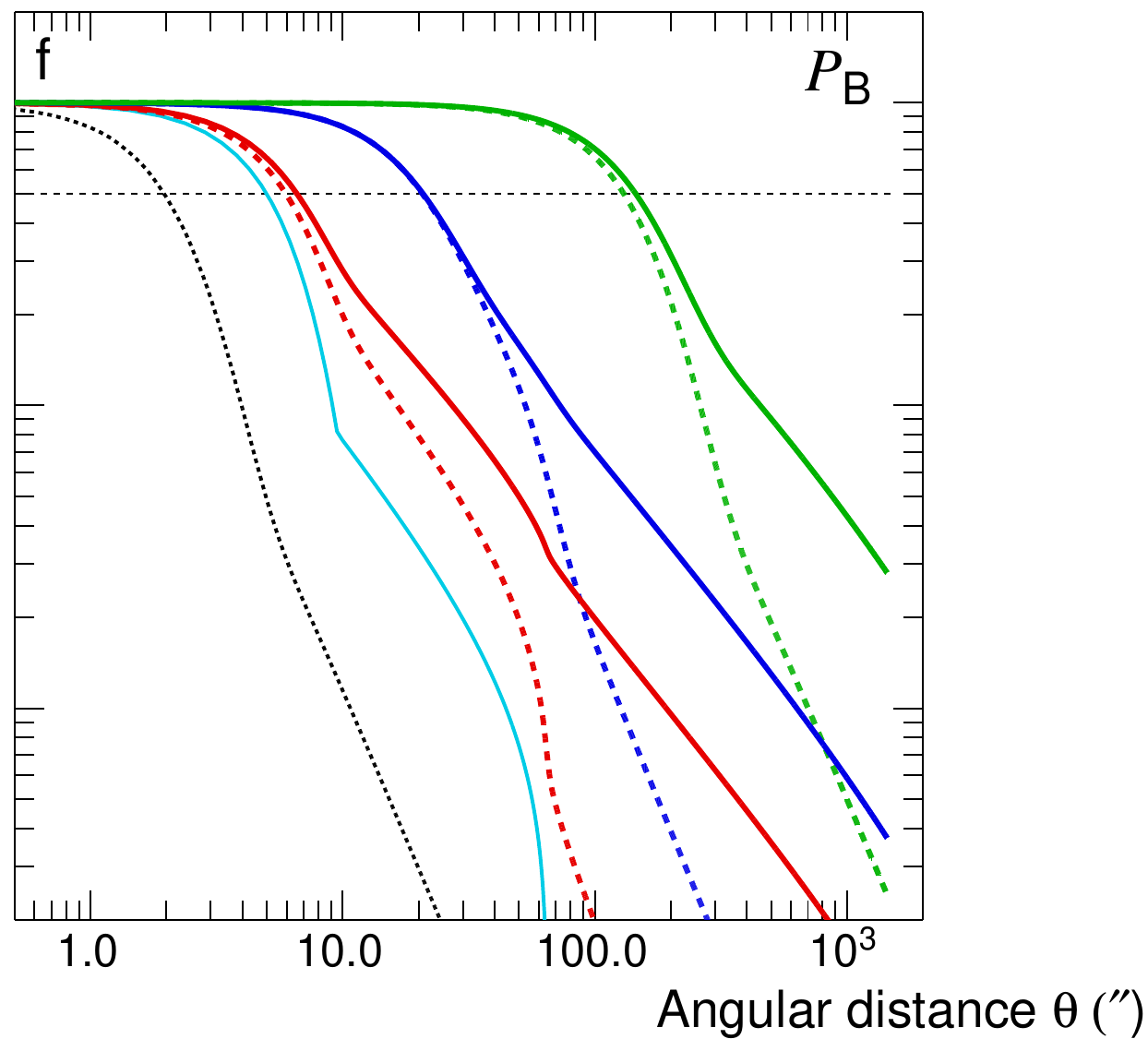}}  \hspace{-11.85mm}
  \resizebox{0.2287\hsize}{!}{\includegraphics{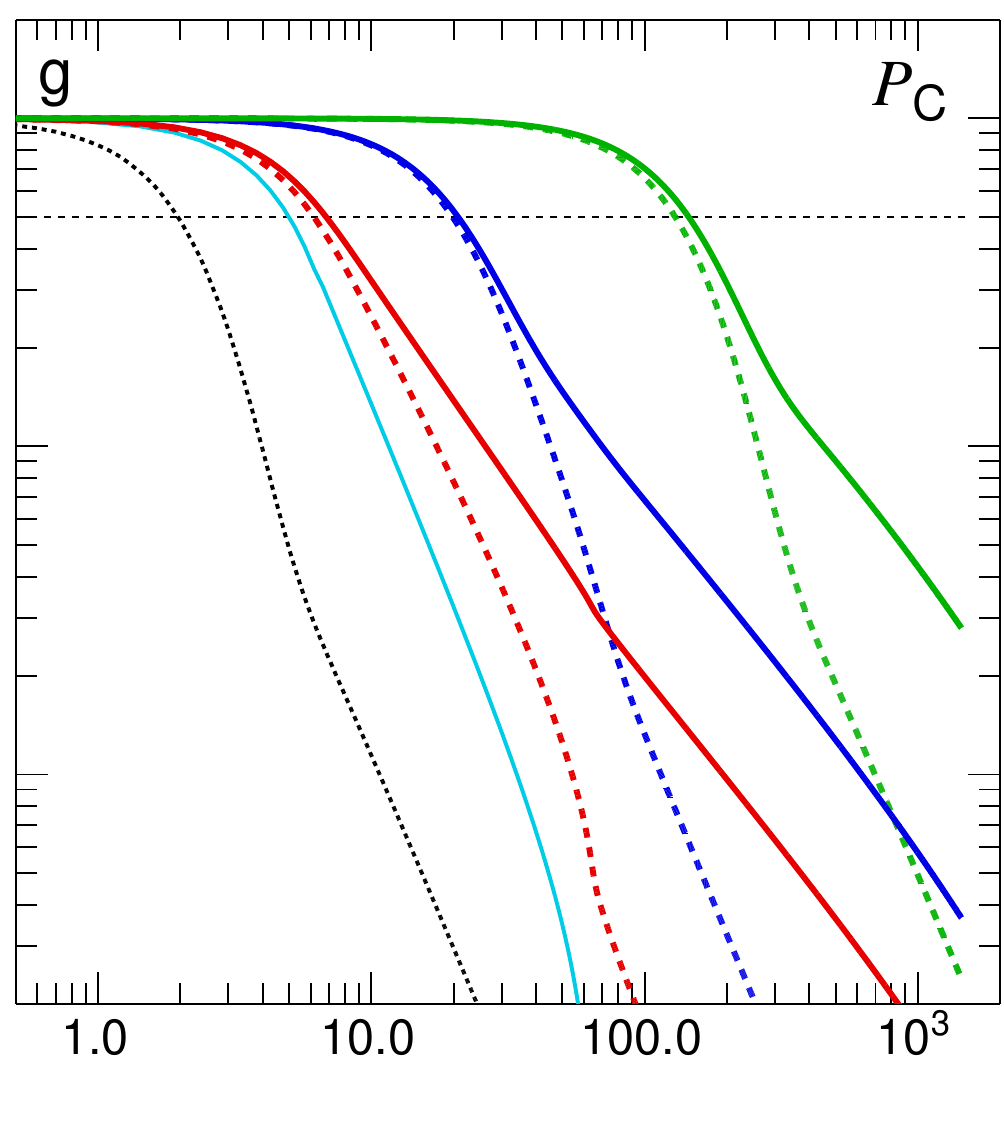}}  \hspace{-1.8mm}
  \resizebox{0.2675\hsize}{!}{\includegraphics{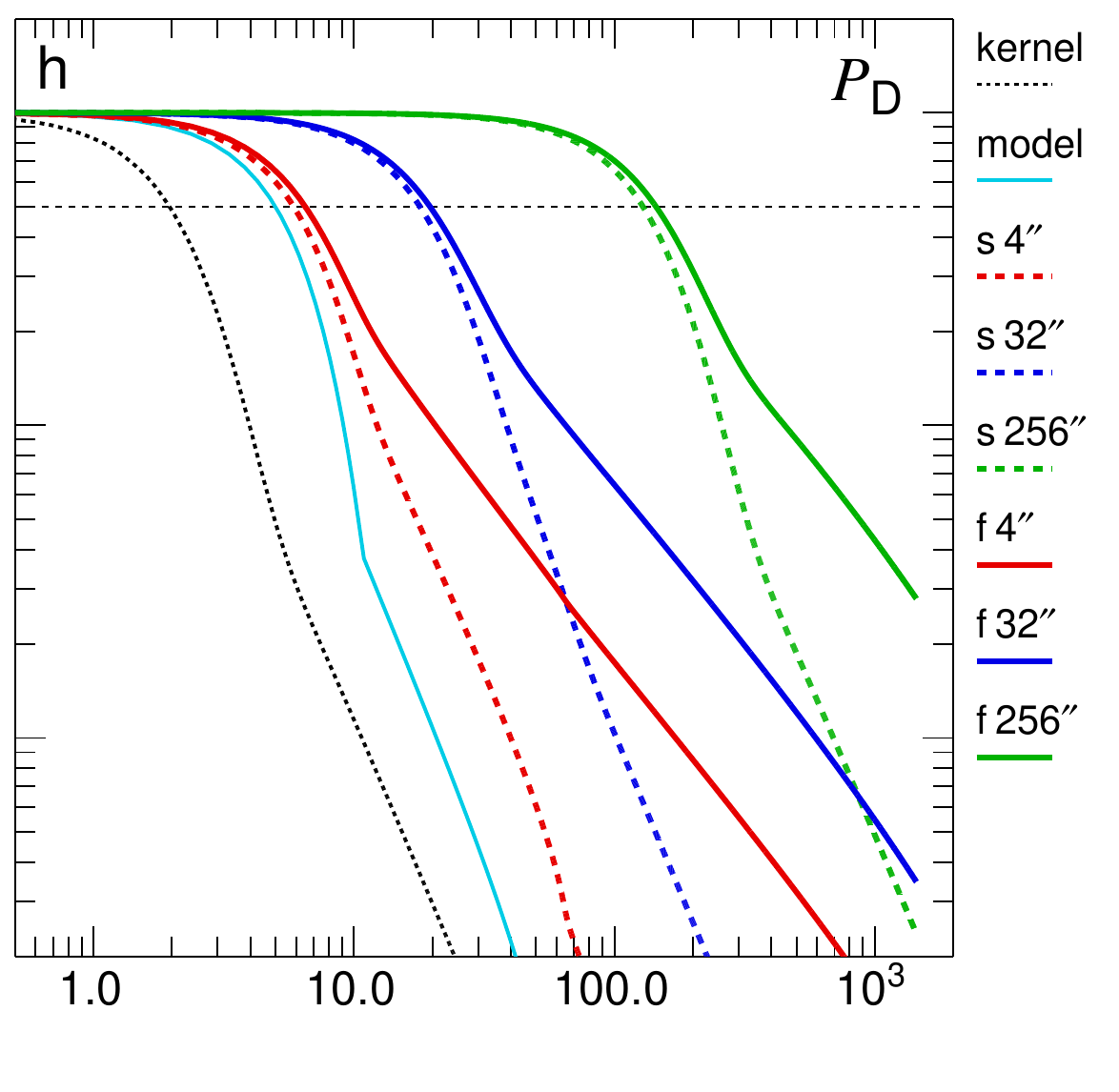}}}
\caption
{ 
Convolution of the finite power-law models (with an outer edge at $\Theta = 64${\arcsec}). The spherical (dashed curves) and
cylindrical (solid curves) models $\mathcal{P_{\rm \{A|B|C|D\}}}$ (Fig.~\ref{modelprofs}) were convolved with two types of kernels
(thin dotted curves) of the half maximum sizes of \{$4, 32$, $256$\}{\arcsec}. Shown are the results for the pure Gaussian kernels
$\mathcal{O}_{j}$ (\emph{top}) and power-law kernels $\mathcal{K}_{j}$ (\emph{bottom}). The dashed horizontal lines indicate the
half maximum level.
} 
\label{plawfinconv}
\end{figure*}

\subsection{Convolution effects}
\label{convsrcfil}

There are convolution effects for different geometries of structures that may be important for some studies. For an illustration,
it is sufficient to consider just two types of structures, represented by the round sources and straight filaments, discussed in
this paper. Application of the Gaussian source size deconvolution (Eq.~(\ref{deconvolution})) to the filamentary geometry carries
an implicit assumption that convolution always produces identical profiles for the two types of the morphologically different
structures. This assumption is only valid, when the source, filament, and convolution kernel have perfectly Gaussian profiles and
the filament crest is a straight line. Only in this simplistic case, convolution would produce the same radial profiles (and
deconvolved sizes) for both sources and filaments (Fig.~\ref{gausconv}), as expected.

In the astronomical studies, it is very unlikely that the observed structures have Gaussian profiles, especially after the
subtraction of an inaccurate background. Moreover, most of the observed physical objects are expected to be strongly non-Gaussian
(e.g., volume density $\rho \propto r^{-2}$, surface density $\sigma \propto \theta^{-1}$) and telescope PSFs usually have
non-Gaussian, roughly power-law shapes beyond their central (Gaussian) cores. To illustrate the general behavior of convolution for
different kernels, this section presents results for the Gaussian kernels $\mathcal{O}_{j}$ and simple power-law kernels
$\mathcal{K}_{j}$, consisting of the Gaussian core with the half maximum size $O_{j}$ and power-law wings $I \propto
\theta^{-\varkappa}$ with $\varkappa = 2$ (Fig.~\ref{gausconv}). The simple power law, which is somewhat shallower and more intense
than those shown by the more complex PSFs of orbital telescopes, was chosen to better illustrate the convolution effects.

The spherical and cylindrical power-law models, described in Sect.~\ref{models}, may be referred to as the finite models, because
they have an outer boundary at $\Theta = 64${\arcsec} (Fig.~\ref{modelprofs}) and there are angular resolutions $O_{j} > 2 \Theta$,
at which the entire model becomes unresolved. It is also useful to consider the infinite models with the same power-law radial
density distributions and an outer boundary, imposed by the image edge at $\theta = 1429${\arcsec}, that are resolved at all
angular resolutions. The Gaussian models are always finite, because their steep exponential profile has a negligible contribution
at $\theta \ga 20${\arcsec} (Fig.~\ref{modelprofs}).

Convolution of the structures of different geometry with different kernels can seriously affect both the widths and slopes of the
model profiles. The following discussion refers to the widths as the half maximum sizes (Sect.~\ref{measurements}), and to the
slopes, evaluated numerically at the angular distances around $\theta \approx 1000${\arcsec}, where all convolved power-law
profiles have converged to their asymptotic behavior. Whenever the convolved models do not appear to have power-law profiles, the
slopes are not computed.

\subsubsection{Widths of the spherical and cylindrical models}

Figure~\ref{gausconv} demonstrates that the cylindrical Gaussian model $\mathcal{G}$, convolved with the power-law kernel
$\mathcal{K}_{j}$, is systematically wider than the equivalent spherical model. Conversely, the top panels in
Fig.~\ref{plawinfconv} show that the infinite cylindrical power-law models $\mathcal{P}_{\rm \{A|B|C|D\}*}$, convolved with a
Gaussian kernel $\mathcal{O}_{j}$, are systematically narrower than the equivalent spherical power-law models. However, the bottom
panels in Fig.~\ref{plawinfconv} reveal that the same cylindrical models, convolved with the power-law kernel $\mathcal{K}_{j}$,
are systematically wider than the spherical models, with the exception of models $\mathcal{P}_{\rm \{A|B\}*}$ at the low
resolutions of $32$ and $256${\arcsec} that become narrower than the spherical models (Fig.~\ref{plawinfconv} (\emph{e},
\emph{f})). The reason for the appearance of the two exceptional cases is that they are produced by the infinite models with
the shallowest power-law profile, while the power-law kernel $\mathcal{K}_{j}$ is steeper than the model profile ($\varkappa >
\beta_{\mathcal{M}}$) and wider than the model half maximum size ($O_{j} \ga H_{\mathcal{M}} = 10${\arcsec}). Apparently, the
convolution in such cases with respect to the relative widths of the convolved cylindrical and spherical models behaves as if the
kernel had a Gaussian shape. The transition to the exceptional cases in Fig.~\ref{plawinfconv} (\emph{e}, \emph{f}) takes place at
$O_{j} \approx H_{\mathcal{M}}$, at the border of the resolved domain ($R_{j} \approx 1.4$, Eq.~(\ref{resolstate})).
Table~\ref{infconvol} presents the accuracy $D_{j}/H$ of the deconvolved sizes for the infinite power-law structures
(Fig.~\ref{plawinfconv}) that are especially greatly overestimated for the unresolved sources, reaching a factor of $40$ at $O_{j}
= 256${\arcsec}.


Figure~\ref{plawfinconv} reveals that the steepening of the intensity profile near the outer boundary at $\Theta = 64${\arcsec} in
the finite power-law models $\mathcal{P}_{\rm \{A|B|C|D\}}$ makes the results of their convolution with a Gaussian kernel
$\mathcal{O}_{j}$ very similar in both spherical and cylindrical geometries, the spherical model being slightly wider. However, the
same cylindrical power-law models, convolved with the power-law kernel $\mathcal{K}_{j}$, are systematically wider than the
equivalent spherical model (Fig.~\ref{plawfinconv}), with the exception of model $\mathcal{P_{\rm A}}$ at the resolution of
$32${\arcsec}. The inversion with respect to the relative widths of the convolved cylindrical and spherical models is confined to
the range of resolutions $10 \la O_{j} \la 70${\arcsec}. The endpoints of this range correspond to $O_{j} \approx H_{\mathcal{M}}$
and $O_{j} \approx \Theta = 64${\arcsec}, where the former points to the border of the resolved domain ($R_{j} \approx 1.4$) and
the latter corresponds to the sharp drop of the profile $I \propto \theta^{-1}$ at the boundary of the spherical models
(Fig.~\ref{modelprofs}). Table~\ref{finconvol} presents the accuracy $D_{j}/H$ of the deconvolved sizes for the finite Gaussian and
power-law structures (Figs.~\ref{gausconv} and \ref{plawfinconv}) that are overestimated for the unresolved filaments up to a
factor of $14$ at $O_{j} = 256${\arcsec}.

\begin{figure*}
\centering
\centerline{
  \resizebox{0.2675\hsize}{!}{\includegraphics{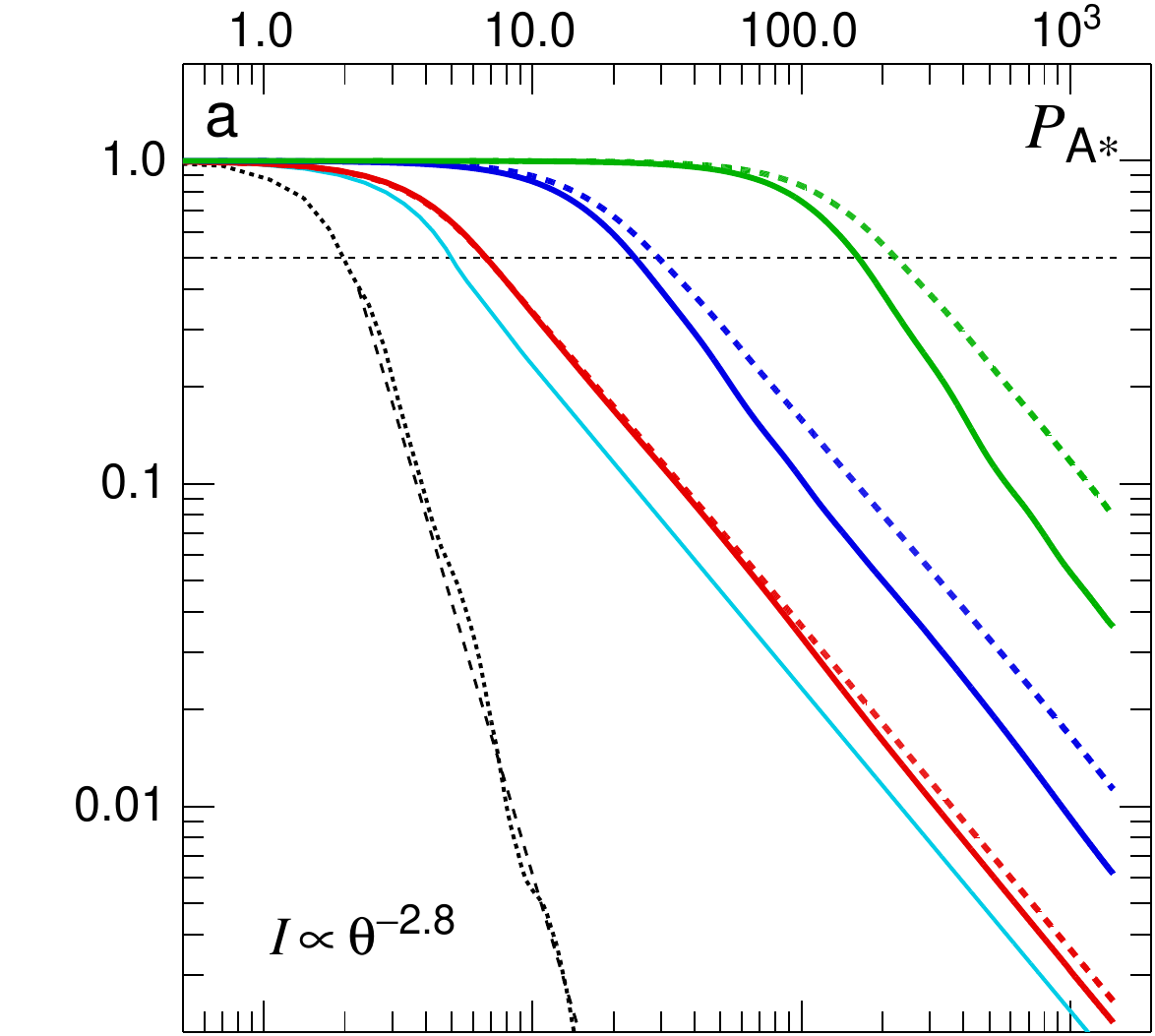}}  \hspace{-1.8mm}
  \resizebox{0.2287\hsize}{!}{\includegraphics{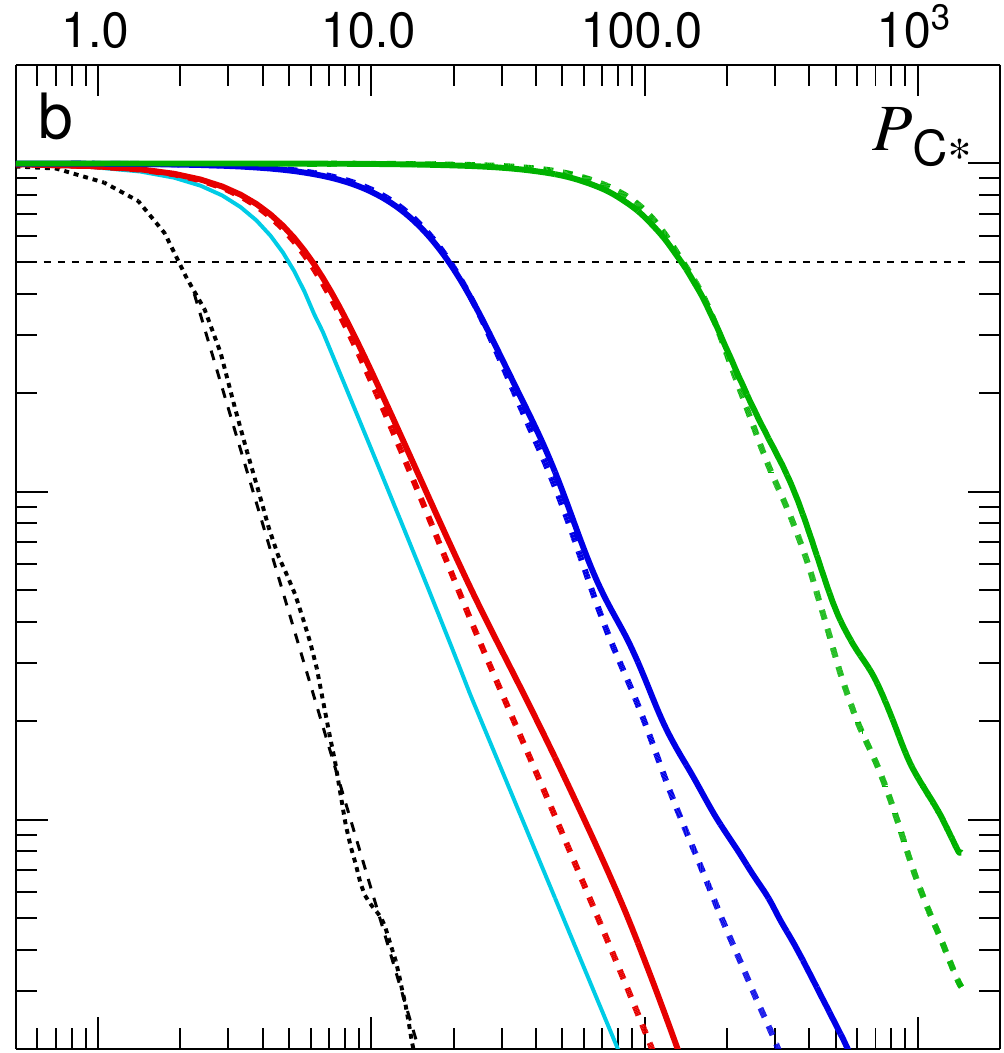}}  \hspace{-1.8mm}
  \resizebox{0.2706\hsize}{!}{\includegraphics{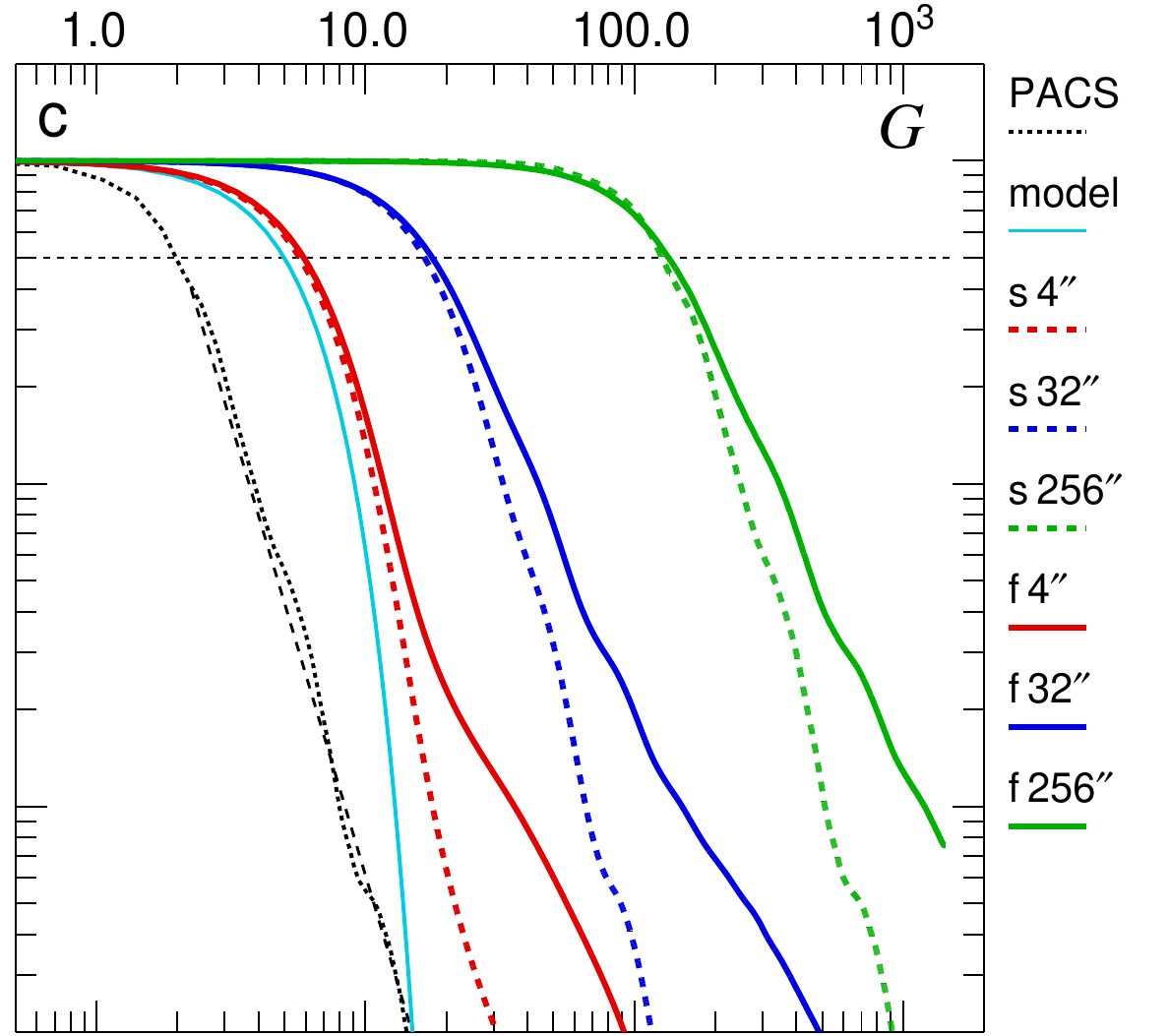}}}
\vspace{-0.94mm}
\centerline{
  \resizebox{0.2675\hsize}{!}{\includegraphics{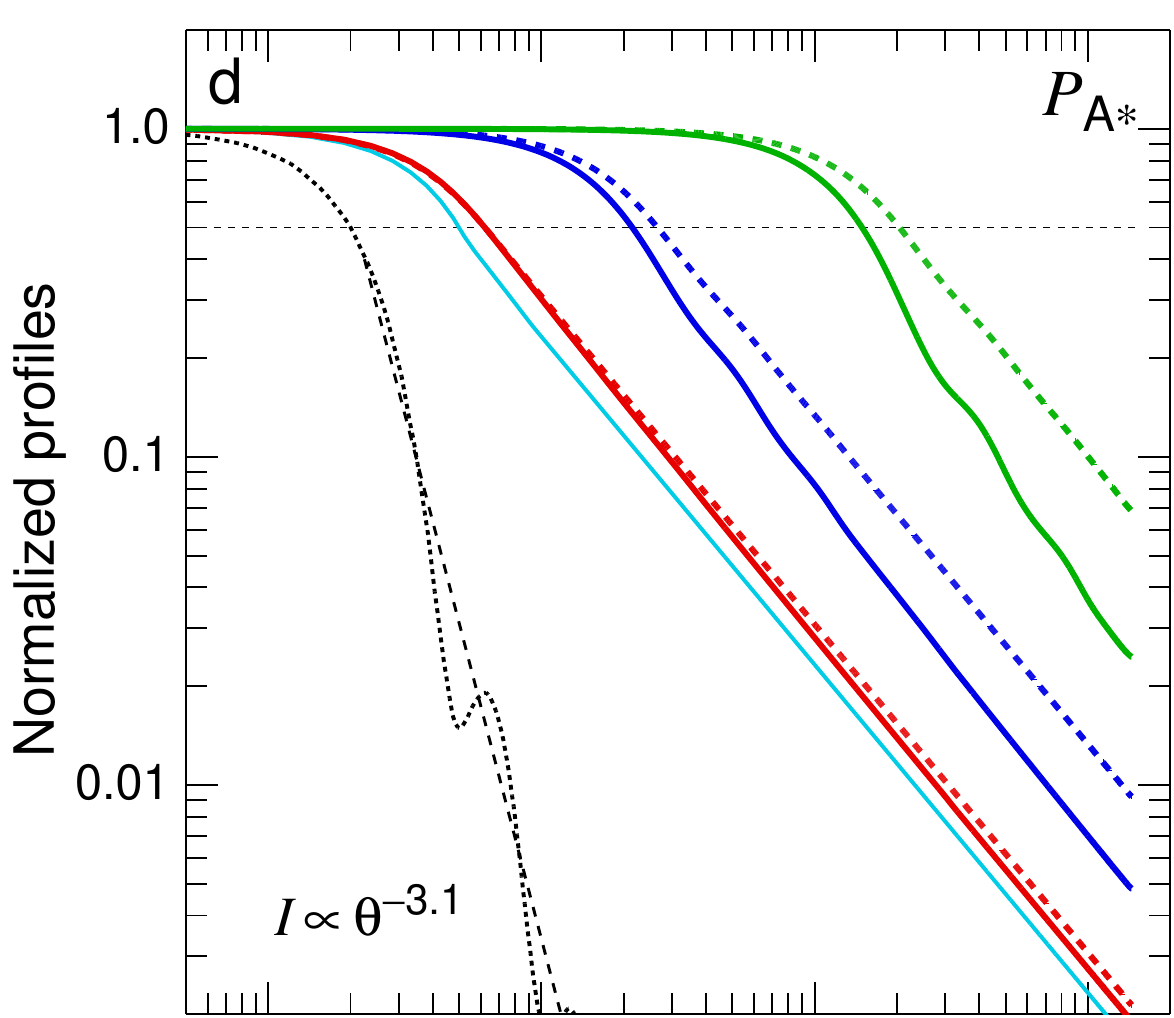}}  \hspace{-1.8mm}
  \resizebox{0.2287\hsize}{!}{\includegraphics{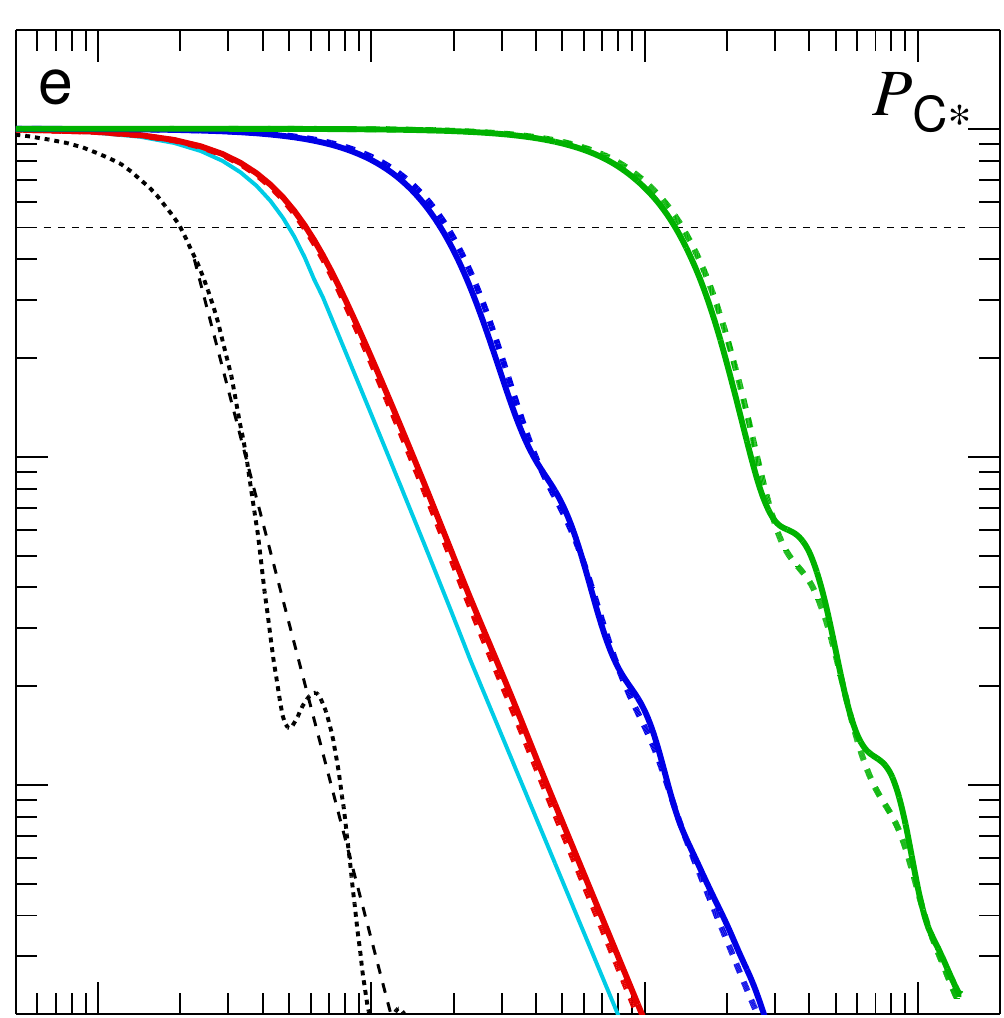}}  \hspace{-1.8mm}
  \resizebox{0.2706\hsize}{!}{\includegraphics{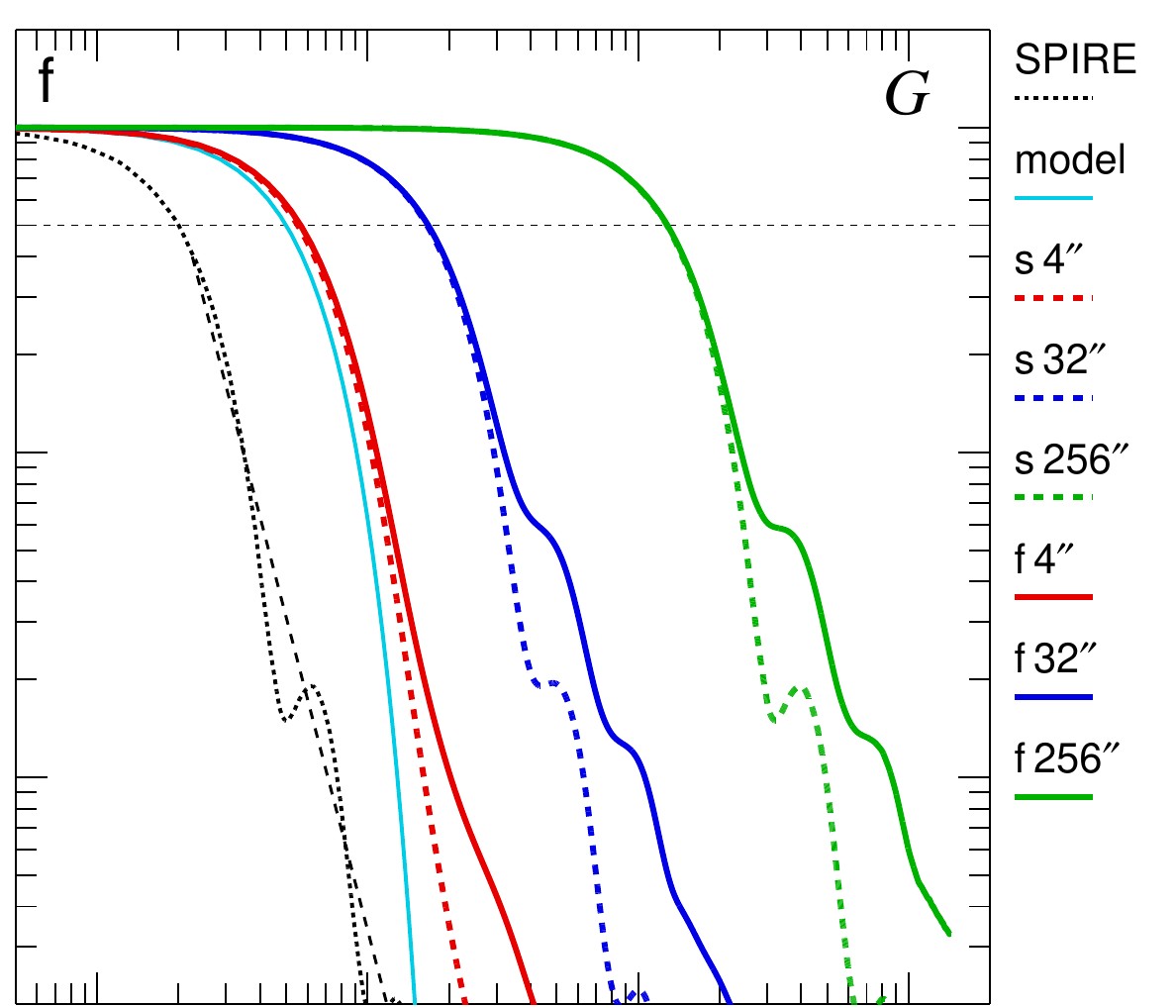}}}
\vspace{-0.46mm}
\centerline{
  \resizebox{0.2675\hsize}{!}{\includegraphics{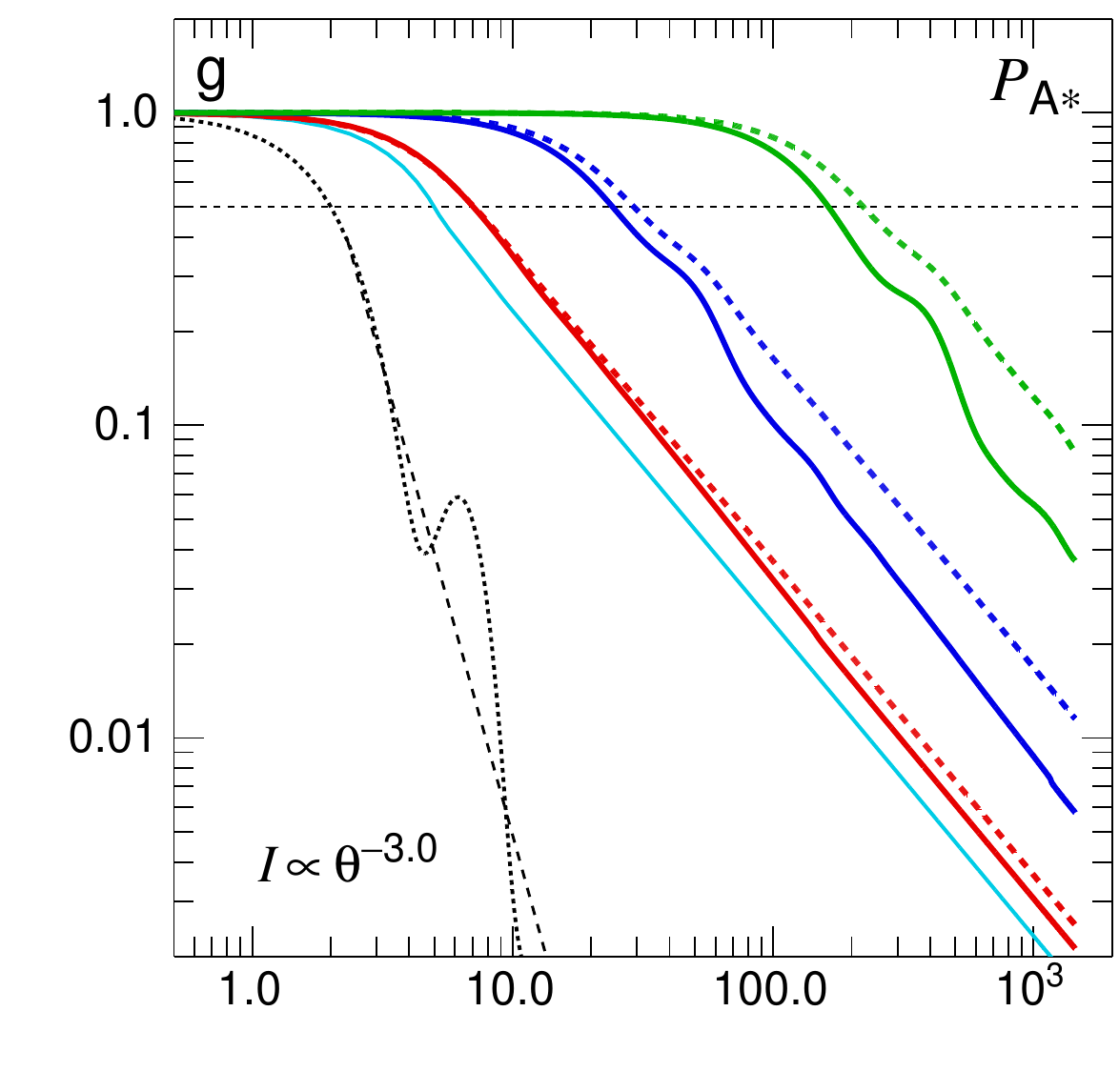}}  \hspace{-1.9mm}
  \resizebox{0.2832\hsize}{!}{\includegraphics{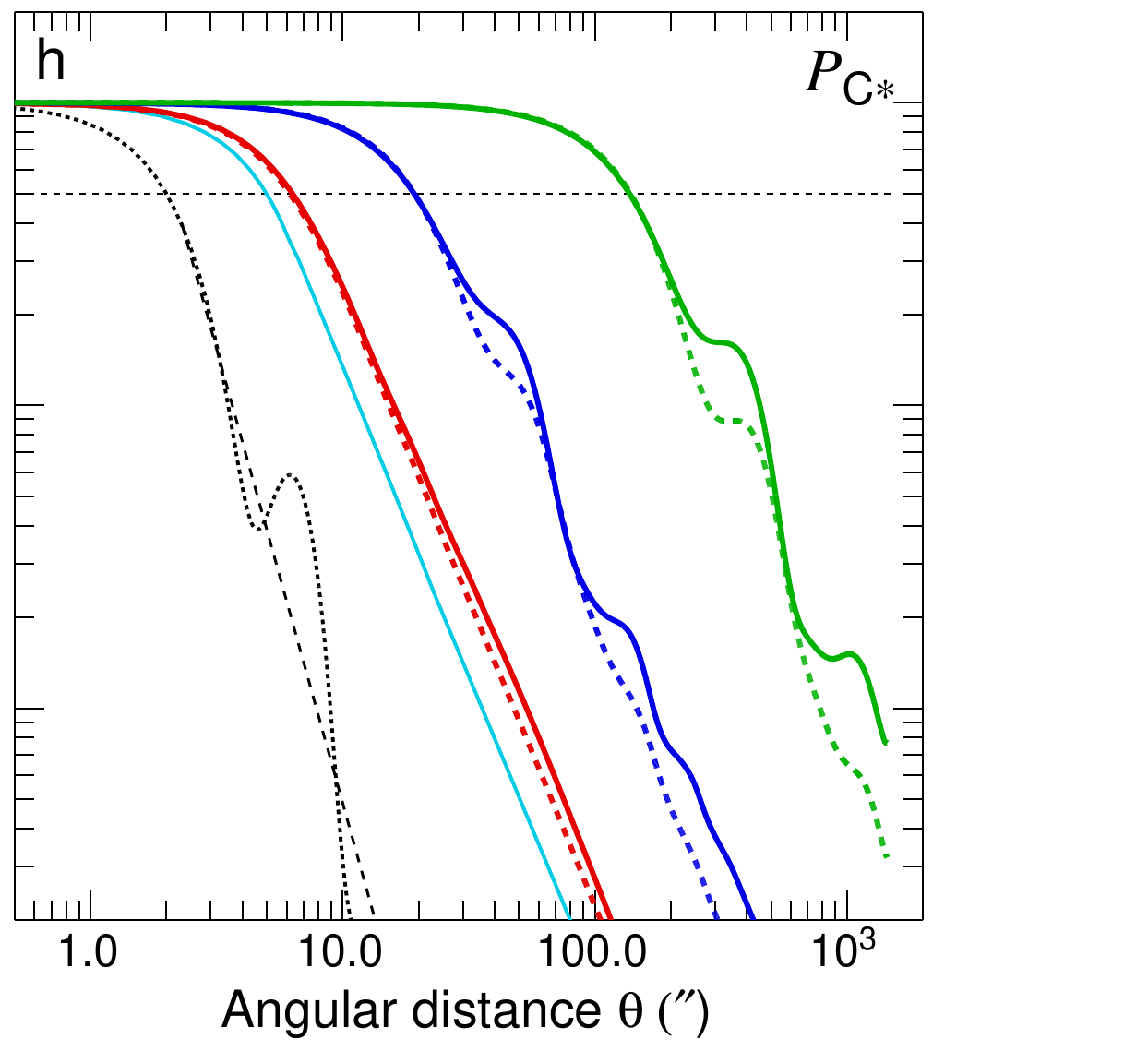}}  \hspace{-11.85mm}
  \resizebox{0.2706\hsize}{!}{\includegraphics{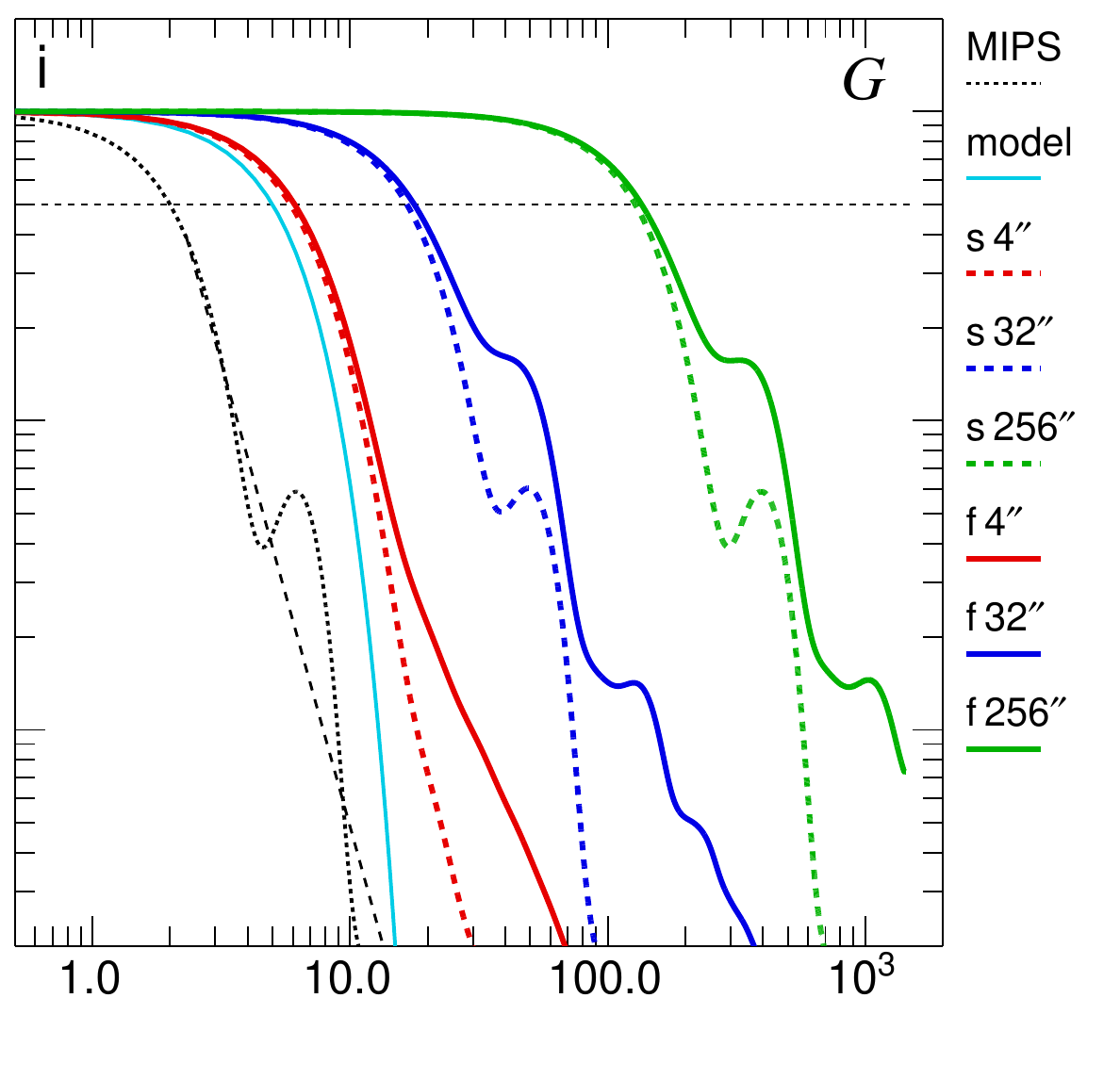}}}
\caption
{ 
Convolution of the Gaussian and infinite power-law models. The spherical (dashed curves) and cylindrical (solid curves) models
from Figs.~\ref{gausconv} and \ref{plawinfconv}, were convolved with two kernels (thin dotted curves) of the half maximum sizes of
\{$4, 32$, $256$\}{\arcsec}, with the PSF shapes \citep{Aniano_etal2011} of \emph{Herschel} PACS at $70$\,$\mu$m (\emph{top}),
\emph{Herschel} SPIRE at $250$\,$\mu$m (\emph{middle}), and \emph{Spitzer} MIPS at $160$\,$\mu$m (\emph{bottom}). The black dashed
lines visualize the approximate power-law slopes of the PSFs, derived by a visual fitting of their profiles over 5 orders of
magnitude. The dashed horizontal lines indicate the half maximum level.
} 
\label{anianoconv}
\end{figure*}

\subsubsection{Slopes of the spherical and cylindrical models}

Figures~\ref{gausconv}\,--\,\ref{plawfinconv} reveal an effect with potential consequences for the observational studies of sources
and filaments. When a convolution kernel has substantially steeper wings than the power-law profile of the model object ($\varkappa
\ge \beta_{\mathcal{M}} + 1$), the resulting slopes $\gamma_{\rm \{S|F\}}$ of the convolved sources and filaments resemble those of
the true model profiles ($\gamma_{\rm \{S|F\}} \approx \beta_{\mathcal{M}}$), at all angular resolutions. This behavior is shown by
all models after the convolution with the Gaussian kernels $\mathcal{O}_{j}$ (Fig.~\ref{plawinfconv} (\emph{a}\,--\,\emph{d})) and
by the shallower power-law models $\mathcal{P}_{\rm \{A|B\}}$ ($\beta_{\mathcal{M}} = 1$) after the convolution with a steeper
kernel ($\varkappa = 2$, Fig.~\ref{plawinfconv} (\emph{e}, \emph{f})).

In contrast, when the power-law wings of the kernel $\mathcal{K}_{j}$ are shallower than the profiles of the model objects
($\varkappa < \beta_{\mathcal{M}}$), the convolved sources tend to resemble the kernel profile ($\gamma_{\rm S} \approx
\varkappa$), whereas the convolved filaments become significantly shallower ($\gamma_{\rm F} \approx \varkappa - 1$). Essentially,
the shallower kernel ``feels'' that the filament geometry has one more dimension than the source geometry, whereas the Gaussian
kernel $\mathcal{O}_{j}$ is too steep to sense the differences. This behavior is displayed by the Gaussian models
(Fig.~\ref{gausconv} (\emph{b}), Table~\ref{finconvol}), by the infinite models $\mathcal{P}_{\rm \{C|D\}*}$
(Fig.~\ref{plawinfconv} (\emph{g}\,--\,\emph{h}), Table~\ref{infconvol}), as well as by the finite models $\mathcal{P}_{\rm
\{A|B|C|D\}}$ (Fig.~\ref{plawfinconv} (\emph{e}\,--\,\emph{h})). Notably, the convolution differences between the profiles of
sources and filaments are present even when the structures are well resolved ($R_{j} \ge 3.4$ at resolutions $O_{j} \le
4${\arcsec}).

Convolutions of models $\mathcal{\{G|P_{*}\}}$ with the power-law kernels $\mathcal{K}_{j}$ ($1.5 \le \varkappa \le 6$) confirm
the geometrical differences as the general property of convolution that depends on the slopes $\varkappa$ of the kernel and $\beta$
of the object,
\begin{equation} 
\gamma_{\rm S} \approx \min \left(\varkappa,\,\beta\right), \,\,\, 
\gamma_{\rm F} \approx \min \left(\varkappa-1,\,\beta\right).
\label{coneffect}
\end{equation} 
For example, when $\varkappa = \beta$, the convolved source reflects the correct slope of the object ($\gamma_{\rm S} = \beta$),
whereas the convolved filament acquires a much shallower shape than that of the object ($\gamma_{\rm F} = \beta - 1$). The kernels
that are too steep with respect to the object being convolved ($\varkappa \ge \beta + 1$) cannot sense the differences in geometry
and, therefore, they produce the same profiles for sources and filaments, consistent with the intrinsic profile of the object
($\gamma_{\rm S} = \gamma_{\rm F} = \beta$).

Figure~\ref{anianoconv} displays similar convolution differences also for the \emph{Spitzer} and \emph{Herschel} PSFs
\citep{Aniano_etal2011}, affected by diffraction patterns, that can be approximated by somewhat steeper slopes ($\varkappa \approx
2.8-3.1$). According to Eq.~(\ref{coneffect}), a filamentary object with an intrinsic slope $\beta = 2$, imaged by telescopes with
the PSF slopes of $\varkappa \ga 2.5$, might appear as having shallower slopes in the range $1.5 \la \gamma_{\rm F} \la 2$.
Convolved sources can also appear with shallower profiles, but kernels must be even shallower ($\varkappa < \beta$). In other
words, a spherical object observed with the above range of the PSF kernels, would always appear with the slope $\gamma_{\rm S} = 2$
of the true intrinsic profile ($\beta = 2$). The convolution effects appear blurred for the PSFs with more strongly oscillating
profiles and steeper power-law wings. Table~\ref{infpsfs} presents the accuracy $D_{j}/H$ of the deconvolved sizes for the
Gaussian and infinite power-law structures convolved with the above PSFs that end up overestimated for the unresolved sources
by factors of up to $37$ at $O_{j} = 256${\arcsec}.

It must be noted that the above discussion assumes no background or no errors in the estimated background. It is known, however,
that subtraction of an inaccurate background can also seriously affect the measured slopes of the observed sources and filaments
(Fig.~\ref{sourcesbgs}).

\section{Conclusions}
\label{conclusions}

This work presents a model-based investigation of the inaccuracies and biases of the simple method of Gaussian size deconvolution
(Eqs.~(\ref{deconvolution}) and (\ref{maximized})), routinely applied in many astrophysical studies to both sources and filaments
extracted from observed images. Simulated images of spherical and cylindrical objects with identical radial profiles include a
Gaussian model as an idealized reference case, in addition to power-law models $I \propto \theta^{-\beta}$ and a critical
Bonnor-Ebert sphere, representing the dense cores and filaments observed in star-forming regions. To simulate observations of the
model objects, the images were convolved to a wide range of angular resolutions to probe various degrees of resolvedness
$\tilde{R}_{j}$ (Sect.~\ref{resolve}), from the unresolved to the well-resolved structures (Eq.~(\ref{resolstate})). Three types of
the model backgrounds (flat, convex, and concave), represent the simplified average shapes of fluctuating backgrounds in molecular
clouds. Following the procedure used in source and filament extraction methods, planar backgrounds across the footprints of the
structures were interpolated and subtracted, then sizes of the sources and filaments were measured and deconvolved.

The results demonstrate that the Gaussian size deconvolution is accurate for all angular resolutions in a single simplistic case --
when the telescope PSF and the background-subtracted structures obey the assumption of their Gaussian shapes. In realistic
situations, when background subtraction happens to be inaccurate, the extracted sources and filaments acquire profoundly
non-Gaussian profiles. When the structures are unresolved ($1 < \tilde{R}_{j} \le 1.1$), the deconvolved half maximum sizes can be
under- or overestimated by factors of up to $\sim$\,$20$ for all models considered. Both the Gaussian model
(Fig.~\ref{decgaussflathillhell}) and critical Bonnor-Ebert sphere (Fig.~\ref{decbesH}) display errors within a factor of
$\sim$\,$2$ for the (partially) resolved structures with $1.2 < \tilde{R}_{j}$ and errors of $\sim$\,$30$\,--\,$20${\%} in the
resolved domain ($1.4 < \tilde{R}_{j} < 2$). For the resolved power-law sources and filaments (Figs.~\ref{decgaussplawH} and
\ref{decgaussplawHfils}), the deconvolution errors can reach a factor of $\sim$\,$6$ in the worst cases, although they are
generally below a factor of $\sim$\,$2$. The deconvolved moment sizes suffer from much higher errors
(Figs.~\ref{decgaussflathillhellmom} and \ref{decgaussplawM}), compared to the half maximum sizes, as intensity moments are too
sensitive to the inaccuracies of background subtraction.

These results demonstrate that Gaussian size deconvolution cannot be applied to unresolved structures, because of the unacceptably
large errors. It makes sense to apply the method only to the Gaussian-like structures, including the critical Bonnor-Ebert sphere,
when they are (partially) resolved (for $1.2 \la \tilde{R}_{j} < 2$), with substantial inaccuracies of $\sim$\,$50$\,--\,$20${\%}.
When the extracted sources or filaments have power-law intensity distributions, the range of deconvolution errors is
$\sim$\,$500$\,--\,$20${\%} for $1.4 < \tilde{R}_{j} < 2$, while the deconvolution method reduces the measured sizes by only
$30$\,--\,$13${\%}. With the errors that are much larger than the deconvolution effect, the method must be considered inaccurate
and not applicable for the power-law sources and filaments with shallow profiles ($1\!\leftarrow \beta \la 2$).


Convolution produces identical results for different geometries only when both the structures and kernel have Gaussian profiles.
When convolved with a power-law kernel, the cylindrical Gaussian models are always wider than the equivalent spherical models
(Fig.~\ref{gausconv}). Conversely, when convolved with a Gaussian kernel, the cylindrical power-law models without an outer
boundary (infinite) are always narrower than the equivalent spherical models (Fig.~\ref{plawinfconv}). For the power-law kernels,
the cylindrical models with power-law profiles are generally wider than the equivalent spherical model, except the unresolved
models with shallow profiles ($\beta = 1$), which are narrower (Fig.~\ref{plawinfconv}). The presence of an outer boundary in the
finite power-law models makes their convolution with a Gaussian kernel similar in both spherical and cylindrical geometries
(Fig.~\ref{plawfinconv}).

Convolution dependence on the object geometry has potential consequences for the observational studies of sources and filaments,
affecting even the well-resolved structures. The non-Gaussian convolution kernels (telescope PSFs) with relatively shallow
power-law wings ``perceive'' that the filament geometry has one more dimension than the source geometry. Such kernels can produce
substantially shallower profiles for the convolved filaments, when the kernel is shallow enough relative to the intrinsic profile
of the observed object (Eq.~(\ref{coneffect})). In principle, this convolution property can cause a physical filament, imaged by
the telescope with a non-Gaussian PSF with power-law wings, to appear substantially shallower than the object is in reality.

\begin{acknowledgements} 
This study used the \textsl{cfitsio} library \citep{Pence1999}, developed at HEASARC NASA (USA) and \textsl{saoimage ds9}
\citep{JoyeMandel2003}. The \textsl{plot} utility and \textsl{ps12d} library, used to draw figures directly in the PostScript
language, were written by the author using the \textsl{psplot} library (by Kevin E. Kohler), developed at Nova Southeastern
University Oceanographic Center (USA), and the plotting subroutines from the MHD code \textsl{azeus} \citep{Ramsey2012}, developed
by David Clarke and the author at Saint Mary's University (Canada). The author thanks Ph.~Andr\'{e} for useful comments and the
anonymous referee for the very detailed and constructive report on the manuscript.
\end{acknowledgements} 


\begin{appendix}

\section{Deconvolution of moment sizes}
\label{resultsmom}

\begin{figure*}
\centering
\centerline{
  \resizebox{0.2675\hsize}{!}{\includegraphics{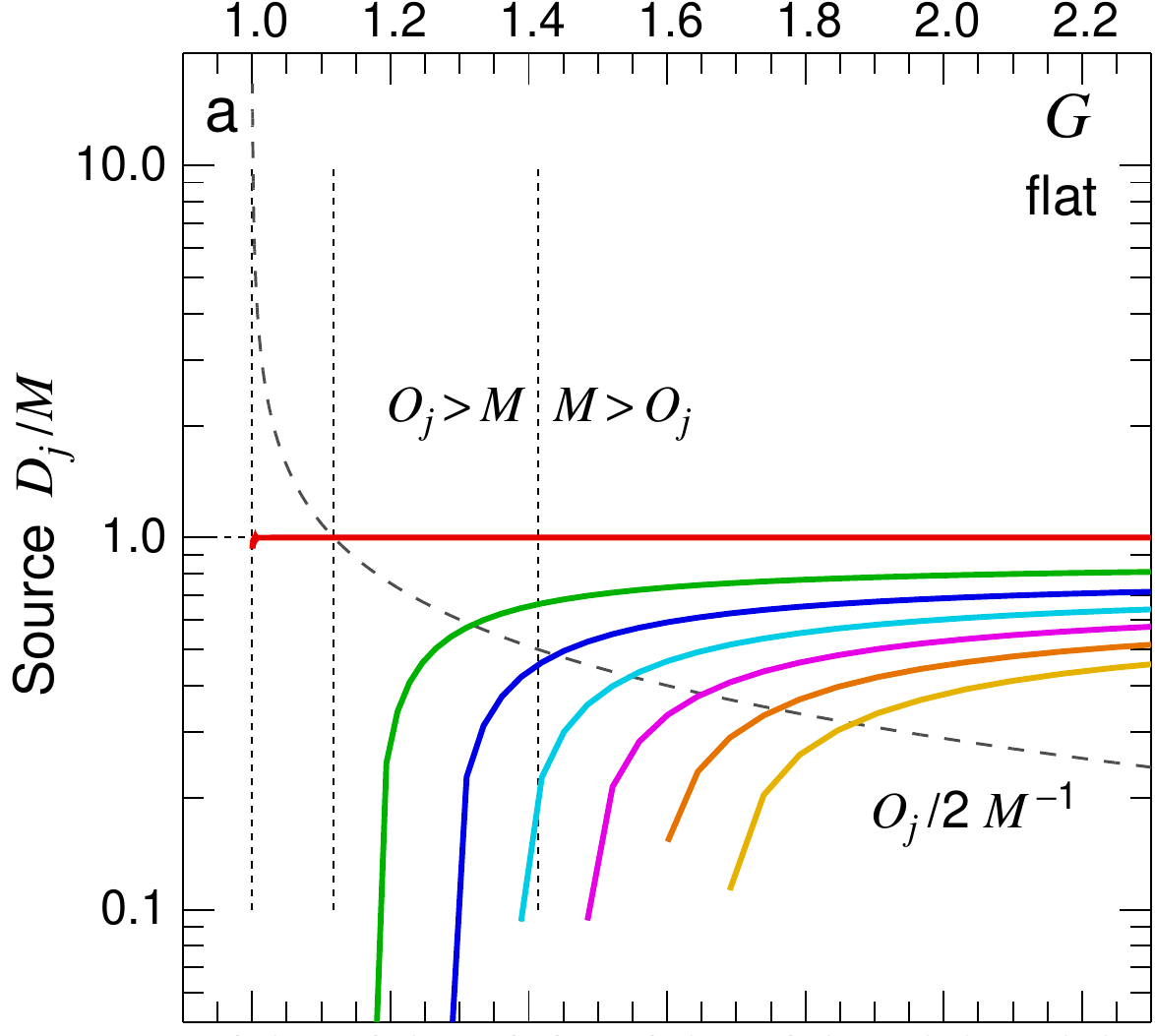}}     \hspace{-1.8mm}
  \resizebox{0.2755\hsize}{!}{\includegraphics{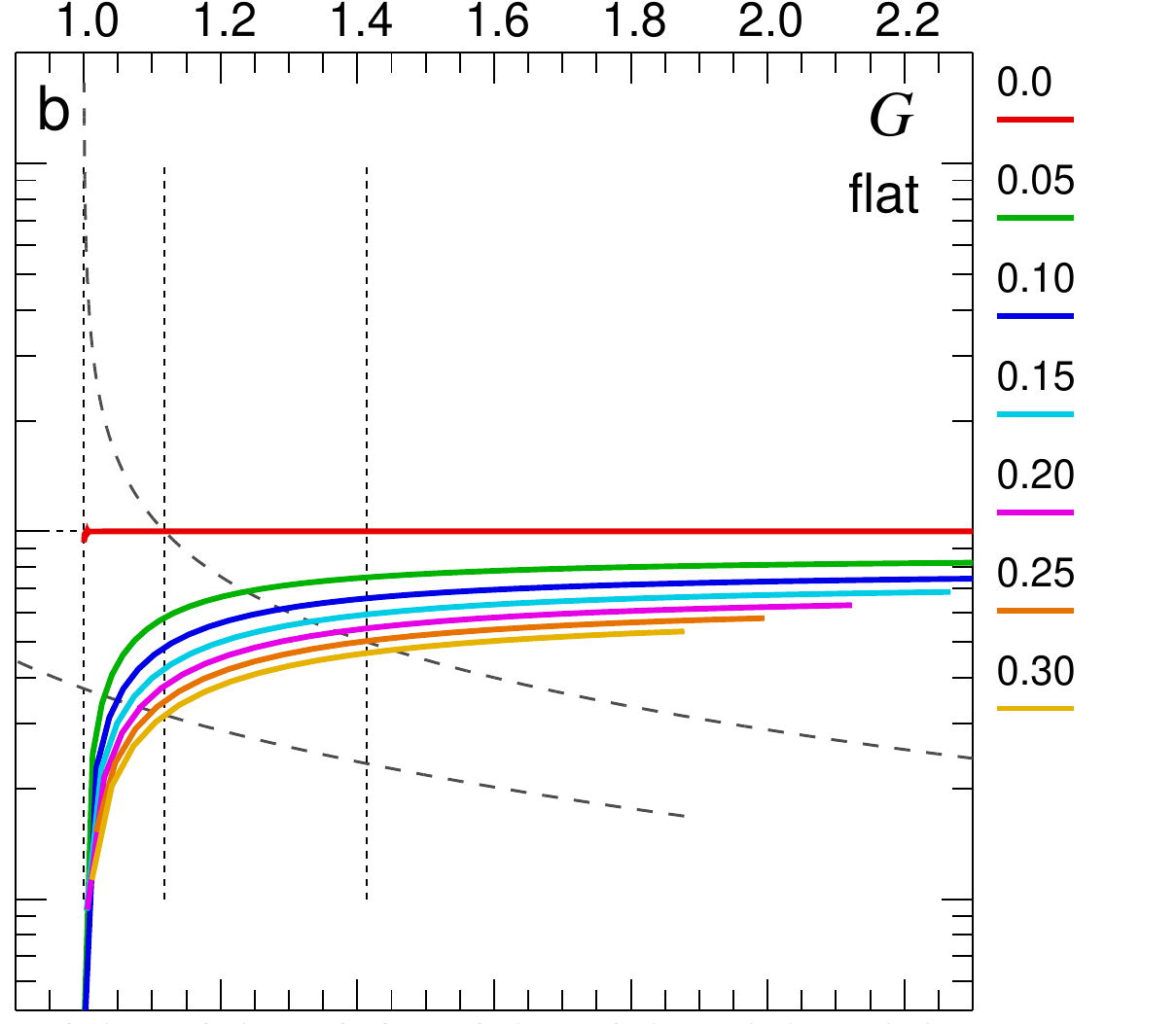}}       \hspace{-1.8mm}
  \resizebox{0.2912\hsize}{!}{\includegraphics{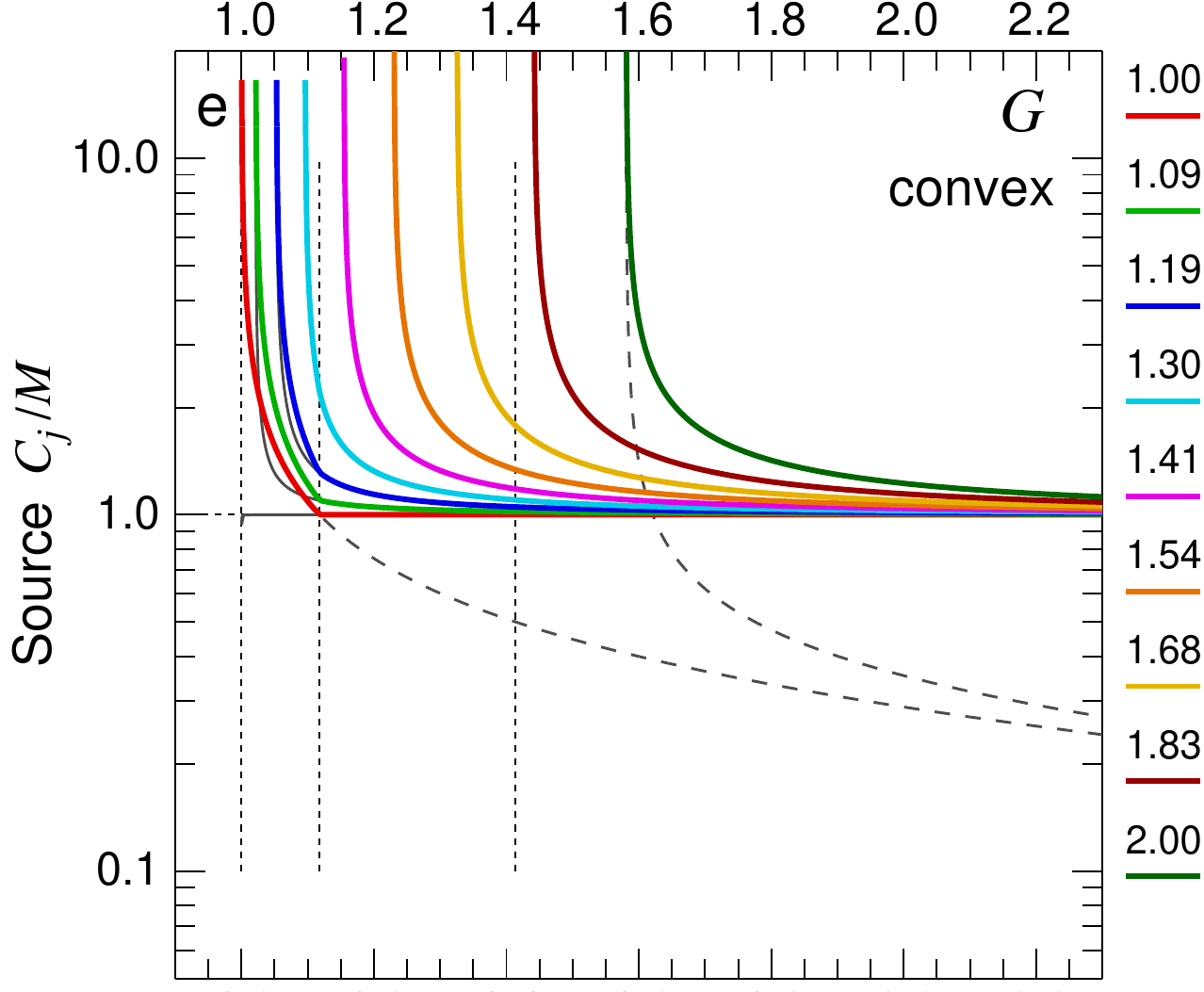}}}
\vspace{-2.54mm}
\centerline{
  \resizebox{0.2675\hsize}{!}{\includegraphics{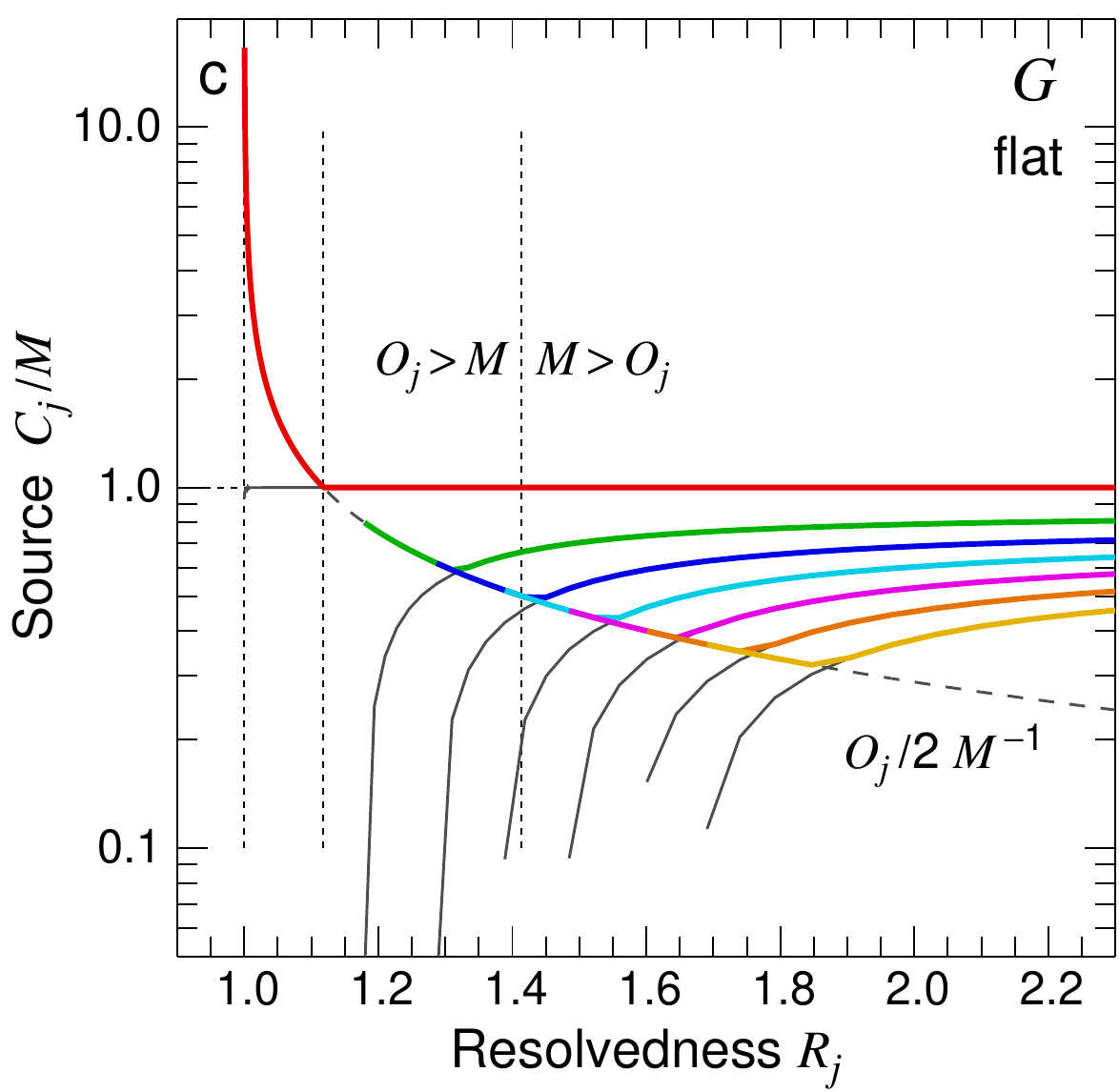}}    \hspace{-1.8mm}
  \resizebox{0.2755\hsize}{!}{\includegraphics{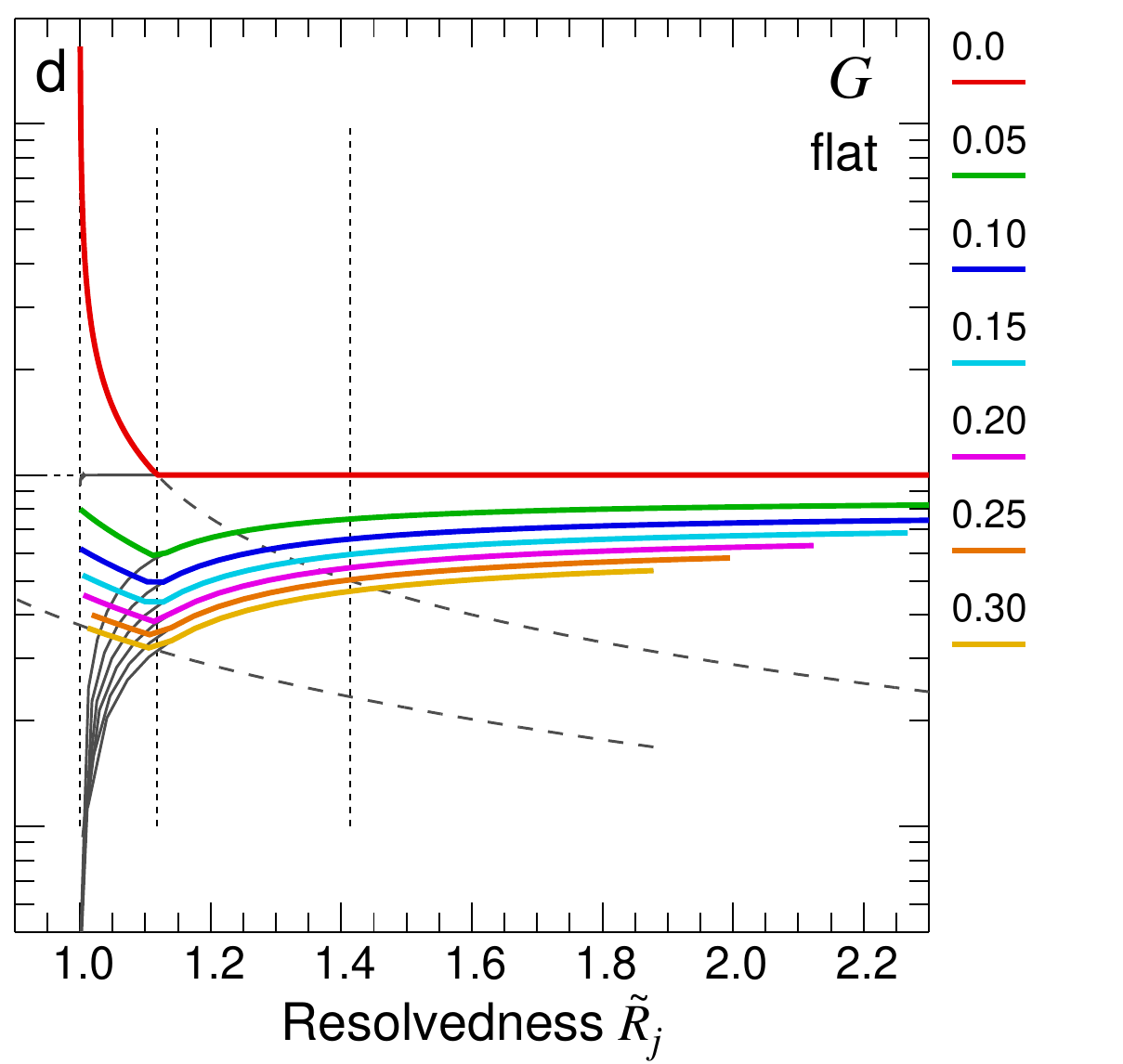}}      \hspace{-1.8mm}
  \resizebox{0.2912\hsize}{!}{\includegraphics{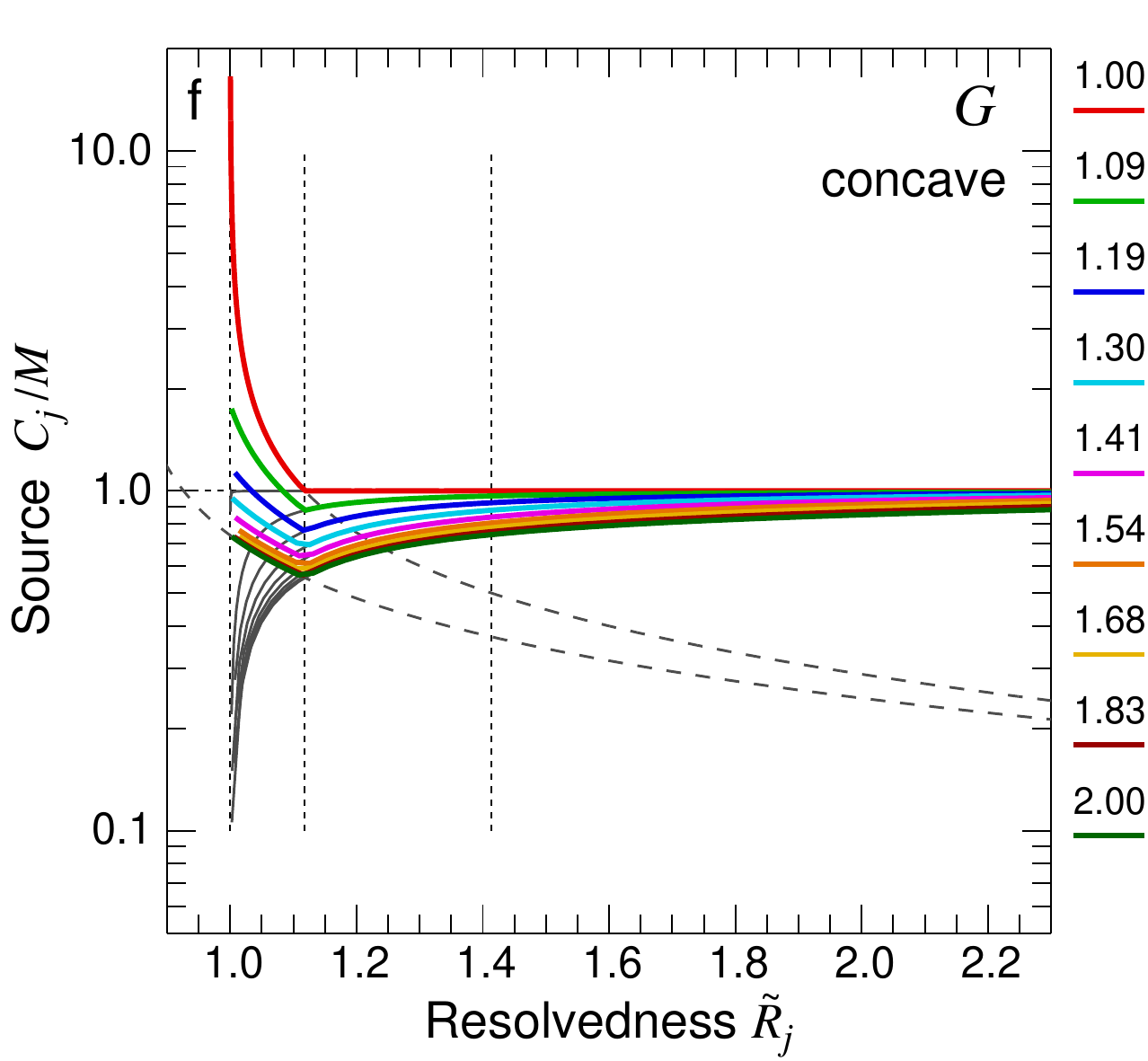}}}
\caption
{ 
Deconvolution accuracy of the moment sizes $\tilde{M}_{j}$ for the Gaussian sources $\mathcal{\tilde{S}}_{{\mathcal{G}}{jl}}$
(Eq.~(\ref{bgsubtraction})), separated from the flat background $\mathcal{B}$, for background over-subtraction levels $0 \le
\epsilon_{\,l} \le 0.3$ and from the convex and concave backgrounds $\mathcal{B}_{{\mathcal{G}}{jk}{\pm}}$, for background size
factors $1 \le f_{k} \le 2$ (\emph{right}). The ratios of the deconvolved sizes $D_{j}$ and $C_{j}$ to the true model size $M$
(Table~\ref{modeltable}) are plotted as functions of the model and source resolvedness $R_{j}$ and $\tilde{R}_{j}$. For reference,
the thin black curves display $D_{j}/M$ from panels \emph{a} and \emph{b} and the dashed curves visualize $(O_{j}/2)/M$ for
$\epsilon_{\,l} = \{0, 0.3\}$. The dashed vertical lines divide the horizontal axis into the unresolved, partially resolved, and
resolved domains (Eq.~(\ref{resolstate})). All curves for $\epsilon_{\,l} > 0$ in panels \emph{b} and \emph{d} are shifted to the
left in comparison with panels \emph{a} and \emph{c} because the resolvedness is underestimated ($\tilde{R}_{j}\!< R_{j}$). The
deconvolution errors in panel \emph{e} for the convex case with $f_{k} \ge 1.54$ reach the values of $24$, $29$, $34$, and $41$.
Corresponding plots for the deconvolved half maximum sizes $\tilde{H}_{j}$ are presented in Fig.~\ref{decgaussflathillhell}.
} 
\label{decgaussflathillhellmom}
\end{figure*}

\begin{figure*}
\centering
\centerline{
  \resizebox{0.2675\hsize}{!}{\includegraphics{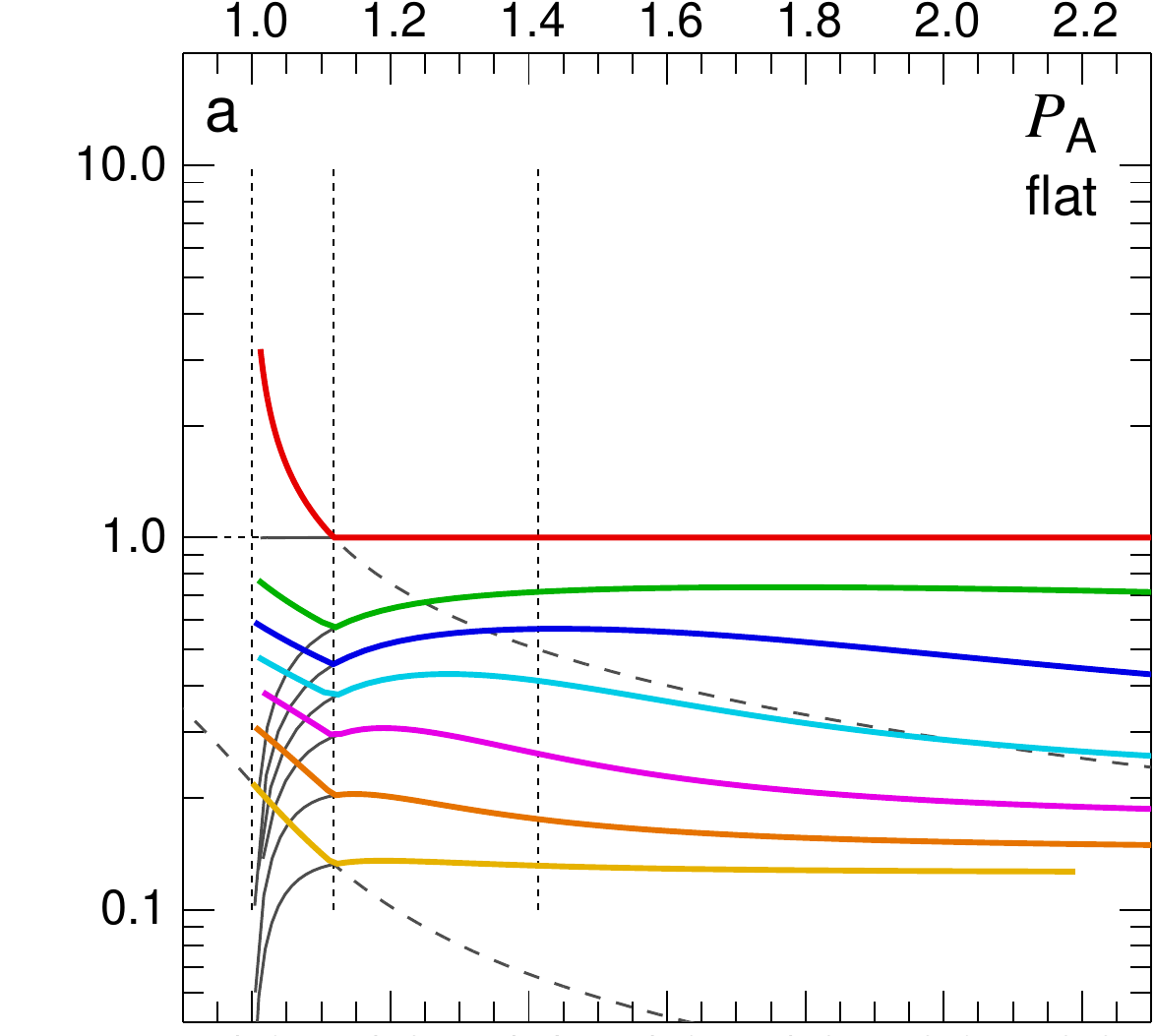}}  \hspace{-1.8mm}
  \resizebox{0.2287\hsize}{!}{\includegraphics{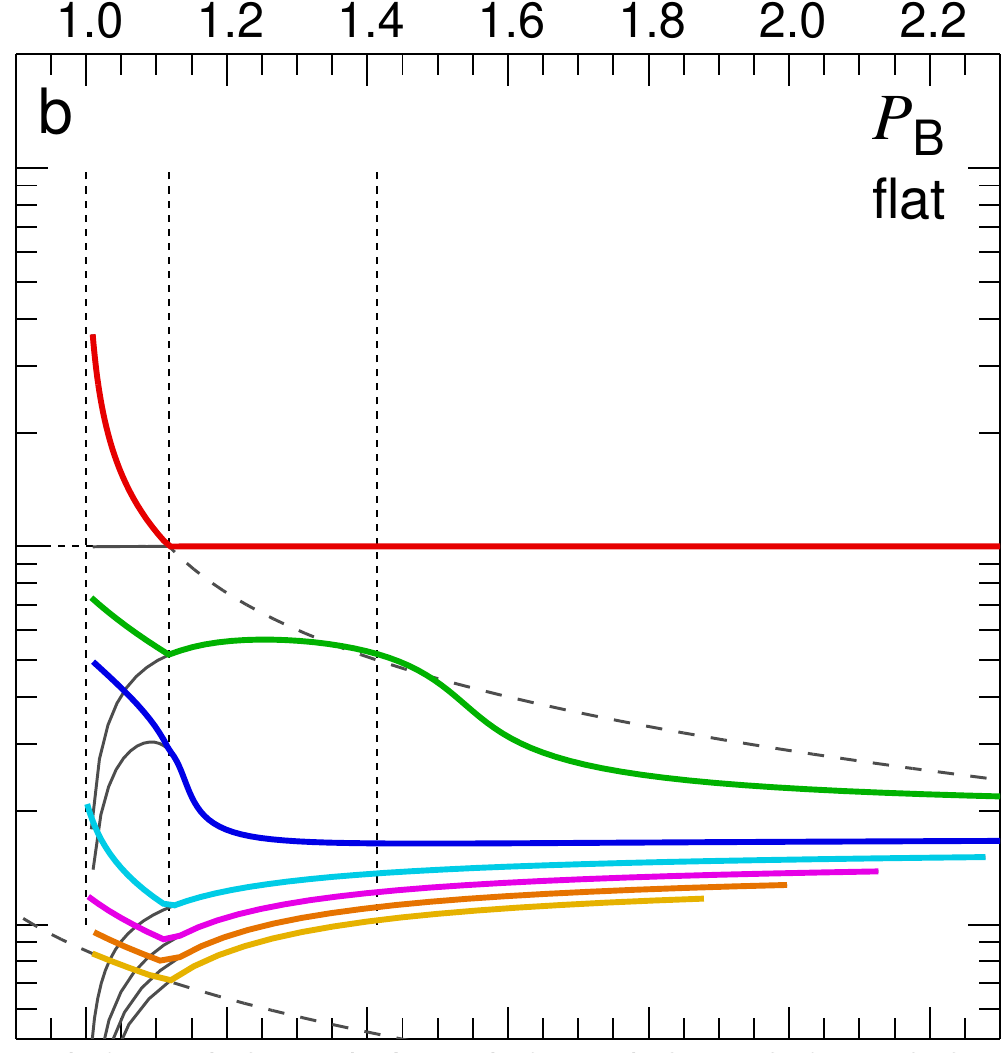}}  \hspace{-1.8mm}
  \resizebox{0.2287\hsize}{!}{\includegraphics{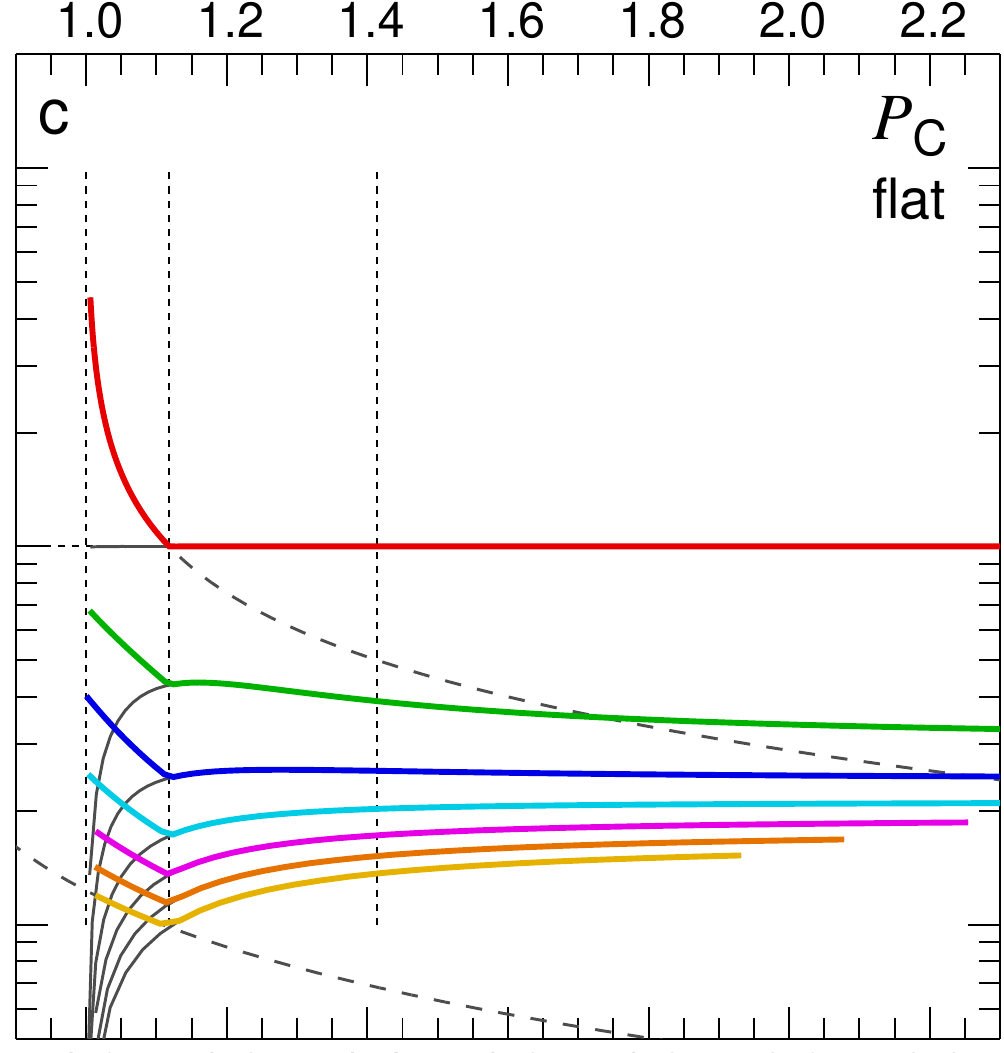}}  \hspace{-1.8mm}
  \resizebox{0.2755\hsize}{!}{\includegraphics{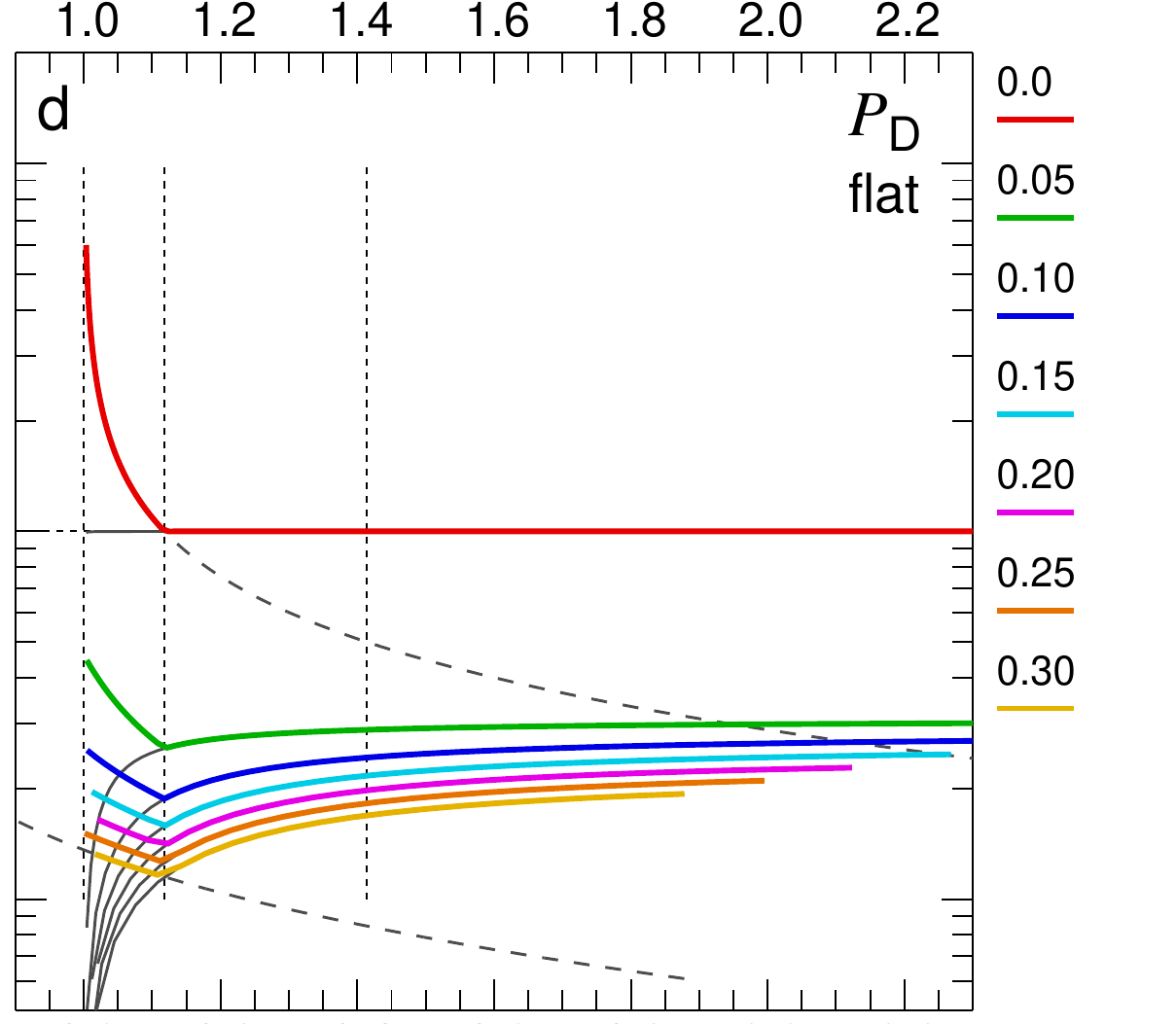}}}
\vspace{-0.94mm}
\centerline{
  \resizebox{0.2675\hsize}{!}{\includegraphics{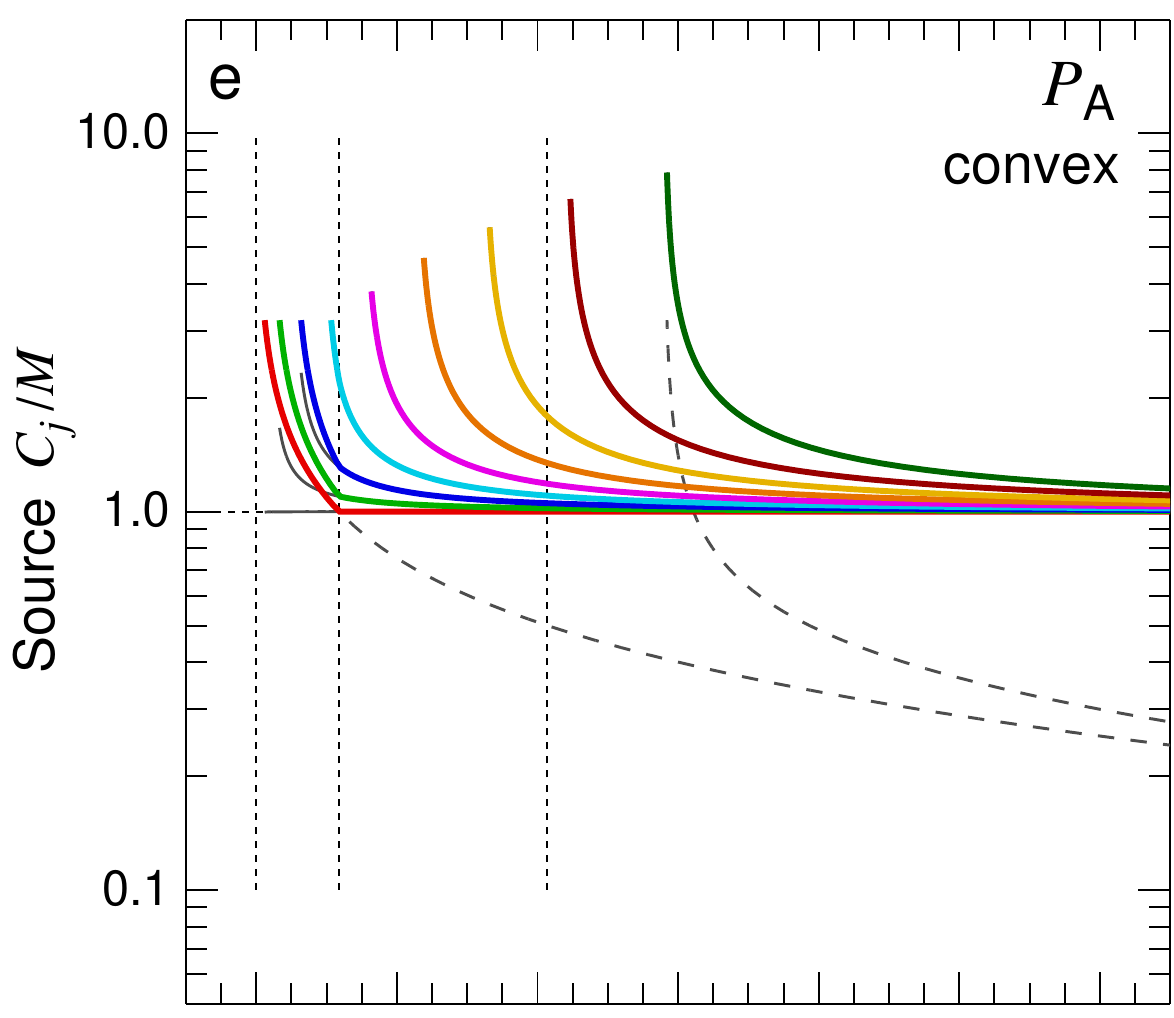}}  \hspace{-1.8mm}
  \resizebox{0.2287\hsize}{!}{\includegraphics{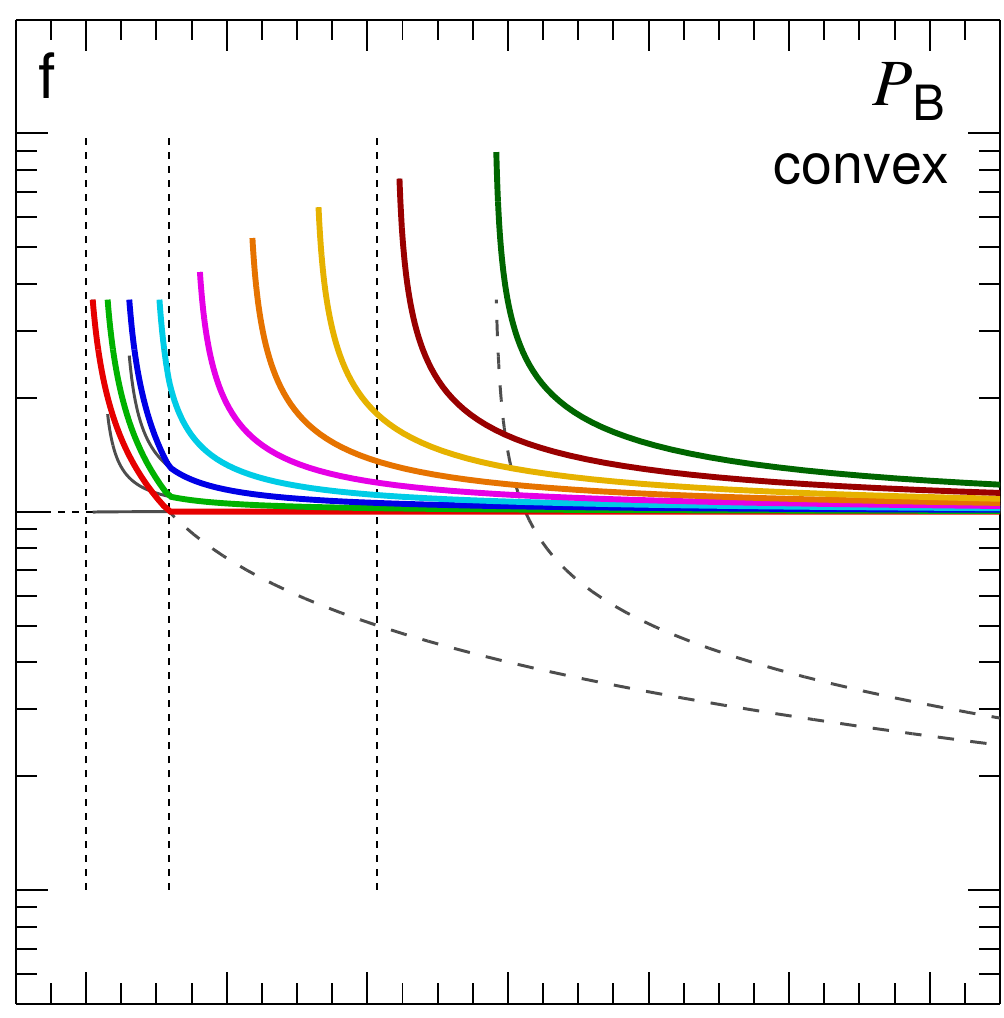}}  \hspace{-1.8mm}
  \resizebox{0.2287\hsize}{!}{\includegraphics{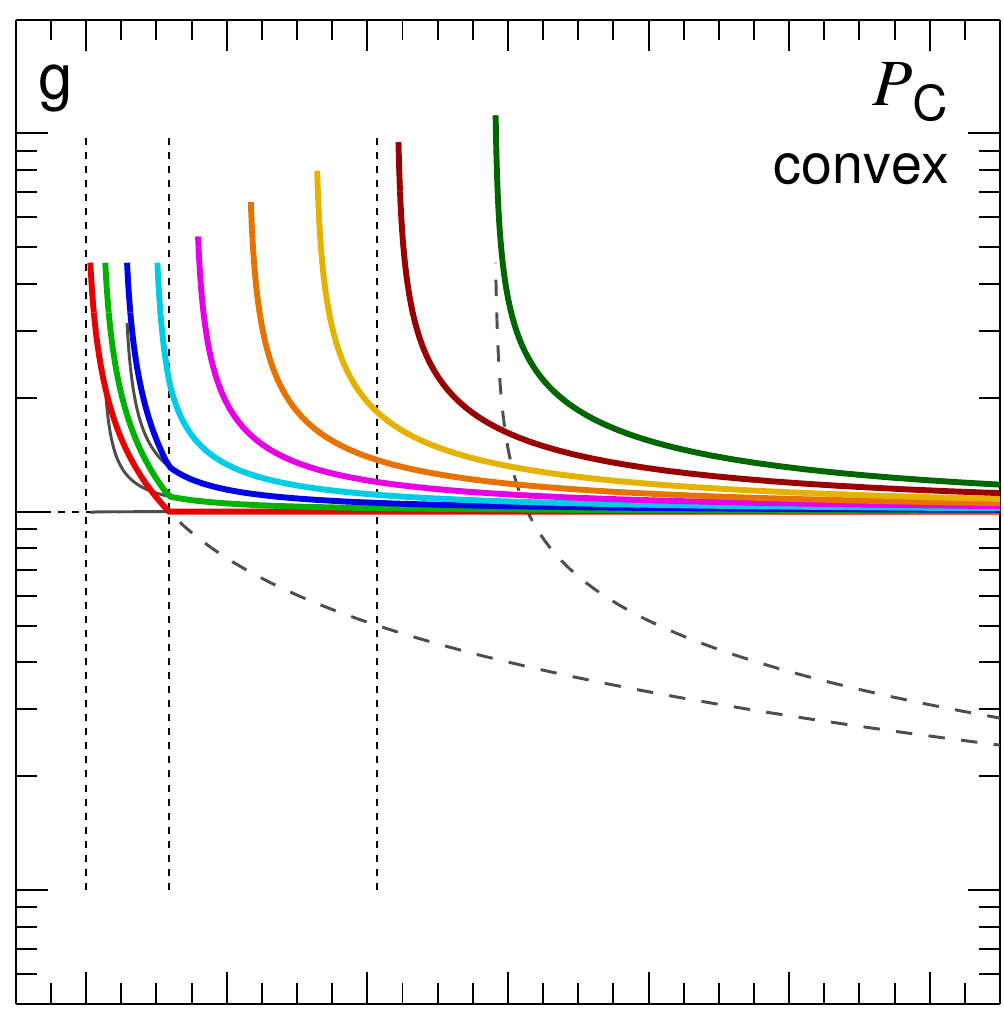}}  \hspace{-1.8mm}
  \resizebox{0.2755\hsize}{!}{\includegraphics{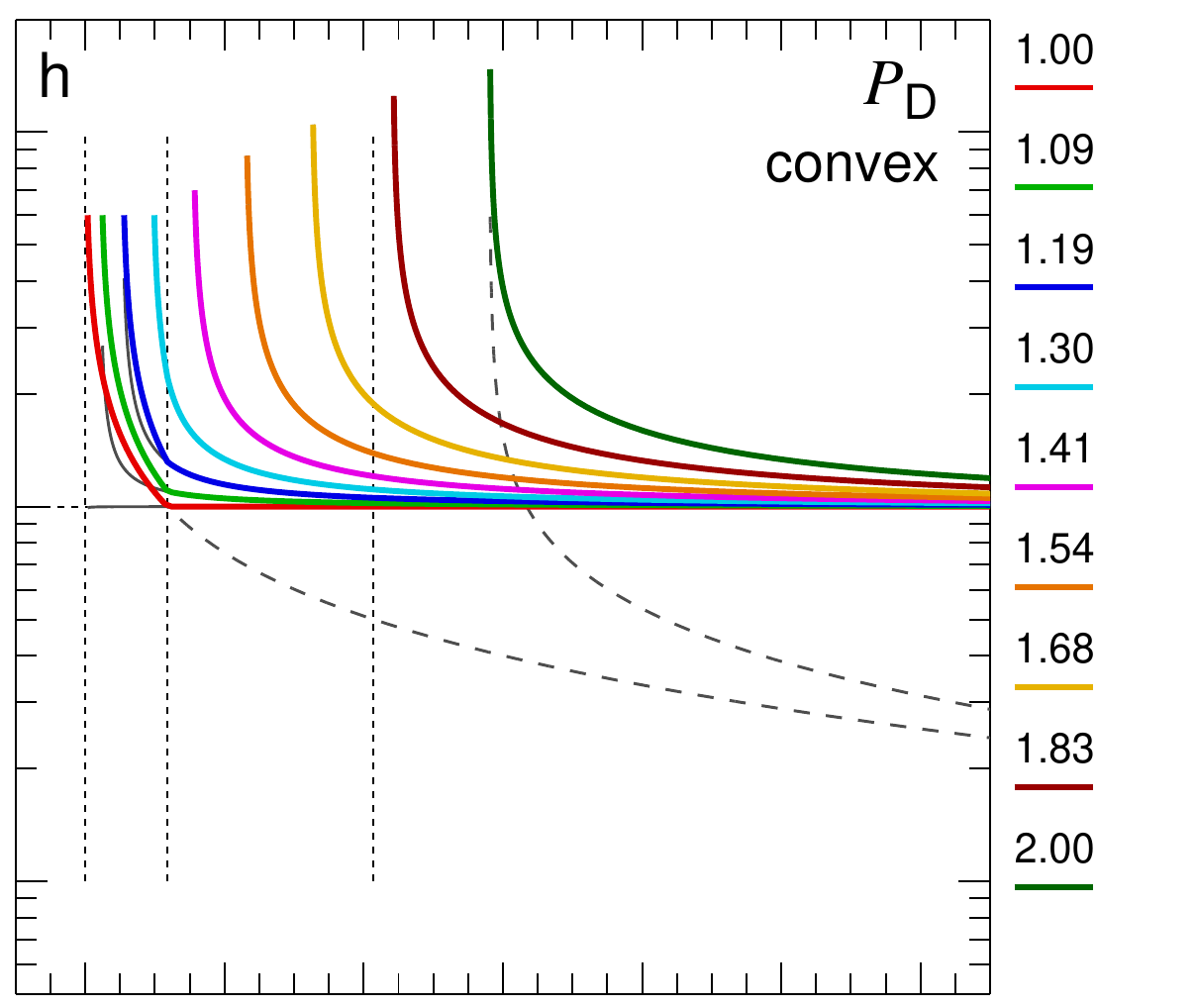}}}
\vspace{-0.94mm}
\centerline{
  \resizebox{0.2675\hsize}{!}{\includegraphics{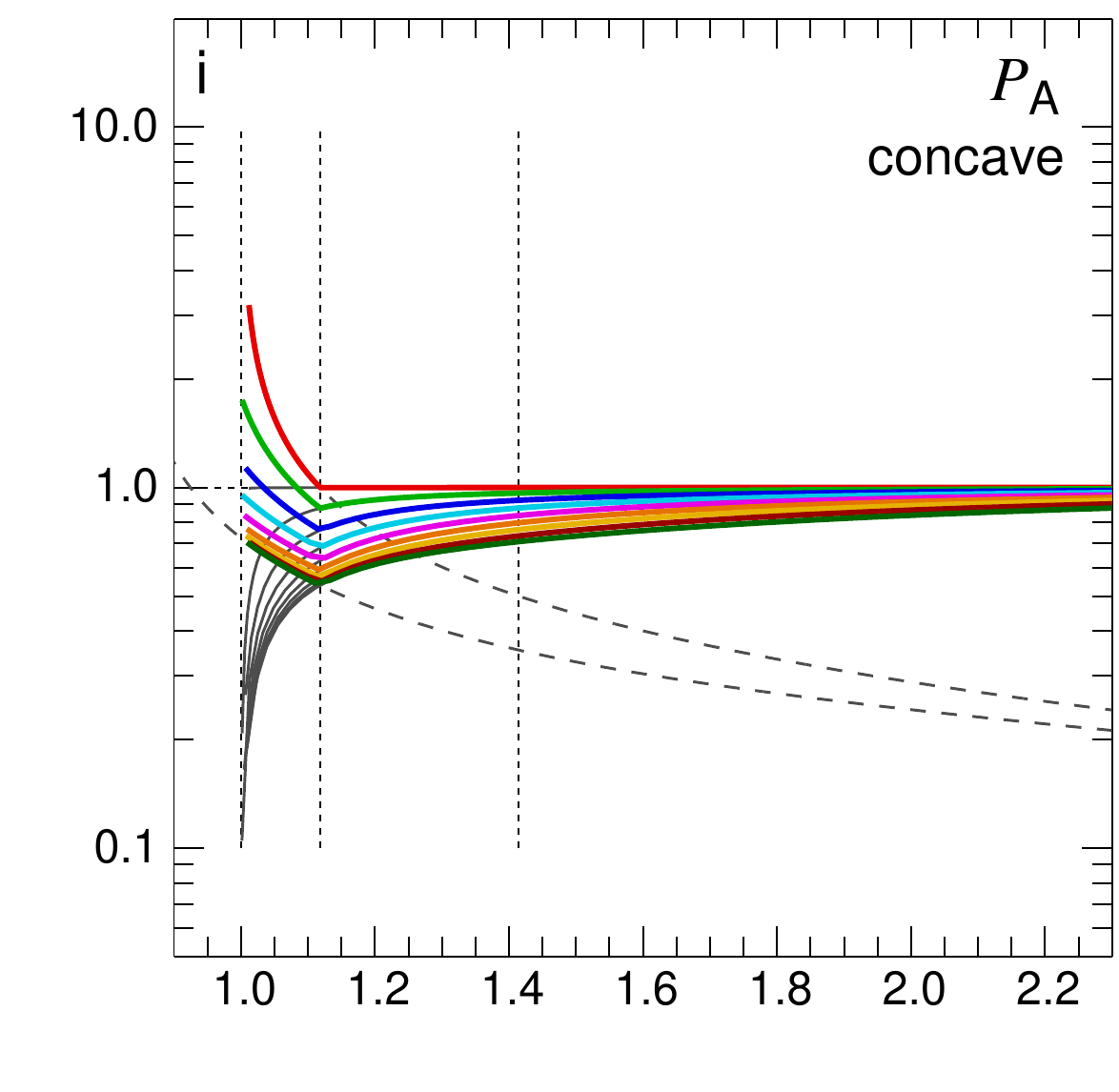}}  \hspace{-1.9mm}
  \resizebox{0.2832\hsize}{!}{\includegraphics{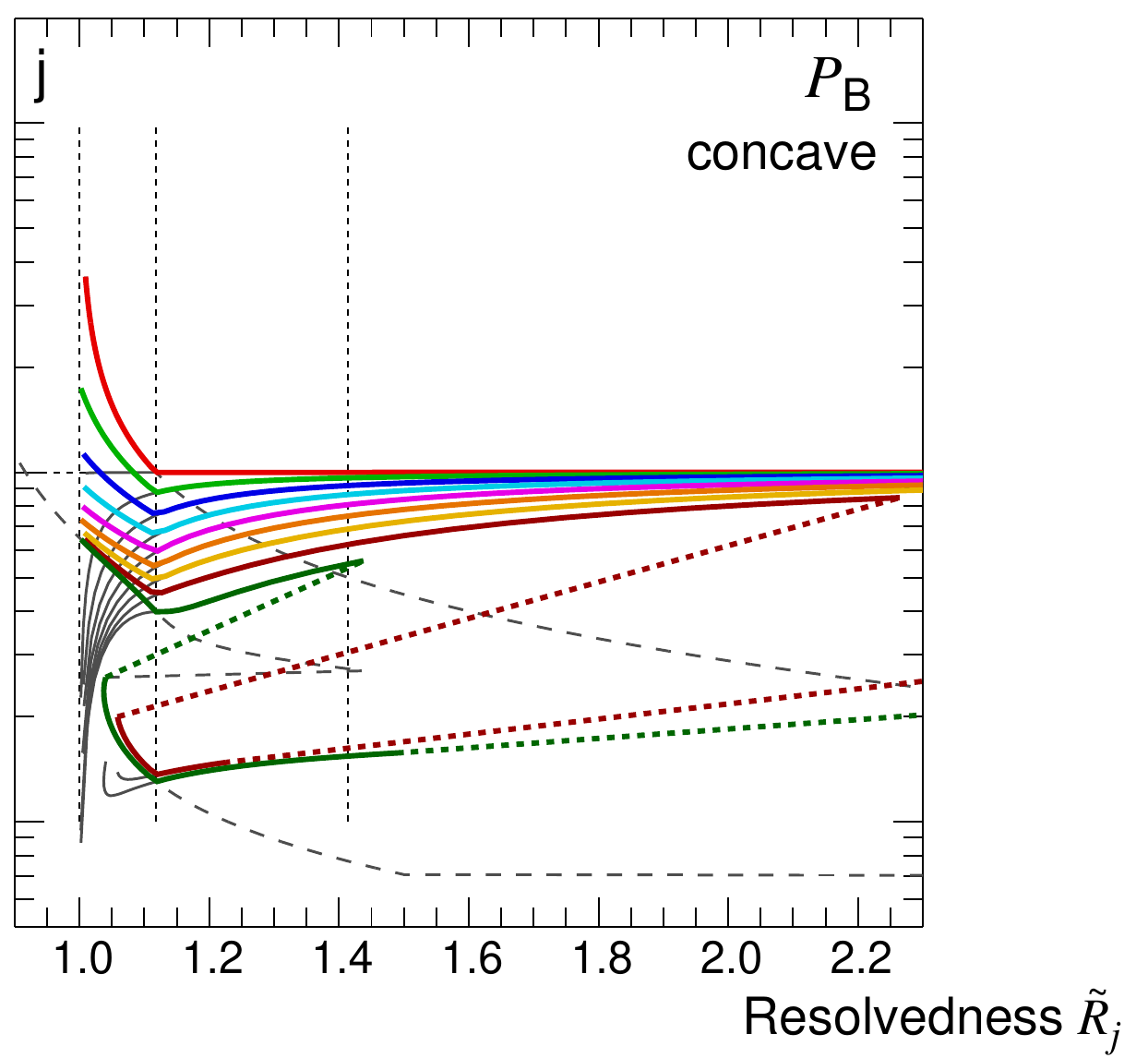}}  \hspace{-11.85mm}
  \resizebox{0.2287\hsize}{!}{\includegraphics{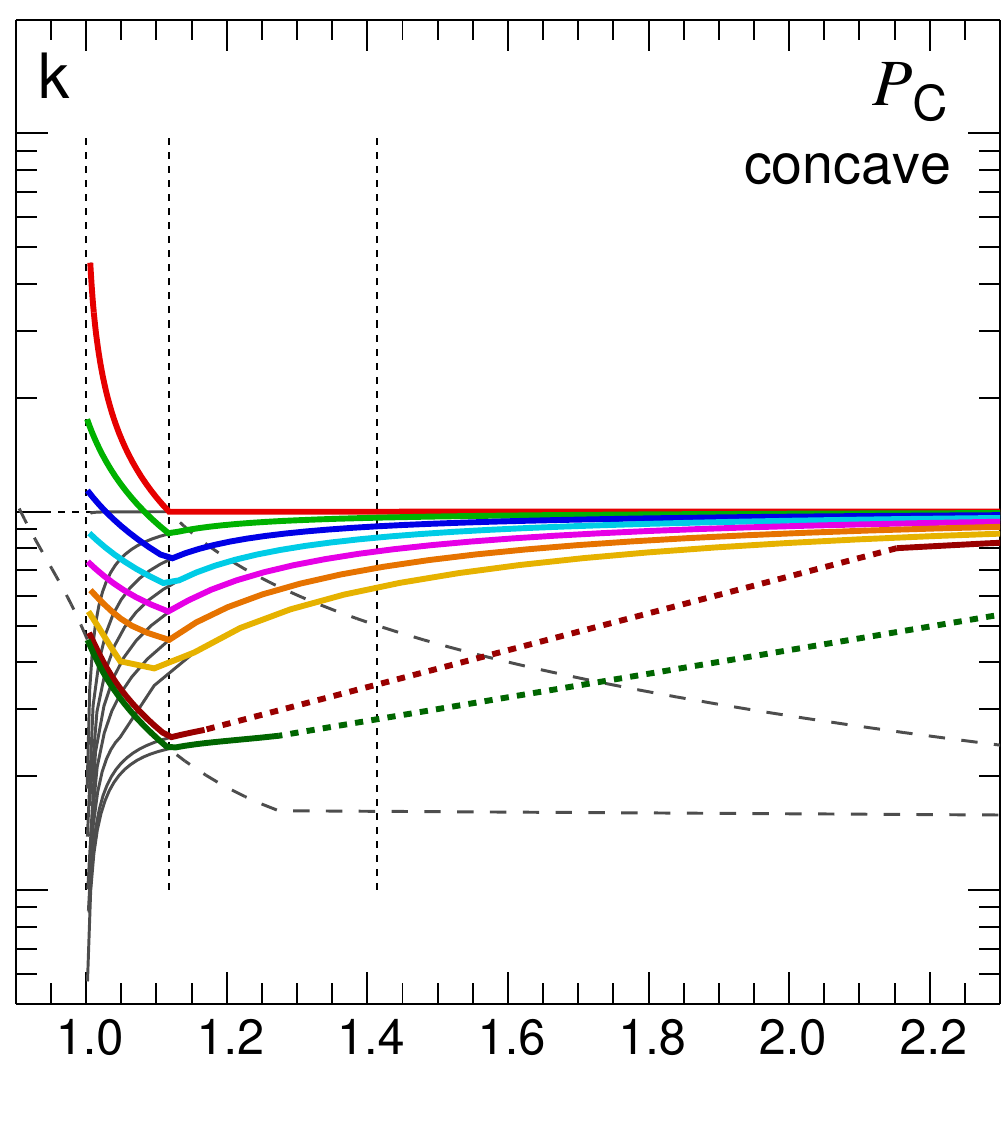}}  \hspace{-1.8mm}
  \resizebox{0.2738\hsize}{!}{\includegraphics{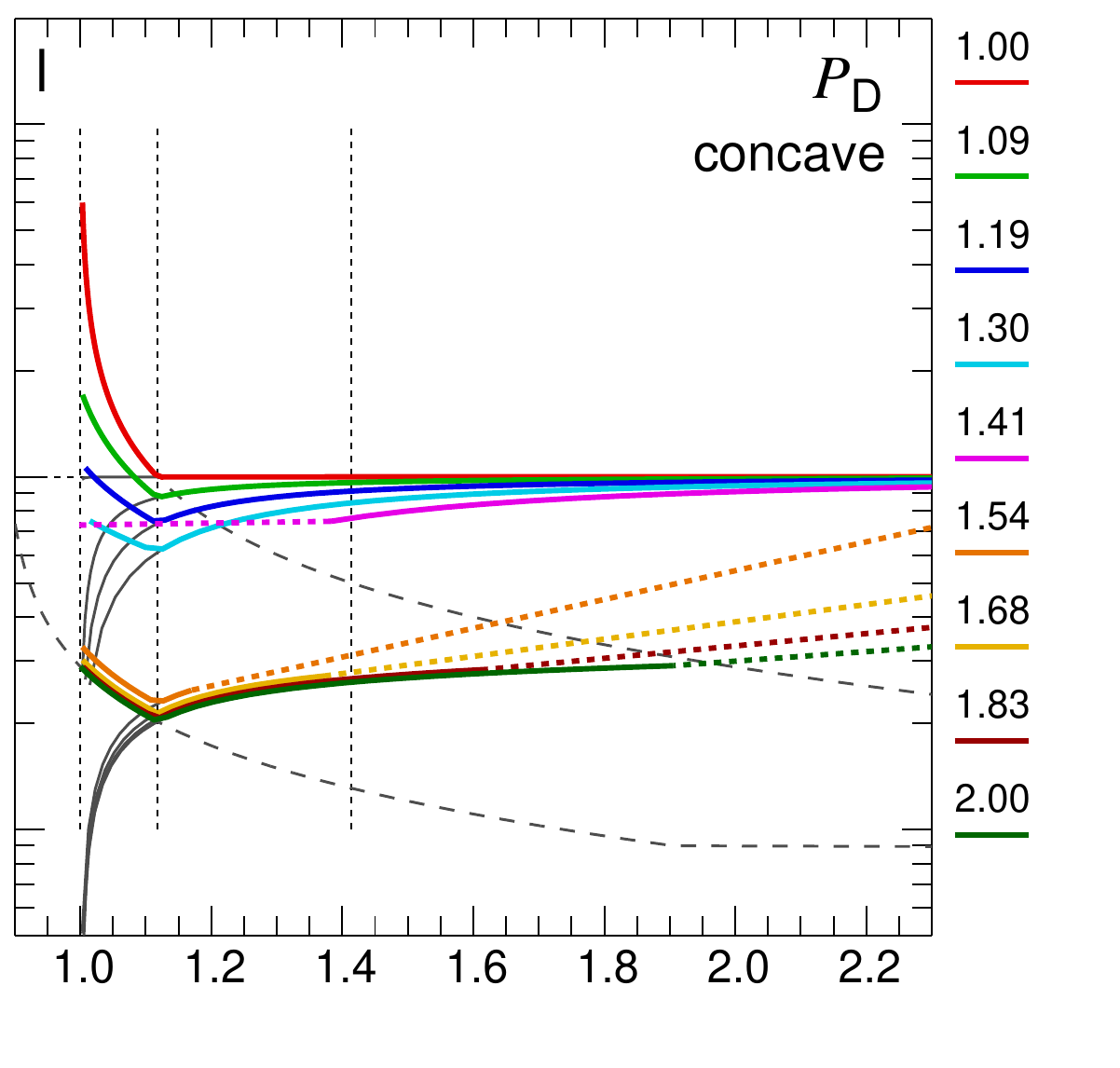}}}
\caption
{ 
Deconvolution accuracy of the moment sizes $\tilde{M}_{j}$ for the power-law sources $\mathcal{\tilde{S}}_{{\mathcal{P}}{jl}}$ and
$\mathcal{\tilde{S}}_{{\mathcal{P}}{jk}{\pm}}$ (Eq.~(\ref{bgsubtraction})), separated from the flat (\emph{top}), convex
(\emph{middle}), and concave (\emph{bottom}) backgrounds, for background over-subtraction levels $0 \le \epsilon_{\,l} \le 0.3$ and
size factors $1 \le f_{k} \le 2$. The ratio of the modified deconvolved sizes $C_{j}$ to the true model size $M$
(Table~\ref{modeltable}) is plotted as a function of the source resolvedness $\tilde{R}_{j}$. The dashed curves indicate
discontinuous jumps in $\tilde{R}_{j}$ between $j$ and $j+1$ (at certain $O_{j}$ values), caused by the jumps in the footprint
radii $\Theta_{{\mathcal{M}}{jk}-}$ (Eq.~(\ref{footprint})) and sizes $\tilde{M}_{j}$ (cf. Fig.~\ref{resolvednM}). For reference,
the thin black curves display $D_{j}/M$ and the dashed curves visualize $(O_{j}/2)/M$ for $\epsilon_{\,l} = \{0, 0.3\}$ and $f_{k}
= \{1, 2\}$. Corresponding plots for the deconvolved half maximum sizes $\tilde{H}_{j}$ are presented in Fig.~\ref{decgaussplawH}.
} 
\label{decgaussplawM}
\end{figure*}

\begin{figure*}
\centering
\centerline{
  \resizebox{0.2675\hsize}{!}{\includegraphics{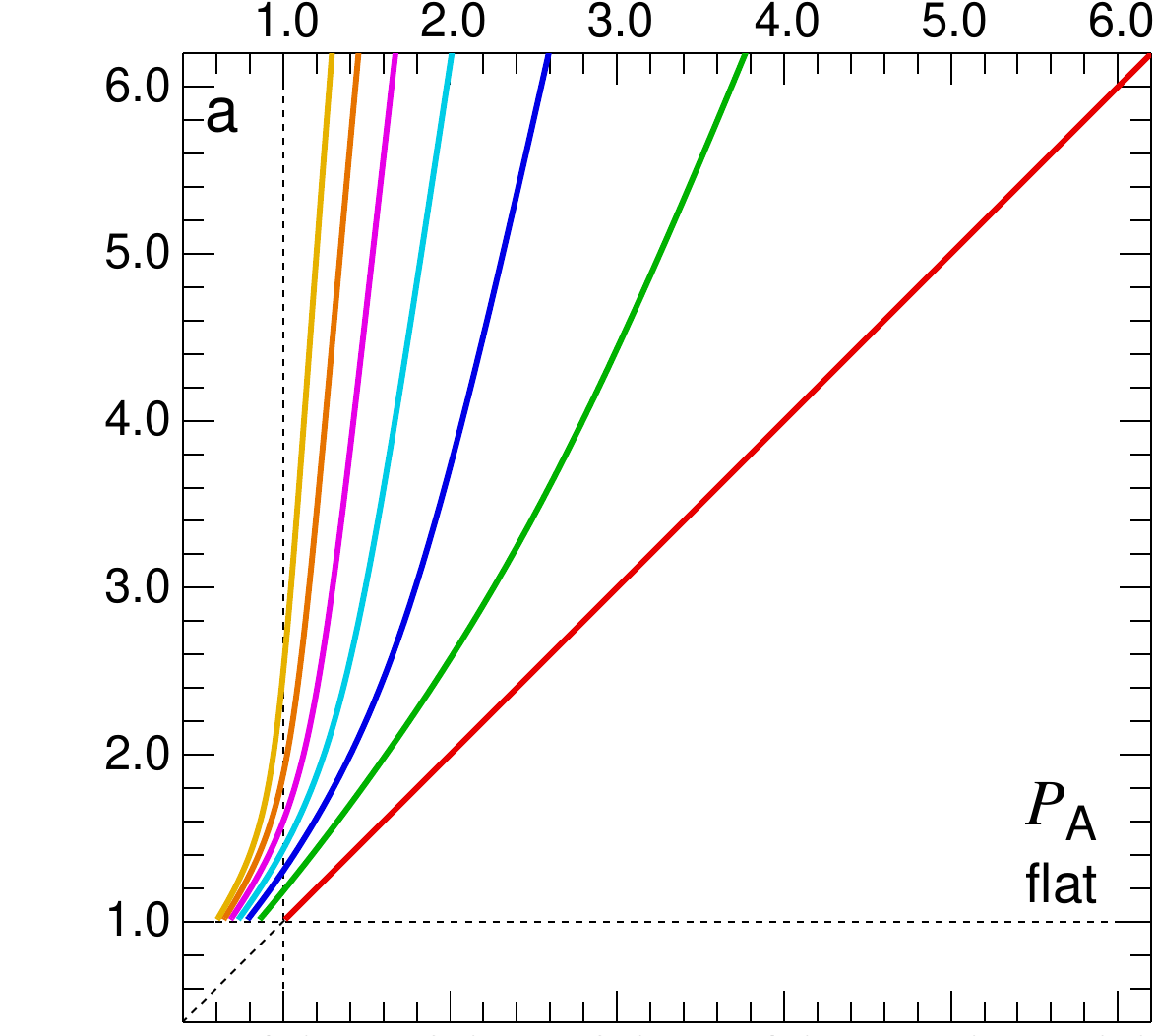}}  \hspace{-1.8mm}
  \resizebox{0.2287\hsize}{!}{\includegraphics{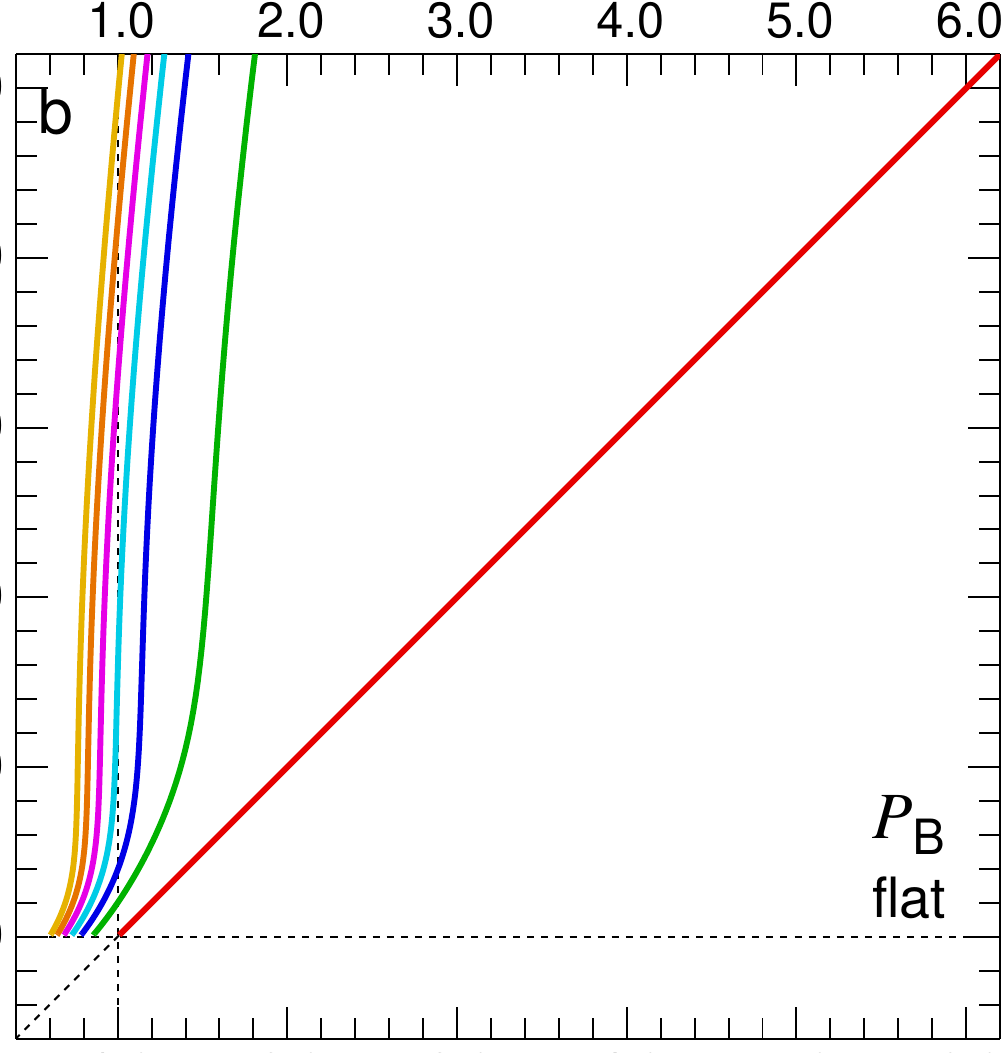}}  \hspace{-1.8mm}
  \resizebox{0.2287\hsize}{!}{\includegraphics{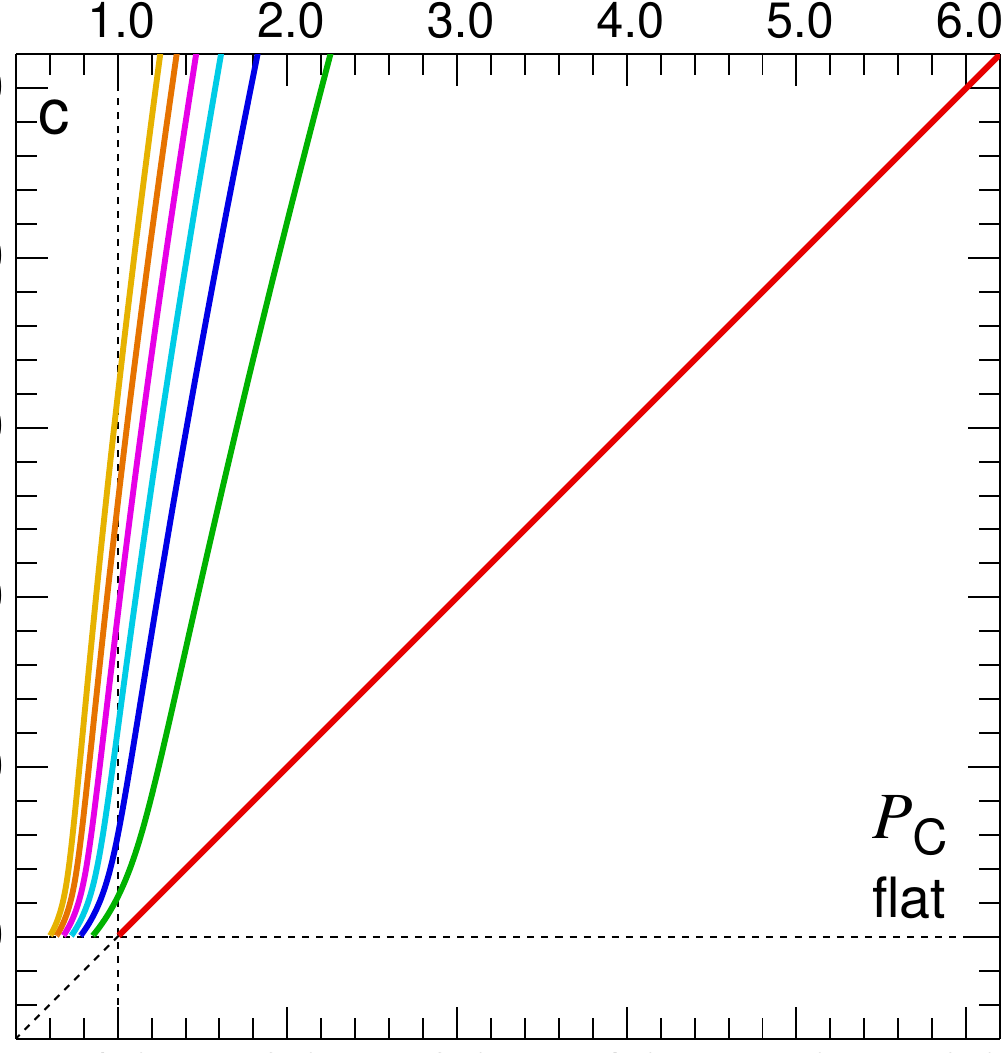}}  \hspace{-1.8mm}
  \resizebox{0.2755\hsize}{!}{\includegraphics{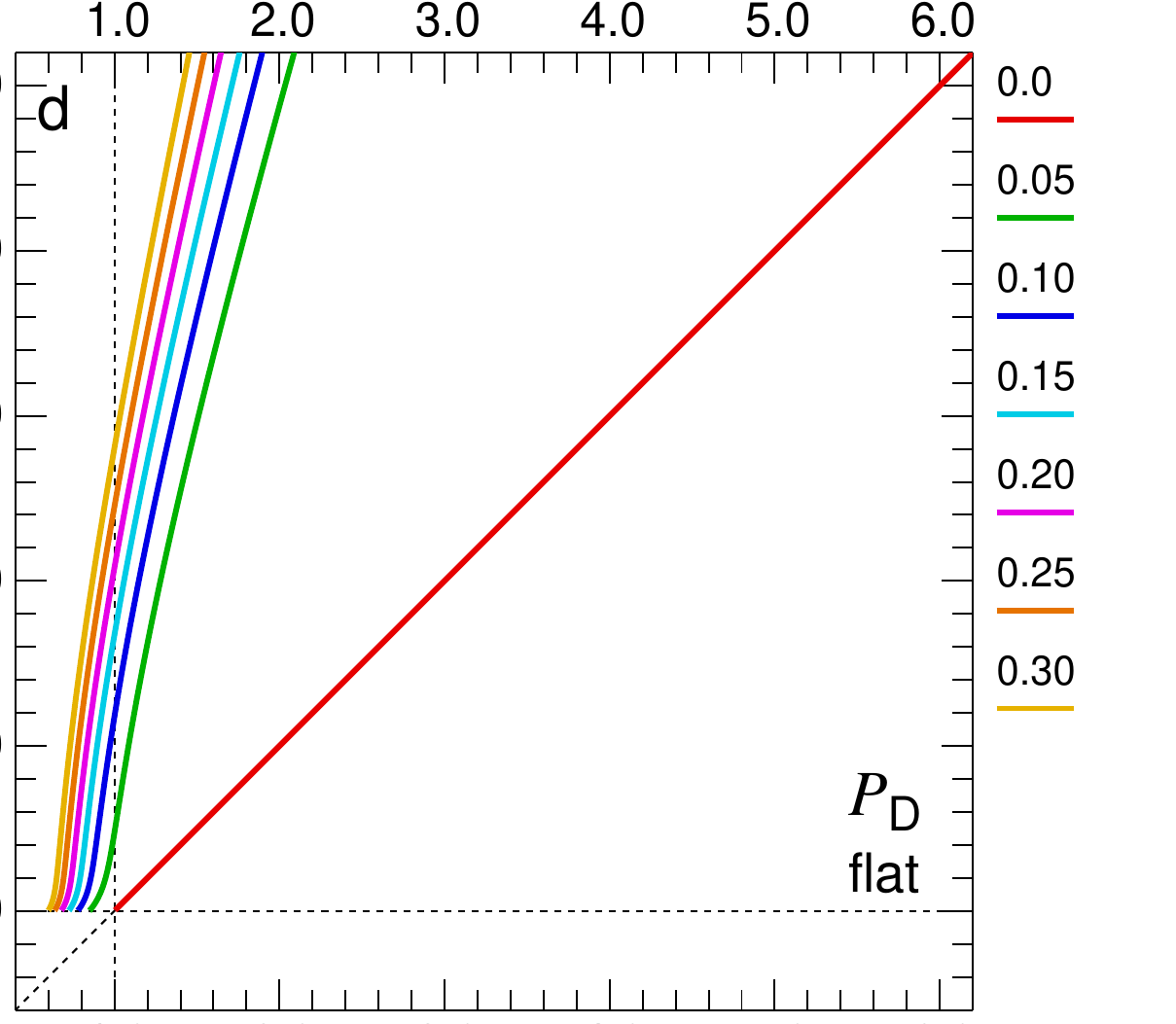}}}
\vspace{-0.94mm}
\centerline{
  \resizebox{0.2675\hsize}{!}{\includegraphics{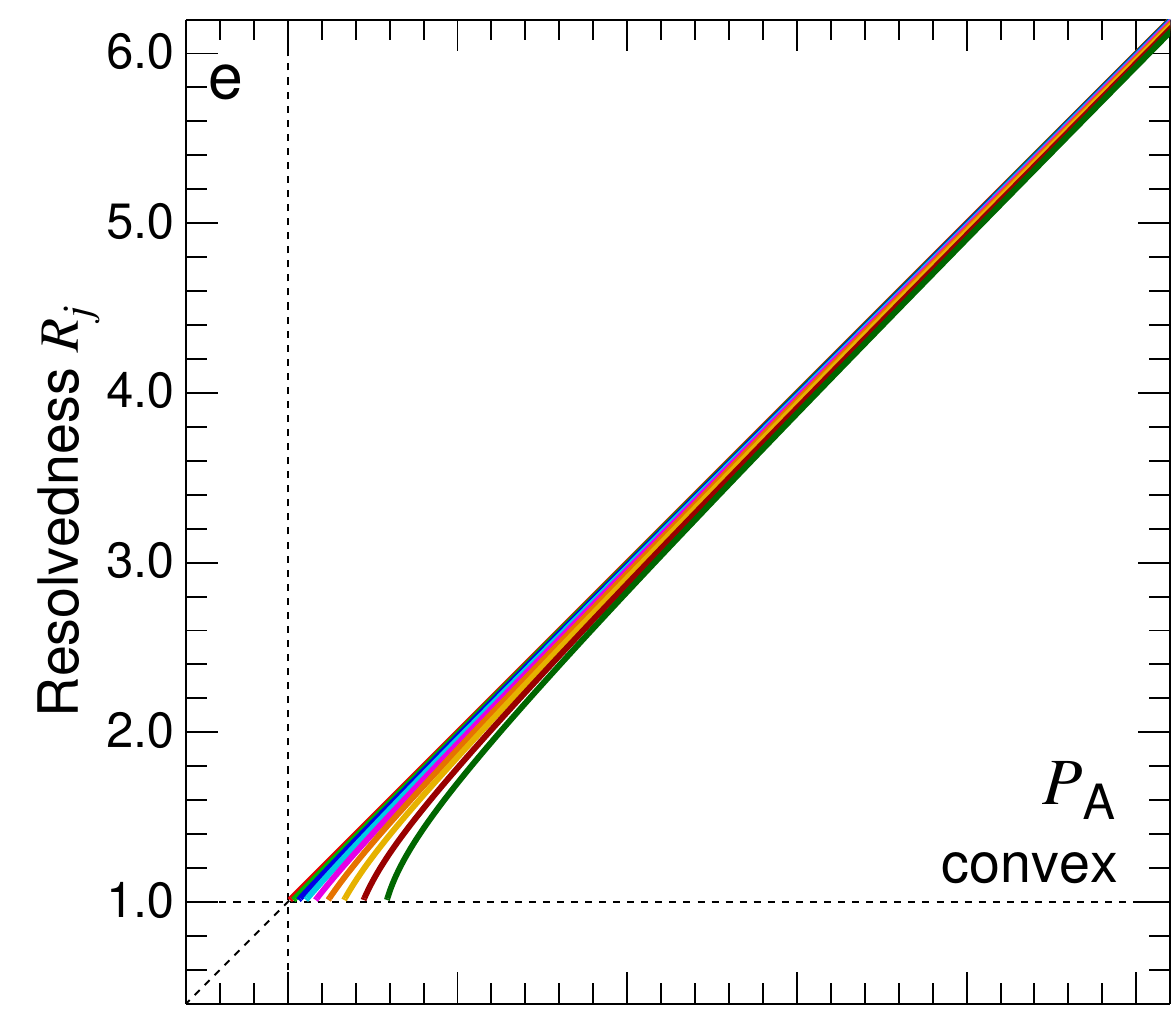}}  \hspace{-1.8mm}
  \resizebox{0.2287\hsize}{!}{\includegraphics{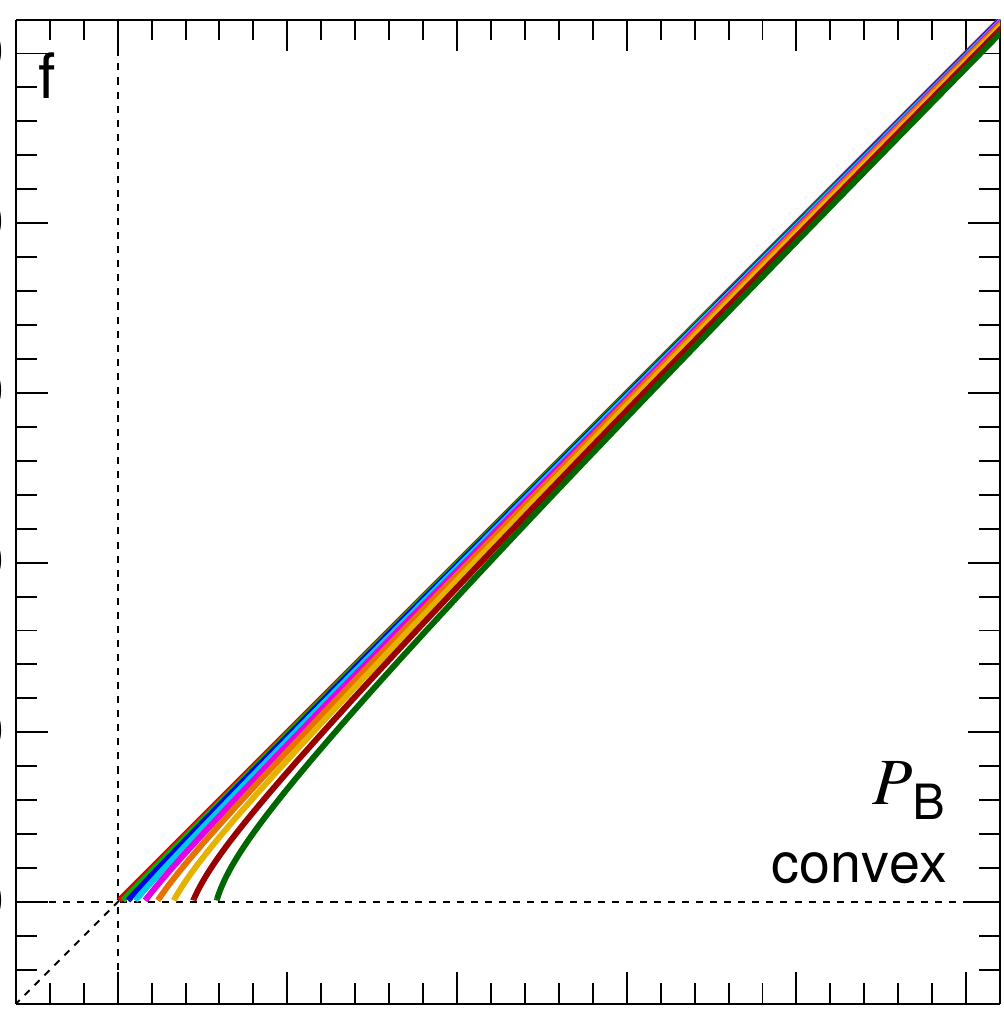}}  \hspace{-1.8mm}
  \resizebox{0.2287\hsize}{!}{\includegraphics{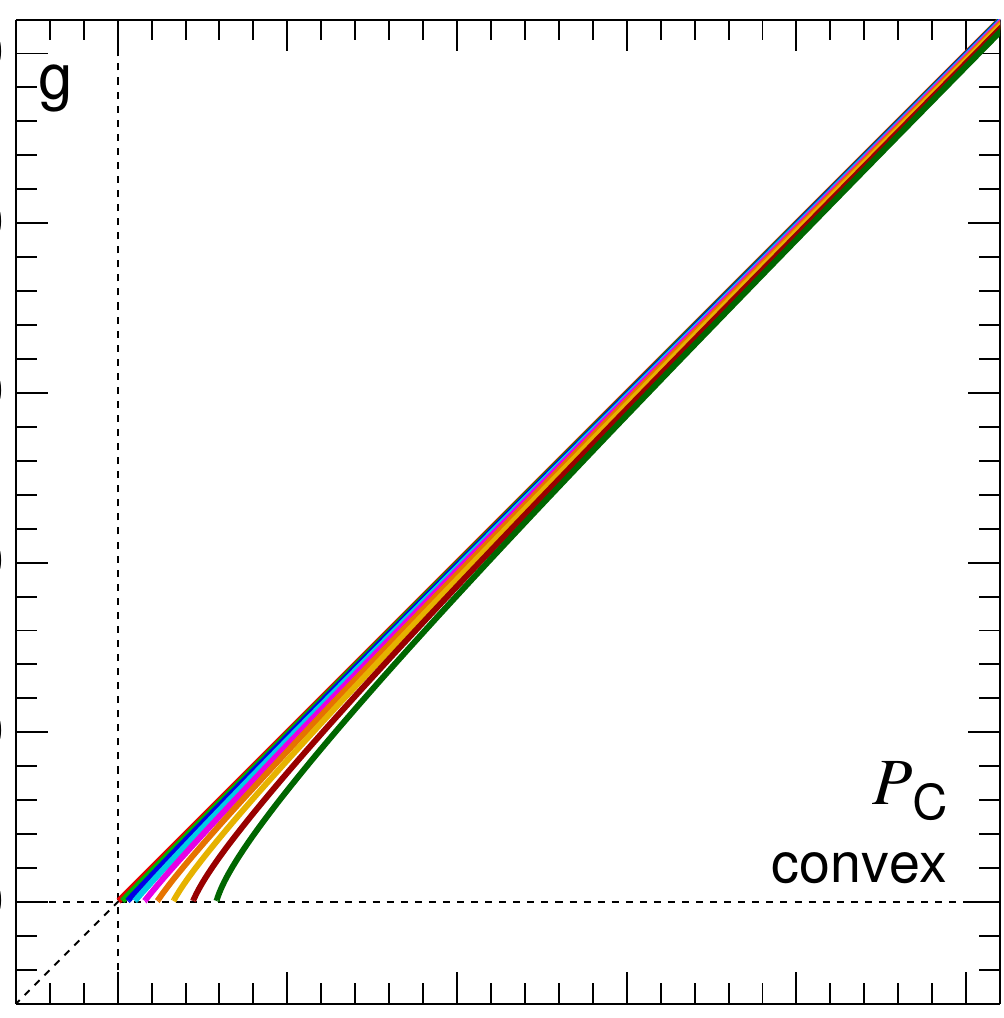}}  \hspace{-1.8mm}
  \resizebox{0.2755\hsize}{!}{\includegraphics{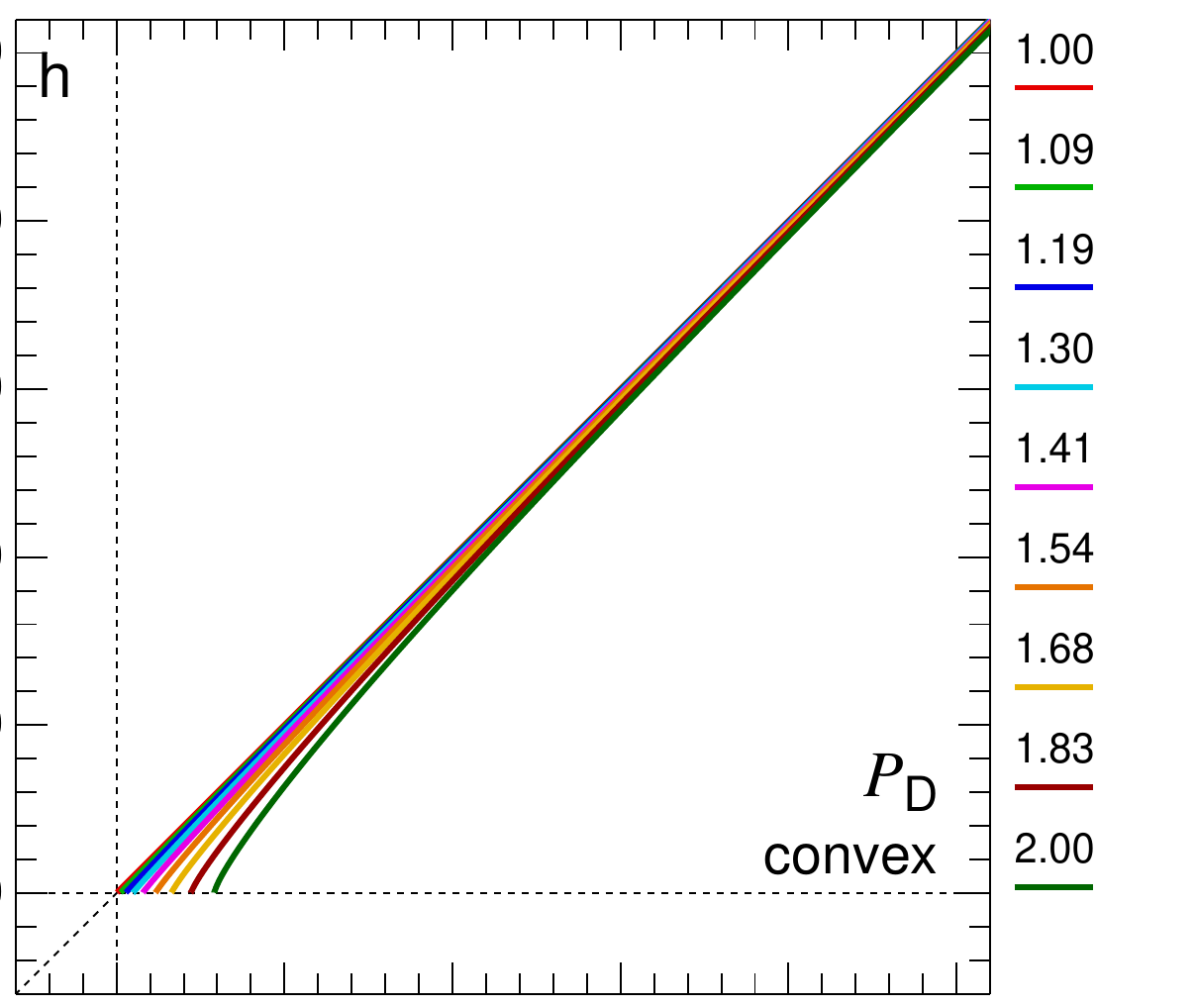}}}
\vspace{-0.94mm}
\centerline{
  \resizebox{0.2675\hsize}{!}{\includegraphics{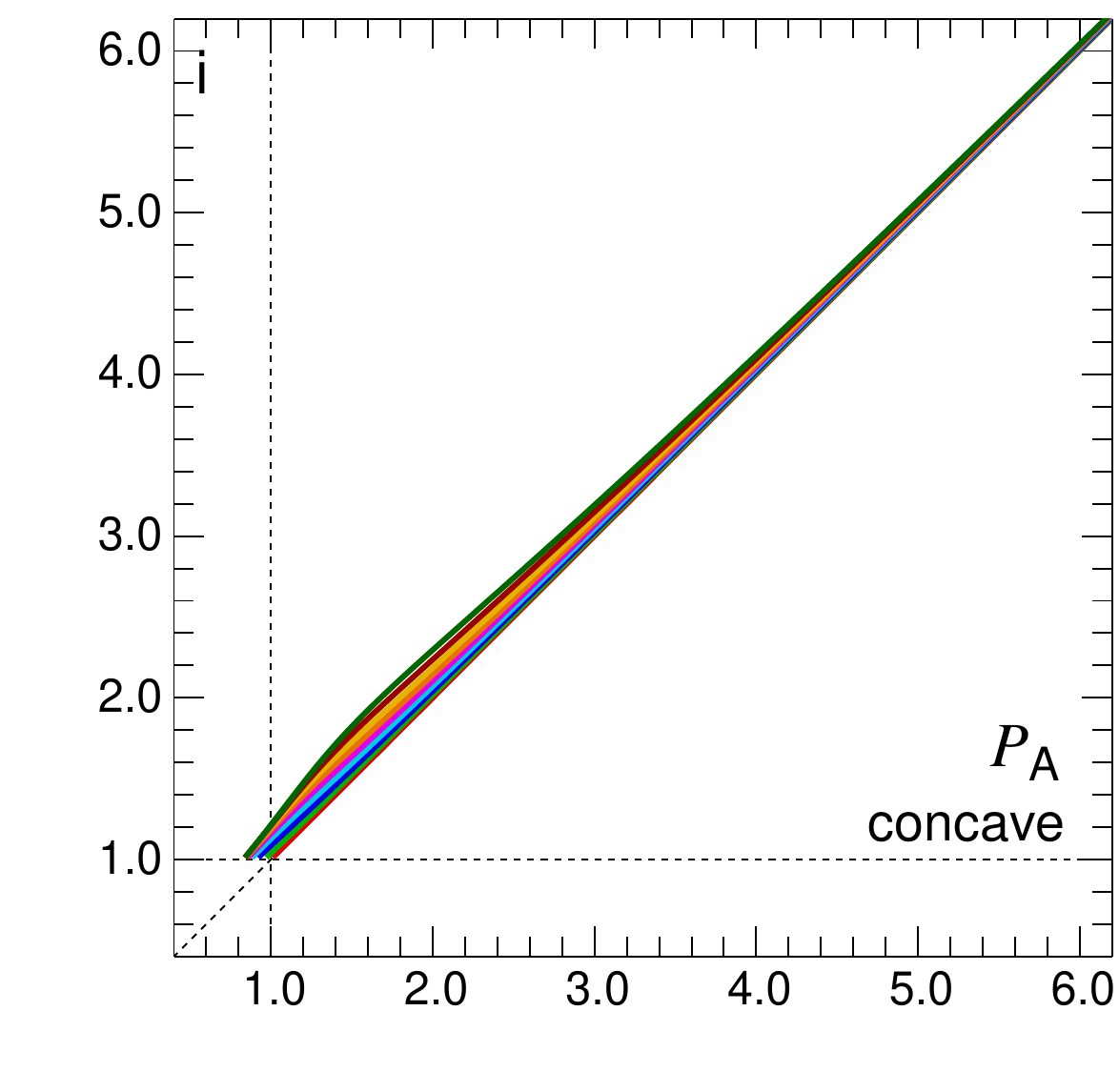}}  \hspace{-1.9mm}
  \resizebox{0.2832\hsize}{!}{\includegraphics{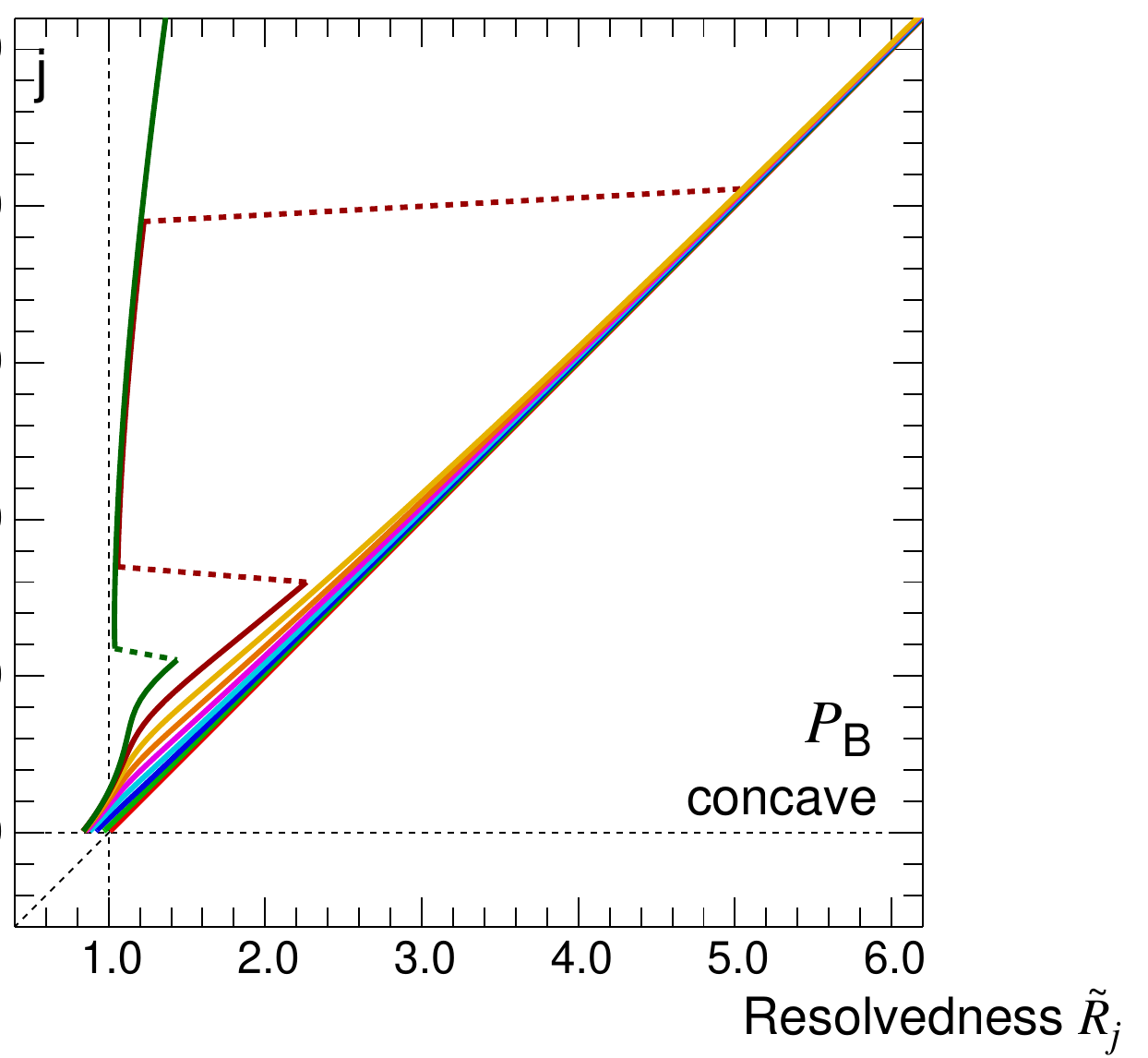}}  \hspace{-11.85mm}
  \resizebox{0.2287\hsize}{!}{\includegraphics{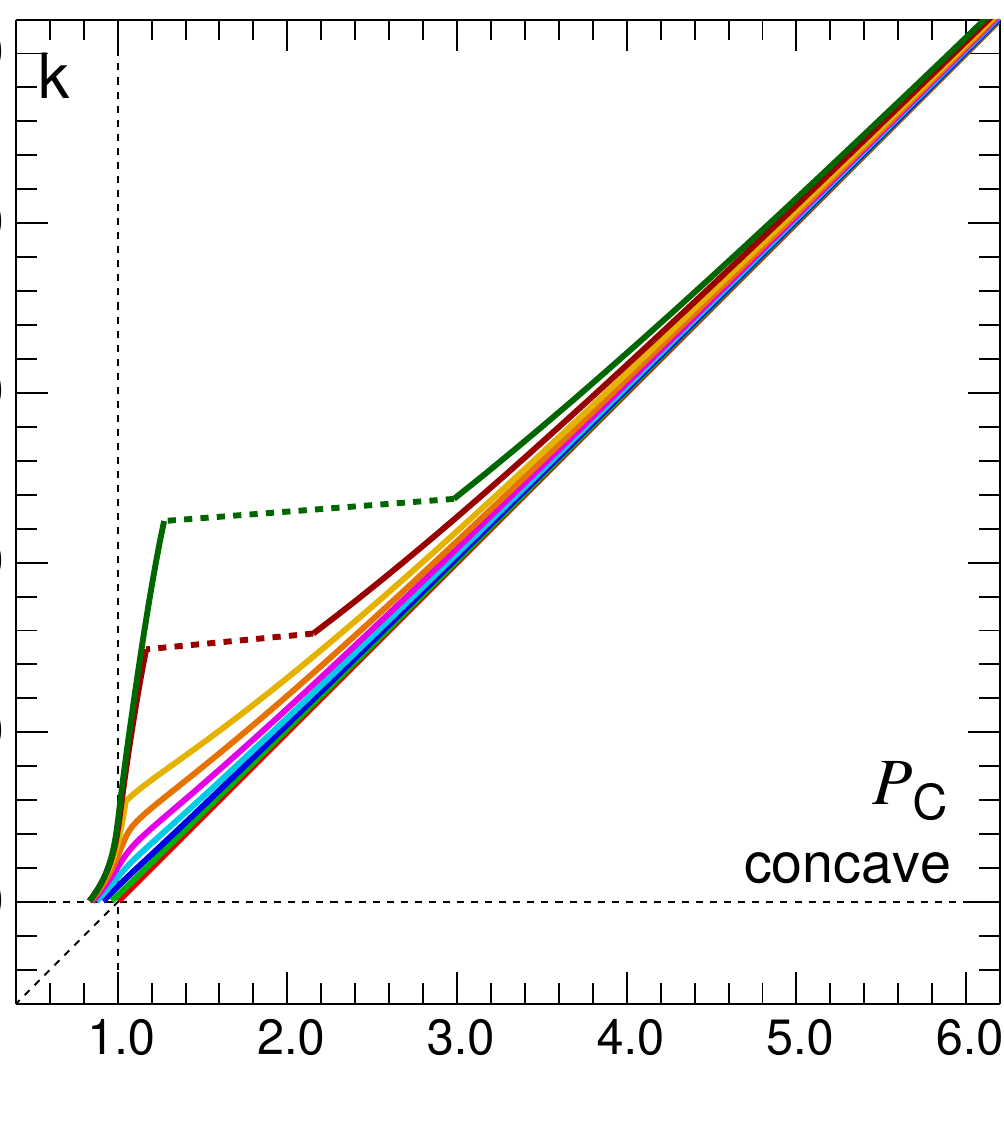}}  \hspace{-1.8mm}
  \resizebox{0.2738\hsize}{!}{\includegraphics{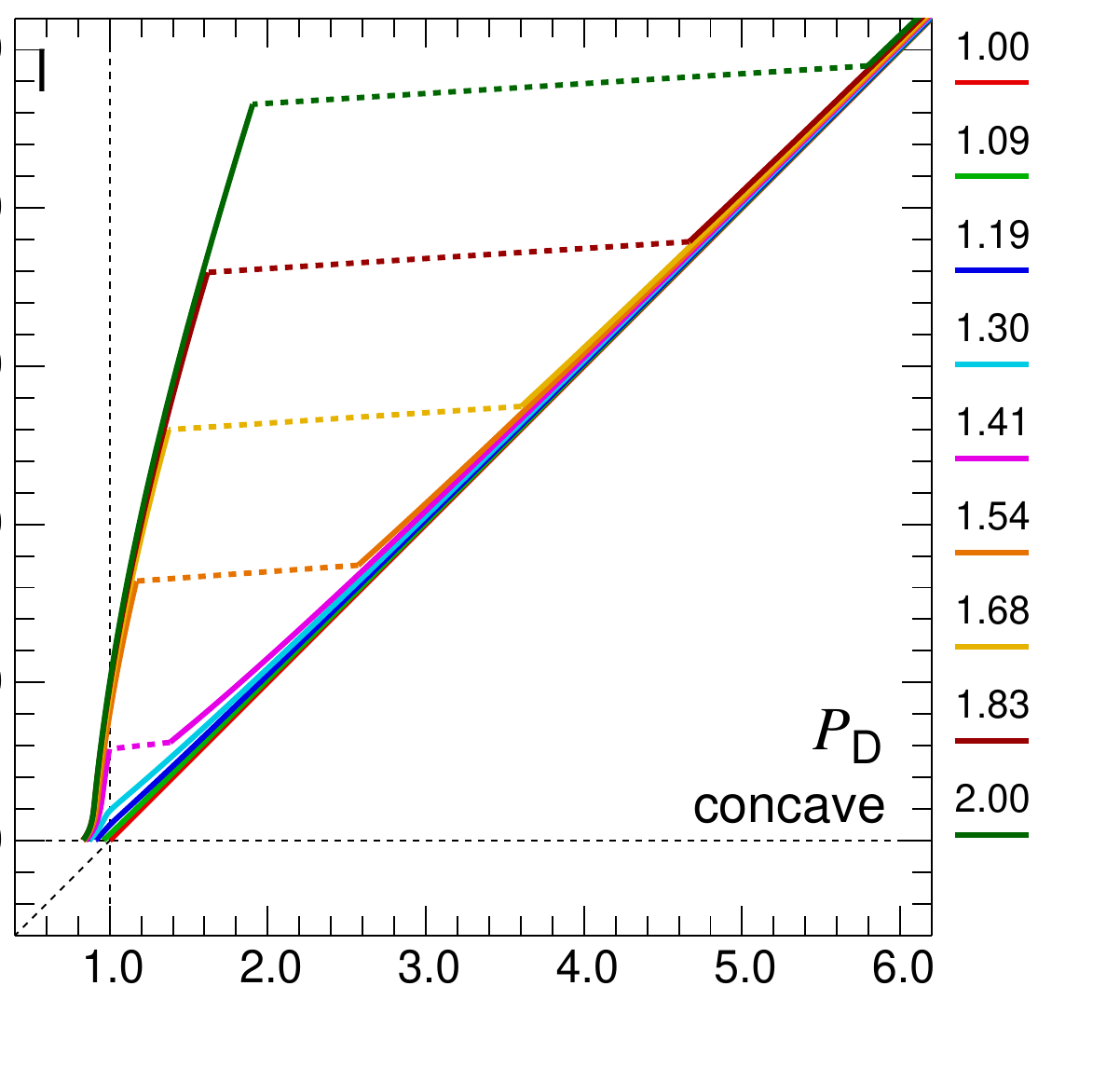}}}
\caption
{ 
Relationships between the model resolvedness $R_{j}$ and $\tilde{R}_{j}$ (Sect.~\ref{resolve}, Eq.~(\ref{resolvedness})) for the
results presented in Fig.~\ref{decgaussplawM}. The plots (with wide ranges of resolvedness values) are useful for a better
comprehension of the deconvolution accuracy plots in Figs.~\ref{decgaussflathillhell}\,--\,\ref{decgaussplawHfils}, \ref{decbesH},
\ref{decgaussflathillhellmom}, and \ref{decgaussplawM}, as well as the discontinuous jumps in Fig.~\ref{decgaussplawM}
(\emph{j}\,--\,\emph{l}).
} 
\label{resolvednM}
\end{figure*}

\subsection{Spherical Gaussian models}
\label{deconvgaussmom}

The deconvolved moment sizes for the Gaussian sources $\mathcal{\tilde{S}}_{{\mathcal{G}}{jl}}$ on the flat backgrounds
$\mathcal{B}$ are shown in Fig.~\ref{decgaussflathillhellmom} (\emph{a}\,--\,\emph{d}). Although the accuracy curves are
qualitatively similar to those presented in Fig.~\ref{decgaussflathillhell} (\emph{a}\,--\,\emph{d}) for the half maximum sizes,
the deconvolution errors become larger by a factor of two at $R_{j} > 1.1$. For the overestimated backgrounds with $\epsilon_{\,l}
> 0$, the moment sizes $\{D|C\}_{j}$ always become more underestimated, because the intensity moments depend on the outer parts of
the intensity distribution that are more distorted by background subtraction (Fig.~\ref{sourcesbgs}).

The deconvolved moment sizes for the Gaussian sources $\mathcal{\tilde{S}}_{{\mathcal{G}}{jl}}$ on the convex and concave
backgrounds $\mathcal{B}_{{\mathcal{G}}{jk}{\pm}}$ are presented in Fig.~\ref{decgaussflathillhellmom} (\emph{e}, \emph{f}). The
resulting accuracies behave qualitatively similar to the ones shown in Fig.~\ref{decgaussflathillhell} (\emph{e}, \emph{f}) for the
half maximum sizes, although with larger errors. For the wider convex backgrounds (with $f_{k} > 1$), the deconvolved moment sizes
$\{D|C\}_{j}$ are strongly overestimated (by factors of $17$\,--\,$41$) within $R_{j} \la 1.7$ (Fig.~\ref{decgaussflathillhellmom}
(\emph{e})), that is even for the resolved sources. In contrast, the overestimation of the half maximum sizes (by a factor of $17$)
was limited to $R_{j} \la 1.2$ (Fig.~\ref{decgaussflathillhell} (\emph{e})). This is because the faint outskirts of the Gaussian
sources are much more affected by the blending with the wider hill-like backgrounds than their peaks around half maximum. For the
concave backgrounds with $f_{k} > 1$, the moment sizes $D_{j}$ are underestimated by $50${\%} at $R_{j} = 1.1$
(Fig.~\ref{decgaussflathillhellmom} (\emph{f})), in contrast to $30${\%} for the half maximum sizes
(Fig.~\ref{decgaussflathillhell} (\emph{f})). At the same time, the moment sizes $C_{j}$ remain strongly biased toward the
unresolved sources ($1\!\leftarrow \tilde{R}_{j} \la 1.1$).


\subsection{Spherical power-law models}
\label{deconvplawsmom}

The deconvolved moment sizes for the power-law sources are markedly different from the half maximum sizes, because the intensity
moments get a major contribution from the extended power-law profiles (cf. Table~\ref{modeltable}). The simple background
$\mathcal{B}$ enables its precise subtraction (with $\epsilon_{\,l} = 0$) and accurate deconvolution for the the moment sizes
$D_{j}$ (Fig.~\ref{decgaussplawM}). For the convex and concave backgrounds that do not distort the source shapes ($f_{k} = 1$), the
deconvolved moment sizes $D_{j}$ are also perfectly accurate. In these simplest cases, the modified sizes $C_{j}$ become steeply
overestimated for the unresolved sources ($1\!\leftarrow \tilde{R}_{j} < 1.1$). For the three types of backgrounds, reduction in
the strength of the power-law profiles of $\mathcal{P}_{\rm B}$, $\mathcal{P}_{\rm C}$, and $\mathcal{P}_{\rm D}$, progressively
worsens the deconvolution results for the moment sizes $\{D|C\}_{j}$ (Fig.~\ref{decgaussplawM}), in contrast to those for the half
maximum sizes (Fig.~\ref{decgaussplawH}). This is because subtraction of an overestimated background makes the source profiles
steepen (drop to zero) toward the outer boundary (Fig.~\ref{sourcesbgs}).

For the power-law models $\mathcal{P}_{\rm \{A|B|C|D\}}$ on the flat background $\mathcal{B}$ that happen to be over-subtracted
($\epsilon_{\,l} > 0$), the deconvolved moment sizes $\{D|C\}_{j}$ become severely underestimated (Fig.~\ref{decgaussplawM}
(\emph{a}\,--\,\emph{d})). For the wider convex backgrounds $\mathcal{B}_{{\mathcal{P}}{jk}+}$ (with $f_{k} > 1$), the moment sizes
are steeply overestimated (Fig.~\ref{decgaussplawM} (\emph{e}\,--\,\emph{h})), even for the resolved sources at $1.4 \la
\tilde{R}_{j} \la 1.7$, which is qualitatively similar to the Gaussian sources (Fig.~\ref{decgaussflathillhellmom} (\emph{e})). For
the wider concave backgrounds $\mathcal{B}_{{\mathcal{P}}{jk}-}$, the deconvolved moment sizes become more underestimated
(Fig.~\ref{decgaussplawM} (\emph{i}\,--\,\emph{l})) than for the Gaussian sources (Fig.~\ref{decgaussflathillhellmom}
(\emph{f})). For certain angular resolutions $O_{j}$ and factors $f_{k}$, models $\mathcal{P}_{\rm \{B|C|D\}}$ exhibit
discontinuous jumps in the $D_{j}$ and $\tilde{R}_{j}$ values between $j$ and $j+1$ (Fig.~\ref{decgaussplawM})
(\emph{j}\,--\,\emph{l}). The jumps are caused by the sudden changes of the footprint radii $\Theta_{{\mathcal{M}}{jk}-}$ triggered
by the footprint determination algorithm of Eq.~(\ref{footprint}) in the perfectly smooth model images of sources and backgrounds.


The footprint changes in Fig.~\ref{decgaussplawM} (\emph{j}\,--\,\emph{l}) are caused by the appearance or disappearance of a
shallow secondary intensity minimum in the source shape, when it is superposed on the concave background (Fig.~\ref{modelsources}).
The profiles of some power-law models become fairly flat over a wide range of $\theta$ (best visible in $\mathcal{P}_{\rm \{B|D\}}$
at $O_{j} = 4${\arcsec}), which leads to the appearance of the secondary minima for some combinations of the model profile
(Fig.~\ref{modelprofs}), angular resolutions (Eq.~(\ref{convol})), and background shapes (Eq.~(\ref{modelback})). The sudden
footprint changes lead to substantial variations in the background-subtracted intensity distributions, hence in the moment sizes
$\tilde{M}_{j}$ and resolvedness $\tilde{R}_{j}$. To clarify the deconvolution results presented in this paper, as well as the
reasons behind the jumps, the relationships between $R_{j}$ and $\tilde{R}_{j}$ for the power-law models are shown in
Fig.~\ref{resolvednM}.

Source extractions in observed images do not show such subtle effects, because the shallow minima in the low-intensity areas are
destroyed by the noise and background fluctuations. The jumpy curves in Fig.~\ref{decgaussplawM} are to be ignored, because the
deconvolution (in)accuracy for the moment sizes, associated with the concave backgrounds, is clearly demonstrated by the continuous
curves.


\section{Tabulated material}
\label{tabresults}

Table \ref{resolvtable} presents the values for the model resolvedness $R_{j}$ (Sect.~\ref{resolve}), corresponding to the half
maximum and moment sizes at selected angular resolutions $O_{j}$. Tables \ref{infconvol}, \ref{finconvol}, and \ref{infpsfs}
present the deconvolution accuracies and power-law slopes, corresponding to the convolved infinite and finite models in
Figs.~\ref{plawinfconv}, \ref{plawfinconv}, and \ref{anianoconv} (Sect.~\ref{convsrcfil}).

\begin{table*} 
\caption{Resolvedness $R_{j}$ for the half maximum and moment sizes $\{H|M\}_{j}$ of the spherical Gaussian and
power-law models $\mathcal{M}$ (Fig.~\ref{modelprofs}, Table~\ref{modeltable}), at selected angular resolutions $O_{j}$.
} 
\begin{tabular}{rlllllllll}
\hline\hline
\noalign{\smallskip}
\noalign{\hspace{0.37cm}$O_{j}$\hspace{2.1cm}from half maximum sizes $H_{j}$\hspace{1.25cm}from moment sizes $M_{j}$}
(\arcsec) & $R_{j\,\,\mathcal{G}}$ & $R_{j\,\,\mathcal{P}_{\rm A}}$ & $R_{j\,\,\mathcal{P}_{\rm B}}$ 
& $R_{j\,\,\mathcal{P}_{\rm C}}$ & \!\!$R_{j\,\,\mathcal{P}_{\rm D}}$ & \,\,\,$R_{j\,\,\mathcal{P}_{\rm A}}$ 
& \!$R_{j\,\,\mathcal{P}_{\rm B}}$ & $\!R_{j\,\,\mathcal{P}_{\rm C}}$ & \!$R_{j\,\,\mathcal{P}_{\rm D}}$ \\
\noalign{\smallskip}
\hline
\noalign{\smallskip}
4   & 2.69    & 2.93  & \!2.69  & \!2.74   & \!\!\!2.69   &      13.0   & \!\!\!\!\!11.5 & \!\!9.17  & \!6.99   \\
8   & 1.60    & 2.02  & \!1.63  & \!1.73   & \!\!\!1.61   & \,\,\,6.54  & \!\!5.79       & \!\!4.67  & \!3.60   \\
16  & 1.18    & 1.68  & \!1.30  & \!1.35   & \!\!\!1.21   & \,\,\,3.38  & \!\!3.02       & \!\!2.49  & \!2.00   \\
32  & 1.048   & 1.49  & \!1.24  & \!1.20   & \!\!\!1.10   & \,\,\,1.90  & \!\!1.74       & \!\!1.52  & \!1.32   \\
64  & 1.012   & 1.25  & \!1.16  & \!1.10   & \!\!\!1.054  & \,\,\,1.29  & \!\!1.23       & \!\!1.15  & \!1.089  \\
128 & 1.0030  & 1.077 & \!1.057 & \!1.036  & \!\!\!1.020  & \,\,\,1.079 & \!\!1.062      & \!\!1.040 & \!1.023  \\
256 & 1.00076 & 1.020 & \!1.015 & \!1.0098 & \!\!\!1.0056 & \,\,\,1.015 & \!\!1.016      & \!\!1.010 & \!1.0058 \\
\noalign{\smallskip}
\hline
\label{resolvtable}
\end{tabular}
\end{table*} 

\begin{table*} 
\caption{Deconvolution accuracy ratios $A = D_{j}/H$ (\emph{left}) and power-law slopes $\gamma$ (\emph{right}) for the
infinite power-law sources $\mathcal{S}$ and filaments $\mathcal{F}$ (Sect.~\ref{convsrcfil}) with profiles, analogous to
$\mathcal{P_{\rm \{A|B|C|D\}}}$ (Fig.~\ref{modelprofs}), convolved with the Gaussian kernels $\mathcal{O}_{j}$ and power-law
kernels $\mathcal{K}_{j}$ of half maximum sizes $O_{j}$ (cf. Fig.~\ref{plawinfconv}), without any background.
} 
\begin{tabular}{rrllllllllllllllll}
\hline\hline
\noalign{\smallskip}
& \!\!\!\!\!\!$O_{j}$ & & \!\!\!\!\!\!\!$\mathcal{P}_{\rm A*}$ & 
& \!\!\!\!\!\!\!\!$\mathcal{P}_{\rm B*}$ & & \!\!\!\!\!\!\!\!$\mathcal{P}_{\rm C*}$ & & \!\!\!\!\!\!\!\!$\mathcal{P}_{\rm D*}$ & 
& \!\!\!\!\!\!\!\!\!$\mathcal{P}_{\rm A*}$ & 
& \!\!\!\!\!\!\!\!\!$\mathcal{P}_{\rm B*}$ & & \!\!\!\!\!\!\!\!\!$\mathcal{P}_{\rm C*}$ & 
& \!\!\!\!\!\!\!\!\!$\mathcal{P}_{\rm D*}$ \\
& ({\arcsec})  
& \,\,\,\,\,\,$A_{\mathcal{O}_{j}}$ & \,$A_{\mathcal{K}_{j}}$ & \,$A_{\mathcal{O}_{j}}$ & $A_{\mathcal{K}_{j}}$ 
& \,$A_{\mathcal{O}_{j}}$ & $A_{\mathcal{K}_{j}}$ & \,$A_{\mathcal{O}_{j}}$ & $A_{\mathcal{K}_{j}}$ 
& \,\,\,$\gamma_{\mathcal{O}_{j}}$ & \!\!$\gamma_{\mathcal{K}_{j}}$ & $\gamma_{\mathcal{O}_{j}}$ & \!\!$\gamma_{\mathcal{K}_{j}}$ 
& $\gamma_{\mathcal{O}_{j}}$ & \!\!$\gamma_{\mathcal{K}_{j}}$ & $\gamma_{\mathcal{O}_{j}}$ & \!\!$\gamma_{\mathcal{K}_{j}}$ \\
\noalign{\smallskip}
\hline
\noalign{\smallskip}
\!$\mathcal{S}_{ }$ &\!\!\!\!4   &\,\,\,\,1.10 &       \!1.44 &       \!1.00 &       \!1.13 &       \!1.02 &       \!1.17 & 1.00 
&       \!1.11 & \,\,\,1.0 & \!\!1.0  & 1.0 & \!\!1.0  & 2.0 & \!\!1.9 & 2.0 & \!\!1.9  \\
  $\mathcal{F}_{ }$ &\!\!\!\!4   &\,\,\,\,1.08 &       \!1.60 &       \!1.00 &       \!1.26 &       \!1.02 &       \!1.31 & 1.00 
&       \!1.24 & \,\,\,1.0 & \!\!1.0  & 1.0 & \!\!1.0  & 2.0 & \!\!1.1 & 2.0 & \!\!1.1  {\smallskip}\\
\!$\mathcal{S}_{ }$ &\!\!\!\!32  &\,\,\,\,4.23 &       \!5.58 &       \!2.62 &       \!3.25 &       \!2.14 &       \!2.42 & 1.49 
&       \!1.67 & \,\,\,1.0 & \!\!1.0  & 1.0 & \!\!1.0  & 2.0 & \!\!1.9 & 2.0 & \!\!1.8  \\
  $\mathcal{F}_{ }$ &\!\!\!\!32  &\,\,\,\,2.77 &       \!4.40 &       \!1.67 &       \!2.87 &       \!1.60 &       \!2.67 & 1.19 
&       \!2.28 & \,\,\,1.0 & \!\!1.0  & 1.0 & \!\!1.0  & 2.0 & \!\!1.1 & 2.0 & \!\!1.1  {\smallskip}\\
\!$\mathcal{S}_{ }$ &\!\!\!\!256 & \,31.91 &\!\!\!\!41.39 &\!\!\!\!29.02 &\!\!\!\!37.31 &\!\!\!\!10.76 &\!\!\!\!12.03 & 8.19 
&       \!9.12 & \,\,\,1.0 & \!\!1.0  & 1.0 & \!\!1.0  & 2.0 & \!\!1.9 & 2.0 & \!\!1.8  \\
  $\mathcal{F}_{ }$ &\!\!\!\!256 & \,14.94 &\!\!\!\!24.60 &\!\!\!\!10.14 &\!\!\!\!19.15 &       \!4.10 &\!\!\!\!14.17 & 2.65 
&\!\!\!\!13.60 & \,\,\,1.0 & \!\!1.0  & 1.0 & \!\!1.0  & 2.0 & \!\!1.1 & 2.0 & \!\!1.1  \\
\noalign{\smallskip}
\hline
\label{infconvol}
\end{tabular}
\end{table*} 

\begin{table*} 
\caption{Deconvolution accuracy ratios $A = D_{j}/H$ (\emph{left}) and power-law slopes $\gamma$ (\emph{right}) for the
finite Gaussian and power-law sources $\mathcal{S}$ and filaments $\mathcal{F}$ (Sect.~\ref{convsrcfil}) with profiles
$\mathcal{\{G|P_{\rm \{A|B|C|D\}}\}}$ from Fig.~\ref{modelprofs}, convolved with the Gaussian kernels $\mathcal{O}_{j}$ and
power-law kernels $\mathcal{K}_{j}$ of half maximum sizes $O_{j}$ (cf. Fig.~\ref{plawfinconv}), without any background. These
results correspond to those presented in Sect.~\ref{results} for a perfectly subtracted flat background ($\epsilon_{\,l} = 0$).
} 
\begin{tabular}{rrlllllllllllllll}
\hline\hline
\noalign{\smallskip}
& \!\!\!\!\!\!$O_{j}$ & & \!\!\!\!\!$\mathcal{G}$ & & \!\!\!\!\!\!\!$\mathcal{P}_{\rm A}$ & 
& \!\!\!\!\!\!\!$\mathcal{P}_{\rm B}$ & & \!\!\!\!\!\!\!$\mathcal{P}_{\rm C}$ & & \!\!\!\!\!\!\!$\mathcal{P}_{\rm D}$ 
& \,\,\,\,\,$\mathcal{G}$ & $\mathcal{P}_{\rm A}$ & $\mathcal{P}_{\rm B}$ & $\mathcal{P}_{\rm C}$ & $\mathcal{P}_{\rm D}$ \\
& ({\arcsec}) & \,\,\,\,$A_{\mathcal{O}_{j}}$ & $A_{\mathcal{K}_{j}}$ 
& \,$A_{\mathcal{O}_{j}}$ & \,$A_{\mathcal{K}_{j}}$ & \,$A_{\mathcal{O}_{j}}$ & $A_{\mathcal{K}_{j}}$ 
& \,$A_{\mathcal{O}_{j}}$ & $A_{\mathcal{K}_{j}}$ & \,$A_{\mathcal{O}_{j}}$ & $A_{\mathcal{K}_{j}}$ 
& \,\,\,\,$\gamma_{\mathcal{K}_{j}}$ & $\gamma_{\mathcal{K}_{j}}$ & $\gamma_{\mathcal{K}_{j}}$ & $\gamma_{\mathcal{K}_{j}}$ 
& $\gamma_{\mathcal{K}_{j}}$ \\
\noalign{\smallskip}
\hline
\noalign{\smallskip}
\!$\mathcal{S}_{ }$ &\!\!\!\!4   &\,\,1.00 &       \!1.10 & 1.10 &       \!1.39 & 1.00 &       \!1.12 & 1.02 &       \!1.17 & 1.00
 &       \!1.11 & \,\,\,\,2.0 & 2.0 & 2.0 & 2.0 & 2.0 \\
  $\mathcal{F}_{ }$ &\!\!\!\!4   &\,\,1.00 &       \!1.23 & 1.08 &       \!1.55 & 1.00 &       \!1.26 & 1.02 &       \!1.31 & 1.00
 &       \!1.24 & \,\,\,\,1.0 & 1.3 & 1.3 & 1.3 & 1.3 {\smallskip}\\
\!$\mathcal{S}_{ }$ &\!\!\!\!32  &\,\,1.00 &       \!1.10 & 3.53 &       \!4.01 & 2.35 &       \!2.63 & 2.11 &       \!2.35 & 1.47
 &       \!1.63 & \,\,\,\,2.0 & 2.0 & 2.0 & 2.0 & 2.0 \\
  $\mathcal{F}_{ }$ &\!\!\!\!32  &\,\,1.00 &       \!2.10 & 2.50 &       \!3.72 & 1.59 &       \!2.67 & 1.59 &       \!2.65 & 1.19
 &       \!2.27 & \,\,\,\,1.0 & 1.3 & 1.3 & 1.3 & 1.3 {\smallskip}\\
\!$\mathcal{S}_{ }$ &\!\!\!\!256 &\,\,1.00 &       \!1.10 & 5.16 &       \!5.69 & 4.52 &       \!4.98 & 3.60 &       \!3.96 & 2.71
 &       \!2.98 & \,\,\,\,2.0 & 2.0 & 2.0 & 2.0 & 2.0 \\
  $\mathcal{F}_{ }$ &\!\!\!\!256 &\,\,1.00 &\!\!\!\!13.24 & 4.04 &\!\!\!\!14.09 & 2.82 &\!\!\!\!13.64 & 2.34 &\!\!\!\!13.50 & 1.62
 &\!\!\!\!13.33 & \,\,\,\,1.0 & 1.2 & 1.2 & 1.2 & 1.2 \\
\noalign{\smallskip}
\hline
\label{finconvol}
\end{tabular}
\end{table*} 

\begin{table*} 
\caption{Deconvolution accuracy ratios $A = D_{j}/H$ for the Gaussian and infinite power-law sources $\mathcal{S}$ and
filaments $\mathcal{F}$ (Sect.~\ref{convsrcfil}) with profiles, analogous to $\mathcal{\{G|P_{\rm \{A|C\}}\}}$ from
Fig.~\ref{modelprofs}, convolved with kernels with the PSF shapes \citep{Aniano_etal2011} of \emph{Herschel} PACS at $70$\,$\mu$m,
\emph{Herschel} SPIRE at $250$\,$\mu$m, and \emph{Spitzer} MIPS at $160$\,$\mu$m (the kernel profiles are shown in
Fig.~\ref{anianoconv}).
} 
\begin{tabular}{rrlllllllll}
\hline\hline
\noalign{\smallskip}
& \!\!\!\!\!\!$O_{j}$ & & \,\,$\mathcal{G}$ & & & \,\,$\mathcal{P}_{\rm A*}$ & & & \,\,$\mathcal{P}_{\rm C*}$ \\
& ({\arcsec}) 
& \,\,\,\,$A_{\rm PACS}$ & \!\!$A_{\rm SPIRE}$ & \!\!$A_{\rm MIPS}$ 
& \,\,\,\,$A_{\rm PACS}$ & \!\!$A_{\rm SPIRE}$ & \!\!$A_{\rm MIPS}$ 
& \,\,\,\,$A_{\rm PACS}$ & \!\!$A_{\rm SPIRE}$ & \!\!$A_{\rm MIPS}$ \\
\noalign{\smallskip}
\hline
\noalign{\smallskip}
\!$\mathcal{S}_{ }$ &\!\!\!\!4   &\,\,\,\,\,1.07 &\!\!1.03 &\!\!1.11 &\,\,\,\,\,1.28 &       1.18 &       1.36 &\,\,\,\,\,1.12 
&       1.06 &       1.17 \\
  $\mathcal{F}_{ }$ &\!\!\!\!4   &\,\,\,\,\,1.11 &\!\!1.06 &\!\!1.15 &\,\,\,\,\,1.28 &       1.17 &       1.35 &\,\,\,\,\,1.15 
&       1.08 &       1.20 {\smallskip}\\
\!$\mathcal{S}_{ }$ &\!\!\!\!32  &\,\,\,\,\,0.90 &\!\!0.88 &\!\!0.91 &\,\,\,\,\,4.93 &       4.24 &       4.88 &\,\,\,\,\,2.21 
&       2.04 &       2.15 \\
  $\mathcal{F}_{ }$ &\!\!\!\!32  &\,\,\,\,\,1.54 &\!\!0.99 &\!\!1.54 &\,\,\,\,\,3.58 &       2.87 &       3.62 &\,\,\,\,\,2.13 
&       1.62 &       2.14 {\smallskip}\\
\!$\mathcal{S}_{ }$ &\!\!\!\!256 &\,\,\,\,\,2.34 &\!\!0.39 &\!\!0.88 &     \,\,36.87 &\!\!\!31.81 &\!\!\!36.21 &     \,\,10.90 
&\!\!\!10.16 &\!\!\!10.69 \\
  $\mathcal{F}_{ }$ &\!\!\!\!256 &\,\,\,\,\,8.16 &\!\!1.81 &\!\!8.90 &     \,\,20.20 &\!\!\!15.45 &\!\!\!20.14 &\,\,\,\,\,9.50 
&       4.36 &       9.97 \\
\noalign{\smallskip}
\hline
\label{infpsfs}
\end{tabular}
\end{table*} 

\end{appendix}


\bibliographystyle{aa}
\bibliography{aamnem99,deconv}

\begin{thebibliography}{30}
\expandafter\ifx\csname natexlab\endcsname\relax\def\natexlab#1{#1}\fi

\bibitem[{{Andr{\'e}} {et~al.}(2022){Andr{\'e}}, {Palmeirim}, \&
  {Arzoumanian}}]{Andre_etal2022}
{Andr{\'e}}, P., {Palmeirim}, P., \& {Arzoumanian}, D. 2022, \aap, 667, L1

\bibitem[{{Andr{\'e}} {et~al.}(2016){Andr{\'e}}, {Rev{\'e}ret}, {K{\"o}nyves},
  {Arzoumanian}, {Tig{\'e}}, {Gallais}, {Roussel}, {Le Pennec}, {Rodriguez},
  {Doumayrou}, {Dubreuil}, {Lortholary}, {Martignac}, {Talvard}, {Delisle},
  {Visticot}, {Dumaye}, {De Breuck}, {Shimajiri}, {Motte}, {Bontemps},
  {Hennemann}, {Zavagno}, {Russeil}, {Schneider}, {Palmeirim}, {Peretto},
  {Hill}, {Minier}, {Roy}, \& {Rygl}}]{Andre_etal2016}
{Andr{\'e}}, P., {Rev{\'e}ret}, V., {K{\"o}nyves}, V., {et~al.} 2016, \aap,
  592, A54

\bibitem[{{Aniano} {et~al.}(2011){Aniano}, {Draine}, {Gordon}, \&
  {Sandstrom}}]{Aniano_etal2011}
{Aniano}, G., {Draine}, B.~T., {Gordon}, K.~D., \& {Sandstrom}, K. 2011, \pasp,
  123, 1218

\bibitem[{{Arzoumanian} {et~al.}(2011){Arzoumanian}, {Andr{\'e}}, {Didelon},
  {K{\"o}nyves}, {Schneider}, {Men'shchikov}, {Sousbie}, {Zavagno}, {Bontemps},
  {di Francesco}, {Griffin}, {Hennemann}, {Hill}, {Kirk}, {Martin}, {Minier},
  {Molinari}, {Motte}, {Peretto}, {Pezzuto}, {Spinoglio}, {Ward-Thompson},
  {White}, \& {Wilson}}]{Arzoumanian_etal2011}
{Arzoumanian}, D., {Andr{\'e}}, P., {Didelon}, P., {et~al.} 2011, \aap, 529,
  L6+

\bibitem[{{Arzoumanian} {et~al.}(2019){Arzoumanian}, {Andr{\'e}},
  {K{\"o}nyves}, {Palmeirim}, {Roy}, {Schneider}, {Benedettini}, {Didelon}, {Di
  Francesco}, {Kirk}, \& {Ladjelate}}]{Arzoumanian_etal2019}
{Arzoumanian}, D., {Andr{\'e}}, P., {K{\"o}nyves}, V., {et~al.} 2019, \aap,
  621, A42

\bibitem[{{Bertin} {et~al.}(2002){Bertin}, {Mellier}, {Radovich}, {Missonnier},
  {Didelon}, \& {Morin}}]{Bertin_etal2002}
{Bertin}, E., {Mellier}, Y., {Radovich}, M., {et~al.} 2002, in Astronomical
  Society of the Pacific Conference Series, Vol. 281, Astronomical Data
  Analysis Software and Systems XI, ed. {D.~A.~Bohlender, D.~Durand, \&
  T.~H.~Handley}, 228

\bibitem[{{Bonnor}(1956)}]{Bonnor1956}
{Bonnor}, W.~B. 1956, \mnras, 116, 351

\bibitem[{{Bontemps} {et~al.}(2010){Bontemps}, {Motte}, {Csengeri}, \&
  {Schneider}}]{Bontemps_etal2010a}
{Bontemps}, S., {Motte}, F., {Csengeri}, T., \& {Schneider}, N. 2010, \aap,
  524, A18

\bibitem[{{Cox} {et~al.}(2016){Cox}, {Arzoumanian}, {Andr{\'e}}, {Rygl},
  {Prusti}, {Men'shchikov}, {Royer}, {K{\'o}sp{\'a}l}, {Palmeirim}, {Ribas},
  {K{\"o}nyves}, {Bernard}, {Schneider}, {Bontemps}, {Merin}, {Vavrek}, {Alves
  de Oliveira}, {Didelon}, {Pilbratt}, \& {Waelkens}}]{Cox_etal2016}
{Cox}, N.~L.~J., {Arzoumanian}, D., {Andr{\'e}}, P., {et~al.} 2016, \aap, 590,
  A110

\bibitem[{{Dewangan} {et~al.}(2023){Dewangan}, {Bhadari}, {Men'shchikov},
  {Chung}, {Devaraj}, {Lee}, {Maity}, \& {Baug}}]{Dewangan_etal2023}
{Dewangan}, L.~K., {Bhadari}, N.~K., {Men'shchikov}, A., {et~al.} 2023, \apj,
  946, 22

\bibitem[{{Hacar} {et~al.}(2018){Hacar}, {Tafalla}, {Forbrich}, {Alves},
  {Meingast}, {Grossschedl}, \& {Teixeira}}]{Hacar_etal2018}
{Hacar}, A., {Tafalla}, M., {Forbrich}, J., {et~al.} 2018, \aap, 610, A77

\bibitem[{{Joye} \& {Mandel}(2003)}]{JoyeMandel2003}
{Joye}, W.~A. \& {Mandel}, E. 2003, in Astronomical Society of the Pacific
  Confe\-rence Series, Vol. 295, Astronomical Data Analysis Software and
  Systems XII, ed. H.~E. {Payne}, R.~I. {Jedrzejewski}, \& R.~N. {Hook}, 489

\bibitem[{{K{\"o}nyves} {et~al.}(2015){K{\"o}nyves}, {Andr{\'e}},
  {Men'shchikov}, {Palmeirim}, {Arzoumanian}, {Schneider}, {Roy}, {Didelon},
  {Maury}, {Shimajiri}, {Di Francesco}, {Bontemps}, {Peretto}, {Benedettini},
  {Bernard}, {Elia}, {Griffin}, {Hill}, {Kirk}, {Ladjelate}, {Marsh}, {Martin},
  {Motte}, {Nguy{\^e}n Luong}, {Pezzuto}, {Roussel}, {Rygl}, {Sadavoy},
  {Schisano}, {Spinoglio}, {Ward-Thompson}, \& {White}}]{Ko"nyves_etal2015}
{K{\"o}nyves}, V., {Andr{\'e}}, P., {Men'shchikov}, A., {et~al.} 2015, \aap,
  584, A91

\bibitem[{{Ladjelate} {et~al.}(2020){Ladjelate}, {Andr{\'e}}, {K{\"o}nyves},
  {Ward-Thompson}, {Men'shchikov}, {Bracco}, {Palmeirim}, {Roy}, {Shimajiri},
  {Kirk}, {Arzoumanian}, {Benedettini}, {Di Francesco}, {Fiorellino},
  {Schneider}, {Pezzuto}, {Motte}, \& {Herschel Gould Belt Survey
  Team}}]{Ladjelate_etal2020}
{Ladjelate}, B., {Andr{\'e}}, P., {K{\"o}nyves}, V., {et~al.} 2020, \aap, 638,
  A74

\bibitem[{{Larson}(1969)}]{Larson1969}
{Larson}, R.~B. 1969, \mnras, 145, 271

\bibitem[{{Louvet} {et~al.}(2021){Louvet}, {Hennebelle}, {Men'shchikov},
  {Didelon}, {Ntormousi}, \& {Motte}}]{Louvet_etal2021}
{Louvet}, F., {Hennebelle}, P., {Men'shchikov}, A., {et~al.} 2021, \aap, 653,
  A157

\bibitem[{{Men'shchikov}(2021{\natexlab{a}})}]{Men'shchikov2021b}
{Men'shchikov}, A. 2021{\natexlab{a}}, \aap, 654, A78

\bibitem[{{Men'shchikov}(2021{\natexlab{b}})}]{Men'shchikov2021a}
{Men'shchikov}, A. 2021{\natexlab{b}}, \aap, 649, A89

\bibitem[{{Men'shchikov} {et~al.}(2012){Men'shchikov}, {Andr{\'e}}, {Didelon},
  {Motte}, {Hennemann}, \& {Schneider}}]{Men'shchikov_etal2012}
{Men'shchikov}, A., {Andr{\'e}}, P., {Didelon}, P., {et~al.} 2012, \aap, 542,
  A81

\bibitem[{{Men'shchikov} {et~al.}(2001){Men'shchikov}, {Balega}, {Bl{\"o}cker},
  {Osterbart}, \& {Weigelt}}]{Men'shchikov_etal2001}
{Men'shchikov}, A.~B., {Balega}, Y., {Bl{\"o}cker}, T., {Osterbart}, R., \&
  {Weigelt}, G. 2001, \aap, 368, 497

\bibitem[{{Men'shchikov} \& {Henning}(1997)}]{Men'shchikovHenning1997}
{Men'shchikov}, A.~B. \& {Henning}, T. 1997, \aap, 318, 879

\bibitem[{{Men'shchikov} {et~al.}(1999){Men'shchikov}, {Henning}, \&
  {Fischer}}]{Men'shchikov_etal1999}
{Men'shchikov}, A.~B., {Henning}, T., \& {Fischer}, O. 1999, \apj, 519, 257

\bibitem[{{Molinari} {et~al.}(2011){Molinari}, {Schisano}, {Faustini},
  {Pestalozzi}, {di Giorgio}, \& {Liu}}]{Molinari_etal2011}
{Molinari}, S., {Schisano}, E., {Faustini}, F., {et~al.} 2011, \aap, 530, A133+

\bibitem[{{Motte} {et~al.}(2003){Motte}, {Schilke}, \& {Lis}}]{Motte_etal2003}
{Motte}, F., {Schilke}, P., \& {Lis}, D.~C. 2003, \apj, 582, 277

\bibitem[{{Nguyen Luong} {et~al.}(2011){Nguyen Luong}, {Motte}, {Hennemann},
  {Hill}, {Rygl}, {Schneider}, {Bontemps}, {Men'shchikov}, {Andr{\'e}},
  {Peretto}, {Anderson}, {Arzoumanian}, {Deharveng}, {Didelon}, {di Francesco},
  {Griffin}, {Kirk}, {K{\"o}nyves}, {Martin}, {Maury}, {Minier}, {Molinari},
  {Pestalozzi}, {Pezzuto}, {Reid}, {Roussel}, {Sauvage}, {Schuller}, {Testi},
  {Ward-Thompson}, {White}, \& {Zavagno}}]{NguenLuong_etal2011}
{Nguyen Luong}, Q., {Motte}, F., {Hennemann}, M., {et~al.} 2011, \aap, 535, A76

\bibitem[{{Padoan} {et~al.}(2023){Padoan}, {Pelkonen}, {Juvela},
  {Haugb{\o}lle}, \& {Nordlund}}]{Padoan_etal2023}
{Padoan}, P., {Pelkonen}, V.~M., {Juvela}, M., {Haugb{\o}lle}, T., \&
  {Nordlund}, {\r{A}}. 2023, \mnras, 522, 3548

\bibitem[{{Palmeirim} {et~al.}(2013){Palmeirim}, {Andr{\'e}}, {Kirk},
  {Ward-Thompson}, {Arzoumanian}, {K{\"o}nyves}, {Didelon}, {Schneider},
  {Benedettini}, {Bontemps}, {Di Francesco}, {Elia}, {Griffin}, {Hennemann},
  {Hill}, {Martin}, {Men'shchikov}, {Molinari}, {Motte}, {Nguyen Luong},
  {Nutter}, {Peretto}, {Pezzuto}, {Roy}, {Rygl}, {Spinoglio}, \&
  {White}}]{Palmeirim_etal2013}
{Palmeirim}, P., {Andr{\'e}}, P., {Kirk}, J., {et~al.} 2013, \aap, 550, A38

\bibitem[{{Pence}(1999)}]{Pence1999}
{Pence}, W. 1999, in Astronomical Society of the Pacific Conference Series,
  Vol. 172, Astronomical Data Analysis Software and Systems VIII, ed. {D.~M.
  Mehringer, R.~L.~Plante, \& D.~A.~Roberts}, 487--+

\bibitem[{{Pouteau} {et~al.}(2022){Pouteau}, {Motte}, {Nony},
  {Galv{\'a}n-Madrid}, {Men'shchikov}, {Bontemps}, {Robitaille}, {Louvet},
  {Ginsburg}, {Herpin}, {L{\'o}pez-Sepulcre}, {Dell'Ova}, {Gusdorf},
  {Sanhueza}, {Stutz}, {Brouillet}, {Thomasson}, {Armante}, {Baug}, {Bonfand},
  {Busquet}, {Csengeri}, {Cunningham}, {Fern{\'a}ndez-L{\'o}pez}, {Liu},
  {Olguin}, {Towner}, {Bally}, {Braine}, {Bronfman}, {Joncour}, {Gonz{\'a}lez},
  {Hennebelle}, {Lu}, {Menten}, {Moraux}, {Tatematsu}, {Walker}, \&
  {Whitworth}}]{Pouteau_etal2022}
{Pouteau}, Y., {Motte}, F., {Nony}, T., {et~al.} 2022, \aap, 664, A26

\bibitem[{{Ramsey} {et~al.}(2012){Ramsey}, {Clarke}, \&
  {Men'shchikov}}]{Ramsey2012}
{Ramsey}, J.~P., {Clarke}, D.~A., \& {Men'shchikov}, A.~B. 2012, \apjs, 199, 13

\end{thebibliography}

\end{document}